# PRINCIPLES OF
# A SECOND QUANTUM MECHANICS,
# INDETERMINISM, NON-LOCALITY,
# UNIFICATION

**Mioara MUGUR-SCHÄCHTER**

**24 april 2018**

http://www.mugur-schachter.net/

## AKNOWLEDGEMENT

This work would not have been possible without the life-long support of Sully Schächter.

## GRATITUDE

I am grateful to my sons François and Vincent for their constant support.
I express my deep gratitude to those who have encouraged me.
I feel particularly indebted to my friends
Jean-Paul Baquiast, Henri Boulouet, Jean-Marie Fessler, Geneviève Rivoire.

## DEDICATION

This work is dedicated to

Louis de Broglie

whose deep unconventional work has founded Quantum Mechanics
and 90 years later permits to re-found it.

# SUMMARY

# PART II

# CRITICAL-CONSTRUCTIVE PRELIMINARY GLOBAL EXAMINATION OF THE HILBERT-DIRAC QUANTUM MECHANICS, BY REFFERENCE TO *IQM*

## INTRODUCTION TO PART II

## CONCLUSION ON THE PART II

# PART III

# THE PRINCIPLES OF A SECOND QUANTUM MECHANICS

***rooted into the microphysical factuality in a physical-operational way***

## INTRODUCTION TO PART III

# ABSTRACT

This work is *not* a 'reinterpretation' of the nowadays Hilbert-Dirac quantum mechanics $QM_{HD}$. It exposes the principles of a new representation of microstates called *a second quantum mechanics* and denoted *QM2*. This representation is rooted directly into the a-conceptual physical reality wherefrom a representation of the microstates is reconstructed *bottom-up*, conceptually and formally, and in uninterrupted relation with factuality.

*QM2* emerges as a fully intelligible, consensual, predictive and verifiable representation *of the **factually** generated microstates*.

First a qualitative but formalized representation of the general characteristics of *any* physical theory of the microstates is developed, quite independently of the quantum mechanical formalism and outside it, under exclusively the [operational-methodological] constraints entailed by the requirement of a consensual, predictive and verifiable description of entities that – radically – cannot be perceived directly by human conceptors-observers. This representation is called *infra-(quantum mechanics)* and is denoted *IQM*. The specific purpose of *IQM* is to offer a *reference-and-imbedding-structure* for the construction of any acceptable theory of the microstates: only a pre-structure of this sort could permit to overcome the inertial ties that immobilize the minds inside an out-dated theory. Indeed *IQM* insures by construction comparability with $QM_{HD}$ thereby endowing with criteria for estimating from various and definite points of view the adequacy of its mathematical formalism. In this sense *IQM* can be regarded as a first realization from a class of structures conceived in order to act as ***infra-(scientific disciplines)***.

Then, by systematic reference to *IQM*, is worked out a preliminary critical examination of $QM_{HD}$. Thereby it appears that: ***(a)*** $QM_{HD}$ is devoid of any general formal representation of the ***physical, individual*** *entities and operations* that are quite essentially involved: the *whole* level of individual conceptualization of the microstates is lacking, massively; only abstract statistics of results of measurements are clearly defined inside $QM_{HD}$. ***(b)*** The mathematical formalism from $QM_{HD}$ *does* involve – and in a fundamental role – a definite *model of a specimen of a microstate*; but this fact remains implicit, so its consequences are not systematically recognized and made use of. ***(c)* $QM_{HD}$ *is simply devoid of an acceptable representation of the quantum measurements.***

These lacunae are then compensated constructively.

First a new representation of the quantum measurements is elaborated for any *unbound* microstate, without quantum potential or containing a quantum potential (while the category of bound states does not raise questions of principle from the point of view of *IQM*). The elaboration involves incorporation of a second central feature from Louis de Broglie's interpretive approach, beside his model. The obtained representation of the quantum measurements is founded upon Gleason's theorem on the possibility of a Hilbert-space expression of any *mathematical* measure, so in particular also probability-measures. The predictive probability-measures on results of outcomes of quantum measurements, are ten constructed factually, via measurements, *just like in the case of the verification of predictions of this kind*. Thereby the constituted representation emerges *independent* of the Schrödinger equation of the considered problem. So the use of this equation – when it is available – is quite generally *duplicated* by a factual-formal procedure of establishing the predictions. This offers possibilities of control of the output of the Schrödinger equation when idealizations or/and approximations are involved. And when the equation cannot be solved or even cannot be defined, this offers the possibility of a total replacement of such an output. A posteriori, the surprising peculiarities of such a situation inside a fundamental theory of theoretical physics appear as quite consonant with the new possibilities and trends generated by the progresses in informatics and in nanotechnology.

Finally, around the core constituted by the representation of the quantum measurements indicated above, is structured a very synthetic outline of the whole of the Second Quantum Mechanics, *QM2*

# GENERAL INTRODUCTION

> « The book will, therefore, draw a limit to thinking, or rather, not to thinking, but to the expression of thoughts; for, in order to draw a limit to thinking we should have to be able to think both sides of this limit (we should therefore have to be able to think what cannot be thought). »
>
> Wittgenstein, Preface of the Tractatus

The first attempts at a representation of microscopic physical entities started in terms of usual 'objects' endowed with delimited spatial volumes. Therefrom classical models and ways of reasoning were more and more deeply lowered into the domain of small space-time dimensions. This process however has come to a clear crisis around 1900: The connections with classical physics ceased being compatible with the experimentally established facts. Therefore Bohr and Plank introduced *non*-classical but ad hoc "principles". Thereby the intelligibility dissolved.

And then, de Broglie’s 'corpuscular-wave' *model* fractured the evolution: *It changed the* ***origin*** *on the vertical that connects knowledge of macroscopic physical entities, to knowledge concerning microscopic entities*. Indeed de Broglie’s model is placed just upon the extreme frontier between the microscopic, still *a-conceptual* factual physical reality, and the realm of the already conceptualized. And therefrom it tried to proceed upward toward the previously conceptualized in classical terms, and to connect to this, intelligibly.

> So the *direction* and the nature of the actions of construction of knowledge along the mentioned vertical of conceptualization were *reversed*.

Instead of continuing to try to guess top-down starting from the classical level and advancing 'downwards' into the realm of microscopic space-time dimensions via mental extrapolating procedures that were unconsciously trussed up into inertial strings developed since millennia inside the classical knowledge and thinking, there timidly began to emerge a new, fluctuating tendency to construct representations bottom-up, by a sort of conceptual climb in the dark guided by operational-observational-formal requirements.

The direction of constructive thought – where it begins and how it acts in order to reach a definite representational purpose – is quite determinant. The succession of the acts of conceptualization is tied with specific questions and reactions to these. So the mentioned inversion of construction of representation involved quite fundamental changes in the process of conceptualization and these in their turn induced obscure and strong mental confrontations between ancestral habits of thought and new procedures that still lacked definite and stable contours, but of which the necessity had become obvious and the consequences were strikingly sensed though feebly understood. The *method* of constructing scientific representations of physical reality was undergoing mutation.

The mathematical representations of Schrödinger and their results – directly initiated by de Broglie's model – and on the other hand Heisenberg's algorithms that were founded on different principles but for bound states offered equivalent results, led to impressing first successes, and these, for a while, neutralized the conceptual disquietudes.

Meanwhile Bohr, strongly aware of the radically new characters of the emerging theory, but of which the source and nature withstood identification inside his mind, tried to protect these characters from any premature restriction, via a preventive interdiction of *any* model of a microsystem. Furthermore, as it is well known, he founded this interdiction upon the assertion of the general philosophical requirement of a strictly 'positivistic' attitude in science, consisting of the acceptance of – *exclusively* – purely operational basic procedures, free of any interpretive assumption.

*But this was an impossible requirement.*

When the entity to be studied is quintessentially un-observable and is unknown, if strictly no model is assigned to it any criteria are lacking for deciding *what sort* of operation deserves being considered to be a 'measurement-interaction' between *'that'* entity and a *given* qualifying quantity; and also for deciding what *value* of the involved qualifying quantity is entailed by the observable marks obtained by one given 'measurement-interaction' of a chosen sort. One cannot even know in advance where in space-time the entity to be studied 'is', nor what extension and contours the space-time support of this entity possesses; its inside and outside keep non-*conceived*; nothing insures even that such classical delimiting notions possess meaning with respect to what, a priori, is called 'microsystem' and 'state of a microsystem'. No specifically adequate *language* has been constructed as yet, nor criteria for constructing such a language. So a fortiori there is no intuitive basis for *beginning* to construct the desired knowledge. When one wants to enter upon a bottom-up process of conceptualization of physical entities, as de Broglie conceived to attempt, the perspective of a whole implicit order of *constructability* opens up necessarily like a ladder from the *as yet never conceptualized* toward the sky of classical knowledge. This ladder has to be constructed and climbed step by step. If this is attempted in a purely formal-algorithmic way, void of any explanation, the procedure cannot but seem arbitrary and the result cannot be endowed with intelligibility.

And precisely this happened indeed.

In a certain very warped way Bohr's interdiction of any model of a microstate protected indeed the development of the emerging Schrödinger-Heisenberg mathematical representation, and later its mutation into the nowadays Hilbert-Dirac reformulation. But on the other hand this interdiction led to hidden violations of certain laws of thought that – remarkably – *do* irrepressibly work inside the human constructive processes of conceptualization. And this entailed non-intelligibility of the achieved formalism. Moreover, it nourished a hidden inner contradiction. Namely, de Broglie's 'corpuscular-wave' model that had triggered Schrödinger's contributions, though rejected by Bohr's positivistic philosophical diktat, *remained* quite essentially involved in the quantum mechanical formalism as well as in the current language that accompanied its manipulations. But it remained there in an only minored way, masked inside mathematical forms and superficially utilized words, so immobilized in atrophy by absence of a declared and definite conceptual status. In consequence of this – up to this very day – this model keeps acting *most fundamentally* inside the formalism without being exposed to overt control and optimization.

This circumstance led to the occultation, inside the quantum mechanical formalism, of also many other features, factual, operational and conceptual, that irrepressibly *do act*, but without being mutually distinguished, named, and genuinely dominated from a semantic point of view. The most massive such occultation is that of the radical difference of nature and role between *individual* representations and *statistical* ones.

Therefore since 90 years our representation of microstates irrepressibly nourishes endless questionings and fumbling that pulverize systematically against a paradoxical negative dike of *absence of definite criteria* for defining the exact contents and the adequacy of this or that mathematical representation. The mathematical representations proliferated densely and they still do so. While in their core there subsists a deleterious semantic magma. There is an urgent need *to overtly organize* ***meaning****, to generate intelligibility*. We are not yet robots. We are still human beings that need to understand in order to optimise with depth, generality and precision, in the full light of rationality. A powder of purely algorithmic, 'technical' ad hoc solutions, amorphous with respect to rationality, does not yet fully satisfy everybody.

> What lacks – dramatically – for organizing meaning is *a structure of insertion-and-reference* constructed independently of the quantum mechanical formalism and outside it, that offer a clear and thorough understanding of the non-classical specificities of the process of bottom-up construction of a human representation of non-perceptible microscopic entities.

Only this could permit an explicit, exhaustive and coherent specification of the way in which a mathematical representation of micro-phenomena can be brought to signify in a controlled and adequate way.

In the first part of this work I construct such a structure of insertion-and-reference [1]. It is the very first one of this kind and it might open the way toward many others of the same type but tied with other representational aims.

In the second part, by reference to this structure, are identified the main lacunae that vitiate nowadays quantum mechanics. Moreover some of these are immediately compensated locally thus offering a cleansed ground for a new construction.

In the third part are outlined the main contours of *a second quantum mechanics*, baptized in advance '*QM2*', that is constructed bottom-up and emerges freed of interpretation problems and fully intelligible, via a step by step explicit identification of the incorporated semantic contents and a constant control of the semantic-syntactic consistency[2]. And this entails an unexpected and major reward:

> *QM2* permits to discern a natural and – a posteriori – an obvious path toward the unification between nowadays microphysics and the theory of gravitation.

The approach practised in this work is not usual. So it will surprise, and many readers might feel rejected. I take the liberty to express that I am deeply aware of this. But I have not been able to find a less singular way toward a fully intelligible representation of the modern microphysics

---

[1] For Maxwell's classical electromagnetism, because "fields" are not directly perceptible, a fully new syntax of specifically adequate field-descriptors has been independently created before the formulation of the theory itself.

[2] For the sake of effectiveness throughout the whole following work all the involved descriptional elements are posited finite, so discrete.

# PART I

# INFRA-QUANTUM MECHANICS

**A qualitative but formalized structure of reference-and-insertion,**

**built outside the Hilbert-Dirac mathematical formalism**

**for guiding the construction of a fully intelligible Quantum Mechanics**

"To reach the truth, once in the life one has to unbound oneself from all the received opinions and to reconstruct the whole system of knowledge, starting from the ground".
René Descartes

# INTRODUCTION TO PART I

A human being who wants to construct knowledge concerning states of microsystems – ‘microstates’ – makes use of physical entities to which he associates this denomination, of instruments and operations, and he introduces representational *aims* and corresponding *methods* of acting and thinking. Thereby the human observer introduces severe constraints that structure the process of construction of knowledge. It is not possible to preserve the process from such constraints. They are precisely what ‘forms’ it. Nor is it possible to eliminate a posteriori the effects of theses constraints from the constructed knowledge, these are essentially incorporated to the achieved form to which they have led. *Any* piece of knowledge is a construction and this construction remains irrepressibly relative to its whole genesis. So, if the observer-conceptor wants to stay in control of the knowledge that he has generated, to be able to understand and to freely optimize it – he has to be thoroughly aware of the conceptual-operational-methodological weft of this knowledge.

In what follows – quite independently of the mathematical formalism of quantum mechanics – is elaborated a structure of the necessary and sufficient features of *a procedure* – not a 'description', nor a 'theory', but a method for reaching an a definite aim – that is specifically appropriate for creating *scientific knowledge*, i.e. communicable, consensual, and verifiable knowledge, on a 'microstate', so on a physical entity that is radically non-perceivable by, directly, the human biological sensorial apparatuses. By comparison with the processes of construction of the classical conceptualization, this new procedural structure involves a deliberate change of the *origin*, on the vertical of conceptualization, of the processes of construction of knowledge: deliberately, it starts at the bottom, upon the very limit between the volume of what has already been drawn before inside the volume of the actions of conceptualization and what we imagine to be the as yet a-conceptual universal physical substance.

So the *order* of conceptual constructability is inversed, the construction progresses bottom-up. This entails a fundamental change in the content of the classical concept indicated by the historically introduced word 'microstate': this content transmutes into that of a *factually* defined concept, because *definitions in the classical sense cannot be realized any more*.

And this in its turn is what entails the emergence of the famous 'problem' of the 'essentially' probabilistic character of the modern microphysics.

In order to bring into maximal evidence this pivotal feature I have kept in use, unchanged, the word 'microstate'. This introduces a key-connection with the classical top-down historical evolution of the scientific conceptualization toward microscopic space-time dimensions, inside the molecular and atomic physics, but at the same time it acts as a reference for comparability between a top-down and a bottom-up conceptualization. Which entails intelligibility. Whereby it becomes possible to identify how have germinated and developed the basic misunderstandings that since a century plague Physics, and to dissolve them. While the narrow guides that emerge progressively lead to a second Quantum Mechanics that is itself fully intelligible and thereby brings forth the methodological unity of Physics.

The approach proposed in the first part of this work is structured in qualitative but explicit, formalized [3] and finite *effective* terms. The result is called in advance *Infra-(Quantum Mechanics)* and is denoted *IQM*.

I would like to convey to the reader from the start what follows.

*Nothing* – throughout the construction elaborated below – is conceived as an assertion of 'objective intrinsic factual *truth*'. Just a succession of *methodological* steps is figured out, each one of which is imposed with necessity by the global aim to construct a guiding structure for the elaboration of a satisfactory representation of physical entities that – radically – *cannot be directly perceived*, by the local aim of the considered step, and by the corresponding cognitive situation. In order to instil intelligibility, each step is explicitly referred to the structure of our classical thought-and-languages that have emerged and settled in our minds by interactions with entities that *are* perceived. But on the other hand each constructive methodological step transgresses our classical forms of thought by definite features commanded by the radical novelty of the conceptual situation that characterizes it, and these features are explicated.

*IQM* is the global procedural whole that is obtained when the methodological steps indicated above are put together under constraint of logical coherence. It is a coherent procedural reference-and-hosting-structure for building a specifically appropriated representation of factually defined micro-entities.

---

[3] We employ the word 'formalized' in the sense that: The posits are explicitly stated; all the specific basic terms are endowed with explicit and finite definitions; and the elements introduced in this explicit way are constructed as general and syntactically related void loci for receiving in them particular unspecified semantic data. Furthermore, once posited or constructed, the elements are explicitly connected in full agreement with current logic, i.e. with the usual syllogistic.

I think that in the absence of such a structure it simply is not possible to construct for such entities a fully appropriate and fully intelligible mathematical representation of scientific knowledge.

*IQM* is organically tied with a general method for constructing scientific human knowledge – consensual, predictive and verifiable –, the Method of Relativized Conceptualization, *MRC* (MMS [4], [2002A], [2002B], [2006]). *MRC* offers the general framework for constructing in a unified way *any* desired infra-discipline. The Infra Quantum Mechanics *IQM* is the very first such 'infra-discipline' and it leads to a second Quantum Mechanics *QM2*.

This second Quantum Mechanics developed inside the Infra-Quantum Mechanics that is an application of the general Method of Relativized Conceptualization, brings into clear evidence a fundamental ***methodological unity*** between all the domains of the modern physics. In particular it brings into explicit and detailed perceptibility in *what a sense, and how* Quantum Gravitation and the Modern Microphysics belong organically to one same basic constraint of a radically relativized reorganization of the scientific representations of matter: in the modern scientific approaches, the constantly increasing distance between direct sensorial perceptibility of that what is represented, and a scientific representation of this, *obliges* to relativize with method and rigor the scientific ways of constructing knowledge.

This fact, implicitly, burgeons already everywhere inside the sciences of matter. The Method of Relativized Conceptualization *MRC* and the Infra-(Quantum Mechanics) *IQM* only offer an explicit perception and a coherent expression of this now ubiquitous fact.

[4] MMS is to be read "M. Mugur-Schächter".

**PROLOGUE**

The extract reproduced below from the volume "*Einstein 1879- 1955 (6-9 juin 1979), colloque du centenaire, Collège de France*, Editions du Centre National de la Recherche Scientifique" – is useful for reminding of the state of mind concerning the fundamental problems in Quantum Mechanics in 1979, *that still persists today*.

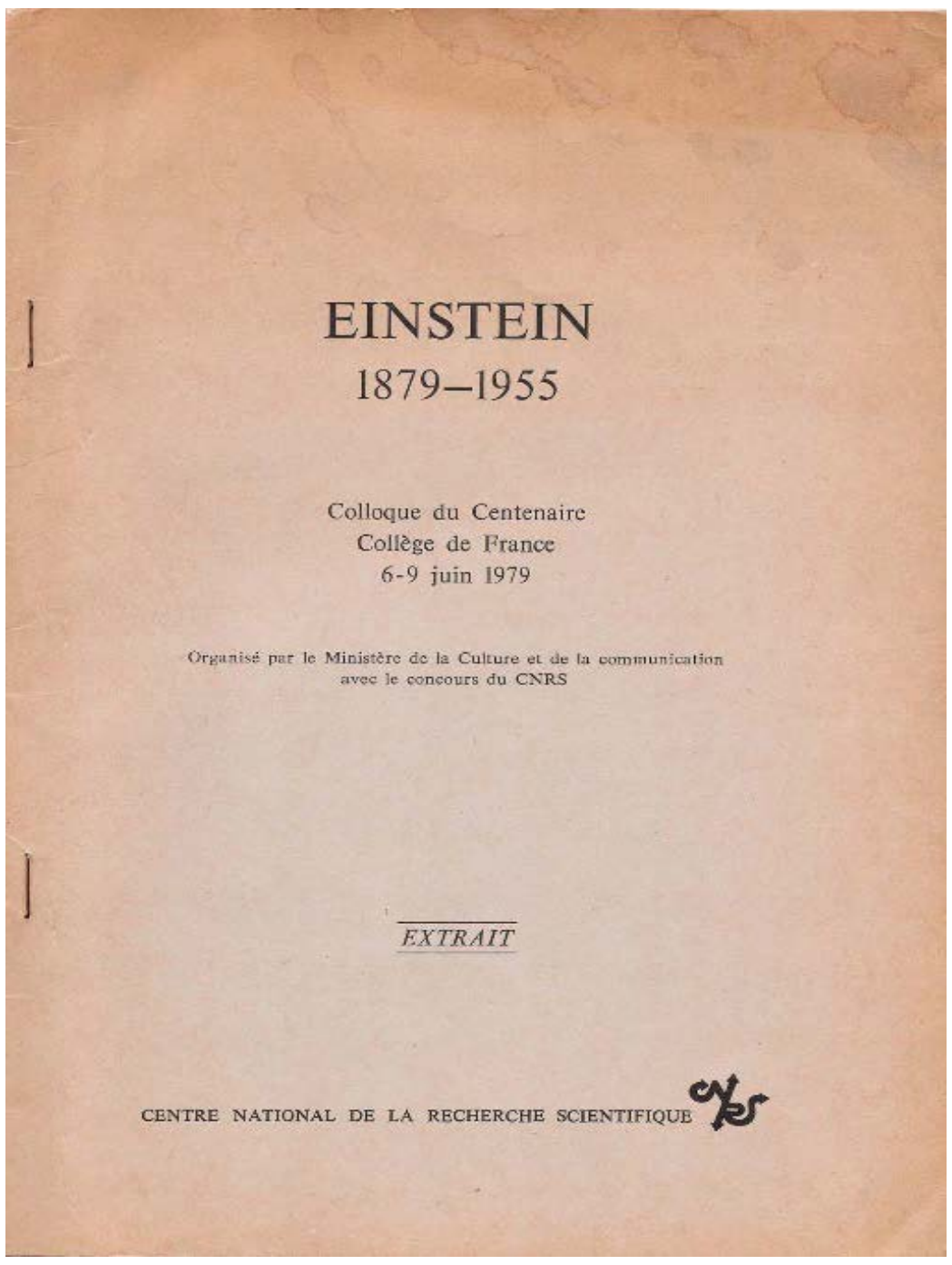

•

I reproduce the original French version

# (EXTRAIT)
# REFLEXION SUR LE PROBLEME DE LOCALITE

**M. Mugur–Schächter**

**UNIVERSITE DE REIMS**
**B.P 347 51062 REMS CEDEX**

## But

Depuis huit ans ce que l'on appelle le problème de localité retient de plus en plus l'attention. Des théoriciens, des expérimentateurs, des penseurs pluridisciplinaires investissent des efforts importants pour élucider ce problème. Les aspects techniques – mathématiques et expérimentaux – ont été déjà examinés dans un grand nombre de travaux et ils sont bien connus de ceux qui font à ce sujet des recherches spécialisées. Mais la configuration conceptuelle qui est en jeu me paraît avoir des contours beaucoup moins définis. Le but de l'exposé qui suit est d'examiner cette configuration conceptuelle. J'essaierai de procéder à cet examen d'une manière aussi simple et frappante que possible, presque affichistique, à l'aide de schémas et de tableaux. Ces moyens me paraissent être les plus adéquats pour donner le maximum de relief aux insuffisances que je perçois dans la définition même du problème de localité.

## Bref rappel

***Le paradoxe EPR (I935)***. Le problème de localité est soulevé par un théorème bien connu de J. Bell (1) qui se rattache à un raisonnement formulé en 1935 par Einstein, Podolsky et Rosen (2). Ce raisonnement, connu sous la dénomination de "paradoxe EPR", et été construit pour démontrer que le formalisme de la Mécanique Quantique ne fournit pas une description complète des microsystèmes individuels. Les hypothèses qui constituent la base de départ du paradoxe EPR sont indiquées dans le tableau suivant (où des notations abrégées leur sont associées):

| | |
|---|---|
| Toutes les prévisions de la Mécanique Quantique sont vraies. | $\forall$ MQ |
| La Mécanique Quantique fournit une description complète des microsystèmes. | C (MQ) |
| La réalité physique existe indépendamment de l'observation. Elle est "déterministe" et locale (ou"séparable")* | $\exists$(r.d.l.) |

Le "paradoxe EPR" consiste dans la démonstration du fait que les hypothèses énumérées ne sont pas compatibles.

L'interprétation proposée par Einstein, Podolsky et Rosen, de cette démonstration, a été la suivante:
Les prévisions du formalisme quantique se montrent correctes. Il n'existe donc aucune base pour abandonner l'hypothèse $\forall$MQ.

L'hypothèse ∃(r.d.l.) exprime un credo métaphysique que l'on est libre d'accepter ou de rejeter. Mais *si* on l'accepte, alors il faut l'adjoindre aux prévisions de la Mécanique Quantique. En ce cas la démonstration de l'incompatibilité du système d'hypothèses [∀MQ + C(MQ) + ∃(r.d.l.)] oblige à abandonner hypothèse de complétude C(MQ). En d'autres termes cette démonstration oblige alors à accepter la possibilité d'une théorie déterministe et locale (TDL) des microphénomènes, où le formalisme quantique sera complété par des éléments descriptifs additionnels, des paramètres cachés (par rapport au formalisme quantique) déterministes et locaux (p.c.d.l.) qui permettent d'accomplir une description complète des microsystèmes individuels. Cette description complète fournie par TDL doit être compatible avec la Mécanique Quantique – en vertu de l'hypothèse ∀MQ – et avec la Relativité, en vertu de l'hypothèse ∃(r.d.l.) qui se trouve intégrée dans la théorie de la relativité. Cette structure d'idées peut être représentée par le schéma suivant:

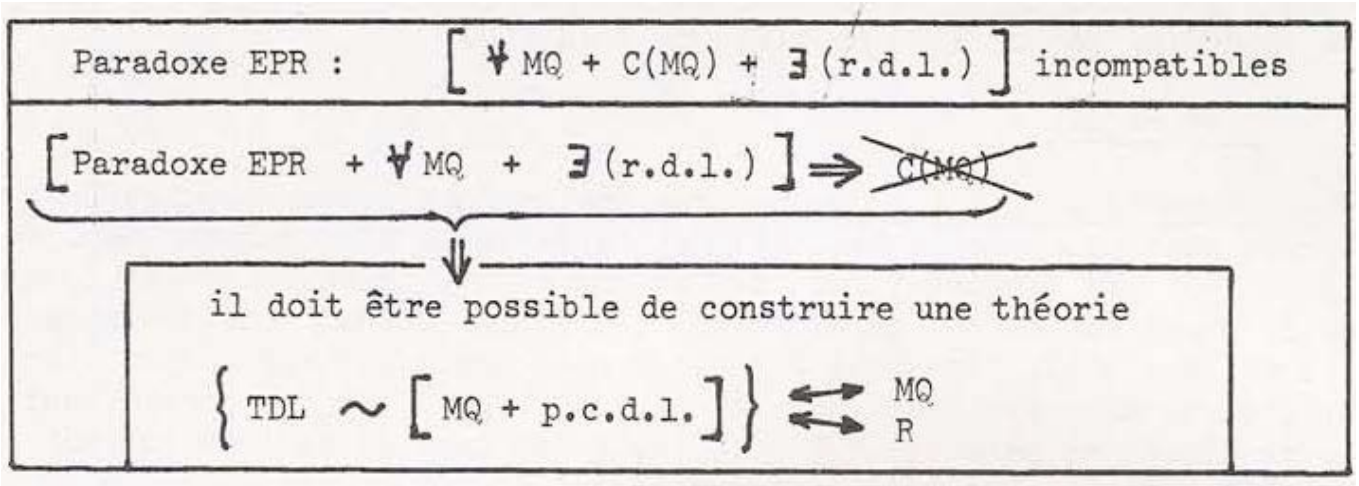

•

***Les réactions pendant 30 ans.*** Les réactions ont été diverses. Pourtant la note dominante a été nettement celle du positivisme: l'hypothèse "réaliste" ∃(r.d.l.) est dépourvue de toute signification opérationnelle. Elle est donc essentiellement métaphysique, extérieure à la démarche scientifique. L'incompatibilité dénommée "paradoxe EPR" n'existe que par rapport à cette hypothèse non scientifique, et donc elle ne constitue pas un problème scientifique. Pour la science il s'agit là d'un faux problème.

***Le théorème de J. Bell (1964).*** Trente années plus tard J.Bell a démontré un théorème qui semble contredire l'interprétation associée par Einstein Einstein, Podolsky et Rosen à leur propre démonstration. La conclusion du théorème de Bell peut s'énoncer ainsi (ou de manières équivalentes): il n'est pas possible, à l'aide de paramètres cachés déterministes et locaux, d'obtenir dans tous les cas les mêmes prévisions que la Mécanique Quantique ; en certains cas, de tels paramètres conduisent à d'autres prévisions. Si alors on veut rétablir l'accord avec les prévisions de la Mécanique Quantique, il faut supprimer le caractère local des paramètres cachés introduits, ce qui contredira l'hypothèse ∃(r.d.l.), que la théorie de la Relativité incorpore. Par conséquent la théorie déterministe TDL compatible à la fois avec la Mécanique Quantique et la Relativité, dont Einstein Podolsky et Rosen ont cru avoir établi la possibilité, est en fait impossible.

La démonstration repose sur la production d'un exemple. On considère deux système $S_1$ et $S_2$ à spins non nuls et corrélés, créés par la désintégration d'un système initial S de spin nul. On envisage des mesures de spin sur $S_1$ selon trois directions a, b, c, à l'aide d'un appareil $A_1$, et des mesures de spin sur $S_2$ selon ces mêmes directions, à l'aide d'un appareil $A_2$ qui peut se trouver à une distance arbitrairement grande de $A_1$. *L'hypothèse ∃(r.d.l.) est ensuite formalisée*: des paramètres cachés sont introduits et ils sont soumis à des conditions telles qu'elles fournissent une traduction mathématique des qualifications de "déterministes" et "locaux". Ainsi la conceptualisation introduite auparavant au niveau d'une sémantique claire, mais qualitative, est élevée jusqu'à un niveau sémantique syntaxisé. Un tel pas est souvent important, car il peut permettre des déductions mathématiques à conclusions quantitatives. Et en effet Bell a démontré que l'hypothèse ∃(r.d.l.) ainsi formalisée entraîne nécessairement une certaine inégalité concernant les corrélations statistiques entre les résultats de mesures de spin enregistrés sur les appareils $A_1$ et $A_2$. Or, cette inégalité n'est *pas* satisfaite par les corrélations statistiques prévues par la Mécanique Quantique. On pourrait retrouver les corrélations quantiques en supprimant la condition qui traduit mathématiquement le caractère "local" des paramètres cachés introduits, c'est-à-dire en renonçant à une partie de l'hypothèse ∃(r.d.l.). On exprime ceci en disant que, dans la circonstance considérée, "la Mécanique Quantique est non-locale" ou "implique des effets non-locaux" qui la rendent incompatible avec ∃(r.d.l.). Schématiquement, on peut résumer l'apport de Bell ainsi (en notant $(p.c.d.l.)_B$ les paramètres cachés soumis aux conditions de Bell).

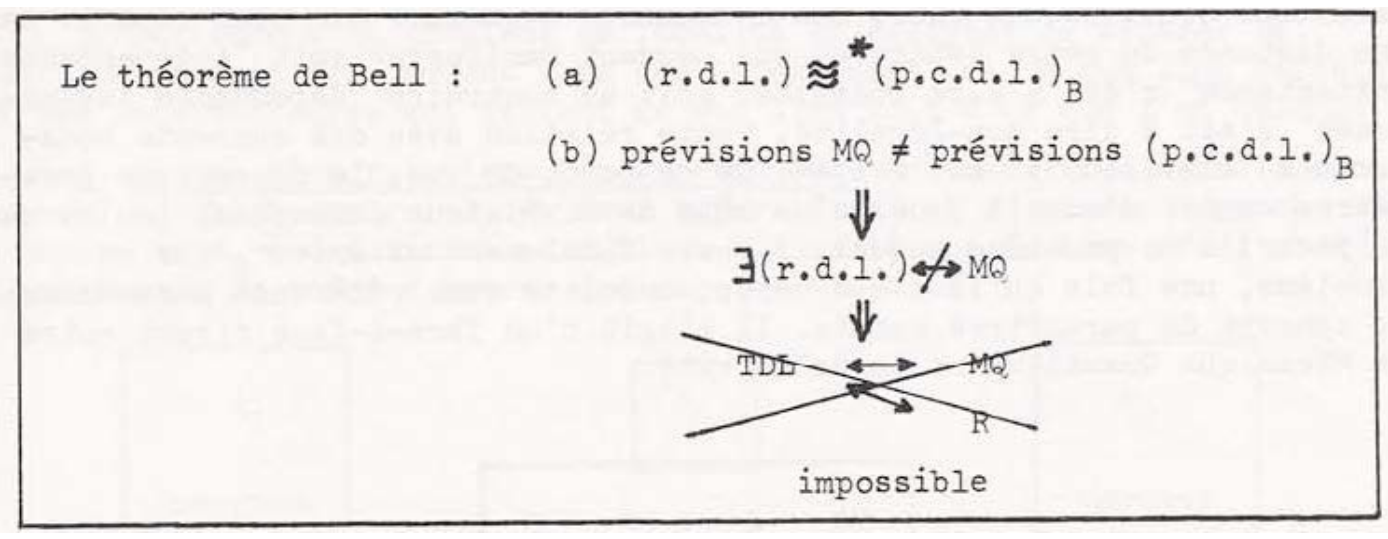

Comme les statistiques dont il s'agit sont observables, il est en principe possible d'établir expérimentalement si les faits physiques correspondent aux prévisions de la Mécanique Quantique ou à celles entraînées par les paramètres cachés déterministes et locaux au sens de Bell. C'est l'un des traits les plus forts du théorème de Bell.

Si l'expérience infirmait la Mécanique Quantique, la situation conceptuelle créée paraîtrait claire. On devrait admettre la possibilité d'une théorie déterministe et locale des microphénomènes, mais différente de celle envisagée par Einstein, Podolsky et Rosen, car elle n'obéirait pas à l'exigence d'identité prévisionnelle avec la Mécanique Quantique, pour tous les cas.

Mais un certain nombre d'expériences de vérification a déjà été fait et il se trouve que les résultats obtenus à ce jour – bien qu'ils ne tranchent pas encore définitivement – étayent fortement la supposition que la prévision de la Mécanique Quantique s'impose comme correcte.

Il s'agit donc de comprendre la situation conceptuelle qui semble s'établir et que l'on dénomme "problème de localité".

**Interprétations**

Le problème de localité est ressenti diversement. Je distinguerai en gros trois interprétations, en omettant ou en bousculent beaucoup de nuances.

***I- Interprétations de refus***. Un certain de nombre de physiciens semble considérer cette fois encore qu'il s'agit d'un problème métaphysique qui n'existe que par rapport au concept non opérationnel de paramètre cachés, mais qui se dissout dès qu'on refuse ce concept. D'autres physiciens considèrent que le problème n'existe parce qu'il est faussement posé (3).

***2- Interprétation minimale***. Selon d'autres physiciens (4), (5), (6), (7), etc.…, le problème satisfait cette fois aux normes positivistes les plus draconiennes, parce qu'il conduit à des testes expérimentaux. Toutefois, ils refusent de conceptualiser au-delà de ce que ces tests mettent en jeu. Ils ne prennent en considération strictement que des corrélations statistiques entre des évènements de mesure qui sont séparés par une distance du genre espace et qui peuvent manifester soit "indépendance instantanée" c'est-à-dire localité, soit au contraire "dépendance instantanée" c'est-à-dire non-localité. Toute relation avec des concepts sous-jacents "explicatifs" est évitée. De ce point de vue, le concept de paramètres cachés n'aurait qu'un rôle de révélateur conceptuel (ou de catalyseur) d'un problème auquel il reste finalement extérieur. Car ce problème, une fois qu'il a été perçu, subsiste sans référence nécessaire au concept de paramètres cachés. Il s'agit d'un face à face direct entre la Mécanique Quantique et de Relativité.

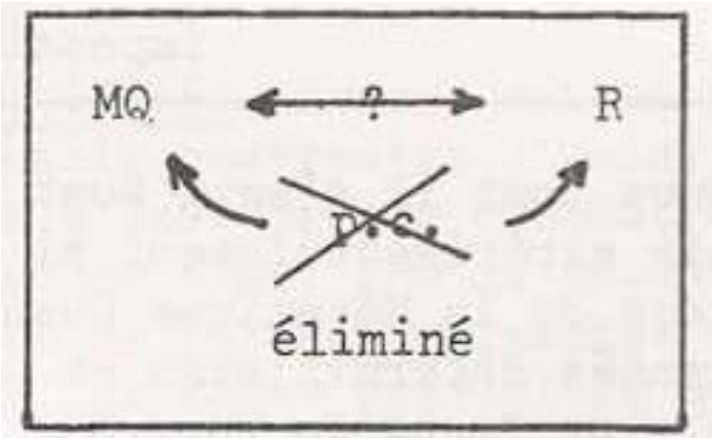

•

*3- L'interprétation épistémologique*. Il existe enfin une tendance (8) à connecter le problème de localité à notre conceptualisation la plus courante de la réalité, qui postule l'existence d'objets isolés possédant des propriétés intrinsèques et permanentes. La violation des inégalités de Bell serait incompatible avec ces suppositions. Il s'agirait donc en dernière essence d'un face-à-face entre la Mécanique Quantique et – à travers le concept de paramètres cachés et à travers la Relativité – des postulats épistémologiques fondamentaux.

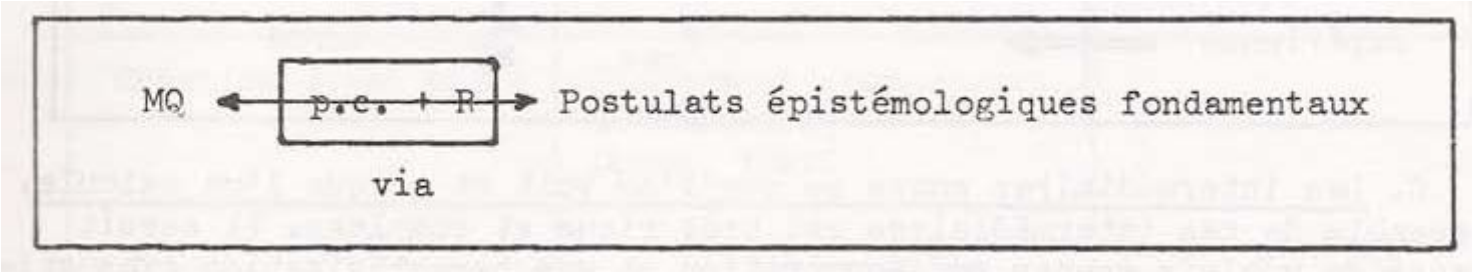

•

Je n'examinerai pas l'interprétation de refus, car elle ne peut conduire à aucun élément nouveau.

Quant aux deux face-à-face impliqués par les deux autres interprétations, aucun d'eux ne me semble s'imposer dans la phase actuelle du débat. Seule une question ressort clairement:

Qu'est ce qui est en jeu – au juste – dans le problème de localité?

L'examen qui suit montrera que, pour fixer une réponse, les conceptualisations existantes et les tests sur l'inégalité de Bell ne peuvent pas suffire. Inévitablement d'autres conceptualisations encore, et les tests correspondants, devront être abordés. Sinon, aucune conclusion définitive ne pourra être tirée, même si l'inégalité de Bell est clairement violée.

**Le problème de localité et le terrain conceptuel sous-jacent**

Reconsidérons le problème de localité en essayant de séparer ce que l'on perçoit directement lors des expériences, de ce que l'on calcule, et des intermédiaires qui relient ce que l'on voit à ce que l'on calcule.

***A. Ce qu'on voit lors des expériences***. On voit (tous les détails mis à part) un objet central $A_o$ et deux appareils $A_I$ et $A_2$ placées à gauche et à droite de $A_o$ à des distances égales. Sur certaines parties de $A_I$ et $A_2$ apparaissent de temps à autres des marques visibles.

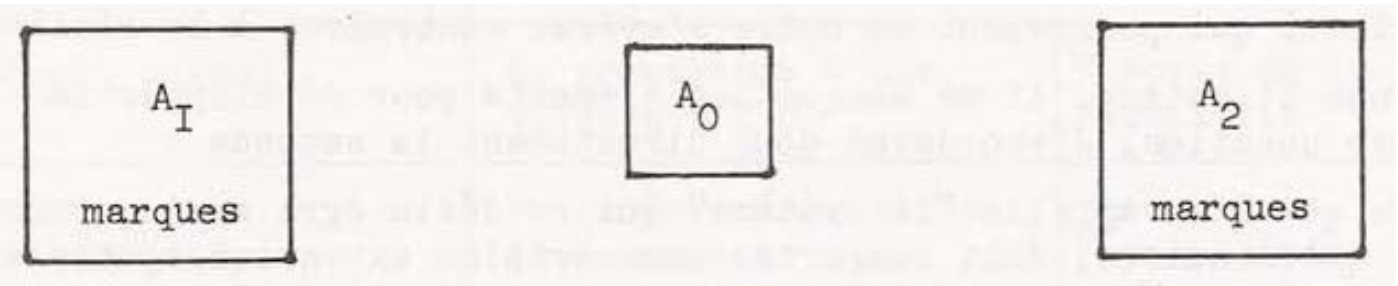

- 

***B. Ce qu'on calcule***. On calcule des corrélations statistiques en employant trois sortes de distributions de probabilités conduisant à trois fonctions de corrélation, une fonction $F_{(TDL)B}$ caractéristique d'une théorie déterministe locale au sens de Bell, une fonction $F_{MQ}$ obéissant aux algorithmes de la Mécanique Quantique, et une fonction $F_{obs}$ correspondant aux statistiques observées. L'inégalité de Bell distingue $F_{(TDL)B}$ de $F_{MQ}$. L'expérience doit montrer si la réalité observée reproduit $F_{MQ}$ ou $F_{(TDL)}$.

| $F_{(TDL)_B}$ , $F_{MQ}$ , $F_{obs}$ | |
|---|---|
| Inégalité de Bell | $\Longrightarrow F_{(TDL)_B} \leqslant$ borne , $F_{MQ} >$ borne |
| Expérience | $\Longrightarrow$ $F_{obs} \overset{?}{\gtrless}$ borne |

- 

***C. Les intermédiaires entre ce qu'on voit et ce qu'on calcule***. L'ensemble de ces intermédiaires est très riche et complexe. Il serait insensé de vouloir donner une énumération et une caractérisation déterministes et locaux de Bell violent la pudeur sémantique dictée par le positivisme. Alors autant aller jusqu'au bout et avouer l'ensemble des questions sémantiques liées aux interprétations 2 et 3 du problème de localité telles que je les ai distinguées plus haut.

Je commence par l'interprétation minimale. Je perçois deux questions.

En premier lieu, les contenus sémantiques assignés aux qualificatifs "déterministes" et "locaux", tels qu'ils sont impliqués par la modélisation mathématisée de Bell, permettent-ils la représentation la plus générale concevable d'un processus d'observation d'un "microétat" à l'aide d'un "appareil" macroscopique?

En second lieu, en supposant que la modélisation de Bell d'un processus d'observation n'introduit vraiment aucune restriction non nécessaire, quelle sorte de non-localité, exactement, la violation des inégalités de Bell démontrerait-elle? La non-localité que la théorie de la Relativité interdit clairement, ou bien des prolongements spontanés et encore flous de celle-ci qui pourraient en outre s'avérer contraire à la réalité?

Pour l'instant, il me manque les éléments pour développer la première question. J'aborderai donc directement la seconde:

Ce qu'on appelle "le système" qui se désintègre en $A_o$, pour autant qu'il existe, doit comporter une certaine extension spatiale non nulle de départ $\Delta x_s(t_o)\neq 0$ (ce qui peuple ce domaine d'espace, est-ce un "objet" ou un "processus", ou les deux à la fois? les définitions même manquant pour répondre). Ce qu'on désigne par les termes "désintégration" ou "création d'une paire $S_1$ et $S_2$", comment le concevoir? Les mots indiquent dans le substrat conceptuel l'hypothèse d'un processus, d'une entité réelle en cours de changement. Pour exister, ce processus doit se produire quelque part et il doit durer, il doit occuper un certain domaine non nul d'espace-temps $\Delta s_c(t).\Delta t_c \neq 0$ (l'indice c: création) à l'intérieur duquel "le système de départ S" existe encore mais change, cependant que $S_1$ et $S_2$ n'existent pas encore mais se forment.

| MOTS | CONCEPTS | ORGANISATIONS SYNTAXIQUES |
|---|---|---|
| 1 système | Macroobjet, objet | Logique des classes d'objets et des prédicats |
| Création d'une paire | Changement, processus, SUCCESSIONS, DUREE, TEMPS | ? |
| 2 systèmes corrélés | Objets ⇄ corrélés / isolés | |
| Appareil | Macroobjet capable de réagir d'une manière significative avec un microsystème | |
| Spin | Propriété d'observation | |
| Mesure | Processus d'observation<br>Evénement de connexion entre microobjets et macroobjets | |
| Paramètres cachés | Propriétés intrinsèques et permanentes (d'objets ou de processus) | |
| Déterministe | Prévisible?<br>A partir de quoi? | |
| Local, signal | Se propageant à une vitesse v < c | Théorie de la Relativité |
| Statistiques Probabilités | Phénomènes aléatoires (événements, processus, objets) | Théorie des Probabilités |
| Prévisions quantiques | Vecteurs d'états, algorithmes quantiques | Mécanique Quantique |
| Influence à distance | Changement transporté à vitesse v < c | ? |

•

Dans l'écriture qui désigne ce domaine d'espace-temps, le facteur de durée $\otimes t_c = t_{12,\ o} - t_o$ s'étend – par définition – d'une certaine "valeur initiale de temps" $t_o$ où le changement de création commence, jusqu'à une "valeur finale de temps" $t_f \equiv t_{12,\ o}$ à partir de laquelle "la paire $S_1$, $S_2$ de systèmes corrélés" commence à exister (des objets? des processus eux aussi? les deux à la fois?). Quand au facteur d'extension spatiale $\Delta s_c$ (t), il semble obligatoire de concevoir, puisqu'il s'agit d'un processus, qu'il change en fonction de la "valeur de temps" t, avec ($t_o < t < t_f$ ), en restant toutefois métastablement connexe tant que $t < t_f$ (c'est-à-dire tant que S subsiste encore et que $S_I$ et $S_2$ ne sont pas encore créés). Pour tout $t > t_f$, toutefois, ce domaine spatial devrait être devenu non connexe *via* une scission plus ou moins "catastrophique" conduisant à cette nouvelle forme de stabilité à laquelle on rattache l'expression "la paire $S_I$, $S_2$ de deux systèmes corrélés".

Je m'arrête un instant et je regarde ce que je viens d'écrire. Quel mélange de "nécessités" et d'arbitraire, de signes et de mots qui ont l'air de pointer vers un désigné précisé et sous lesquels pourtant on ne trouve que des images floues et mouvantes accrochées à ces mots et ces signes de manière non séparée. J'écris entre guillemets "valeur de temps", par exemple, parce qu'à chaque fois que je réfléchis au degré d'inexploration où se trouvent encore les concepts de durée et de temps et leur relation, je ressens une réticence à écrire quoi que ce–soit en dehors d'un algorithme qui fixe une règle du jeu. Le paramétrage de la propriété fondamentale de durée à l'aide de la variable de temps t, telle que cette paramétrisation est pratiquée dans les théories existantes et même dans la Relativité, est encore certainement très simplificatrice et souvent falsificatrice, rigidifiante, mécanisante en quelque sorte. Les changements ne sont pas toujours des déplacements d'entités stables intérieurement. Pour pouvoir rendre compte pleinement de l'entière diversité des types et des intensités de changements, il faudrait une sorte de grandeur vectorielle, un *champ* de temps processuel défini en chaque point de l'espace abstrait encadré par l'axe de durée et par les axes des changements envisagés.

Mais un tel temps se transformerait-il selon Lorentz? Quel rôle joue la vitesse d'un "signal" lumineux face aux vitesses de propagation "d'influences" (?) dans un tel espace processuel? Qu'est ce que la Relativité impose véritablement aux processus *quelconques* et qu'est-ce qu'elle laisse en blanc? Lorsqu'il s'agit de processus très "intenses" localement, "catastrophiques", comme l'est probablement la "création d'une paire", que devient "le temps"?

En théorie relativiste générale de la gravitation, par exemple, un gradient non nul du champ de gravitation est lié à une impossibilité de définir un temps unique, pour les observateurs d'un même référentiel, si ces observateurs sont spatialement distants l'un de l'autre. Quand à l'invariance de la vitesse *de la lumière* elle-même (et non la vitesse d'autres sortes "d'influences") lorsqu'on passe d'un référentiel à un autre, elle n'est postulée que localement, car il n'existe aucune définition uniforme des distances et des temps dans des champs gravitationnels variables (9) (espace-temps courbes). Comment savoir quelle sorte de "courbure" locale de l'espace-temps est produite (ou non) par un processus – essentiellement variable – de création d'une paire?

Enfin, la Relativité n'introduit aucune quantification au sens de la Mécanique Quantique, sa description est continue. Lorsqu'on écrit [vitesse = distance/temps], le temps est un paramètre continu.

Si ensuite on se demande comment on trouve la valeur de t, on s'aperçoit qu'elle est de la forme $NT_H$ où N est un entier et $T_H$ une période "d'horloge" (supposée constante !) ce qui ramène au discret. En macroscopie cela peut être négligeable aussi bien sur le plan du principe que sur le plan numérique. Mais lorsqu'on considère des processus quantiques et relativement très brefs, quel est le degré de signifiance d'une condition comme

$$v = \frac{\text{distance}}{\text{temps}} = \frac{\text{distance}}{NT_H} = \text{const?}$$

•

Quelle horloge faut-il choisir, avec quel $T_H$, et comment par ailleurs s'assurer que lorsqu'on écrit $\otimes t = 10^{-x}$, on fait plus qu'un calcul vide de sens?

On comprend, devant de telles questions, les prudences positivistes et les normes qui conseillent de se maintenir dans la zone salubre de l'opérationnellement défini et du syntaxisé, où la pensée circule sur des voies tracées et consolidées. Au dehors, on s'enfonce dans une véritable boue sémantique. Pourtant ce n'est que là, dans cette boue, et lorsqu'on force le regard à discerner les formes mouvantes, que l'on peut percevoir du nouveau. De toute façon le problème de localité nous y force: c'est un problème très fondamental où tout comportement inertiel, inanalysé ou approximatif, conduit inéluctablement à l'arrêt de la capacité de raisonnement, ou à des problèmes et perspectives illusoires. On ne peut pas cette fois suivre un chemin parce qu'il est construit. On est obligé de choisir et de suivre la direction qui convient.

Je reviens donc sur la création d'une paire corrélée $S_1$, $S_2$. J'imagine ce processus comme ayant des analogies avec la formation de gouttes. (Ceci peut être faux, mais ce n'est pas *a priori* impossible, et *on n'a besoin que d'un exemple de possibilité*). Je dessine donc ainsi la projection spatiale (en deux dimensions) du domaine d'espace- temps $\otimes s_c(t).\otimes t_c$, $t_o < t < t_f$, pour 4 époques:

* $t = t_o$;
* $t_o < t < t_f$
* $t_o < t < t_f^-$ (c'est-à-dire immédiatement *avant* $t_f$) ;

et $t = t_f^+$ (c'est-à-dire immédiatement *après* $t_f$)

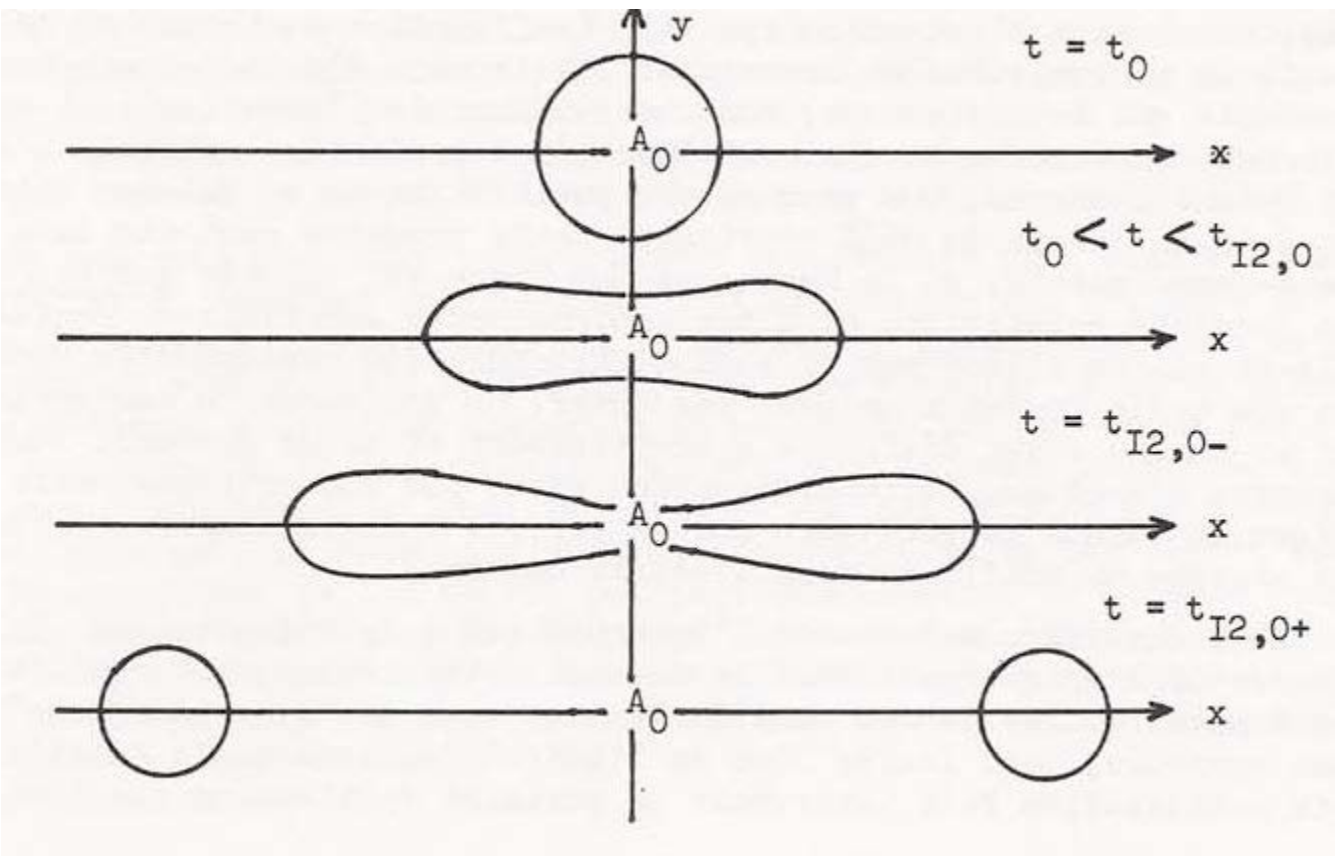

Supposons maintenant que la distance $d_{12}$ entre les appareils $A_1$ et $A_2$ est plus petite que la projection maximale de $\otimes s_c(t)$ sur l'axe x correspondant à $t = t_f^{-}$.

Les appareils $A_1$ et $A_2$ seront donc atteints non pas par "$S_1$" et "$S_2$" respectivement, mais par "S en cours de désintégration", qui *peut* néanmoins déclencher des enregistrements sur $A_1$ et $A_2$. Supposons encore que la durée des évènements de mesure se trouve être telle par rapport à $d_{12}$, que la distance d'espace-temps entre les événements de mesure soit du genre espace. Enfin, supposons que, en dépit de tout cela, les évènements de mesure ne soient pas indépendants au sens de Bell, c'est-à-dire supposons qu'un changement de $A_2$ puisse agir à une vitesse v > C sur le résultat de l'un des enregistrements de $A_2$. Les statistiques de résultats d'enregistrements sur $A_1$ et $A_2$ seront alors "non localement corrélées" et l'inégalité de Bell sera violée. Mais serait-il en ce cas justifié de conclure qu'on a démontré une contradiction avec la théorie de la Relativité? La théorie de la Relativité ne statue *que* sur des "signaux" (quelle est exactement la définition?) se propageant "dans le vide". Elle ne statue rien du tout concernant la transmission "d'influences" (définition?) à travers un "système" (objet? processus?). En particulier, elle n'impose rien du tout concernant "l'ordre temporel" (?) ("causal" ou "non causal") (?) d'événements placés à des endroits spatiaux différents d'"un même système". L'exemple imaginé – un modèle de "création d'une paire" – n'appartient tout simplement pas au domaine de faits que la Relativité décrit. Aucune théorie constituée ne le décrit encore. Pourtant cet exemple, quelles que soient ses inadéquations face à la réalité inconnue, caractérise certainement d'une manière en essence acceptable ce qui mérite la dénomination de processus de création d'une paire: un tel processus doit occuper un domaine non nul d'espace-temps, dont la projection spatiale, connexe au départ, évolue, devenant non connexe.

Cet exemple de possibilité me semble suffire comme base pour la conclusion suivante: les tests destinés à vérifier l'inégalité de Bell, même s'ils violaient définitivement l'inégalité, ne pourront jamais établir à eux seuls que le principe einsteinien de localité a été enfreint. Pour préciser ce qui est en jeu, la modélisation de Bell et le test correspondant devront être associés à d'autres modélisations et à d'autres tests, concernant l'extension d'espace-temps des évènements qui interviennent, non observables ("création") et observables (mesures). La minimalité de l'interprétation minimale n'est en fait qu'une prudence, une peur encore positiviste de se laisser entraîner trop loin en dehors du déjà construit. Cette prudence cantonne dans un face-à-face indécis, où la Mécanique Quantique est opposée *indistinctement* à la localité relativiste et à des prolongements inertiels et confus de celle-ci qui ne s'insèrent en aucune structuration conceptuelle constituée. Mais une telle prudence ne peut pas durer. Un processus de conceptualisation en chaîne s'est déclenché subrepticement et aucun obstacle factice ne pourra l'arrêter. Cette affirmation n'est pas une critique, elle désigne la valeur la plus sûre que je perçois dans la démarche de Bell, et elle exprime ma confiance dans l'esprit humain.

Je considère maintenant l'interprétation épistémologique. Celle-ci s'avance déjà précisément dans le sens de cette inéluctable modélisation supplémentaire. Les termes considérés sont ceux de "1 système" et "2 systèmes corrélés mais isolés l'un de l'autre" (au sens de la Relativité). La modélisation supplémentaire mentionnée fait intervenir le postulat épistémologique courant d'existence de propriétés intrinsèques pour des entités réelles isolées. On déduit de ce postulat des inégalités du même type que celle de Bell, concernant des statistiques de résultats de mesures sur des entités supposées *isolées*. On établit donc une connexion entre des tests sur des inégalités observables d'une part, et d'autre part le postulat épistémologique d'existence de propriété intrinsèques pour des objets isolés au sens de la Relativité. Sur cette base on admet (il me semble?) que la violation de l'inégalité de Bell infirmerait à elle seule la signifiance de la conceptualisation en termes d'entités isolées possédant des propriétés intrinsèques. Or j'ai montré ailleurs (10) (en termes trop techniques pour être reproduits ici) que cela n'est pas possible. Ici je ne ferai à ce sujet que quelques remarques qualitatives.

Tout d'abord, les considérations faites plus haut concernant la création d'une paire peuvent aussi se transposer d'une manière évidente au cas de l'interprétation épistémologique. Mais prolongeons encore autrement ces considérations: plaçons-nous cette fois d'emblée à l'instant $t=t_o$ où $S_I$ et $S_2$ sont créés. Pour $t>t_o$, $S_1$ et $S_2$ occupent maintenant deux domaines d'espace disjoints $\Delta s_1(t)$ et $\Delta s_2(t)$ qui s'éloignent l'un de l'autre et qui rencontrent ensuite respectivement les appareils $A_1$ et $A_2$, produisant des interactions de mesure. L'interaction de mesure de $S_1$ avec $A_1$ est elle-même un évènement qui occupe un domaine non nul d'espace-temps $\Delta s_{m1}(t_{m1})$. $\Delta t_{m1} \neq 0$ (l'indice m se ł mesure) où $t_{m1} \in \Delta t_{m1}$ et le facteur de durée $\Delta t_{m1}$ *dépend* de l'extension spatiale $\Delta s_{m1}(t_{m1})$ liée à l'époque $t_{m1} \in \Delta t_{m1}$ (en supposant que cette extension spatiale reste constante au cours de l'époque $t_{m1} \in \Delta\ t_m$). Il en va de même pour l'évènement de mesure sur $A_2$ dont l'extension d'espace-temps est $\Delta s_{m2}(t_{m2}).\Delta t_{m2} \neq 0$. Comment définir maintenant la distance d'espace-temps entre ces deux évènements de mesure? Quelle que soit la distance spatiale fixée entre $A_1$ et $A_2$, comment savoir si la distance correspondante d'espace-temps entre les évènements de mesure est ou non du genre espace? Car c'est cela qui décide si oui ou non la condition cruciale "d'isolement" réciproque de ces évènements de mesure, se réalise, et c'est sur la base de cette condition que l'on s'attend à l'inégalité de Bell pour les statistiques des résultats enregistrés. Que la distance d'espace-temps entre les évènements de mesure soit ou non du genre espace, cela dépend évidemment (entre autres) des facteurs d'extension spatiale $\Delta s_{m1}(t_{m1})$ et $\Delta s_{m2}(t_{m2})$. Or, que savons-nous de la valeur de ces facteurs? $S_1$ et $S_2$ se déplacent-ils "en bloc", "mécaniquement", comme le suggèrent le modèle de Louis de Broglie et le concept récent de soliton, ou bien s'étalent-ils comme le suggère le concept quantique courant de paquet d'ondes à évolution linéaire Schrödinger?

On pourrait peut-être espérer avoir une réponse plus claire dans le cas où $S_1$ et $S_2$ seraient des photons "dont la vitesse est C". Mais la vitesse de quoi? Du front de l'onde photonique, oui, mais que penser du "reste" du photon? Comment est fait un photon, comme un microsystème de Louis de Broglie, avec une singularité et un phénomène plus étendu autour? Le comportement manifesté par des ondes radio le laisse supposer. De quelle extension alors? Dans la phase actuelle, que savons nous, exactement et individuellement sur ces entités que l'on dénomme "photons"? La Mécanique Quantique newtonienne ne les décrit pas ; l'électromagnétisme ne les décrit pas individuellement. La théorie quantique des champs a été marquée, au cours des années récentes, par des essais "semi-classiques" dont le but est d'éliminer tout simplement la notion de photon afin d'éviter les difficultés conceptuelles liées aux algorithmes de re-normalisation (11).

On peut donc conclure en toute généralité que, quelle que soit la distance spatiale fixée entre $A_1$ et $A_2$ (qu'il s'agisse de microsystèmes à masse non nulle au repos ou de photons), pour savoir si les évènements de mesure sur ces microsystèmes sont séparés ou non par une distance d'espace-temps du genre espace, il faudrait connaître (entre autres) l'extension spatiale des états de ces microsystèmes, en fonction du temps.

Sans détailler plus des enchaînements logiques non essentiels, ces seules remarques suffisent pour indiquer la base de l'affirmation suivante.

A eux seuls, les tests de l'inégalité de Bell ne permettront jamais de conclure concernant la signifiance de l'assignation de propriétés intrinsèques à des entités réelles isolées au sens de la Relativité d'Einstein. Donc pour l'instant aucun face-à-face n'est encore défini entre la Mécanique Quantique et les postulats épistémologiques de notre conceptualisation courante de la réalité. Seule une direction de pensée est tracée, qui suggère l'intérêt de recherches nouvelles sur la structure d'espace-temps de ce que l'on appelle des microsystèmes individuels. Cette direction de pensée me paraît courageuse et très importante, mais dans la mesure où elle se reconnaît et s'assume. Elle s'associe alors naturellement à des recherches récentes sur l'extension des microsystèmes à masse non nulle au repos (12), (13) et sur le concept de photon (11). Il est très remarquable de voir que toutes ces recherches se concentrent sur les phénomènes et concepts d'interférence. En effet c'est là qu'à travers le statistique peut apparaître l'individuel. C'est là que peut se trahir – si on l'y cherche – la confusion entre des interférences *mathématiques* de statistiques standard et d'autre part des statistiques d'interférences physiques d'une entité individuelle qui se superpose avec elle- même (14), (15).

A travers le problème de localité, j'ai dirigé volontairement les regards sur la couche sémantique qui se trouve sous les mots qu'on emploie. L'état de celle-ci est en quelque sorte l'objet principal de ces remarques. La boue sémantique au dessus de laquelle nous voltigeons salubrement d'algorithme en algorithme, accrochés à des cordes de mots, me paraît mériter d'être connue de plus près. Il faudra bien y plonger pour forger les concepts nouveaux qui manquent et en fixer les contours d'une manière qui permette de s'élever jusqu'à des syntaxisations.

Le concept d'objet au sens macroscopique de ce terme est cerné avec rigueur – bien que qualitativement – à l'intérieur de la logique des classes d'objets et de prédicats. Celle-ci est par essence une théorie des objets macroscopiques explicitement structurée et de généralité maximale. Mais cette théorie est *foncièrement* inapte à une description non restreinte des changements. En effet, la logique des classes d'objets et des prédicats est fondée sur la relation d'appartenance $\in$: si pour l'objet x le prédicat f est vrai, alors x appartient à la classe $C_f$ définie par f: $f(x) \rightarrow x \in C_f$. Mais cette relation fondamentale d'appartenance $\in$ est conçue au départ d'une manière statique, hypostasiée. Aucun aménagement ultérieur ne peut compenser les rigidités introduites ainsi au départ. La théorie des probabilités d'une part et d'autre part les différentes théories physiques (la mécanique, la thermodynamique, les théories des champs, la Mécanique Quantique, la Relativité) sont arrivées à combler cette lacune à des degrés différents. Mais chacune pour une catégories particulière de faits et par des méthodes implicites et diversifiées. Une théorie générale et spécifique des évènements et des processus, une logique des changements absolument quelconques, à méthodologie explicite et unifiées, n'a pas encore été construite*.

Considérons maintenant de nouveau la logique des classes d'objets et de prédicats. Elle transgresse foncièrement l'individuel, puisqu'elle décrit des classes. Elle semblerait donc être vouée naturellement à une quantification numérique de type statistique ou probabiliste, à l'aide d'une mesure de probabilité définie sur les classes. Pourtant, à ce jour, une telle quantification numérique de la logique n'a pas pu être accomplie. Les "quantificateurs" logiques $\exists$, $\forall$, Ø, sont restés *qualitatifs* !

Complémentairement en quelque sorte, à ce jour, la théorie des probabilités n'a pas encore développé explicitement un traitement classificateur. Le concept fondamental employé est celui d'espace de probabilité $[U,\tau, p(\tau)]$ où $p(\tau)$ désigne une mesure de probabilité posée sur une tribu d'événements $\tau$, définie sur l'univers $U=\{e_i, i=1,2,....\}$ d'événements élémentaires $e_i$. Cette tribu peut refléter, en particulier, une classification des événements élémentaires $e_i$ commandée par un prédicat f et en ce cas des propriétés spécifiques "logiques" s'ensuivent pour l'espace $[U,\tau, p(\tau)]$. *Via* ces propriétés classificatrices, la connexion entre logique et probabilités pourrait être amorcée. Mais ceci n'a pas été tenté, et la connexion reste pour l'instant non élaborée.

Considérons maintenant la Mécanique Quantique. Elle introduit des espaces de probabilité. Pourtant les relations entre ces espace sont telles que certains mathématiciens affirment que "la Mécanique Quantique n'est pas une théorie de probabilités". La connexion entre la théorie des probabilités et la Mécanique Quantique reste pour l'instant elle aussi très obscure.

---

* J'ai pu prendre connaissance d'une tentative originale et courageuse de formaliser la durée (I6). Jusqu'ici seules les valeurs associables à la durée ("le temps") ont fait objet de certaines formalisations.

D'autre part les relations de la Mécanique Quantique avec les divers concepts suggérés par le langage qu'elle introduit – 1 système, 1 système de 2 systèmes corrélés, etc. – restent elles aussi très obscures. La Mécanique Quantique n'indique en fait strictement rien concernant ces concepts tels que l'on pourrait vouloir les imaginer en dehors de l'observation. Même la probabilité de présence n'est qu'une probabilité de résultats d'interactions d'observation: il est permis par la Mécanique Quantique d'imaginer qu'un "système" qui fait une marque sur un écran à un moment t, se trouvait, en lui-même, aussi loin que l'on veut de cette marque, aussi peu que l'on veut avant le moment t. La Mécanique Quantique laisse parfaitement non conceptualisée en elle-même, "la réalité" dont elle codifie de manière si riche et détaillée les manifestations observables à travers les interactions de mesure.

Considérons enfin la théorie de la Relativité. Cette théorie est, à sa base, individuelle, non statistique, et continue, non quantifiée. En outre, elle décrit "ce qui est", bien que relativement à l'état d'observation. Sa connexion avec les espace de probabilité à évènements foncièrement *observationnels* et quantifiés de la Mécanique Quantique, soulève des problèmes bien connus et très résistants.

Ainsi nous sommes actuellement en possession de plusieurs structurations syntaxiques constituées, chacune très complexe, riche et rigoureuse. Mais ces structurations sont comparables à des icebergs émergeant de la mer de boue sémantique, sous le niveau de laquelle les bords et les bases disparaissent. Quand à l'ensemble des concepts liés à la propriété fondamentale de durée, les concepts de processus, d'évènement, de changement, de permanence, de succession, de TEMPS, ils n'agissent librement qu'à l'état épars, primitif et subjectif, tels que l'expérience et le langage les a diversement induits dans les esprits. Car les organisations auxquelles ces concepts ont été soumis à l'intérieur de la théorie de la Relativité, de la théorie des probabilités, ou à l'intérieur de telle ou telle autre théorie physique, sont toutes particularisantes et amputantes. La situation est encore telle que la décrivait Bergson: «La déduction est une opération réglée sur les démarches de la matière, calquée sur les articulations mobiles de la matière, implicitement donnée, enfin, avec l'espace qui sous-tend la matière. Tant qu'elle roule dans l'espace ou dans le temps spatialisé, elle n'a qu'à se laisser aller. C'est la durée qui met des bâtons dans les roues' » (17).

Je résume une fois encore par un schéma:

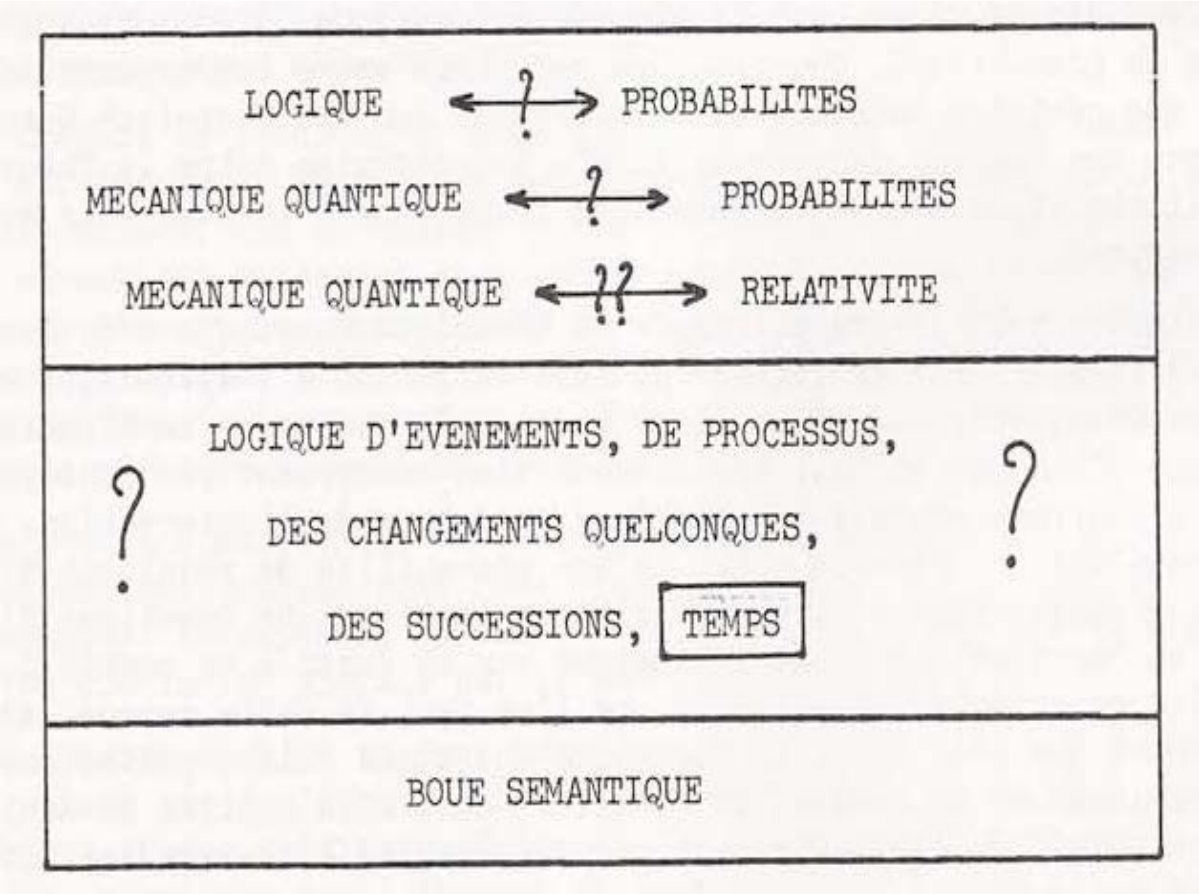

Quand il n'existe encore aucune unification entre la démarche statistique, discrète, observationnelle, orientée vers le microscopique, de la Mécanique Quantique, et d'autre part la démarche individuelle, continue, réaliste, orientée vers le cosmologie, de la Relativité, quant tout ce qui touche à la durée et au temps est encore si peu élucidé, quand tout ce qui touche à la manière d'être de ces entités que l'on appelle des microsystèmes – ou plus encore, de microétats – est encore tellement inexploré, quel sens cela peut-il bien avoir d'affirmer qu'on se trouve – sur la base de tests de "non-localité" – devant un face-à-face *contraignant*, direct ou pas, entre la Mécanique Quantique et la Relativité? Ou bien entre la Mécanique Quantique et notre conceptualisation du réel?

**Conclusion**

Je ne puis qu'écarter, pour ma part, les face-à-face que les autres physiciens pensent percevoir. Pour moi la valeur du théorème de Bell réside ailleurs: ce théorème, et l'écho qu'il soulève, illustrent d'une manière frappante la puissance d'action des modélisations *mathématisées*, lorsqu'elles sont connectables aux tests expérimentaux. Pendant des dizaines d'années, les tabous positivistes ont fait obstacle aux modèles. Le résultat est ce vide vertigineux de modèles syntaxiques, et même seulement qualitatifs, que l'on découvre maintenant sous les algorithmes quantiques. Or, la modélisation de Bell a déclenché une dynamique de conceptualisation et de syntaxisation. Cette dynamique atteindra peut-être l'attitude positiviste. Elle ébranlera peut être la Mécanique Quantique et la Relativité. Car elle attire et maintient longuement l'attention sur l'état du milieu conceptuel dans lequel les théories actuelles sont immergées. De ce contact prolongé sortiront peut-être des théorisations nouvelles, plus unifiées, plus étendues et plus profondes. Je perçois (ici comme en théorie de l'information) les premiers mouvements de formalisation de l'épistémologie, les premières ébauches, peut-être, d'une méthodologie mathématisée de la connaissance. Et cela pourrait s'avérer plus fertile que toute théorie particulière d'un domaine donné de réalité.

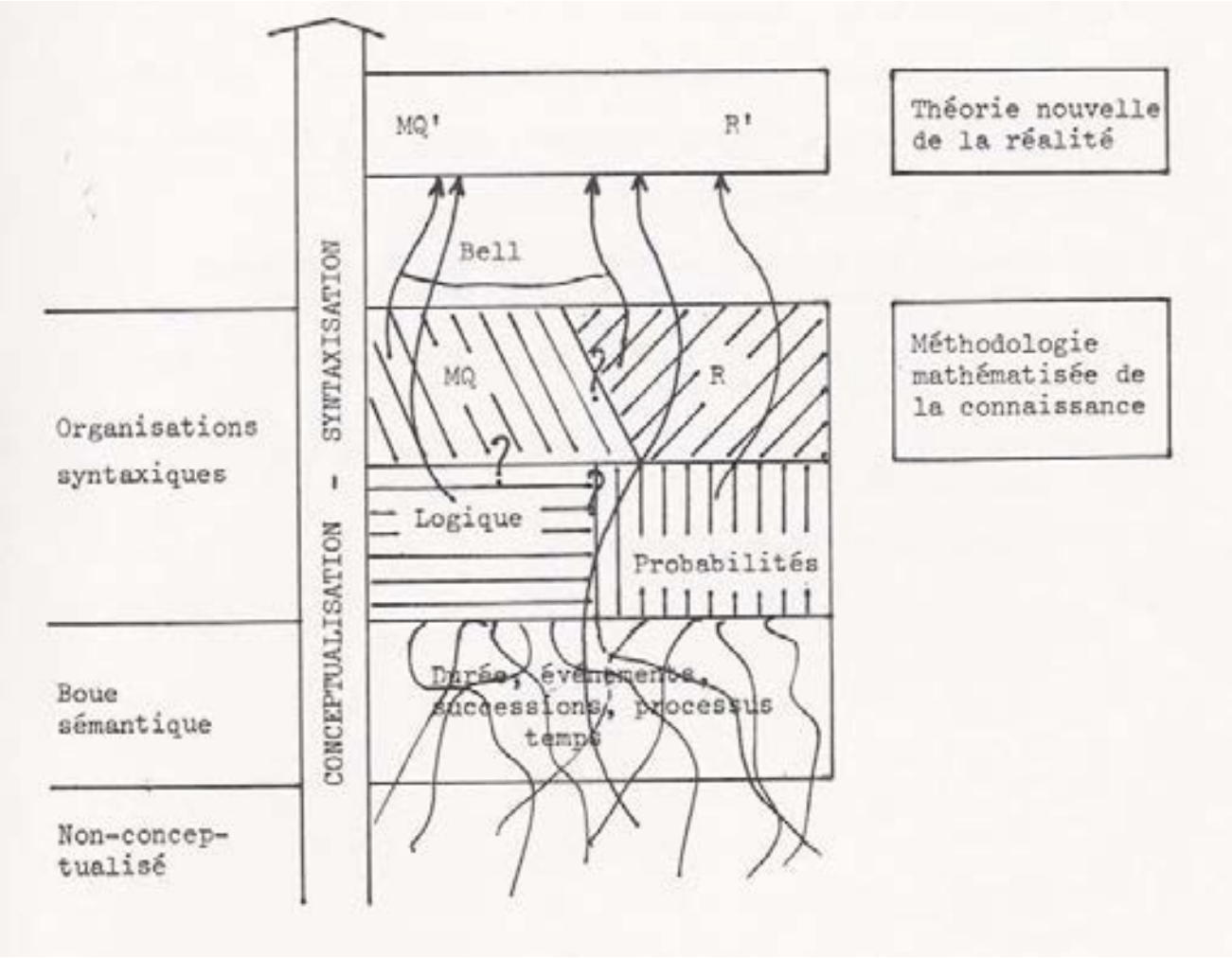

•

**REFERENCES**

(1) Bell, Physics, I, I95, (I964)
(2) Einstein, Podolsky, Rosen, Phys. Rev. 47, 777 (I935)
(3) Lochak, Found. Phys. 6, I73 (I976).
(4) Costa de Berauregard, Found. Phys.6, 539 (I976), Phys. Lett. 67. A, I7I.
(5) Selleri, Found. Phys. 8, I03 (I978).
(6) Stapp, Phys. Rev. DI3, 947 (I976).
(7) Vigier, Nuovo Cimento Lett. 24, 258 (I979).
(8) d'Espagnat, Phys. Rev. DII, I454 (I975) et DI8,
(9) Weinberg, Gravitation and Losmology, J. Wiley Sons, N.Y. (I975).
(I0) Mugur-Schächter, Espistemological Letters (I976).

(I1) Cohen Tannoudji, Exposé au Collège de France, juin I979.

(I2) Mugur–Schächter, Evrad, Tieffine, Phys. Rev. D6, 3397 (I972).

(I3) Evrard, thèse, Univ. de Reims (I977).

(I4) Mugur–Schächter, Quantum Mechanics a Half Century Later (eds. J Leite Lopes and M. Paty) D. Reidel (I977).

(I5) Mugur–Schächter, *Etude du caractère complet de la Mécanique Quantique,* G. Villars (I964)

(I6) Schneider, la Logique self-référentielle de la temporalité (non publié).

(17) Bergson, l'Evolution Créatrice (1907 ).

**************

The most striking in this account from 39 years ago is that the public conceptual situation concerning microphysics did not notably change in its essence.

As for the author of the present work, she believes that by precisely the work exposed below – and from her own point of view – she has finally accomplished in its essence the program delineated in the Conclusion reproduced above, that has been in work since 1979.

# 1.I [5]

# THE FIRST GERM OF DESCRIPTION OF A MICROSTATE: GENERATION OF A MICROSTATE AND QUALIFICATION OF ONE SPECIMEN OF A MICROSTATE

## 1.I).1. OPERATION *G* OF GENERATION OF A MICRO-ENTITY-TO-BE-STUDIED AND A BASIC METHODOLOGICAL DECISION

## COMPOSED OPERATION OF GENERATION *G(G1, G2,...Gk)*

In agreement with Dirac we distinguish between stable characteristics assigned to a 'micro*system*' (mass, spin, etc.), and unstable dynamical characteristics assigned to a 'micro*state*' (position, momentum, etc.). So far this is just a verbal sign posited to point toward a physical thing that is entirely unknown as to all its *specificities*.

By its definition the concept of knowledge means qualification of something-to-be-qualified. In this first part of the present work we want to establish the general a priori features of any process of creation of scientific knowledge on microstates, that is, of communicable, consensual and verifiable knowledge tied with microstates, when one wants to start at the extreme 'bottom' and to proceed down-up. So – once given the cognitive situation that is at work – we have to establish how it is possible to produce out of the as yet never qualified, a microstate in the role of entity-to-be-qualified, and how and in what a sense it is possible to qualify this in a scientific way.

### (1.I).1.1. A basic question

In current languages and in classical grammars an object-to-be-qualified is usually supposed to pre-exist, as such. It just "is" there. Its definition is realized by use of grammatical predicates (“bring me the brown thing from that drawer”, etc., and look in a dictionary). The predicates also are considered to pre-exist – in the air of thought, platonically – or expressed by verbal pointers of location ('there', etc.), or even by just pointing physically toward the object-to-be-qualified. In the classical logic these assumptions are sanctified. The objects-to-be-qualified are represented by a set of letters *('x', 'y', 'z',....)* and the functional expressions like $f_P(x)$ that contain such a letter (*x* in this case) and where *P* designates a 'predicate', are called propositional functions and they become true or false according to whether *x* satisfies the predicate *P* or not, which is a physical fact that is perceived by the human observer. All this is founded upon the naïvely realist postulate that the objects-to-be-qualified are perceived 'such as they really are' via their intrinsic 'properties' represented by predicates, and upon the fact that, classically, most of the objects-to-be-qualified are directly perceived. This last fact however has increasingly many exceptions and this wraps up the central view in a thickening ball of procedures for reaching perception.

[5] To be read ‘chapter 1 from Part I’.

But how can a radically non-perceivable and unknown microstate be introduced as that-what-is-to-be-studied, when in general it does not even pre-exist? (In this respect the unbound microstates are the most striking example). How can a microstate be obtained in this role, and in a way endowed with some sort of stability so as to be kept available for further cognitive action concerning it, that permit also verifiability, so scientificity?

Of course as soon as we presuppose an unknown microstate and we indicate it by some word or label we already have presupposed that it is tied with something that pre-exists and out of which that microstate can be brought into the role of an object-to-be-qualified. But in order to effectively bring into this role a given sort of micro-entity-to-be-studied, some definite macroscopically controllable physical *operation of generation* of *this* should be realized accordingly to some previously established knowledge, and on some specified space-time support: If not we cannot even think of this micro-entity, so a fortiori we cannot study it. Furthermore this operation has to be repeatable, for if not no *verifiability* of its consequences can be conceived, so again a scientific study is out of reach.

This problem does not exist with respect to the directly perceivable objects from our current life that – admittedly – just subsist while we cease observing them, and when we want to perceive them again we manage to bring them again into our domain of perception. But for a radically non-perceivable micro-entity this problem emerges basically and dramatically. In the historically realized top-down approach, from a small step to another small step, this problem remained more or less hidden by the classical models and assumptions. But when one wants to start at the extreme 'bottom' and to proceed down up, this problem is gaping and it has to be solved explicitly.

### (1.I).1.2. Operation *G* of generation of a micro-entity-to-be-studied

Then let us focus upon an operation of generation of a micro-entity in the role of micro-entity-to-be-studied, in the scientific sense. We denote it by *G*.

As remarked, the repeatability of *G* is an unavoidable pre-condition for constructing scientific knowledge on microscopic physical entities. But how can we know that when *G* is repeated it emerges the same?

Well, we *cannot* know whether yes or not *G* comes out the same when it is repeated. Nor can we insure factually a positive answer. This is so because the operation *G* is a factual physical process that has to be inter-subjectively specified and communicated, which is possible only by some ***finite*** definition. And any finite factual definition is quite essentially unable to constrain into absolute (or 'total') sameness the whole factual singularity of each realized replica of the operation *G* (Umberto Ecco has said that as soon as we speak or write we conceptualize and thereby we quit and lose irreversibly the infinite singularity of any piece of factual entity).

> Here the unconceivable *in*finity of possible ways of being of any fragment of factual physical reality stays face-to-face with the finiteness of the human capacity to constrain and to control in predefined ways.

However giving up because of this the whole project of establishing how it is possible to create some sort of knowledge on the dynamical states of micro-entities would be an unacceptable weakness from the part of a human mind. We are in presence of a problem of strategy, of method. So we have to conceive an appropriate strategy.

### (1.I).1.3. A methodological decision ('the microstate *corresponding to G*' and 'one specimen' of it)

We organize a first *methodological decision denoted MD* that introduces a global strategy of speaking and thinking on the basis of which it becomes possible to start, to act, and to achieve the bottom-up construction of *IQM*.

***MD***

- Each time that one individual operation denoted *G* of generation of a dynamical state of a micro-entity-to-be-studied, is realized *as such*, in agreement with a definition expressed in terms of a finite number of parameters that are controllable factually from our macroscopic level of existence – which, for us, is the only sort of possible factual definition – this operation *G* itself is admitted to come out the same by construction, *with respect to its factual finite definition*.
- *That* what emerges in consequence of *one* realization of *G* is not directly observable by our bio-psychical apparatuses but it is posited a priori to be observable indirectly via future appropriate operations of qualification; and it is conceived as one *specimen* (or variant) denoted $\sigma(ms_G)$ of something more global than any individual specimen $\sigma(ms_G)$.
- The more global entity posited above will be labelled by '$ms_G$' and we call it 'the microstate *corresponding to G*'.
- This amounts to denote

$ms_G \equiv \{\sigma(ms_G)\}$ ($\sigma$ : specimen)

- On this basis we shall enter upon the bottom-up constructive research of an observable and verifiable, *law*-like one-to-one relation

$$G \leftrightarrow ms_G \qquad (1)$$

The necessarily finite character of the human definition of *G* and the obvious fact that absolute sameness is just nonsense, have imposed inside *MD* a *factual* and ***multiple*** content for the *new* concept called 'the microstate corresponding to *G*' and denoted $ms_G$. Whereas the classical concept of a microstate is defined *abstractly* and the content assigned to it is specified *individually*. So

A microstate '$ms_G$' in the sense of *MD* is essentially different from a microstate in the classical sense.

Nevertheless the word 'microstate' is kept in use[6] because it can play the very useful role of a recurrent element of reference and of comparison between the classical top-down approach specified conceptually via abstract definitions, and the factual bottom-up approach practised here. This word will work as a memento that in this work the *origin* of the process of construction of knowledge has been changed. That we now start from the extreme boundary between the previously conceptualized and the as yet a-conceptual universal physical substratum, of which the existence is unanimously presupposed throughout Physics.

So we start from *local zeros* of previously constructed knowledge on – *specifically* – each individual micro-entity brought in as an entity-to-be-studied, as is the case for a specimen $\sigma(ms_G)$ of the microstate $ms_G$[7]. And therefrom we construct bottom-up. This changes also the *order of constructability of concepts* (abstract concepts or factually defined ones) as well as the place inside this order of each sort of piece of *verifiable*

[6] The absence of an explicit specification that the concept '$ms_G$' is different from the classical concept of 'microstate', has nourished a harmful and years-long misunderstanding with Henri Boulouet.

[7] A specimen $\sigma(ms_G)$ of $ms_G$ is more than $ms_G$ alike to a classical microstate, but it emerges entirely undefined in its individuality, its specificity.

*knowledge*. It can be hoped that the explicit awareness of this new order from a bottom-up approach, when compared with that from the classical top-down approach that started spontaneously from our everyday level of ancestral conceptualization, will bring forth many clarifications concerning the problems of interpretation of the modern microphysics.

The posit (1) – via the definition $ms_G \equiv \{\sigma(ms_G)\}$ – expresses the way in which is infused into microphysics the so much discussed "*essential* indeterminism". Namely by the imperative necessity to introduce a new sort of factually defined concept of microstate, associated with the ineluctable *finiteness* of our capacity to produce effective assertions, so in particular effective definitions.

> The 'essential indeterminism' of the modern microphysics is *factual, observational, and predictive*; whereas the classical postulate of determinism is *abstract, purely conceptual*, and it is devoid of any rigorously attainable observational support; which is explained by the notion of 'imprecision of measurement', unpredictable (chaotic) mathematical development, etc.[8].

But it is very noteworthy indeed that:

> *The whole posit MD and in particular the one-one relation (1) $G \leftrightarrow ms_G$ that found the "the essential* ***indeterminism****" of the modern microphysics, in fact still express a basically deterministic view.*

Indeed, the relation (1) $G \leftrightarrow ms_G$ amounts to the assertion of existence of a probability measure, which still is the assertion of a 'law', of a one-one causal connection "*if* this *G*, *then* that ($ms_G$)": Our human minds – such as they have been modelled by optima of adaptation of our ways of perceiving, thinking and acting – have selected and imprinted upon our minds a principle of causality. This is a mental fact. This principle works so strongly that in order to transgress it, we still *use* it, but in a way that displaces its frontier upon 'probabilistic' factual-operational-observational contours, instead of point-like individual assertions. So what acts in this circumstance is not in the least an "essential indeterminism"; this is ***a factual impossibility to in general insure an in principle rigorously individual prediction***. And the Methodological Decision (1) permits to nevertheless save a global inner coherence founded upon a general deterministic postulate, by *distinguishing* explicitly between: ***(a)*** a general abstract posit of punctual causality, and on the other hand ***(b)*** the *sort* of scientific consensual, predictive-verifiable ***knowledge*** that can be generated in a cognitive situation that is *entirely* founded upon factual-physical operations, in the strict absence of any direct sensorial human perceptibility.

### (1.I).1.4. Mutation of the classical concept of 'definition': a split

We have noted already that in the classical conceptualization the entity-to-be-studied is conceived to pre-exist as a stably available potential support for qualifications achieved by identifying predicates conceived to represent 'properties intrinsically possessed' by this entity. The direct perceptibility permits this confortable ellipsis that *absorbs* in it the necessity of an explicit operation *G* of generation of the considered entity as an entity-to-be-studied. But for microstates this is not possible. And that is why:

> *MD* splits the classical concept of definition into a sequence of two distinct operations that can be achieved only separately, namely; an initial action of generation of the object-entity-to-be-studied, that is already specified inside *MD*;

---

[8] The investigations on "chaos" have brought forth that 'determinism' does not entail observational predictability.

and a *subsequent* act (that still remains to be specified) of qualification of the object-entity-to-be-studied generated before.

### (1.I).1.5. Composed operations of generation $G(G_1,G_2,...G_k)$: a principle of composition of physical operations of generation

From its start, the study of microstates has brought into evidence a class of microstates that have been called '(auto)-interference-states' and that have played a founding role in the emergence of quantum mechanics (the paradigmatic case is Young's two slits experiment). The process of generation of an interference-state permits to distinguish at least two operations of generation $G_1$ and $G_2$ that are involved in the following very peculiar sense: Each one of these two operations *can* be produced separately, and if they are, then two different corresponding microstates $ms_{G1}$ and $ms_{G2}$ do emerge. But when $G_1$ and $G_2$ are 'composed' into only *one* operation – let us denote it $G(G_1,G_2)$ [9] – then, accordingly to (1), there emerges only *one* corresponding microstate $ms_{G(G1,G2)}$ that manifests 'auto-interference effects'.

On this factual basis tied with the just indicated way of speaking, we introduce here an only *qualitative* but nevertheless a general *'principle of composition of operations of generation'* according to which:

> In certain operations of generation of a microstate, two or more operations of generation – deliberately produced by human researchers or brought forth by natural processes – can 'compose' while acting upon *on*e preliminary unspecified microstate, so as to generate together *one* microstate-to-be-studied, in the sense of *MD*.

When this happens we shall speak of one microstate $ms_{G(G1,G2,...Gn)}$ with a composed operation of generation $G(G_1,G_2,...G_k)$ [10]. When this does not happen, for contrast and precision we can sometimes speak of a 'simple' operation of generation.

The operation of 'composition of operations of generation of a microstate' defined above, as well as the corresponding underlying principle of possibility to compose such operations, are only very feebly defined here. But in the Parts II and III of this work this principle will gain more specification and it will entail most essential consequences.

### (1.I).1.6. Universality of *G*

At a first sight it might seem that the concept of operation of generation of an entity-to-be-qualified constitutes a radical novelty of which the necessity is specific of microphysical entities. But a deeper analysis reveals that in fact this mutation only brings into evidence a *universal* phase in the human conceptualization (MMS [2002], [2006]) that acts quite obviously already inside all the 'exceptions' that surround the core of the fully classical conceptualization. Indeed any definition presupposes – more or less implicitly but quintessentially – an operation of initial specification of the entity-to-be-defined. Often this is a specification via a merely psycho-sensorial out-cut from the continuum of the directly perceived surrounding 'exterior reality'; or even only a reflex human gesture (turning the whole head or only the eyes toward some delimited domain of direct perceptibility); or even an exclusively mental selection via a focalization of the attention. But in many 'classical' situations the act of specifying the entity-to-be-qualified consists of a deliberate and laborious *physical* operation of separation and of supply into immediate accessibility (think of medical analyses or geological or archaeological procedures). And sometimes, exactly like in the case of microstates, a 'classical' operation of generation consists of a deliberate radical *creation* of the entity-to-be-qualified (production of prototypes in the industry of artefacts[11], simulated test-situation in a detective research, etc.). In short:

> Strictly always a human being, in order to acquire some knowledge on some thing, somehow singularizes this thing from inside the continuum of 'the reality', explicitly or implicitly[12].

---

[9] This notation stresses that only *one* operation of generation has been effectively achieved by 'composing' other operations of generation that *could* have been achieved separately but have *not* been separately achieved.

[10] We do not try to specify the conditions that restrict the possibility of composing operations of generation (in particular, the space-time conditions) though such conditions do certainly exist. Nor do we try to specify some limit to the possible number of composed operations of generation. These are features that are still unexplored from both a factual and a conceptual point of view because inside nowadays quantum mechanics – together with the concept of operation *G* of generation of a microstate itself – they remain hidden *beneath* what is mathematically expressed, in consequence of a basic confusion between 'superpositions' in the *mathematical* sense, and *factual* superpositions in space-time, of operations or of physical entities. The consequences of this basic confusion will be narrowly surveyed and in the third part of this work they will play a quite essential role.

That is so because a human being can have only finite perceptions and can perform only finite actions, whether these actions are psychical, or psychophysical, or physical. So he is obliged to somehow delimit inside the in-finite whole of what we call 'reality' that what he wants to qualify, to parcel this out in some sense. This inescapable necessity to parcel out induced by the human imprisonment in *finiteness* has very basic and unexpected consequences.

> *This* is what introduces a basic impossibility to assert an *individually* deterministic one-to-one *factual-observational* relation in *MD*, which in the case of microphysical entities-to-be-qualified becomes systematic and obvious and entails the non-classical posit $ms_G \equiv \{\sigma(ms_G)\}$ and a corresponding displacement of the one-to-one deterministic relation (1) upon the probabilistic level of qualification.

The systematic and obvious character mentioned above is indeed specific of *the human cognitive situation* with respect to microstates. But the *presence*, in any act of scientific conceptualization, of an operation *G* of generation of the entity-to-be-studied, that is controlled by the involved human cognitive situation, is *not* specific of the human cognitive situation with respect to microstates. This presence is a universal cognitive fact; a more or less hidden fact, but a quite universal fact. This simply has not been explicitly remarked, precisely because it is universal, but also no doubt because in the current life – historically and during a very long time – inside the domain of physical reality that was accessible to direct perception it has very often been possible to put spontaneously a physical entity in the role of entity-to-be-qualified, or even to realize this in reflex unconscious ways. While inside the global methodical approaches, the act of bringing an entity in the role of entity-to-be-studied got lost in an ocean of other, more complex and more specific norms (think of the global requirement of 'reproducible experimental conditions' in classical physics).

The universal presence of the operation of generation of an entity-to-be-qualified throughout the human conceptualization is *not* an abstract principle like the posit of determinism. It just is the necessarily existent first phase of any processes of construction of knowledge; and in particular of scientific knowledge, consensual, predictive and verifiable. So the abstract principle of determinism on the one hand, and on the other hand the effect in any given process of construction of consensually observable and verifiable knowledge, of the presence of an operation of generation of the entity-to-be-qualified followed by acts of qualification, must be radically distinguished from one another. And the *relation* – in any given cognitive situation – between the general abstract principle of determinism, and on the other hand the different sorts of effects of the involved *pair* of an operation of generation and an operation of qualification, has to be *specified*, in a way that permit clear *comparison between the various sorts of cognitive situations*.

[11] Cf. H. Boulouet [2014].
[12] And probably any living being does this.

It is true that the acts of measurement introduce systematically observational imprecisions with respect to a posited general causal behaviour of the physical reality. But:

> *These imprecisions of measurements are not the unique source of the statistical-probabilistic character of that what can be observed*. This character is also strongly dependent upon the sort of operation of generation that is involved.

And a clear comparability between the effects entailed in different cognitive situations, by the different sorts of pairs

*[(an operation of generation of the entity-to-be-qualified),*
*(an operation of qualification of this entity)]*

cannot be realized – cannot even be *conceived* – without the use of a common, general language defined inside a general methodological framework that organizes a consensual unity of the criteria. While on the other hand precisely these effects are the source of an illusory 'incompatibility' between macroscopic and cosmic physics, and the modern microphysics.

An explicitly constructed common methodological framework is a necessary condition for understanding and dominating the conceptual and factual consequences of the *way* in which, in each given cognitive situation, the pair *[(an operation of generation of the entity-to-be-qualified), (an operation of qualification of this entity)]* that is involved introduces – or not – observational dispersion. This becomes clearer by the following examination of the acts of qualification of a microstate.

## (1.I).2. BASIC FEATURES OF THE GENERAL CONCEPT OF QUALIFICATION OF ONE SPECIMEN OF A MICROSTATE

### (1.I).1.4. Classical qualification

Inside the classical thinking an act of qualification involves more or less explicitly a *genus-differentia* structure. The *genus* can be conceived as a *semantic dimension* (or *space*) and the *differentia* can be regarded as *'values'* from a *spectrum of values* carried by this semantic dimension. The spectrum can be numerical or not, ordered or not, and it can be specified by the help of material samples or otherwise. Let us denote the semantic dimension by $X$ and by $X_j, j=1,2,...J,$ the values from the spectrum posited to be carried by $X$ (for instance $X$ can be 'colour' and then the spectrum of values $X_j$ consists of a finite number of definite colours *{red, green, blue, etc.}* defined by a finite set of material samples (since for effectiveness we consider only finite definitions)).

As already recalled, inside classical thought with its languages, logic and grammars, a given semantic dimension and the spectrum of values carried by it are currently imagined to somehow pre-exist in the realm of ideas, even if only potentially. But here – and even for classical acts of qualification – we conceive them as being constructed more or less deliberately by the human observer who conceptualizes accordingly to his local *aims* of description, and under the general and permanent though ignored control of the irrepressibly restrictive general human ways and possibilities of thinking and doing, and of the cognitive situation that is at work. All this, considered globally, acts like a net of a priori *constraints*.

According to the classical conception again, there also usually just 'exists' some possibility to estimate what value $X_j$ of $X$ has been found for a given entity-to-be-qualified when it has been examined 'via' $X$. This amounts in essence to imagining more or less explicitly a sort of act of *measurement-interaction* – biological or not, spontaneous or scientific – between some *measurement apparatus* $\mathcal{A}(X)$ and the entity to be qualified. Let us denote by *MesX* such an act of measurement-interaction.

> The result $X_j$ of an act of *MesX*, when perceived by the observer, becomes a piece of *knowledge* concerning the examined entity: indeed, by definition, ***knowledge** of some thing is just **qualification** of this thing, so **what is not qualified in any way is not known.***

This apparent triviality is simply ignored by our spontaneous conception on what we call 'reality'. And even inside scientific 'realistic' thought, the aim to know 'how things *truly* are', 'intrinsically', 'in themselves' – so in fact in the *absence* of any qualification that is *known* consensually via well-defined and repeatable examinations – is not yet generally perceived like a self-contradicting aim.

The operation *MesX* – just like *G* – cannot be defined otherwise than by some *finite* specified set of controllable parameters. Unavoidably features and circumstances that cannot be conceived a priori transcend the control entailed by these parameters. So again, just like in the case of *G* and (1), there is no other way than just *admit* that *all* the realizations of *MesX* are the 'same' *with respect to* a necessarily *finite* set of specified parameters[13]. *This is not reducible to 'imprecision'; **it is an essential feature**.*

When the registration of the value $X_j$ of a semantic dimension or 'quantity' *X* that is posited to be able to qualify an 'object' in the classical sense, is performed directly via a human *biological* sensorial apparatus, it generates in the observer's mind a *quale*, a strictly subjective perception of a definite particular 'quality' that cannot be described but of which the subjective existence can usually be communicated by words, gestures, or other signs that label it *consensually* in connection with its exterior source that is publicly perceivable, namely the considered classical 'object'[14]. We denote globally this classical coding-process by $cod.proc(X_j)$ and we represent a classical *grid of qualification (gq)* by writing

$$gq[X, X_j, MesX, cod.proc(X_j)] \qquad (2)$$

### (1.I).2.2. Qualification of one specimen of a microstate

But how can be qualified a microstate $ms_G$ that cannot be directly observed? The answer, if it is thoroughly constructed, appears to be a genuine saga.

Consider a qualifying quantity *A* with 'values' $a_j$. In order to qualify by a value $a_j$ of *A* the the microstate $ms_G$ such as the operation of generation *G* has brought it forth, *G* must be followed *immediately* by a qualifying measurement interaction *MesA* realized inside the space-time neighbourhood of the space-time support of the operation *G*. Indeed each outcome of $ms_G$ is conceived as a *dynamical* state of a changing physical entity. So, even though any specific knowledge of this changing entity is still lacking in our minds, nevertheless – insofar that knowledge on that what *G* has generated is researched (not on something that has evolved out of that) – the measurement interaction *MesA* must follow the operation *G* immediately. For this purpose a whole succession *[G.MesA*] has to be realized in order to obtain one qualification via *A* of a specimen $\sigma(ms_G)$ of $ms_G$. And since a measurement-interaction with a specimen $\sigma(ms_G)$ requires an appropriate *non*-biological apparatus, its result can only consist of some publicly observable marks registered by devices of this apparatus. Furthermore, in general the measurement-interaction destroys the involved specimen $\sigma(ms_G)$ generated by the previously accomplished operation *G* of generation. And so on. All these questions have been already discussed very much indeed and they have suffered heavy trivialization, but without having been genuinely studied.

---

[13] Suppositions of this kind are made everywhere inside science.

[14] For instance – as it is very well known – each one of us experiences the feeling of a quality that *he* has learned to call 'red' while referring to the *source* to which he connects this value of the quality 'colour' (say a flower). Thereby – by learning and via the involved sort of context – that quale and its values acquire common inter-subjective verbal labels that point inside each given mind toward strictly subjective, non-communicable events. So in classical circumstances each very currently arising quale acquires an inter-subjective labelling that is tied with the illusion that it just 'exists', 'objectively', 'outside there', '*in* the object itself', as a 'property *possessed by it*'.

But much more radically, and rather curiously, a huge gap seems to have been unanimously left entirely implicit, namely *the coding problem*. Below our own examination of the process of qualification of one specimen of a microstate $ms_G$, is centred on this problem and, deliberately, it will be exposed in an outrageously explicit way.

### *(1.I).2.2.1. The coding problem* **versus** *model of any specimen* $\sigma(ms_G)$

What *criteria* do permit to define the procedure that deserves being called a measurement-interaction *MesA* between a specimen $\sigma(ms_G)$ of a given microstate $ms_G$ for measuring on this a quantity *A*? What procedure can endow the publicly observable marks produced by one given act of 'measurement-interaction' *Mes(A)*, with *meaning*, and in terms of – precisely – a *given* value $a_j$ of precisely the quantity *A* that one wants to measure? To reformulate this question in summarized terms we shall call such a procedure *a coding procedure in terms of a value* $a_j$ *of A* and we denote it *cod.proc*$(a_j)$. So:

> When the physical characters toward which the symbol '$\sigma(ms_G)$' points are still entirely unknown so that not even the *applicability* to it of qualifications via a given dynamical quantity *A* first defined inside the classical mechanics (position, or momentum, or energy, etc.) can be asserted a priori, *how can one define the coding procedure cod.proc*$(a_j)$?

This is a most fundamental problem. Nevertheless it has been left implicit. So it has been taken into account only intuitively, without generality, nor rigor. Let us stop on this problem.

The general content of a grid for mechanical qualification of a specimen $\sigma(ms_G)$ accepts the same general *form* (2) of a classical grid. But when a specimen of a *factually* defined microstate $ms_G$ is the object of qualification the signs *A*, $a_j$, *MesA*, *cod.proc*$(a_j)$ point toward *contents* – entities and circumstances – that with respect to the human observer involve cognitive constraints that are radically different from those that act in the case of 'mobiles' in the classical sense:

- That what is to be qualified – one specimen $\sigma(ms_G)$ of a microstate $ms_G$ for which the one-to-one relation (1) $G \leftrightarrow ms_G$ is posited – has been extracted by the operation *G* of generation *directly* from the as yet a-conceptual physical reality. It is still radically unknown in its physical *specificities* inside the class $ms_G \equiv \{\sigma(ms_G)\}$. It is only posited to exist and is labelled.

- Every individual specimen $\sigma(ms_G)$ remains constantly and entirely non-perceptible *itself* by the observer. Suppose that a given sort of measurement *MesA* (for instance with *A* meaning 'momentum' *P*) *does* make sense with respect to what the symbol '$ms_G$' represents, and that we know how to perform such a measurement. When an the act *MesA* is performed upon a specimen $\sigma(ms_G)$, *exclusively* groups $\{\mu\}_{kA}$ of some publicly observable marks (with $k_A=1,2,...m_A$) can be obtained on registering devices of some corresponding apparatus $\mathcal{A}_p(P)$ (a spot on a sensitive screen, a sound-registration at a time *t*, etc., some group of such marks).

- Since the registered group $\{\mu\}_{kA}$ of observable marks is the result of a measurement-*interaction MesA* between $\sigma(ms_G)$ and an apparatus $\mathcal{A}_p(A)$, its *meaning* can *not* be conceived in terms of some 'property' assignable to $\sigma(ms_G)$ ***alone***. *The marks* $\{\mu\}_{kA}$ *characterize exclusively the achieved measurement interaction as a whole*. While in the radically incipient cognitive situation that is considered here no criteria are conceivable for separating a posteriori inside $\{\mu\}_{kA}$ the contributions from the two sources $\sigma(ms_G)$ and $\mathcal{A}_p(A)$.

- A fortiori, since $\sigma(ms_G)$ itself is not directly perceivable, no qualia tied with exclusively this entity can be formed and triggered in the observer's mind via *MesA*: The observer gets no inner subjective feeling whatever tied with the nature of *A* and with the specimen $\sigma(ms_G)$.

The characters listed above will be globally indicated as the result of one *primordial transferred qualification of a specimen* $\sigma(ms_G)$ *of a microstate* $ms_G$, which means: a strictly first compact whole of observable marks that are *transferred* on the registering devices of an apparatus, that do *not* entail any sort of qualia tied with – separately – the studied microstate *itself*, and that cannot be analysed further in effects of the involved specimen $\sigma(ms_G)$ and effects of the involved act of *MesA* [15].

We come now back to the central question from this section: How are we to *conceive* an act of measurement-interaction *MesA* in order to found the assertion that the registered marks $\{\mu\}_{kA}$ do qualify the involved specimen $\sigma(ms_G)$ of the studied microstate $ms_G$ in terms of a given value $a_j$ of a given measured quantity *A*? In what a way can an observable group of brute marks be brought to *signify* in terms of one definite value $a_j$ of *A*? ***How can the observable result of an interaction 'MesA' be endowed with a definite meaning***? It seems clear that:

> In the absence of *any* general model of a specimen of a factually defined microstate $ms_G$ it is not conceivable to produce an a priori meaningful definition of the possible results of a measurement-interaction with a specimen $\sigma(ms_G)$ of $ms_G$ [16]. So a consensual study of a 'mechanics' of the microstates cannot even begin. For this purpose a general model of a microstate must be *given* as a basic primary datum.

We hit again the transparent wall that imprisons us inside our human ways of thinking and acting. This is a major *fact* that cannot be transgressed. Then we must let it work freely and take it into account explicitly. We must organize a framework where we are *insured* that working freely accordingly to the specific laws of our thought we develop clearly controllable and meaningful results. Models and formal systems of signs, logical or logical-mathematical, generate knowledge only when they are made use of *together*. If not, instead of genuine inter-subjective *knowledge* – communicable, consensual, predictive and verifiable scientific knowledge – we will construct either purely *mental* representation, or just meaningless heaps of unintelligible signs, verbal, logical, mathematical heaps of signs that will generate in our minds only unease and passive, vile, idolatrous submission to illusory 'results'. In the classical physics we are protected from such a failure by the models that emerge spontaneously from the perceptions generated by our biological sensorial apparatuses (which still nowadays most genuine thinkers, implicitly, identify firmly with 'reality such as it truly is in itself').

---

[15] *Any* very first – primordial – registration of the result of a measurement interaction is 'transferred', even in the case of directly perceived entities like in the classical domain (private exchanges with Henri Boulouet). What is specific here is the fact that *no qualia can be formed in the observer's mind*. As soon as the studied entity accedes to some sort of direct perceptibility via some apparatuses (microscopes, etc., as it happened historically for molecules and atoms), this absence of qualia *ceases*. But this does *not* entail that the qualia that have been produced in this way can be confounded with 'intrinsic properties' of the studied entity. Any material entity is nowadays conceived to merge with the universal 'sub-quantic substance' so that it is devoid of delimiting contours. *Delimitation by some G is a human necessity in the processes of conceptualization*. Furthermore even the absence of spatial delimitation is just a *model* conceived by human mind, not some sort of *unconceivable* representation of a 'true property' of the studied entity such as 'it really is in itself'. *This Fata Morgana notion is* ***self-contradiction*** because *any* ***knowledge*** *is* ***qualification*** *and any qualification is relative to the apparatuses and the physical operations by which it is achieved*, as well as to the conceptual definition of the qualifying quantity.

[16] In MMS [2013] (pp. 117-126) I have constructed a "space-time coding" procedure that *identifies* – so labels a posteriori – the results of an arbitrarily constructed "test-interaction" *T* between a corresponding test-apparatus and the specimens $\sigma(ms_G)$ of a factually defined microstate $ms_G$, but *without endowing these results with any meaning that relates them to some previously achieved conceptualization*. Such a coding-procedure cannot signify in terms that possess some meaning in terms of pre-established conceptualization, so it cannot directly connect to the classical science. But – and this is noteworthy – it can initiate quintessentially new processes of conceptualization that, indirectly, via intuitive substrata, take profit from the already established conceptualization.

But when no direct sensorial perception of the entity-to-be-studied generates models any more this natural resort dissolves, and as long as an efficient model of the entity-to-be-studied is not constructed *conceptually* we are simply blocked in any action for deciding what sort of measurement-interaction can produce information on a definite qualifying concept *A*. And only some connection with a definite cognitive situation where direct perceptibility offers a foundation can suggest such a model; namely, a connection with observable data *and* with the previous classical conceptualization, because this is the unique domain of organized meaning that emerges spontaneously for us and so, that can be used by us as a first ground for starting to model, even if we start by changing this ground [17]. But on the other hand:

> Inside *IQM*, that is deliberately required to define with full generality the features of ***any*** acceptable theory of microstates, no particular model of a microstate can be given without perpetrating vicious circularity: *The coding problem cannot be* ***treate****d inside IQM.*

In the second part of this work we shall identify the model of a microstate that is acting inside the Hilbert-Dirac formulation of quantum mechanics and this model will play a fundamental role in the construction of a fully intelligible second quantum mechanics. But here we just already draw strongly attention upon the existence of the coding problem and upon the unavoidable necessity, in any *given* definite theory of the microstates, to posit some model of a microstate, while knowing that it is just a model and *not* 'intrinsic' factual truth.

> The conceptual situation brought into evidence above refutes the very *possibility* to obey Bohr's positivistic interdiction of any model of a microstate.

Which in its turn proves that in fact this interdiction has never been genuinely taken into account. It has only enormously intimidated the physicists and pushed them, as it will appear, into passive and abstruse acceptance of basic conceptual impossibilities.

---

[17] Notice that this is how nowadays quantum mechanics effectively proceeds for constructing mathematical representations of the qualifying quantities: Bohr's interdiction of models strikes only the entities-to-be-studied.

***(1.I).2.2.2. Graphic representation of one qualification of one specimen of a microstate***

The global content of (1.I).2 are summarized graphically below in the Fig.1.

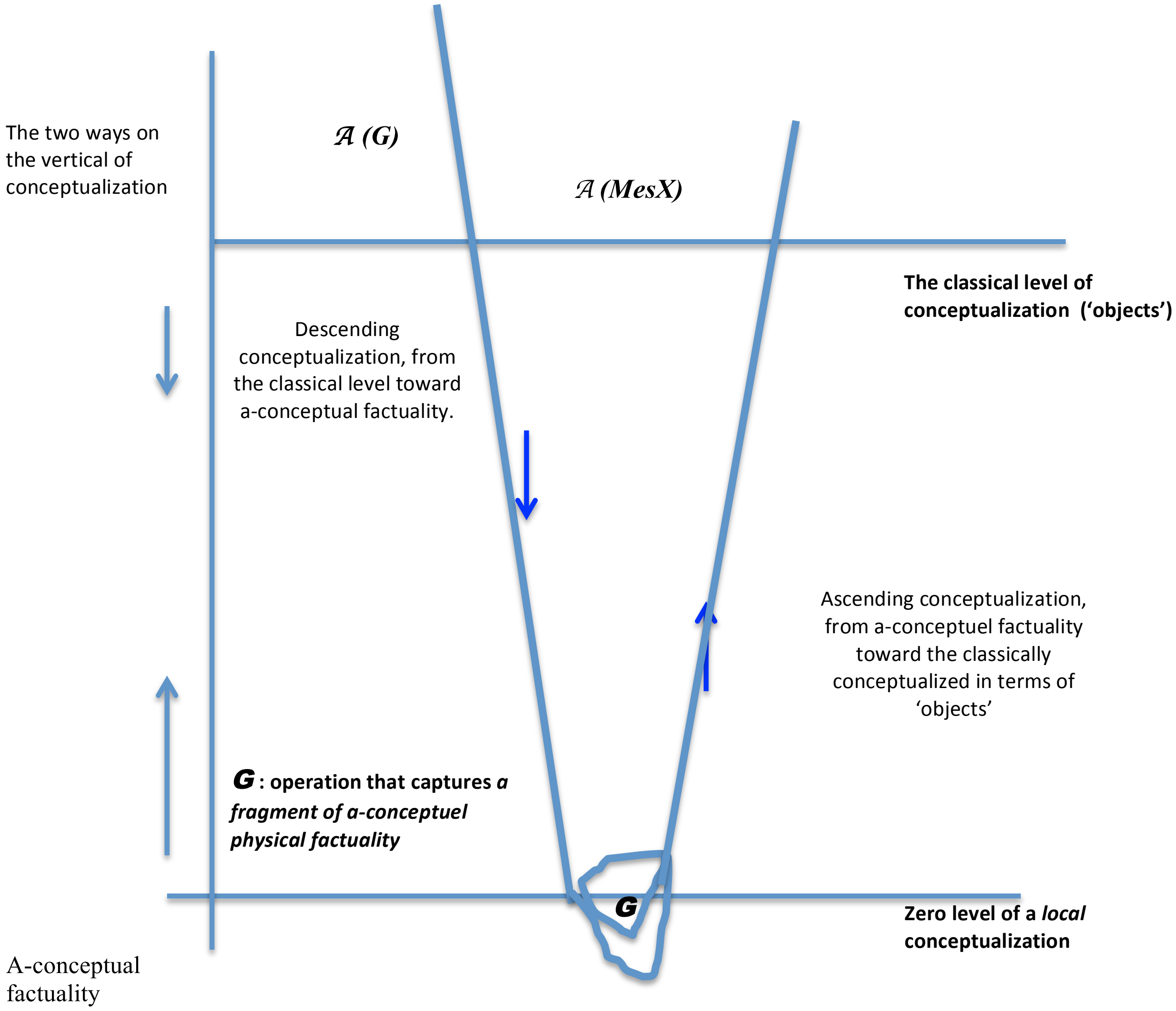

**Fig.1**. *One qualification of one* ***specimen*** *of a microstate: the germ of the structure of a primordial transferred description*

The apparatus for producing the operation of generation $G$ is denoted $\mathcal{A}_{pp}(G)$; the apparatus for producing the measurement interactions for the dynamical quantity $A$ is denoted $\mathcal{A}_{pp}(MesA)$. The basic operational construct that generates the result of only *one* act of measurement-interaction performed upon *one* outcome of one specimen of the microstate $ms_G$ defined in (1) can be represented as a chain:

$[(G\leftrightarrow ms_G)\text{-}[G.MesA]\text{-}\{\mu\}_{kA}$ *coded in terms of one* $a_j)]$, $\quad k_A=1,2,...m_A, \quad j=1,2....J \qquad (3)$

The chain (3) that brings forth just one act of qualification of one specimen $\sigma(ms_G)$ of a factually defined microstate $ms_G$ will be called *a (one) coding-measurement-succession*. It constitutes the very first germ of the factual constructive representation of the process of generation of knowledge on such a microstate. This germ is already endowed with a rather complex inner structure and it already specifies in what a sense the pairs

*[(one operation of generation of the entity-to-be-qualified),*
*(one operation of qualification of this entity)]*

play a basic role in the construction of consensual predictive and verifiable knowledge. A chain (3) acts like a fragile narrow bridge over the frontier between the a-conceptual universal physical substance of which the existence is posited by our minds, and the volume of human conceptualization.

In what follows this germ will be developed into a still far more complex concept, namely a general form of a full *scientific description* of a microstate, a deliberate, consensual, predictive and verifiable piece of stable knowledge on a microstate $ms_G$: *the primordial transferred description* of a factually defined microstate $ms_G$ in the sense of *MD*.

# 2.I

# BOTTOM-UP CONSTRUCTION OF THE TRANSFERRED DESCRIPTION OF A FACTUALLY DEFINED MICROSTATE

## (2.I).1. PRELIMINARY CONSTRUCTION OF LANGUAGE: DEFINITION OF 'MICRO-*SYSTEM*', 'MICRO-*STATE* $ms_G$', '*TYPES* OF MICRO-STATES $ms_G$'

### 2.I).1.1. The general problem

In our current life we begin by embedding structures of thought in structures of some current language that emerged and evolves collectively by an anonymous and spontaneous process. But from a scientific point of view the structures of thought expressed inside a current language are most often beds of Procustes because the aim of the natural languages is to be ***contextual*** in order to maximally permit rapid, allusive, suggestive, approximating transmissions of meaning, of poetic connotations, of humour, etc. The accent falls upon local and contextual efficiency in space and time, and upon the harmonics of the core-meaning. Whereas the aim of a scientific language is to induce *maximally strict and stable consensus* inside some definite group of consensus, via a priori definitions that point as precisely as possible toward a uniquely defined significance; which can be realized – nearly strictly – only via axiomatic constructions. The just mentioned two sorts of aims are opposite to one another. And quantum mechanics, like the majority of the mathematical theories of Physics, is not axiomatic, it is a mathematized representation imbedded in the natural language where one relies on contextual communication. This blurs the significance of many basic words that occur currently in the feebly defined verbal support of the quantum mechanical mathematical representations (to 'prepare' (the 'system', the 'state'); to 'measure'; 'superposition', etc.). Thereby much confusion is induced. In what follows we suppress beforehand the possibility of several such basic confusions.

### 2.I).1.2. The specific problem

Consider a measurement-interaction involving a specimen $\sigma(ms_G)$ generated by the operation $G$ that corresponds to the studied microstate $ms_G$. This produces observable marks that have to be translatable in terms of *one* value $a_j$ of ...... of *what*, exactly? Of one value $a_j$ of only *one* measured dynamical quantity $A$, for *any **sort*** of 'involved microstate', or possibly of *several* such quantities or values of quantities permitted for *some* sorts of microstates? Shall we organize our concepts-and-language so as to require that *one* act of measurement on only *one* specimen $\sigma(ms_G)$ of the studied microstate $ms_G$ brings forth *necessarily* only *one* value $a_j$ of each measured dynamical quantity $A$? Or that it shall necessarily involve – at most – only *one **set*** of 'compatible' quantities (which is not the same thing as in the preceding question)? And, in this case, what exactly does 'compatible' mean? What restrictions are we prepared to accept?

Furthermore, according to (1) *each specimen* of the one micro-*state* tied with one operation of generation $G$ can involve one or more other micro-entities (like when $G$ creates a pair). How can we name these micro-entities? If we call them 'particles' – as it is often done – we suggest a model, which we want to avoid inside *IQM*. Could we then speak of one, or two, or more micro-*systems* involved by each specimen of a given micro-state? This does not contradict the current way of speaking inside quantum mechanics. If then we do call micro-*systems* the component entities from one specimen of a given micro-state, how are we to count them, according to what observational criteria? What presuppositions have to be incorporated in order to stay in clear agreement with the concepts of a factually defined micro-state and of a specimen of it, in the sense of (1), as well as with the current ways of speaking and thinking that accompany the formal quantum mechanical writings?

The answers are not at all obvious. Inside the current languages the word "system" points usually toward a complex whole that contains 'components'. But inside quantum mechanics, on the contrary, the word "system" – "*the* system" – points often just toward *what is studied*, no matter whether it is posited to involve one or several components; moreover the term ‘micro-state’ indicates the dynamical characters of the *whole* studied entity, and the word 'system' points toward exclusively the constant characters of 'a particle'.

All these ways of using words are not severely regulated, while in what follows we want to stay rigorous in order to avoid false problems. So we define a language that stays in agreement with:

*(a)* The general fact that the concept of ‘dynamical state’ [18] designates a variable behaviour that involves an invariant material support (violating such a fundamental slope of natural human conceptualization would uselessly waste energy).

*(b) MD*, that introduces the basic posit (1) $G\leftrightarrow ms_G$, with $ms_G=\{\sigma(ms_G)\}$, according to which one operation of generation $G$ produces factually one 'specimen' $\sigma(ms_G)$ of the micro-state denoted $ms_G$; while the number of the involved ‘systems’ is *not* restricted by (1) because this concept is not involved by *MD*.

*(c)* The hidden consensual assumptions that can be identified inside the moving ways of speaking and writing practised inside quantum mechanics.

***Definition [(micro-state) and (micro-system).*** The concept delimited by the *persistent* characters (mass, charge, etc.) assigned to any element from the set $\{\sigma(ms_G)\}$ of mutually distinct specimens of the micro-*state* $ms_G$ in the sense of *MD* is called a micro-*system* involved by $ms_G$.

***Definition [(*one *micro-system) and (*one *micro-state of* one *micro-system)].*** Consider a micro-state $ms_G$ that is such that *one* act of measurement accomplished upon *one* specimen $\sigma(ms_G)$ *of* $ms_G$ can bring forth only *one* group $\{\mu\}_{kA}$ of observable marks significant in terms of a value of the measured quantity. We shall say that this micro-state $ms_G$ brings in specimens $\sigma(ms_G)$ each one of which consists of *one* micro-*system S* and so we shall call it in short *a micro-**state** of (with) **one** micro-**system***.

***Definition [*one *micro-state of* n *micro-systems].*** Consider now $n>1$ micro-*systems* of a type of which we know that, for each one of them *separately* it is *possible* to generate a micro-state in the sense of the preceding definition; which, if done, would lead to ‘***n*** *micro-states of **one** micro-system*’ in the sense of the preceding definition. But let $G(nS)$ ($nS$ : $n$ systems) denote only *one* operation of generation that, acting upon some physical initial support that relatively to $G(nS)$ is regarded as ‘prime matter’, has generated *one common* micro-*state* for *all* these $n$ micro-*systems*; or even, out of some initial substratum, $G(nS)$ has simultaneously generated the $n$ micro-*systems themselves* that are contained by each specimen of the studied common *one* micro-state[19, 20]. In both these cases we shall say that the micro-state generated by *G(nS)* is *a micro-state of (with) n micro-systems* and we shall denote it by $ms_{G(ns)}$[21].

[18] A somewhat self-contradicting expression.
[19] This is the case, for instance, when $G(nS)$ consists of some interaction with pre-existing elementary particles that brings forth ‘a pair’.
[20] This way of speaking seems convenient in both fundamental quantum mechanics and the fields-theories.

***Definition [complete measurement on* one *micro-state of* n *micro-systems].*** *One* act of measurement performed on one specimen $\sigma(ms_{G(ns)})$ of a microstate $ms_{G(ns)}$ of *n* micro-systems, *can produce at most n distinct groups of observable marks* signifying *n* observable values of dynamical quantities. An act of measurement that effectively realizes this maximal possibility will be called a *complete* act of measurement on the one specimen $\sigma(ms_{G(ns)})$ of the one micro-state $ms_{G(ns)}$ of *n* micro-systems. The quantities *A* and the values $a_j$ to which these *n* distinct groups of marks are tied, are permitted to be either identical or *different*.

***Definition [*****in*****complete measurement on* one *micro-state of* n *micro-systems].*** One act of measurement accomplished upon one specimen $\sigma(ms_{G(ns)})$ of a microstate $ms_{G(ns)}$ of *n* micro-systems that produces *less* than *n* distinct groups of observable marks, will be called an *incomplete* act of measurement on $ms_{G(ns)}$.

Finally, for self-sufficiency of this sequence of definitions, we restate here telegraphically the definition from 1.I of a micro-state $ms_{G(G1,G2,..Gk)}$ generated by a composed operation of generation:

***Definition [*****one** ***micro-*****state** ***generated by a*** **composed** ***operation of generation].*** Consider – indifferently – either *one* micro-state of *one* micro-system, or *one* micro-state of *n>1* micro-systems. If the specimens of this micro-state are generated by a composed operation of generation $G(G_1,G_2,..G_k)$ in the sense defined in 1.I then we call it *a microstate with composed operation of generation*.

***Definition [*****one** ***'bound' micro-*****state** ***of several micro-*****systems*].*** This is the usual verbal designation of the result of a *natural* operation of generation, i.e. accomplished in consequence of the physical laws of nature, before any human aim of investigation (like in the case of the natural realization of an atomic structure). But in principle it can be also thought of in terms of the result of a *composed* operation of generation (so much more so as a bound micro-state of several micro-systems manifests systematically 'interference-effects').

We hold that the preceding definitions insure, both, global coherence relatively to the implications carried by the language practised inside nowadays microphysics, and continuity with the basic principles of the classical conceptualization and language. If one contests the adequacy of some feature from these definitions, he should specify the reasons for the contestation and propose a better usage of words. Meanwhile the definitions from (2.I)1 are adopted throughout what follows.

We now announce the following

***Choice****. In this work the bound microstates will occupy a very marginal position.*

We make this choice on the basis of two reasons. The first one is that a bound state can pre-exist any desired investigation, just as it is supposed for classical 'objects'. The second reason is that furthermore, to a bound state it is possible to assign – in a certain relative sense of course – *a definite spatial delimitation*, again as in the case of a classical mobile. These two features might explain why the mathematical representation of bound microstates has constituted the natural passage from classical physics to quantum mechanics when the practised approach still was top-down. But in this work we want to explicate and stress the radical novelties imposed by a bottom-up representation of microstates. Only these novelties will permit to bring into evidence:

---

[21] The posit (1) entails that the uniqueness of the operation *G(nS)* is to be a priori conceived as a source of certain global observational specificities of each specimen of $ms_{G(ns)}$ and so of $ms_{G(ns)}$ itself.

- To what a degree the scientific representations can become a deliberate consensual construction of which the necessary and sufficient conditions of possibility depend strongly on the involved cognitive situation (that can evolve with the evolution of the sciences and the techniques).

- To what a degree this should modify our conception on scientific representation, and stress the utmost importance of the relativities to the constraints and the aims that act.

And these novelties are brought forth – specifically – by unbound microstates. So here we are quasi exclusively concerned with unbound microstates. The bound microstates will finally be naturally absorbed in the new representation constructed here.

On the basis of the contents from (2.I).1 we enter now upon the construction of the general concept of description of a microstate.

## (2.I).2. *PRIMORDIAL* TRANSFERRED DESCRIPTION OF AN UNBOUND MICROSTATE $ms_G$

What follows is formulated in terms that are valid for *any* microstate.

### (2.I).2.1. Preliminary requirements

We start again from the remark that inside current thinking and speaking the qualifications are in general just asserted freely concerning an object-for-qualification that is conceived to pre-exist *such* as we qualify it (this tree *is* big, today the air *is* cold, etc.); whereas a scientific description is required to be endowed with explicit consensual definitions that are *communicable* with precision and without restriction to co-presence of the members of a specified group of consensus, and to be predictive and verifiable. All these requirements *subsist* when it is recognized like in (1.I) that the qualifications that have been obtained cannot be considered to be properties of the entity-to-be-described alone, isolated from the measurement-interaction. And the requirement of verifiability entails *repeatability* of the involved operations as well as the existence of some definite descriptive *invariant* brought forth by many repetitions of the action of qualification: only such invariants can permit prediction and verification. Now, in the case of microstates these implications of the condition of scientificity entail specific and non-trivial consequences among which the following are the most important

#### *(2.I).2.1.1. Consequences of the requirement of repeatability*

A classical mobile is conceived as an "object" that in general pre-exists to qualifications of it; it stays available "there outside". So in general a measurement operation *MesA* on a classical "mobile" can be conceived separately from an operation of generation *G* of that mobile [22]. But an unbound microstate-to-be-studied *does **not** pre-exist in some known and attainable way*, like a macroscopic "object"; and furthermore in general it is destroyed by the act of qualification. So the observer-conceptor, if he wants to create a germ of knowledge on such a microstate, has to radically generate that microstate before achieving on it an act of qualification, so to realize a whole

[22] This, in fact, is confusion. Indeed – by definition – an operation of generation *G* in the sense of (1) is what brings an entity in the *role* of entity-to-be-studied. And a classical mobile that just is conceived to 'exist' is not thereby automatically in the role of entity-to-be-studied. Always some supplementary act is necessary from the part of the observer-conceptor, even if this consists of just bringing the mentioned mobile inside the domain of perceptibility by the observer-conceptor and focusing attention upon it. As already remarked, the existence of an operation of generation *G* is a universal character of any act of qualification, so of any act of creation of a piece of knowledge. This fact is far from being trivial: it is part of the hidden key that opens up access to a path toward unification of microphysics and quantum gravitation.

'measurement-succession' *[G.MesA]*: The explicit necessity steps in, to realize repeatedly and in a physical-operational way whole pairs

*[(one operation of generation of the entity-to-be-qualified),*
*(one operation of qualification of this entity)]*

And this, for scientific descriptions, entails an arm-wrestling between *IQM* and the classical presuppositions. Indeed:

In classical mechanics the studied mobile is admitted to be publicly observable, and the registration of the result of an act of measurement does not destroy the studied mobile, nor does it necessarily perturb notably its dynamical state. So it has been possible to conceive and to formulate a basic classical mechanical law as an *individual* invariant with respect to repetitions of an act of measurement *MesA*. Furthermore such a law is posited to characterize exclusively the studied entity *itself*, it is regarded as the revelation of a behavioural 'property' of, exclusively, the studied mobile; a classical mechanical law it is not explicitly referred to the whole of the measurement interaction. When the results of repeated measurements on the studied mobile manifest a statistical dispersion this is posited to be due exclusively to imprecisions in the acts of measurement, which withstands the *knowledge* of the exact individual value $a_j$ of the measured quantity $A$ that is "possessed" by the entity-to-be-studied, but does not concern the existence of this value. According to the classical thinking this obstacle on the way toward knowledge, however, is doomed to disappear asymptotically while progress is achieved concerning the techniques of measurement; so one advances toward knowledge of how the studied physical entities "truly are, exactly and in themselves".

This sort of illusory scientific realism is quasi unanimous.

Whereas the factually defined concept of microstate $ms_G$ from *MD* is organically tied with a conceptual segregation of a radically different nature that we recall synthetically:

***(a)*** Since the unavoidably physical operation of generation $G$ can be defined by only a finite set of parameters while the domain of physical reality from which this operation stems, as well as that on which it acts, are endowed with the unlimited singularity of the being, it would be *unconceivable* that repetitions of $G$ bring always forth specimens $\sigma(ms_G)$ of the studied microstate $ms_G$ that are all mutually identical, i.e. the posit $ms_G \equiv \{\sigma(ms_G)\}$ where $\{\sigma(ms_G)\}$ is a set of mutually distinct specimens, is quintessential for microstates; it introduces a basic 'statisticity' that is *not* asserted as a physical truth, but only as an unavoidable operational fact involved by a deliberate human action of construction of consensual, predictive and verifiable knowledge on microstates (cf. (1)).

***(b)*** Since one act of measurement *MesA* also cannot be defined otherwise than by a finite set of macroscopically specified parameters, when it is repeated its own effects equally cannot be conceived otherwise than dispersed, in general [23].

*(c)* The specimens $\sigma(ms_G)$ are not observable, while the observable result of one succession *[G.MesA]* (cf. figure 1) characterizes exclusively this succession *as a whole* in a way that cannot be analysed further.

***(d)*** So repetitions of the *whole* succession *[G.MesA]* are unavoidable, and these lead in general to a statistical distribution of the observable results of the achieved successions that – quintessentially – cannot be removed nor analysed (MMS [2002B], [2006], [2017B]).

*(e)* So:

---

[23] When a unity is defined it sets a conventional lower bound to the dispersion that is taken into account. The nanotechnologies might reduce strongly the dispersion of certain specifically targeted observable effects.

> The researched *law*-like invariant – a concept that is 'deterministic' by definition – can manifest itself observably *only* in terms of *probabilistic convergence of repeated statistical distributions of results of sequences of very numerous repetitions of the whole succession [G.MesA].*

Let us remind now that:

- According to the classical conceptual segregation, the entity-to-be-studied is defined conceptually via individually specifying predicates [24] and it obeys individually specifying laws tied with a posited individual invariance of principle of the results of measurements that is only imperfectly observable because of imprecisions of measurement.

- This classical conceptual segregation ***cannot be transposed to factually defined microstates*** $ms_G$ ***in the sense of MD***, because the existence of an operation *G* of generation of the entity-to-be-studied at the beginning of any process of generation of consensual knowledge, has a universal character ((1.I).1.6), and because in consequence of *(a)* and in terms of the general framework (*G, MesA, [G.MesA]*) this classical segregation holds only ***locally*** inside the set of all the various processes of conceptualization brought in by all the various cognitive situations that can confront a human scientist. Namely:

> The classical segregation holds only in the cases in which the considerations from *(a)* fade out because the operation *G* introduces a dispersion that is negligible in some sense (for instance, for the majority of the macroscopic directly perceived "objects" (MMS [2002B], [2006]) or for the purely *mental* conceptual-mathematical representations of celestial entities-to-be-studied (black holes, galaxies), introduced by a purely mental operation of generation *G* and that can be confirmed or invalidated by – exclusively – verification of consensually observable predictions that have been drawn *deductively* from these representations)[25].

But for consensual predictive and verifiable knowledge on microstates all the requirements *(a),(b),(c),(d),(e)* do hold significantly and the classical segregation breaks down. So let us examine the consequences that these requirements impose upon the existence of a descriptional invariant.

### *(2.I).2.1.2. A consensual, observable, predictive and verifiable descriptional invariant concerning microstates*

Consider now the constraint of existence of some descriptional invariant with respect to repetitions of successions *[G.MesA]*. In general when one given succession *[G.MesA]* is repeated one obtains *different* results $a_j$. So in general a whole statistic of results $\{a_j\}$, $j=1,2,...J$ emerges, notwithstanding that in each succession *[G.MesA]* each one of the two operations *'G'* and *'MesA'* is 'the same' *with respect to the two finite groups of parameters that define it.* This is a fact. We are placed on an observational ground that – with respect to knowledge – has a primordially statistical character. Whereas on the other hand any 'law' that permit predictions and verification of these, is an invariant with respect to repetition. So the unique possible sort of observational invariant consists of a primordially *probabilistic* invariant of the statistical distributions of the possible results $a_j$ of realizations of the succession *[G.MesA]*. Which involves [a big set of [$N$ repetitions of the succession *[G.MesA]* with $N$ very big]] *and the concept of 'probabilistic convergence' of these statistical distributions* introduced by the classical theory of probabilities, that in *MD* shifts us upon a postulated level of ***'deterministically** probabilistic' conceptualization expressed by the one-one relation (1)* $G \leftrightarrow ms_G$ *with* $ms_G = \{\sigma(ms_G)\}$. It might seem counter-intuitive to assert that a probabilistic qualification is a deterministic qualification, but – ***globally*** – it *is* a deterministic qualification, in this sense that the recurrence of the convergence is predictable.

---

[24] A sort of definition assisted by direct perception.

[25] The de Broglie-Bohm formal representation of the Universal Substance introduces a limiting conceptual situation: *both G* and *MesA* are simply absent – basically – and so there is no source of observational dispersion any more, we are in presence of just a global and mathematically expressed metaphysical model that remains to be explicitly connected to this or that local consensual, predictive and verifiable knowledge that – necessarily – involves a superposed specification of local *factual* successions *(G,MesA)* and repetitions of these.

So we consider now this concept of probabilistic convergence.

It is a *non*-effective abstract mathematical concept embodied in the weak law of large numbers. But here we have chosen to develop from the start a strictly effective approach (cf. the introduction to Part I). So we have to specify an *effective* equivalent of the classical concept of probability. For this we proceed as follows.

Consider the weak theorem of large numbers

$$\forall j, \ \forall(\varepsilon,\delta), \quad (\exists N_0 : \ \forall(N \geq N_0)) \Rightarrow [\boldsymbol{\Pi} \, [|n(e_j)/N - \pi(e_j)| \leq \varepsilon]] \ \geq (1-\delta) \qquad (4)$$

(The significance of the notations is well known). From this it is possible to extract explicitly a relativized finite implication that is defined below: The probability $\pi$ and the meta-probability $\boldsymbol{\Pi}$ are ***limit***-(real numbers) toward which, at infinity, converge the corresponding distributions of relative frequencies. Consider a universe of events $U=[e_1,e_2,....e_J]$, $j=1,2,...J$, with $J$ a finite integer. *If* the probability $\pi(e_j)$ of an event $e_j$ is postulated to exist for any $e_j$, *then* (4) asserts that for any pair of two arbitrarily small real numbers $(\varepsilon,\delta)$ there exists an integer $N_o$ such that – for any $N \geq N_0$ and with an uncertainty not bigger than $\delta$ – the meta-probability $\boldsymbol{\Pi}$ of the event $[|n(e_j)/N - \pi(e_j)|) \leq \varepsilon]$ that the relative frequency $n(e_j)/N$ observed for the event $e_j$ inside a sequence of $N$ events from $U$ does not differ from $\pi(e_j)$ by more than $\varepsilon$, is bigger than $(1-\delta)$. This assertion itself, such as it stands, i.e. the passage to the limit being *suppressed* – with $N_0$ chosen freely and with the corresponding pair $(\varepsilon,\delta)$ – will be considered in what follows to define a general and ***factual***, *finite* concept of probability of the event $e_j$, namely *the* $(\varepsilon,\delta,N_0)$*-probability* $\pi(e_j)$ *of the event* $e_j$ that will be called the factual probability law of $e_j$ with respect to the triad $(\varepsilon,\delta,N_0)$ [26].

In our case $U$ consists of the finite spectrum of values $a_j$ assigned to $A$. And we make the *strong* assumption that the systematic repetition, for any $A$, of the corresponding succession *[G.MesA]*, introduces sufficient constraints for entailing a factual $(\varepsilon,\delta,N_0)$-probability law $\pi(a_j)$ for any association between a chosen pair $(\varepsilon,\delta)$ and the relative frequency $n(a_j)/N$ found for a value $a_j$ that is present inside the chosen qualification grid (2) $gq[A,a_j,MesA,cod.proc(a_j)]$ [27], with $j=1,2,...J$. Which amounts to a – conceptual – verification of the posit (1) $ms_G \leftrightarrow G$. So:

Given a definite factually defined microstate $ms_G$, the posit (1) introduces for any couple of pairs $((G,A),(\varepsilon,\delta))$ a corresponding 'factual $(\varepsilon,\delta,N_0)$-probability law'

$$(\varepsilon,\delta,N_0)\text{-}\{\pi(a_j), \ \forall j\}_G, \quad A \text{ fixed} \qquad (5)$$

[26] In (MMS [2014B]) this *factual* probability law has been constructed from an *interpretive assumption* on the concept of probability and *it has been proved **compatible** with the weak theorem of large numbers* (cf. also (Wasserstein&Lazar [2016], Leek&Penn, [2015] concerning the conceptual status of – merely – a *statistic*, with respect to the conceptual status of a *probability law*).

[27] The event $a_j$ being identified from a group of observable physical marks, via the utilized coding-procedure that inside *IQM* cannot be defined but that is supposed to have been defined inside the employed theory of microstates.

***(2.I).2.1.3. Compatibility of quantities versus specificity of the 'knowledge' on a microstate***

The initial factual and methodological definition (1) of the microstate-to-be-studied amounts to merely label this unknown and unobservable microstate '$ms_G$' by the operation $G$ that is supposed to have produced it; the final purpose is to substitute to the mere label '$G$', a 'description' of the microstate-to-be-studied in terms of predictive and verifiable knowledge tied with – specifically – this entity *itself*. Now, does a factual probability law (5) constitute such knowledge?

No, not yet, because nothing entails that only one probability law (5) established for $ms_G$ relatively to only one dynamical quantity $A$, cannot be observed also for *another* microstate different from $ms_G$, i.e. generated by another operation of generation $G'\neq G$. The law (5) alone might not be specific of $ms_G$.

It seems likely however that two probability laws (5) corresponding to two mutually different dynamical quantities $A$ and $A'\neq A$ – considered conjointly – might already constitute an observational factual specificity associable to the considered particular microstate $ms_G$ generated by $G$. While a fortiori – in as far as the language introduced in *MD1* resists to the observable facts – *all* the mutually different laws (5) that are defined for $ms_G$ are certainly specific of this microstate.

But what sort of difference between two dynamical quantities $A$ and $A'\neq A$ is determining in this context?

Consider two distinct dynamical quantities $A$ and $A'\neq A$ and a given type of microstate $ms_G$, in the sense of the definitions from (2.I)1.

> We shall say that $A$ and $A'\neq A$ are mutually compatible with respect to the microstate-to-be-sudied iff it is possible to measure them simultaneously *on one specimen* of this microstate.

Suppose then that the microstate-to-be-studied $ms_G$ is a microstate of *one* micro-system. In this case each specimen of $ms_G$ consists of only one system and – with respect to $ms_G$ – the requirement posited above amounts to the possibility to achieve for both $A$ and $A'$ a physically unique common measurement-interaction upon a specimen that consists of this system. So the common interaction has to cover a unique *common* space-time support and to finish by the registration of a unique group $\{\mu\}_{(AA'),k}$, $k=1,2,...m_{AA'}$ of brute observable marks. Then in this case a 'difference' between $A$ and $A'$ can be worked out only *after* the realization of this unique common physical-operational interaction, by exclusively conceptual definitions and calculi that construct two conceptually distinct values $a_j$ and $a_j'$ to be assigned, respectively, to $A$ and to $A'\neq A$**[28]**. If the condition required above can be realized we shall say that $A$ and $A'$ are *mutually compatible quantities with respect to a microstate of one microsystem*; if this cannot be realized we shall say that $A$ and $A'$ are *mutually incompatible quantities with respect to a microstate of one microsystem*. In the first case the two factual probability laws (5) constructed for $A$ and $A'$ introduce a poorer factual constraint than in the second case. So the corresponding knowledge is less specific and a maximally specific knowledge on the studied microstate is obtained by establishing the probabilistic behaviour of this microstate with respect to *all* the groups of mutually in-compatible dynamical quantities that are defined for the studied microstate.

But suppose now that $ms_G$ is a microstate of two (or more) microsystems. In this case two (or more) mutually distinct measurement-interactions can be accomplished on different systems from a unique specimen $\sigma(ms_G)$ of the studied microstate $ms_G$ (cf. the definitions from (2.I)1 and also the future point (3.I)2). So in this case we shall say that *any two different quantities A and to A'≠A* ***can*** *be compatible with respect to a microstate of two or more microsystems*. And the maximally specific knowledge on the studied microstate is obtained by establishing its probabilistic behaviour with respect to *all* the dynamical quantities that are defined for it.

---

[28] This happens, for instance, for the classical quantities $\boldsymbol{p}$ and $p^2/2m=T$ for which it is possible to first determine in a physical-operational way the numerical value of the common basic quantity $|p|=m(v_x+v_y+v_z)$, and out of this basic operational determination, to work out afterward, conceptually, the two results '$\boldsymbol{p}$' (a vector) and '$p^2/2m$' (a scalar) that are mutually distinct from a conceptual point of view as well as by their numerical values).

This settles the question of specificity with respect to the studied microstate, of the knowledge on this microstate captured in a factual probability law (5).

Let us note that: The concept of compatibility of dynamical quantities that has been defined here in connection with the question of specificity of the knowledge created concerning the studied microstate, is essentially ***relative*** to:

- the concept of one ***individual*** specimen of the studied microstate;
- the *sort* of considered microstate, in the sense of the definitions from (2.I)1.
- the *coding procedure* that is involved, so also the *model* of a microstate that is presupposed in the theory that is made use of;
- the available techniques for measuring, which in general *vary* while time passes.

This conclusion is striking when it is compared to the concept of compatibility of qualifying quantities defined in the nowadays Hilbert-Dirac formulation of the quantum mechanics [29].

### (2.I).2.2. Primordial description of a microstate.

The considerations from the preceding point lead us to posit by definition that – even though the laws (5) do *not* concern *exclusively* the studied microstate $ms_G$ itself, i.e. separately from the measurement interactions from the successions *[G.MesA]*, $\forall A$ that led to them – nevertheless:

The set

$$\{(\varepsilon,\delta,N_0)\text{-}\{(\pi(a_j)\},\ \forall j)\}_G\},\ \ \forall A \quad (5')$$

[30]

of *all* the factual $(\varepsilon,\delta,N_0)$-statistical-probabilistic laws (5) established with respect to *one* given operation of generation *G* and *all* the dynamical quantities *A* defined for a microstate, can be regarded as a *mechanical description 'of* $ms_G$*'*. Indeed, it is the maximally specifying characterization that can be realized for the considered microstate $ms_G$ in the sense of *MD1*, and it is a characterization that is specifying with a strong degree of certainty. So, to the initial definition (1) of the microstate $ms_G$ that only labels this microstate by the operation *G* that generates it, and then, to one chain (3) that endows us with a very first unstable dot of qualification tied with this microstate itself, (5') substitutes finally:

- a characterization of $ms_G$ in terms of a whole *stable* and dense structure of communicable, consensual, predictive and verifiable pieces of observable factually-probabilistic data,
- that exhausts the defined possibilities to qualify this microstate,
- and that are all tied with this particular microstate *itself*, with effects of ***its*** interactions with measurement procedures.

Moreover, via the coding-procedures *cod.proc*$(a_j)$, $\forall A$, posited to be necessarily involved by the definitions of the measurement interactions *MesA*, $\forall A$, from the theory of microstates that is employed, the information contained in (5') is intelligible in this sense that it is connected to the already previously constructed classical mechanics.

---

[29] In the nowadays quantum mechanics the concepts of mutual compatibility or incompatibility of dynamical quantities are given a *statistical* definition that *do not* ***reach*** *the level of* ***individual*** *conceptualization*, and they are uncritically assigned an *absolute, intrinsic nature* embodied in a posited algebra of operators. The correlative 'principle of complementarity' has instilled many considerations devoid of any clear and intelligible feature of factual or logical necessity.

[30] From now on, for the sake of simplicity, for a usual repetitive index like '*j*' in $a_j$ we shall write $\forall j$ instead of $j=1,2,....J$, keeping in mind that the cardinal *J* is finite.

So (5') finally installs the concept of a microstate $ms_G$ as a scientific concept that is endowed with a definite, stable and specific, intelligible 'own' content.

Nevertheless the sort of knowledge represented in (5') violates strongly the current classical ways of thinking in terms of "objects" that – as delimited wholes – are endowed with a delimited and stable global space-time location entailing a definite inside and a corresponding outside, as well as an *inner* organization conceived in terms of *properties* that these objects would *possess*.

Moreover the genesis and the content assigned to (5') violates surreptitiously but radically the clear-cut conventional views on 'objective' facts. The set of relativities that mark (5') concerns characters of the human observer-conceptor (his ways of conceiving, thinking and acting and his technical possibilities) at least as much as it concerns the studied microstate. One is led to speak now much more cautiously, namely in terms of only inter-subjective consensus on predictions and verifications of outcomes of human methodological ways of operating. Thereby the classical notion of knowledge of some 'thing', recedes.

#### *(2.I).2.2.1. Notations, denominations, comments*

Let us now immediately organize and denote in detail the new sort of knowledge involved by (5'). In order to deal efficiently with all the unusual descriptional elements introduced here we shall now improve and summarize the names and notations associated with this knowledge[31].

- The grid of qualification introduced by a dynamical quantity *A* defined for microstates will be called *the aspect-view A*. The definition of each aspect-view *A* is assumed to contain the explicit specification of a coding-rule, in order to compensate the absence of direct perceptibility and of qualia assignable to the studied microstate itself. This is what insures a way to associate a meaning in terms of a definite value $a_j$ of *A,* to the group of brute observable marks $\{\mu\}_{kA},\ k_A=1,2,...m_A$ produced by one act of measurement-interaction from a succession *[G.MesA]*.

- The whole set of all the dynamical quantities defined for a microstate will be called *the mechanical view defined for a microstate* : $\{A\}\approx V_M$ ('$_M$': mechanical)

- A pair *(G,A)* that founds the operational succession *[G.MesA]* is called an *epistemic referential*; the pair $(G,V_M)$ is called *the mechanical epistemic referential.*

- A triad

$$(G,\ ms_G,\ A) \qquad (6)$$

of the basic genetic elements from (5') will be called a *genetic triad of (5')*. It can be regarded like a sort of inorganic physical-conceptual string of *DNA*.

- The whole set

$$\{[G.MesA]\},\quad \forall A\in V_M \qquad (7)$$

of repeated successions of operations of the general form *[G.MesA]* achieved by the use of all the genetic triads (6) realized inside the process (5') will be called the *genesis* of (5').

[31] These insert *IQM* explicitly in the general Method of Relativized Conceptualization, *MRC*, both conceptually and verbally.

- The brute result of the genesis $\{[G.MesA]\}$, $\forall A \in V_M$ of (5') consists exclusively of the set-of-sets of observable marks

$$\{\{\mu_{kA}\},\ k_A=1,2,...m_A,\ \forall A \in V_M\}\} \qquad (8)$$

These are the factual data produced by (5').

The totality (8) of all the factual data emerges at very dispersed moments, and also very dispersed spatially, on various registering devices of possibly various apparatuses. Observationally, this totality consists of just heaps of traces of vanished interactions, transmuted into meaning by a man-made operational-conceptual-methodological machine[32]. These heaps of traces however hide inside them a very elaborate unity of *human* curiosity, project and method. In a still non-expressed way, the factual data from (8) are already marked in their inner content by all the organizing relativities that inside (5') have been endowed with an explicit, intelligible and consensual final expression via the use of some definite model of a microstate. Nevertheless the factual data from (8) and their explicitly meaningful final expression (5') are devoid of any own space-time organization, as well as of any qualia assignable to the studied microstate $ms_G$ alone. This, of course, is a striking feature of *any* probabilistic description. But here, in consequence of total non-perceptibility of the entity to be studied, it acquires a limiting degree of purity.

The definitions (5) and (5') of the primordial probabilistic predictive laws concerning $ms_G$ – separated from their geneses (7) – will be *re*-noted now, respectively, as:

$$(D/A)(ms_G) \equiv \{(\varepsilon,\delta,N_0)\text{-}\pi(a_j),\ \forall j\}_G,\quad A \text{ fixed} \qquad (9)$$

$$D_M(ms_G) \equiv \{\{(\varepsilon,\delta,N_0)\text{-}\pi(a_j),\ \forall j\}_G,\quad \forall A \in V_M \qquad (9')$$

The notation $(D/A)(ms_G)$ from (9) will be called *the primordial transferred description of the microstate* $ms_G$ *with respect to the mechanical qualification A* (a description entirely 'transferred' on registering devices of apparatuses). It is the basic concept of transferred description.

The notation $D_M(ms_G)$ from (9') will be called *the primordial transferred mechanical description of the microstate* $ms_G$.

The writings

$$(D/A)(G,ms_G, A) \qquad \text{or} \qquad D_M(G,ms_G, V_M) \qquad (10)$$

can replace the expressions from the first members from (9) and (9'), respectively, when one wants to recall the geneses of, respectively, the laws from (9) and (9'): They stress that in the case of microstates the gained knowledge and the conceptual-physical-operational generation of this knowledge by the human observer-conceptor, constitute an intimate unity wherefrom the intelligibility stems.

Considered globally, this whole point *(2.I).2.2.1* is an application of the general Method of Relativized conceptualization *MRC* (MMS [2002A], [2002B], [2006]).

[32] Let us stop a moment to realize how simplistic it would be to assert that this knowledge pre-existed and has been 'discovered', when so obviously it has been invented and constructed.

# 3.I

# THE PROBABILITY TREE OF THE PRIMORDIAL TRANSFERRED DESCRIPTION OF AN UN-BOUND MICROSTATE

A primordial transferred description is a radically basic and new cognitive concept. But this, in spite of all the specifications and comments from the section (2.I)2, , still remains too abstract for triggering an intuitive and sufficiently detailed as well as integrated perception of the whole novelty of the concept of a primordial transferred description. Therefore we shall now construct graphic representations of the contents carried by the written representations [(1)→(10)]. We shall do this only for the two main sorts of unbound microstates defined in (2.I)1, namely a microstate of one micro-system and a microstate of two (or several) micro-systems. This will suffice for bringing forth that this concept involves a genuine revolution of classical probabilities and logic [33].

## (3.I).1. THE PROBABILITY TREE OF AN UNBOUND MICRO-*STATE* OF *ONE* MICRO-*SYSTEM* WITH NON-COMPOSED OPERATION *G* OF GENERATION

Throughout what follows we distinguish clearly between distinct levels of conceptualization.

We begin with the basic case of one unbound microstate of one micro-system. For this case we shall be able already to reveal non-classical specificities involved by (9) and (9').

### (3.I).1.1. Individual level of conceptualization

By definition the very numerous successions of operations *[G.MesA]*, $\forall A \in V_{Mec}$ involved in a genesis (7) start all with one same operational realization of a 'trunk'-operation of generation *G*. But afterward – in consequence of *individual* and *relative* compatibilities and incompatibilities between dynamical quantities in the sense defined in (2.I)2 – the set of all the individual space-time supports of these successions of operations *[G.MesA]* falls apart, in general, in distinct space-time genetic 'branches'. So in general there emerges a tree-like graphic structure. For simplicity we presuppose here only two non-compatible quantities *A* and *B*. The generalization is obvious.

The two mutually incompatible dynamical quantities *A* and *B* introduce respectively the two grids of qualification of form (2)

$$gq[A, a_k, MesA, cod.proc(a_k)],\ j=1,2,....M; \quad gq[B, b_r, MesB, cod.proc(b_r)],\ r=1,2,....M \qquad (2')$$

For simplicity we have endowed them with the same number *M* of possible values $a_j$ and $b_r$, respectively, and accordingly to the note attached to (5') we shall write only $\forall j$ or $\forall r$.

Let $[d_G.(t_G\text{-}t_o)]$ denote the invariant space-time support of each one realization of the operation *G* of generation of the studied microstate $ms_G$; this plays the role of a common 'rooting' into the microphysical factuality. Let $[d_A.(t_{MesA}\text{-}t_G)]$ and $[d_B.(t_{MesB}\text{-}t_G)]$, respectively, denote the mutually distinct space-time supports of a measurement-operation *MesA* and a measurement-operation *MesB*, the time origin being re-set on zero after each time-registration (obvious significance of the notations). So each realization of one whole succession *[G.MesA]* covers a same global space-time support

$$[d_G.(t_G\text{-}t_o)+d_A.(t_{MesA}\text{-}t_G)]$$

and it produces a group of observable marks $\{\mu_{kA}\}_j$, $k_A=1,2,...m_A$, $\forall j$, that is coded in terms of a value $a_j$ accordingly to (2'); while each realization of a succession *[G.MesB]* covers another same global space-time support

$$[d_G.(t_G\text{-}t_o)+d_B.(t_{MesB}\text{-}t_G)]$$

and produces a group of observable marks $\{\mu_{kB}\}_r$, $k_B=1,2,...m_B$ that is coded in terms of a value $b_r$ of the quantity *B*.

---

[33] Inside *MRC* it appears that this revolution reaches and incorporates also Shannon's theory of information and the representations of complexity.

Thereby for the considered case the genesis (7) from the level of *individual* conceptualisation involved by the representation (9), is achieved. This individual phase has a dominant *physical*-operational character.

### (3.I).1.2. Probabilistic level of conceptualization

Let us now start from the final result of the phase of individual conceptualization: values $a_j$ of *A*. The coding values $a_j$ are stored. Mutatis mutandis, the same holds for a succession *[G.MesB]*.

Suppose now that a sequence of a very big number *N* of realizations of a succession $[G.MesA]_n$, $n=1,2,....N$, has been realized. The relative frequencies $n(a_j)/N$, $\forall j$ (where the symbol $n(a_j)$ is to be read 'the number *n* of values $a_j$ of *A*') have been established *and* by global repetitions of the whole process an $(\varepsilon,\delta,N_0)$-convergence in the sense of (5) has been found to emerge indeed for these relative frequencies. In these conditions the primordial transferred description (9) has been factually specified *fully*, operationally *and* numerically. Furthermore on the top of the branch we have *effectively* constructed for the pair *(G,A)* a relativized Kolmogorov-like factual $(\varepsilon,\delta,N_0)$-probability-space. The universe of elementary events from this probability space is $U=\{a_j\}$, $\forall j$, and the probability law from this space is the primordial transferred description relatively to *A*, (9) $(D/A)(ms_G)\equiv\{(\varepsilon,\delta,N_0)\text{-}\pi(a_j),\ \forall j\}_G$, *A* fixed, (we do not yet consider explicitly the algebra on the universe of elementary events).

Mutatis mutandis, the same holds for the quantity *B* and its values $b_r$.

Thereby the primordial transferred description (9) relatively to *B*, $(D/B)(ms_G)\equiv\{(\varepsilon,\delta,N_0)\text{-}\ \pi(b_r)\}_G$, $\forall r$ is also effectively constructed.

So we have the transferred description (9') for the considered case: Out of the brute observable factual data $\{\mu_{kA}\}_j$, $k_A=1,2,...m_A$, $\forall j$, and marks $\{\mu_{kB}\}_r$, $k_B=1,2,...m_B$ we have worked out factually for the qualifying quantities *A* and *B* a purely *numerical* probabilistic content, via individual genetic, physical-operational actions (7). So when this second level of conceptualization is also achieved, the probability laws obtained on it – considered *separately* from their geneses [34] – possess a purely abstract *mathematical* character [35].

[34] We stress this because inside quantum mechanics the asserted probability laws *are* indeed considered separately from the corresponding probability spaces, so in particular separately from the universe of elementary events that generate these laws. Furthermore *they are **not** defined factually for the purpose of prediction, their factual (re)production serves exclusively the purpose of verification of the predictive statistics*. This circumstance deserves being noted immediately and kept in mind because it plays a major role in the parts II and III of this work.

[35] Notice how, out of the qualitative and physical operational approach practised here, the factual $(\varepsilon,\delta,N_0)$-probability laws induce spontaneously a promontory into the realm of the mathematized, because they express exclusively the results of effective counting.

### (3.I).1.3. A meta-probabilistic level of conceptualization

But the geometric representation from the Fig.2 does not permit to stop here, it pushes further. Indeed the striking awareness of the role of the *unique* operation $G$ of generation of the specimens of the studied microstate $ms_G$ from *both* branches hinders to stop because it strongly stresses that the two different effective probability laws $(D/A)(ms_G)\equiv\{(\varepsilon,\delta,N_0)\text{-}\pi(a_j), \forall j\}_G$, with $A$ fixed, and $(D/B)(ms_G)\equiv\{(\varepsilon,\delta,N_0)\text{-}\pi(b_r), \forall r\}_G$, with $B$ fixed, that crown the space-time branches from the zone of individual conceptualization stem both from one same trunk-operation of generation $G$, i.e. they concern one same microstate $ms_G$. So it seems unavoidable to *posit* that there exists some sort of meta-probabilistic correlation between these two probability laws $\{(\varepsilon,\delta,N_0)\text{-}\pi(a_j), \forall j\}_G$, and $\{(\varepsilon,\delta,N_0)\text{-}\pi(b_r), \forall r\}_G$. Such a correlation accepts an expression of the general form

$$\pi(a_j)=\mathbf{F}_{aj,B}\{\pi(b_r), \forall r\}_G, \qquad \forall A,\ \forall B \tag{11}$$

$$\mathbf{F}_{AB}(G)=\{\mathbf{F}_{aj,B}\{\pi(a_j), \forall j\}_G, \qquad \forall AB\in V_M \tag{11'}$$

where $\mathbf{F}_{aj,B}\{\pi(b_r), \forall r\}_G$ and $\mathbf{F}_{AB}(G)$ are two functionals that represent, respectively, the *individual* probability $\pi(a_j)$ in terms of the *whole* probability law $\{(\varepsilon,\delta,N_0)\text{-}\pi(b_r), \forall r\}_G$, and the global correlation between the two whole laws $\{(\varepsilon,\delta,N_0)\text{-}\pi(a_j), \forall j\}_G$, and $\{(\varepsilon,\delta,N_0)\text{-}\pi(b_r), \forall r\}_G$. Together the relations (11) and (11') will be called *the meta-probabilistic correlations involved by (1)* $G\leftrightarrow ms_G$ with respect to $(A,B)$ and will be symbolized by $(M\pi c(G))_{AB}$ ($M\pi c$: 'meta-probabilistic correlation') [36]. So the description (9') of the studied microstate has to be explicitly completed:

$$D_M(ms_G)\equiv\{[(\varepsilon,\delta,N_0)\text{-}\ \pi(a_j), \forall j\}_G,\ (M\pi c(G))_{AB}\,]\},\quad \forall A,\ \ \forall AB\in V_M, \tag{9''}$$ [37]

(in $(M\pi c(G))_{AB}$ the indexes $j$ and $r$ remain implicit).

In order to distinguish clearly between the factual probability-laws $\{(\varepsilon,\delta,N_0)\text{-}\pi(a_j), \forall j\}_G$, $A$ fixed, from (9), (9') and the meta-probabilistic correlations $(M\pi c(G))_{AB}$, $\forall AB\in V_M$ defined by (11), (11'), we shall say by definition that (9), (9') contain exclusively *probabilistic qualifications of the first order* whereas $(M\pi c(G))_{AB}, \forall AB\in V_M$ from (9'') expresses also *probabilistic qualifications of the second order* [38, 39].

---

[36] The two functionals $\mathbf{F}_{aj,B}\{\pi(b_r), \forall r\}_G$, $\forall A, \forall B$ and $\mathbf{F}_{AB}(G)$ can acquire a precise numerical definition only inside a theory of microstates where are specified the general model posited for a microstate and the corresponding general concept of an act of measurement $MesA$, $\forall A$, with the involved coding procedure.

[37] Mackey [1963], Suppes [1966], Gudder [1976], Beltrametti [1991], and probably quite a number of other authors also, have tried – directly by purely mathematical means – to establish a satisfactory formulation of a meta-probability law associable with a quantum mechanical state-vector. The tree-like structure constructed here explicates the qualitative and semantic foundations of such a law. This, in the future, should much facilitate the specification of a consensual mathematical expression for what is here denoted $M\pi c(ms_G)$.

[38] We note that the whole process of description (9'') has been developed inside an a priori given *cell for conceptualization*, namely the pair $(G,V_M)$, that acted like a l*ocal 'epistemic'* referential.

[39] For the sake of brevity, from now on we cease to always write explicitly the specification *'$(\varepsilon,\delta,N_0)$'*; but it will be constantly presupposed: we consider exclusively factual, effective probability laws.

### (3.I).1.4. The global result of the preceding genesis

All what precedes is represented on the Fig.2:

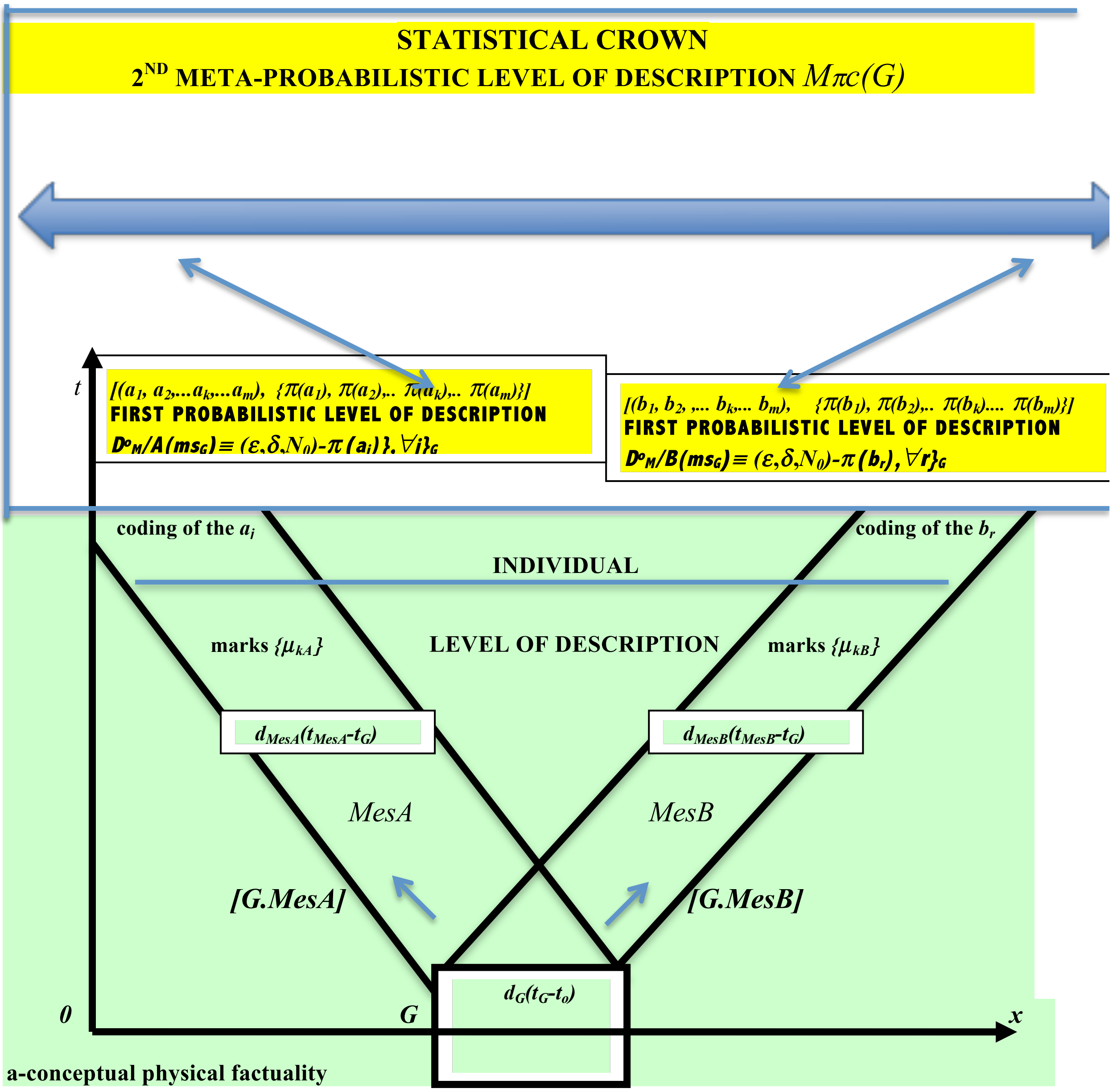

**Fig.2**. *The probability-tree T(G,(A,B) of an unbound microstate* $ms_G$

We have remarked that – in contradistinction to its purely numerical results – the genesis of these results does possess a definite space-*time* structure. However the temporal character of progressive emergence via successively registered individual results is abstracted away, of course. Accordingly to the so mysterious volatility of time it evaporates from the structure while it is progressively accomplished. So, in the exclusively spatial tree-like representation that persists, only the existence of distinct branches still just recalls the genetic temporal classifying features entailed by the mutual compatibilities or incompatibilities between the measured dynamical quantities, with respect to the considered type of microstate [40].

---

[40] I perceive this as a troubling hint that time *could* perhaps be conceived as just a very basic artefact of Nature brought forth inside human thought by the biological evolution, as a basic feature of our fitness to subsist as a species that is *constrained to parcel in order to conceptualize* (other authors also reach a similar notion, for instance Carlo Rovelli [2015] and Donald D. Hoffmann&Ananda Gefner [2016]). In such a perspective, in the scientific conceptualization of

Let us denote by *T(G,(A,B))* (*T*: 'tree') this entirely geometrized residual structure of the genetic process of a description (9") [41]. The green zone of genetic conceptualization – purely individual and *physical*-operational – is clearly separated from the posterior superposed yellow zone of exclusively abstract conceptualization.

### (3.I).1.5. More detailed probabilistic examination of *T(G,(A,B)*

The concept of probability-tree of a microstate involves significances that are far from being trivial: they have already helped us to expand Kolmogorov's purely abstract, mathematical concept of a probability-space – where in particular the distributions of probability remain an only *general* pure concept that is *not* specified numerically – into a new and much more complex tree-like probabilistic whole where each element is defined, while the probability measure (5) $(\varepsilon,\delta,N_0)$-$\{\pi(a_j)\}, \forall j)$, emerges endowed with a finite conceptual definition *and* a factual numerical specification. Let us explicate this a little more.

**- *Random phenomenon.*** The classical theory of probabilities offers no formalization of the notion of random phenomenon. It just makes use of the word 'experiment'. Whereas on the fFig.2 one literally sees how – from nothingness – a whole group of Kolmogorov probability-spaces emerges for a microstate $ms_G$, mutually connected by the corresponding operation of generation *G*, *and* by meta-probabilistic correlations between these. Thereby the basic concept of 'random phenomenon' acquires for a detailed inner structure, expressed in definite terms, namely [*G*, *MesA* or *MesB*, etc., marks $\{\mu\}_{kA}$ or marks $\{\mu_k\}_{kB}$, etc., code $\{a_j\}$ or code $\{b_r\}$, etc.], wherefrom factual finite Kolmogorov probability-spaces are then constructed. Inside these mutually connected probability spaces are lodged *numerically* specified factual $(\varepsilon,\delta,N_0)$-probability laws that are *effective and **relativized to all the actions and features that determine them***.

This result can be generalized to any physical entity and it can be induced in a strongly enlarged abstract theory of probabilities that accepts naturally a deep-set unification with *a relativized and extended logic* (MMS [2002A], [2002B], [2006], [2013], [2014]). Thereby these two fundamental structures of the human thought merge into a unification that has been shown to include Shannon's informational approach (MMS [1980], [1982], [2006]).

**- *Probabilistic dependence.*** The factual Kolmogorov-type probability spaces that crown the two branches from the Fig. 2 admit, respectively, the denotations

$$[U(a_j),\ \tau_A,\ \{\pi(a_j,\forall j\}_G], \qquad [U(b_r),\ \tau_B,\ \{\pi(b_r,\forall r\}_G],$$

where $\tau_A$ and $\tau_B$ are the involved algebras of events. Let us consider now explicitly these algebras also. Inside the classical theory of probabilities the concept of probabilistic dependence is defined *only* for events from the algebra from *one* given space. Kolmogorov ([1950], p.9) has written:

> «…..one of the most important problems in the philosophy of the natural sciences is – in addition to the well known one regarding the essence of the concept of probability itself – to make precise the premises which would make it possible to regard any given real events as independent. »

And he has posited – just posited by definition – that two events $a_1$ and $a_2$ from the algebra $\tau$ of a probability space are mutually independent from a probabilistic point of view if the numerical product $\pi(a_1).\pi(a_2)$ of the probabilities $\pi(a_1)$ and $\pi(a_2)$ of their separate occurrences is equal to the probability $\pi(a_1 \cap a_2))$ of their (set)-product-event

---

*physical* 'reality', space would be more basic than time, notwithstanding that according to Kant's postulate space and time, "equally", are both a priori forms of the *human* intuition.

[41] The expression "probability tree" is already much made use of, with various significances. All these should be very carefully distinguished from the particular significance represented in the Fig.2.

$a_1 \cap a_2$ from $\tau$ ; whereas if this is not the case, then $a_1$ and $a_2$ are tied by a probabilistic dependence. But inside the classical theory of probabilities the concepts of probabilistic dependence or independence are ***not*** defined for elementary events from *one same* universe *U*. Such a sort of 'dependence' can be apprehended only indirectly, by comparison with the probability law that acts upon a universe of elementary events defined as a Cartesian product of two other universes, one of which is *U*. But this involves *another* random phenomenon, distinct from the random phenomenon that generates the space where *U* is the universe of elementary events [42] and a rigid *juxtaposition* of these two random phenomena. Whereas inside *IQM* this limitation will be circumvented in (3.I).2 via the definition (1) of an operation of generation *G* combined with the definition of one microstate of two or several micro-systems permits to circumvent this limitation.

In the case of a microstate of one micro-system, the classical definitions are sufficient only if each one of the two probability spaces that crown the two branches from the Fig.2 is considered *separately* from the other one. But consider now an elementary event $a_j$ from the space that crowns the branch *MesA*, and an elementary event $b_r$ form the space that crowns the branch *MesB*. *Observationally* these two events *are* mutually in-dependent in the sense of Kolmogorov: Since the quantities *A* and *B* are mutually in-compatible in the sense defined in (2.I).2, the two measurement-operations *MesA* and *MesB* cannot be realized together, simultaneously, for only *one* specimen $\sigma(ms_G)$ of the studied microstate $ms_G$, so *the elementary events $a_j$ and $b_r$ cannot even coexist in an **actualized** way*. But nevertheless, the events $a_j$ and $b_r$ concern the *same* microstate $ms_G$ – in the sense of (1) – generated by one same operation of generation *G*. And even though inside our approach a microstate in the sense of (1) is distinct by definition from any specimen $\sigma(ms_G)$ of it, the considerations that led to (11)+(11') entail with a sort of necessity the assertion of a *meta*-probabilistic correlation *(Mπc(G))* and of the corresponding extension (9") of (9), because both spaces that are considered stem from one same operation of generation *G*. This argument amounts to the assertion of a sort of 'probabilistic dependence' of second order that knits into one whole all the distinct branch-random phenomena of which the common operation of generation *G* from the trunk of the tree introduces a priori the ***potentiality***.

The classical theory of probabilities also defines the general concept of probabilistic correlations, quite explicitly. But it does not *singularize* inside it a class of meta-probabilistic correlations that manifests specifically the fact that one same basic concept of a physical entity ($ms_G$ or *G*) is involved in different and *separately actualized* random phenomena [43]. This however is obviously an important case because it can be extremely frequent and it can entail subtle explanations for queer but observable behaviours.

In short, considered globally, *the probability-tree of a microstate constitutes **one whole of potential knowledge**, a closed cell of potentially possible fabrication of different but interconnected sub-wholes of actualized knowledge* [44].

---

[42] This detour could stem from the desire to stay inside the domain of the *actualized*. But let us notice that outside an algebra of events, the Kolmogorov concept of probabilistic dependence between two elementary events from the universe *U* is also a mere potentiality when these elementary events are not compatible in our sense, while in *this* case inside an algebra it can be regarded as actual only because the concept of event from an algebra involves potentiality by construction.

[43] K.J. Jung has introduced a concept of 'synchronicity' that seemed rather mysterious and has much struck Pauli, possibly because quantum mechanics – via the "principle of exclusion" – had suggested to him implicitly similarities with the behaviour of microstates, and this has been discussed in the correspondence Jung-Pauli (MMS [2002B], note pp. 279-281).

[44] Human psychic "time" is strongly populated by potentialities, by the virtual; so the representation of probabilities should fully encompass also such 'states' of events. A probability tree in our sense is overtly constructed as a *potential-and-actual* structure that spreads out freely inside the whole domain of possibilities where develop the individual inner times of human beings, wherefrom the public time is constructed (MMS [2006]).

Whereas the Kolmogorov conceptualization – even though it makes verbal use of the concept of 'experiment' – remains still entirely blind toward the probabilistic specificities entailed by the concept of operation of generation $G$ in the sense of (1) when moreover this is combined with the definitions from (2.I).2. Consequently the classical theory of probabilities reveals imprisonment in fragmenting delimitations when it is compared to the concept of probability that works in the case of a transferred description.

Why is this so? The answer is striking: because *Kolmogorov's concept of a probability space does not reach the factual root of the probabilistic whole that works inside the definition (9") of a transferred description*; it has been conceived entirely on the classical level of conceptualization. Whereas the bottom-up approach practised here starts much deeper, on the frontier between the already conceptualized and the a-conceptual physical factuality, wherefrom it proceeds upward via the methodological decision (1) $G\leftrightarrow(ms_G\equiv\{\sigma(ms_G)\}$ that introduces the factual-operational definition of precisely this factual root of a new probabilistic whole. The entire probabilistic output of this root – with respect to an arbitrary but *given* collection of mutually incompatible branch-qualifying mechanical quantities – can be represented inside *one* new sort of enriched probability space:

$$[\ U_T(e_{jb})=\cup_b U(e_{jb}), \qquad \tau_T=\cup_b \tau_b\,, \qquad \{\pi_T(e_{jb}\,,\forall jb\}_G = \cup_b\{\pi_b(e_{jb},\forall j\}_G\ ]$$

where the index '$T$ ' labels the considered probability-tree; the index '$b$' labels the considered branch from $T$; the index '$jb$' labels the elementary event $e_{jb}$ from the branch labelled by $b$, and $\tau_T$ designates the *total* algebra of events from this enriched probability space.

### (3.I).1.6. The particular case of a one-trunk probability tree

What happens if no sort of relative mutual incompatibility does act in the considered circumstance? In this case the space-time domain covered by the involved operation of generation $G$ leads to only one 'branch' that is common to all the considered mutually compatible mechanical quantities; which amounts to saying that the common trunk-and-branch of the tree is crowned by a set of probability spaces – one for each quantity $A$ – that, inside (9") – are only conceptually distinguished from one another and then meta-correlated to one another, as indicated below.

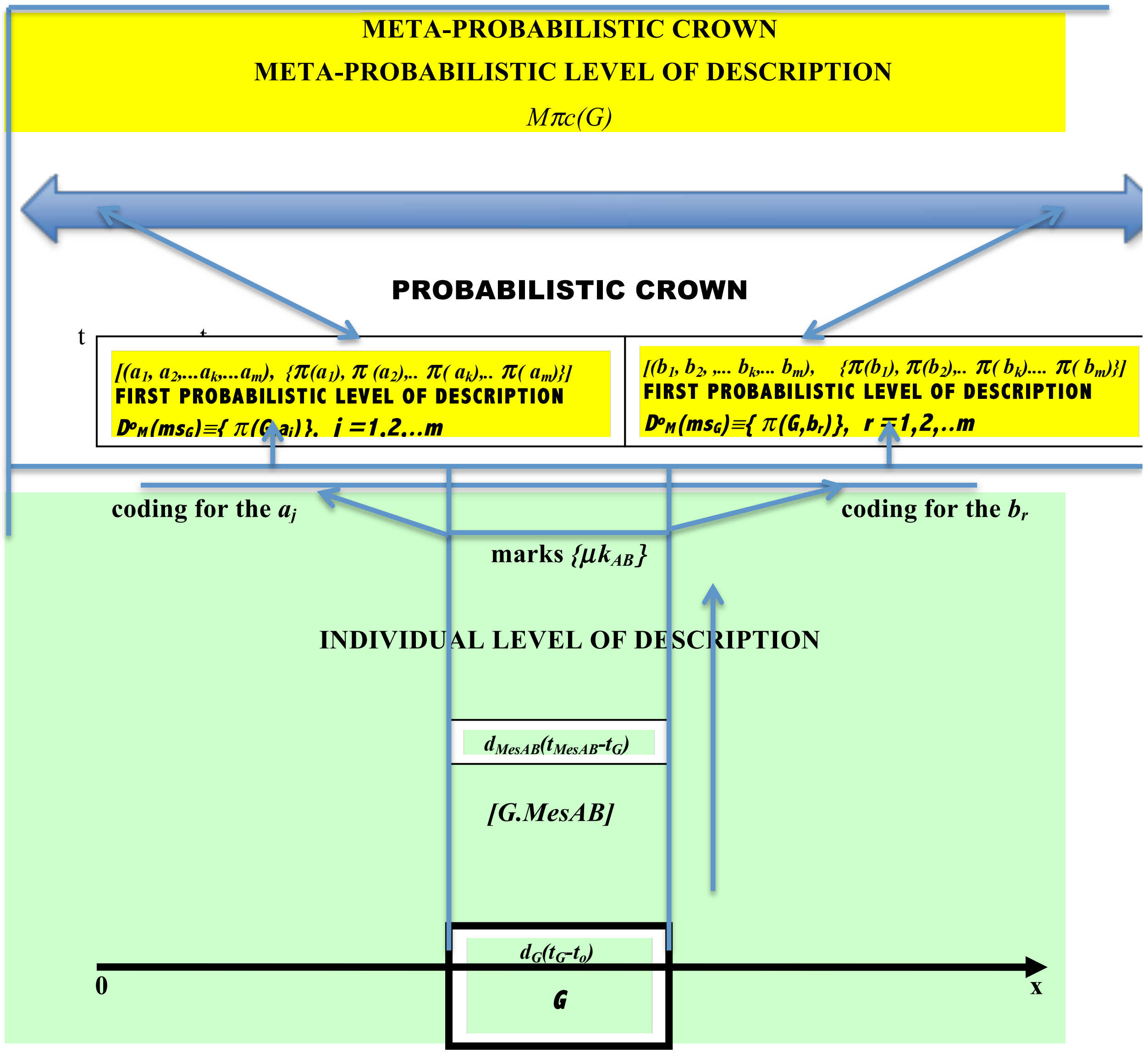

**Fig. 3**. *The probability-tree T(G,(A,B) of two mutually compatible observables*

Here the capacity of the case of an unbound microstate of *one* microsystem to reveal non-classical probabilistic contents of (9"), comes to exhaustion. But the most surprising such contents appear just below.

### (3.I).2. PROBABILITY TREE OF *ONE* UNBOUND MICRO-*STATE* OF *TWO OR MORE* MICRO-*SYSTEMS*

We now enter upon the case of micro-states of two or several micro-systems. Thereby we come face-to-face with what is called *the problem of non-locality*. This case brings strikingly forth to what a degree the concepts-and-language introduced by the definitions from (2.I)2 and by the basic concept of a probability tree defined in (3.I)1 introduce a structure of conditions of inner coherence that entails intelligibility.

Consider *one* progressive micro-state $ms_{G(2S)}$ of *two* micro-systems $S_1$ and $S_2$, in the sense of the definitions from (2.I)2). How shall we construct the probability tree of $ms_{G(2S)}$ ?

According to (1) one microstate $ms_{G(2S)}$ is generated by one corresponding operation of generation *G(2S) to which it is tied in the sense of (1)* and of the identity $ms_{G(2S)} \equiv \{\sigma(ms_{G(2S)})\}$.

According to the definitions from (2.I)2) in this case *one* complete operation of measurement-interaction on *one* specimen of the factually defined microstate $ms_{G(2S)}$ involves *two* partial measurement-interactions, a partial measurement-interaction *MesA* with $S_1$, *and* a partial measurement-interaction *MesB* with $S_2$ (in particular the quantities *A* and *B* can identify, but in general they are permitted to be different). For maximal graphic clarity instead of *A*, $a_j$ and *B*, $b_r$ we shall exceptionally write in this case, respectively, *A1*, $a1_j$ and *B2*, $b2_r$. So a complete act of measurement will be denoted *Mes(A1,B2)*.

In (3.I)1, for the case of one micro-state of *one* micro-system, we have assigned by construction an *own* branch to *each* given sort of 'complete' act of measurement that is involved i.e. which involves fully one specimen of the microstate-to-be-studied (this happens always for a microstate of only one micro-system). In order to stay in agreement with all the constructive definitions from (2.I)2) and (3.I)1, here we must apply this same procedure: [one given sort of complete act of measurement involving fully one specimen of the microstate-to-be-studied] corresponds to [one branch of the tree]. So to each sort of complete act of measurement of the same form as *Mes(A1,B2)* we assign *one* branch from the probability tree of *G(2S)*. Then the *two* partial measurements *MesA1* and *MesB2* from one complete act of measurement *Mes(A1,B2)* operated respectively upon the two micro-systems $S_1$ and $S_2$ from each *one* specimen $\sigma(ms_{G(2S)})$ of the studied micro-state $ms_{G(2S)}$, are ***both*** lodged inside ***one*** *same branch* of the probability tree of *G(2S)*. So we must assign another branch of this tree to the complete measurements that involve another pair of quantities denoted for instance *(C1,D2)* with values, respectively, $c1_k$ and $d2_z$, where *at least* either *C1* is *in*-compatible with *A1* or *D2* is *in*-compatible with *B2* in the sense defined in (4), or where both these possibilities are realized; *there is no condition then concerning the compatibility of C1 and D2*.

So a two-branches-tree from the figure 4 founded upon the operation of generation *G(2S)*, can be denoted *T(G(2S),(A1,B2; C1,D2))*.

Let us now focus upon the following fact: For one micro-state of *two* micro-systems the two dynamical quantities *A1* and *B2* that are involved in *one* complete act of measurement *Mes(A1,B2)≡(MesA1 and MesB2) are **always** compatible* in the sense defined at the point 3 from (2.I)2, because the measurements *MesA1* and *MesB2* are performed, respectively, upon the *two* mutually distinct systems *S1* and *S2* that are involved in any *one* specimen of the microstate $ms_{G(2S)}$ and so no incompatibility between these space-time supports of these two partial acts of measurement comes in necessarily [45] (if in some circumstance these two space-time supports tend to overlap it should be easily possible to eliminate the problem).

Since *one* complete act of measurement *Mes(A1,B2)* contains by definition an act of measurement *MesA1* and an act of measurement *MesB2*, the corresponding pair of observable marks $(\{\mu\}_{kA1},\{\mu\}_{kB2})$ – let us denote it $\{\mu_{kA1B2}\}$ – once it has been coded in terms of a pair of values $a1_j, b2_r$, $j,r=1,2,...M$, constitutes ***one*** elementary event from the universe of elementary events $U=\{a1_j,b2_r\}$, $j,r=1,2,...M$ from the probability-space that in the Fig.4 crowns the *unique* branch of the complete measurements *Mes(A1,B2)*; while the factual probability distribution on the universe of elementary events from this probability space consists of the transferred description (9) with respect to the pair of quantities *(A1,B2)* and has to be denoted as

$$D/(A1,B2)(ms_{G(2S)}) \equiv (\varepsilon,\delta,N_0)\text{-}\{\pi(a1_j,b2_r), j, r=1,2,...M\}_{G(2S)}, \quad j, r=1,2,...M$$

---

45 We recall that inside the approach developed here the compatibility or incompatibility of two dynamical quantities is defined only for *one specimen* of the studied microstate and it is *relative* to both the nature of these quantities and to the type of the microstate that is considered, in the sense of the definitions from (2.I)2.

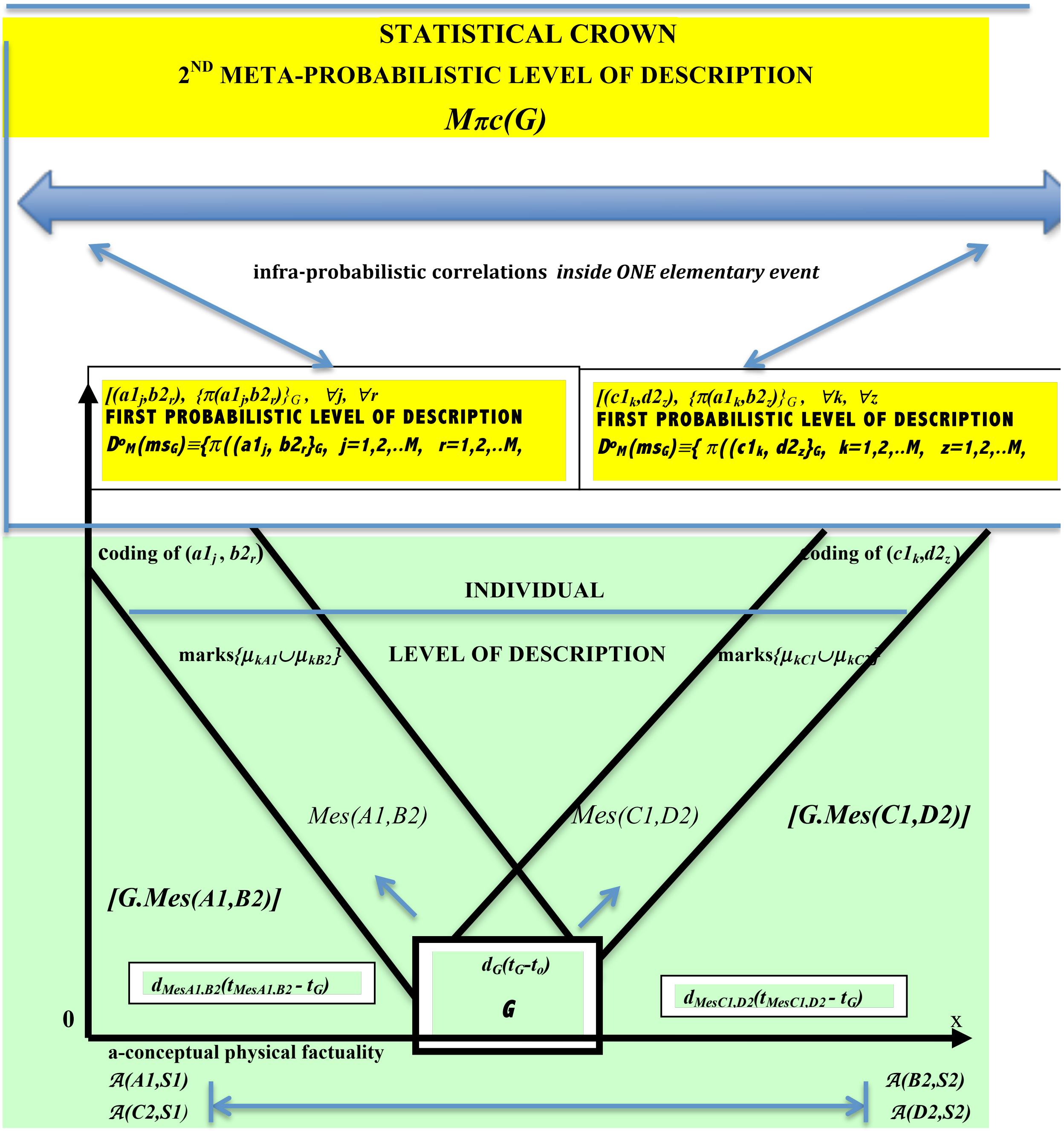

**Fig. 4**. *The probability-tree* $T(G(2S),(A1,B2;\ C1,D2))$ *of a microstate* $ms_{G(2S)}$*:*
*the case of the 'problem of non-locality'*

So the *two* quantities *(A1,B2)* of which one qualifies the system *S1* and the other one the system *S2* are involved *both* in each *one* – *and* ***elementary*** – *'event'* in the probabilistic sense, that concerns only *one* complete act of measurement *Mes(A1,B2)* on *one* specimen of the studied micro-*state*. And nevertheless the *here-now's* of the corresponding *two* observable and registered events – in-the-***physical***-sense this time – namely [the observation by a human observer, of marks $\{\mu\}_{kA1}$ coded by a value $a1_j$ that qualifies *S1*] and [the observation by a human observer, of marks $\{\mu\}_{kB2}$ coded by a value $b2_r$ that qualifies *S2*], *can be separated from one another by an arbitrarily big space-time* *distance*. While *the corresponding description (9) – **one** factual probability law – is itself* ***devoid of a defined space-time structure***.

The Fig.4 represents graphically a most explicit analysis of the *inner* texture of the 'problem'. This is the 'problem' of non-locality re-expressed according to the genuine *algorithm for probabilistic conceptual organization* involved by the probability-tree of the studied microstate, such as this algorithm is entailed by the concept (1) of operation of generation *G* of the considered sort of microstate-to-be-studied, and by the definitions from (2.I)2 [46].

The present way of reaching this problem out of nothing conceptualized before, inside a radically first and merely qualitative bottom-up approach, brings clearly into evidence that what is called 'non-locality' is tied with preconceived classical, so particularizing assumptions, and with a general conceptualization of the 'microstates' inside the nowadays microphysics that is unachieved from various points of view. Indeed:

- Consider the *two* "micro-*systems*" $'S_1'$ and $'S_2'$ from *one* specimen $\sigma(ms_{G(2S)})$ of the studied microstate $ms_{G(2S)}$. In the absence, inside modern microphysics, of an explicit use of a general model of a microstate, these two micro-systems have been spontaneously and implicitly imagined more or less like two spatially delimited small balls radically *exterior* to one-another, so mutually 'separated' by a ***void*** 'distance' that in its turn is also 'exterior' to these entities themselves; which raises strongly and intuitively the question of *what* 'exists' and 'happens' ***outside and between*** $'S_1'$ and $'S_2'$ [47, 48]. Whereas the experimentally registered *time*-distance between $'S_1'$ and $'S_2'$ seems to be quasi null, or in any case smaller than is entailed by the Einstein-velocity of a 'light-signal'. This conceptual situation acts as a strident common call for a general model of a microstate, in spite of the orthodox interdiction imposed by the Copenhagen school.

- But also other presuppositions are involved. For instance, the non-locality problem emerges in a particularly striking way because it is explicitly and essentially *lodged inside the space-time frame **of the human observers with their apparatuses***. One complete act of measurement $Mes_{12}(A1,B2)$ involves two blocks of macroscopic apparatuses $\mathcal{A}(A1,S1)$ and $\mathcal{A}(B2,S2)$ that – themselves – are *perceived* with delimited volumes and are endowed with registering devices that pre-structure classes of possible space-time locations of the observable results of measurements coded $a1_j$ *and* $b2$, which can define perceived spatial and temporal distances between potential locations *of **these** pre-constructed observable space-time locations of the perceivable marks*. This entails an inextricable mixture between: a mathematical formalism; implicit expectations induced by the classical human macroscopic conceptualization; and cognitive human actions, registering 'objects', and observable pre-organized events, that are unavoidably involved by the acts of measurement. While obviously, such heterogeneous features with their respective conceptual roles have to be strictly distinguished from one another via a well defined methodological structure imposed upon the study.

---

[46] Is it not remarkable that an approach like that one developed here – so general, and only qualitative – brings forth so rapidly this whole analysis, in a way so deeply tied with the basic tree-like representation of a microstate and *independently of any mathematical formulation*?

[47] The question of 'separability' has been much discussed, but via mere words and undefined subtleties. While any primordial transferred description (9") of *that* what here, in the reference-and-imbedding structure that we are now constructing in a rigorous way, is called a 'microstate' in the sense of (1) *simply cannot as yet entail any sort of space-time specifications*, neither inner ones nor exterior ones, since it emerges still radically ***devoid*** of any definite inner space-time structure: only later, in a subsequent phase of conceptualization, such specifications *might* become possible (or not) inside a theory of microstates where a general *model* of a microstate, necessarily, is defined.

[48] Descartes held that void space does not *exist*. Which I understand as the assertion that only void space with respect to some definite aspect can exist, because 'space' ***is** exclusively* the universal bearer posited a priori in human mind for any given quality that 'exists' with respect to some given physical entity.

- The purely conceptual probabilistic situation is also unintelligible, from a very basic probabilistic point of view. The space-time distances, *together* with the observed correlations, emerge in relation with only ***each one-branch***-probability distribution

$$(\varepsilon,\delta,N_0)\text{-}\{\pi(a1_j,b2_r,\ \forall(j,r)\}_{G(2S)} \quad \text{or} \quad (\varepsilon,\delta,N_0)\text{-}\{\pi(c1_k,d2_z),\ \forall(k,z)\ \}_{G(2S)}$$

***not*** inside the meta-probabilistic correlations denoted by us *'Mπc(G)'* between the ***two*** branch-probability distributions that are involved (cf. the Fig.2). In this case, the considered 'correlations' appear as tied with a sort of 'probabilistic dependence' that stems from the ***insides*** of the observable events $(a1_j,b2_r), \forall(j,r)$ or $(c1_k,d2_z), \forall(k,z)$ from ***one*** probability law $(\varepsilon,\delta,N_0)\text{-}\{\pi(a1_j,b2_r,\forall(j,r)\}_{G(2S)}$ or, respectively, $(\varepsilon,\delta,N_0)\text{-}\{\pi(c1_k,d2_z),\forall(k,z)\}_{G(2S)})$. While in the probabilistic sense, these are ***elementary*** *events*.

> The classical concept of probabilistic dependence defines exclusively a concept of mutual probabilistic dependence for two distinct events from the algebra posited on the universe of elementary events from one probability space, an algebra that does not even necessarily contain the elementary events. This classical concept of probabilistic dependence cannot deal with features of the inner structure of elementary events.

Here the classical probabilistic conceptualization is literally overwhelmed.

- Finally, let us consider also the *direction* of conceptualization, top-down or a bottom-up. This direction also plays an essential role in this circumstance, but via a quite general feature. Historically the human conceptualization has been developed top-down on the vertical that connects the macroscopic level of conceptualization, to the microscopic one, and this entailed that the notion of a *common* trunk *G* of a possible probability-tree from which stem distinct branches, had not yet been conceived at the time when Kolmogorov elaborated his theory of probabilities. The general genetic concept of operation of generation *G* of the entity-to-be-studied had not yet emerged itself. So Kolmogorov has defined only probability spaces entirely separated from one another, each one of which tops only one 'experiment' (or 'random phenomenon').

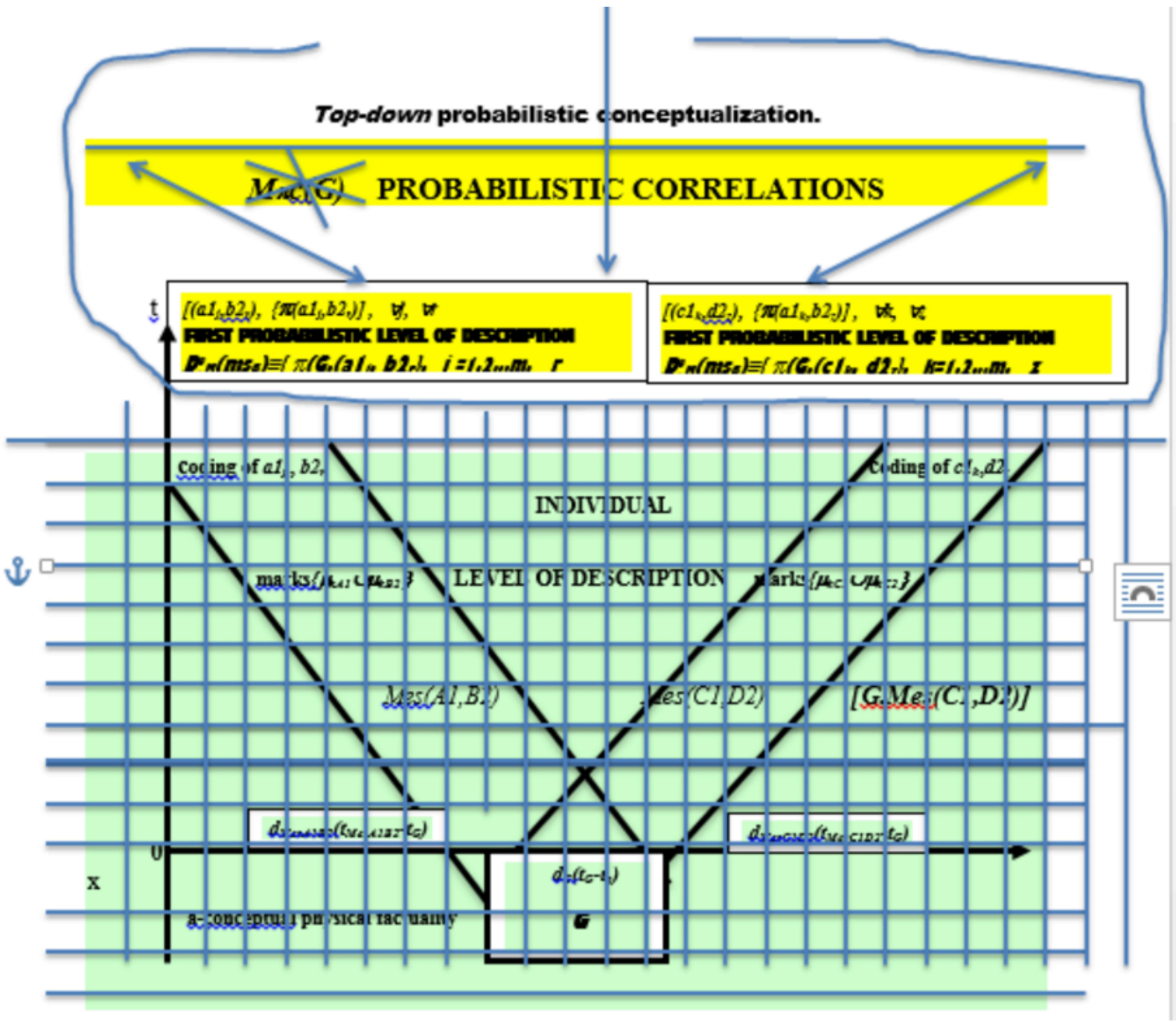

**Fig. 5**. *A probability-tree T(G(2S),(A1,B2; C1,D2)) as encountered by a top-down approach that installs Kolmogorov's classical concept of probability spaces and then* ***stops*** *its progression downward.*

And then Kolmogorov stopped, of course. He had already realized a very remarkable progress with respect to the preceding Bernoulli-von Mises concept of – directly and alone – the mathematical concept of a probability measure. But one isolated probability space gave no *access* to possible common roots of different probability spaces.

For human beings that start their conceptualization on what we call 'the macroscopic level', common roots can stay hidden a very long time with respect to a top-down approach. While in order "to make precise" the premises of probabilistic dependence (cf. the Kolmogorov-quotation from (3.I).1)), a sine qua non condition is to be aware of the existence – of the *general* existence – and the basic role of the 'operation of generation *G* of the entity-to-be-studied' [49], and the mentioned existence and role, though they always exist in some particular avatar, physical or sensorial or only mental, become striking and are endowed with general contours only inside microphysics, and only if one is attentive to the consequences of the *physically* operational character of the operation of generation *G* when the entity-to-be-qualified is a 'microstate'.

- But even if the concept of operation *G* of generation *is* taken into account, *still* the Bell-case can stay probabilistically non-intelligible, because the notion that for unbound microstates the whole successions *[G.MesA]* are a condition for obtaining consensually perceptible marks, if it is active *alone*, entails only a monolithic concept of probabilistic correlation that cannot distinguish between correlations interior to one elementary event in the probabilistic sense, and correlations between distinct probabilistic events. *This distinction, as stressed, requires explicit recourse to also the definitions from* (2.I)2.

In short, the case of the probability tree of one microstate of two or several microsystems is paradigmatic from various points of view, and very basic ones. It is not surprising that it has raised, and still raises, so many researches and considerations. The *IQM*-analyses of this case illustrate strikingly the utility and the forces of an explicitly constructed structure of reference.

## (3.I).3. PROBABILITY TREE OF ONE MICROSTATE WITH *COMPOSED* OPERATION OF GENERATION

Consider now a *composed* operation of generation $\boldsymbol{G(G_1,G_2)}$ (cf. (1.I),(2.I).1) of a microstate in which only two simple operations of generation $G_1$ and $G_2$ are involved, like in the Young two-slits experiment. And consider an effectively realized microstate $ms_{G(G1,G2)}$. Let us compare its factual description (9') with the factual descriptions (9') of the two microstates $ms_{G1}$ and $ms_{G2}$ that *would* be obtained, respectively, *if* the two operations of generation $G_1$ and $G_2$ were each one ***fully realized separately***. According to our present knowledge on microstates such a comparison would bring forth the physical *fact* that in general, between the probability $\pi(G(G_1,G_2),a_j)$ of realization of the value $a_j$ of a dynamical quantity *A* via acts of measurement *MesA* performed on one outcome of $ms_{G(G1,G2)}$, and the probabilities $\pi(G_1,a_j)$ and $\pi(G_2,a_j)$ of this same value $a_j$ established, respectively, via measurements *MesA* performed onthe microstates $ms_{G1}$ and $ms_{G2}$, there holds an *in*-equality

$$\pi_{12}(\mathbf{G}(G_1,G_2),a_j) \neq \pi_1(G_1,a_j) + \pi_2(G_2,a_j) \qquad (12)$$

[49] Dirac's "theory of transformations" – that obviously involves probabilistic correlations – does not assert them explicitly. It is presented as exclusively an algorithm of Cartesian type for passing from one system of coordinates to another one. While it might come out in the future that ***any*** probabilistic correlation can be assigned to a certain class of distinct branches from a huge probability-(meta-tree)-of-probability trees). This would found in a very toned way Gustav Jung's concept of 'synchronicity' that has interested Pauli. (MMS [2002B], the note pp. 279-281). (In particular, it is not a priori absurd that certain subconscious psychical perceptions of 'synchronicity' of physical events come out to be connected with some sort of instinctive, reflex reactions to physical events from a same probability tree that are separated from one another by an arbitrarily big spatial distance *internal* to some basic sort of physical substance (like that from the de Broglie-Bohm view) relatively to which Einstein's 'limit-velocity' of, specifically, **light**-'signals', simply does not exist).

This circumstance deserves being noticed. It suggests that a microstate tied with $\mathbf{G}(G_1,G_2)$ belongs to a domain of phenomena that is of another nature than the domain of phenomena entailed by $G_1$ and $G_2$ *when* these are realized separately. But 'different' *in what sense*, exactly? In this preliminary stage of the conceptualization of the microstates this question remains *open*. Nevertheless we can already formulate the following important remark.

The inequality (12) is usually expressed verbally in *positive* terms by saying that '$ms_{G1}$ and $ms_{G2}$ *interfere* inside $ms_{G(G1,G2)}$'. But inside the present approach – according to the one-to-one relation (1) between a given operation of generation and the corresponding microstate – this expression is misleading from a conceptual point of view. The relation (1) entails that *only* the *one* microstate $ms_{G(G1,G2)}$ is effectively generated when the operation of generation $\mathbf{G}(G_1,G_2)$ is performed. So $\mathbf{G}(G_1,G_2)$ cannot be coherently conceived to generate *also* the two microstates $ms_{G1}$ and $ms_{G2}$ when the microstate $ms_{G(G1,G2)}$ has been generated. When the microstate $ms_{G(G1,G2)}$ has been generated, the microstates $ms_{G1}$ and $ms_{G2}$ have to be conceived as somehow *non*-achieved or non-'completed' microstates that, by construction, can at most possess the status of partial effects of two a priori *possible* full operational state-individualizations via $G_1$ and $G_2$, but that in fact have *not* been fully actualized when $G_1$ and $G_2$ are composed inside $\mathbf{G}(G_1,G_2)$.

The symbol $\mathbf{G}(G_1,G_2)$ adopted here suggests that – if and when this seems useful – the mentioned effects can be *referred* two the unachieved microstates $ms_{G1}$ and $ms_{G2}$. So, in terms of probability-trees, the trees $T(G_1)$ and $T(G_2)$ are only two reference-trees, ghost-trees, only the one tree $T(\mathbf{G}(G_1,G_2))$ is factually realized.

And, since $ms_{G1}$ and $ms_{G2}$ have not been both and separately effectively realized by $\mathbf{G}(G_1,G_2)$, they do not 'exist' inside $ms_{G(G1,G2)}$, as it is implied by the assertion that 'they' *interfere* inside $ms_{G(G1,G2)}$. Such a language – *like also the mathematical writing (12)* or some equivalent one – are misleading inside the present approach: the natural language involves many shades, and these work inside the minds, so they have to be carefully dominated.

The preceding considerations can be generalized in an obvious way to the case of an operation of generation $\mathbf{G}(G_1,G_2,...G_m)$ that composes several operations of generation.

So inside the present approach the mathematical representation of the whole category of 'microstates with *composed* operation of generation' will have to be openly considered in a critical state of mind and in the third part of this work it will play the role of a discriminating test of the construction submitted there.

This point (3.I)3 closes our exploration on probability trees of progressive microstates[50]. Indeed, for the reasons expressed at the end of (3.I)1 the concept of probability tree is not useful for bound microstates. Therefore in what follows we only add a brief but essential remark on the evolution of an unbound microstate.

---

[50] We mention that the concept of a probability-tree has been worked out in (MMS [2002A], [2002B], [2006], [2014]) *in quite general terms* (not only for the case of microstates). In the second part of this work it will appear that the nowadays quantum mechanics also implies it, via the Hilbert-space representations and Dirac's calculus of transformations.

### (3.I).4. ON THE EVOLUTION OF ANY UNBOUND MICROSTATE

Is it possible, inside this qualitative and general approach, to assert something concerning the evolution of a progressive microstate? The answer is yes, and again it brings into evidence the crucial role of the concept of operation $G$ of generation of a microstate.

Imagine the *final* moment $t$ assigned to an operation of generation $G$ from (1) that introduces the microstate $ms_G$ to be studied. In contradistinction to what has been assumed before, let us admit that during some time interval $\Delta t_1=t_1-t$ subsequent to $t$ the human observer does *not* act upon the microstate $ms_G$. Nevertheless during $\Delta t_1=t_1-t$ the initial microstate $ms_G$ can be *posited* to 'evolve' in the 'exterior' conditions $EC$ that it encounters (exterior known macroscopic fields or obstacles). Indeed it would seem weird to posit that '$ms_G$' remains immobilized from any conceivable point of view. Now – formally – this evolution *can be integrated in (1)*, in the following way.

Nothing hinders to posit in full logical coherence with the preceding development, that the association of the initially conceived and realized operation of generation $G$, with what happens with $ms_G$ during $\Delta t_1=t_1-t$, act together like *another operation of generation* – let us denote it $G_{t1}=F(G,EC,(t_1-t))$ ($F$ : some functional) that generates in the factual sense from *MD1* another microstate $ms_{G1}$ corresponding to $G_{t1}$ via (1). As stated before for any factually defined microstate $ms_G$, the microstate $ms_{G1}$ can be studied via sequences of successions $[G_{tk}.MesA], k=1,2,...M, \forall A\in V_M$. The time interval $t_1-t$ can be chosen with any desired value, the external conditions $EC$ being kept unchanged. So – given the initial operation $G_{to}$ – one can study successively a set of mutually 'distinct' microstates $ms_{Gk}$ that correspond respectively to the set of successive operations of generation:

$$G_o,\ G_1=F(G,EC,(t_1-t_o)),\ldots G_k=F(G_o,EC,(t_k-t_{k-1})),\ldots.G_f=F(G_o,EC,(t_f-t_{K-1}));\quad k=1,2,\ldots K \qquad (13)$$

($K$: an integer; '$f$': final; $f\equiv K$). For each operation of generation $G_k$ from this set one can construct the corresponding probability tree $T(G_k,A)$, $\forall A\in V_M$, i.e. also the corresponding descriptions (9') and (9''). So – with $G_t=F(G_o, EC, (t-t_o))$ and $G_t\leftrightarrow ms_{Gt}$ – in general and simplified re-notations we have:

$$[G_o.(t-t_o)] \equiv G_t \qquad (13')$$

The relations (13') *absorbs* into the general concept of operation of generation $G$, the phase of ***individual*** *evolution* ***before*** *measurement* of any involved specimen $\sigma(ms_G)$ of the studied microstate $ms_G$.

So – by definition – inside a succession $[G_t.MesA]$ the 'initial' state of the involved specimen of $ms_G$ is to be understood as the state of this specimen when *begins* the act of measurement *MesA*. This permits to replace $G$ by $G_t$ in (9), (9'), (9''). Then (9''), for instance, becomes

$$D_M(ms_{Gt}) \equiv \{ (D/A(ms_{Gt}), (M\pi c(G_t))_{XY} \},\ \forall A,\ \forall AB,\ \forall t,\ \forall j, \qquad (9''')$$

Together the relations (13') and (9''') express an essential new concept, namely ***a factual*** *statistical law of evolution* associated with the studied factual microstate.

## (3.I).5. CONSTRUCTION *VERSUS* VERIFICATION OF THE DESCRIPTION OF A MICROSTATE

What follows here is very brief to be stated, but becomes so important in the Part III of this work that it deserves this separate sub-section:

How can we *verify* a description (9) of any sort – (9') or (9'') or (9''') ?

The answer is obvious: *Only by reconstructing it* as many times as one wants and by modifying the parameters $(\varepsilon,\delta,N_0)$ from the definition (5) $(\varepsilon,\delta,N_0)$-$\{\pi(a_j)\}$, $\forall j$ until one observes the desired degree of stability. If this is not possible the verification of the prediction is not realized.

*Inside IQM the sequence of operations [G.MesA] from (7) or the variant [$G_t$.MesA] of this sequence in the sense of (13') constitute the basic operator for* ***both*** *the* ***construction*** *and the* ***verification*** *of a description of a microstate.*

## (4.I)

# INFRA QUANTUM MECHANICS

We have organized a methodological pre-structure of reference-and-embedment for constructing a fully intelligible mathematical theory of a mechanics of 'microstates' named a priori Infra-Quantum Mechanics and denoted *IQM*.

*IQM* has been developed independently of any mathematical formalism.

In order to insure explicit control on all the levels of conceptualization we have started on the level of *zero* pre-accepted knowledge concerning the individual physical and fully singular specimens of any microstate-to-be-studied[51], and therefrom we have proceeded *bottom up*.

> So the origin and the order of progression on the vertical of conceptualization have been changed, with respect to the classical science inside which, historically, the generation of knowledge on microphysical entities has progressed *top-down*. And this has entailed a basic modification of the concept of 'microstate'.

Indeed, via an unavoidable methodological decision *MD*, the definition of the concept of microstate – that classically is a precise, an *individual*, and an *abstrac*t definition – has been transmuted into a *factual* definition that involves a *physical operation of generation G*. And this operation, being a human instrument, cannot itself be defined otherwise than by a finite number of controllable parameters. Whereas any fragment of a-conceptual physical being introduced by a factually defined concept of a microstate entails a priori an unlimited set of unpredictable possible effects. And, with respect to the constructability of scientific predictive and verifiable knowledge starting from a local zero of specific knowledge, this contrast entails a 'primordially statistical' character. In short, we have started with a modified concept of a factually defined sort of microstate that entails primordially statistical scientific knowledge.

So, starting from local zeros of specific knowledge, and according to general criteria of factual or logical necessity or of declared methodological choices, *IQM* has been composed bottom-up. In essence, it consists of a network of symbolizations of classes of conceptual moulds of different sorts (methodological procedures, physical operations, probabilistic laws, etc.) in each one of which – later, inside a given theory of the microstates – will have to be lodged a semantically more specified element from the same class. We have endowed this basic construct with all the foundational elements, formatted by all the constraints that are required for achieving *any* acceptable scientific theory of the microstates, and the whole has been organized in a logically coherent way; elements that are not generally necessary, as well as arbitrary a priori restrictions, have been excluded.

The final result can be regarded as a structural definition of the general concept of *a theory of the microstates*. It can be characterized as follows.

***1.*** The core of *IQM* consists of a primordially probabilistic *transferred description* constructed inside a representational cell delimited by an a pair *(G,A)* where *G* symbolizes a physical operation of generation of specimens of the studied microstate, in the sense of (1), and $A$ indicates a mechanical grid of qualification (2) that is defined for the studied sort of microstate (in the sense specified in (1.I).2). The most basic form of transferred description is

$$D/A(ms_G) \equiv \boldsymbol{[}(\varepsilon,\delta,N_0)\{\pi(a_j, \forall j\}_{Gt}\boldsymbol{]}, \qquad \forall A \in V_M \qquad (9)$$

[51] Unavoidably, the concept of a microstate itself must be *given* – as a receptacle where to pour future specifications – because from nothing, nothing more can be drawn.

It is written time-*in*dependent, involves only one qualifying quantity $A$ and it stops on the first probabilistic level. The most comprehensive form of transferred description involves all the qualifying quantities $A$ that, inside the epistemic referential $(G_t, V_M)$, are defined for the studied microstate and all the levels of conceptualization, and it can be represented by the writing

$$D_M(ms_{Gt}) \equiv \boldsymbol{[}(\varepsilon,\delta,N_0)\{\pi(a_j), \forall j\}_{Gt}, (Mpc(G_t))_{AB}\boldsymbol{]}, \qquad (G_t, V_M) \qquad (9''')$$

Throughout *IQM* the physical operation $G$ of generation of the individual specimens of the microstate to be studied – never noticed before – reveals a basic and central role.

The descriptional structure (9''') is marked by very remarkable peculiarities:

- It is strongly *relative* to a triad (6) $[G_t, ms_G, A]$ of genetic elements, where the cell of conceptualization delimited by a given pair (epistemic referential) $(G_t, A)$ is formed in adequacy with a particular descriptional aim.

- The global basic genetic process of type (7) $\{[G_t.MesA]\}$, $\forall A$, that by repetitions of all the successions of the same general form brings forth a description (9'''), involves explicitly the fact that *each* one act of measurement performed on a microstate requires in general a *previous* corresponding realization of also the operation of generation of a specimen of the microstate $ms_G$ to be studied, because in general a measurement-interaction with a specimen of the studied microstate $ms_G$ destroys this specimen of the involved micro-*state* even if the involved micro-*systems* do persist.

- The brute observable result (8) $\{\mu\}_{kA}$, $k_A=1,2,...m_A$ of each one genetic succession $[G_t.MesA]$ from (7) – a group of publicly observable physical marks – is entirely meaningless by itself because it carries no perceivable qualities (qualia) associable with, separately, the involved specimen of the studied microstate, nor with the qualifying quantity $A$. In order to gain indirectly for the observable marks $\{\mu\}_{kA}$, $k_A=1,2,...m_A$, a meaning in terms of a value $a_j$ of the previously defined quantity $A$ and that be somehow tied also with the involved specimen of the studied microstate, *the measurement-evolution MesA has to incorporate an adequate coding-procedure*.

In its turn such a coding-procedure, in order to be definable in a non-arbitrary way, requires a *general* ***model*** of a microstate as well as recourse to a corresponding and explicit *re-definition of the qualifying quantity A for – specifically – the sort of studied microstate with its particularized model*. So:

> Any acceptable *theory* of microstates *must* introduce a generic model of a microstate as well as its variants with respect to the considered *sort* of microstate (in the sense of the definitions from (2.I).1.2), as well as re-definitions of the qualifying quantities $A$ for, specifically, the sort of considered model, *and* corresponding laws of evolution.

- A description (9''') cannot be assigned to the studied microstate itself considered separately, but only – globally – to the whole measurement interactions from the successions $\{[G_t.MesA]\}$, $\forall A \in V_M$ that generated the description (where $G_t$ is reducible to $G_o$ in the sense defined in (13'). So *a description (9''') is strongly and indelibly tied to its genetic process*. The classical notion of 'object' is not yet extracted from a description (9'''). This feature is intimately tied with the absence of any defined space-time structure assigned to the description (9''') *itself*[52].

***2.*** In contradistinction to a description (9''') itself, the genetic successions of operations (7) *[$G_t$.MesA]* that are achieved by the observer-conceptor in order to construct such a description *are* quite essentially endowed with a specific space-time structure. This fact is manifested by the tree-like geometrical structure of the graphic representation from the Fig.2. And this structure entails *non-classical extensions of the classical concept of probability*. These extensions:

* Require a deeply modified and enriched concept of probabilistic dependence that involves an explicit distinction between what is presupposed to be only potential and what is presupposed to have been actualized, as well as incorporation of both these ways of 'existing' conceived by the human mind.

* They vary according to whether one micro-*state* of *one* micro-*system* is involved, or *one* micro-*state* of *several* micro-*systems*. And in the second case the entailed probabilistic extensions violate brutally the classical probabilistic ways of thinking because in this case their significance concerns the *interior of elementary events*, in the probabilistic sense [53, 54].

Thereby inside *IQM* the 'problem' of non-locality becomes intelligible. (As for Bell's theorem of non-locality, its structure involves also the mathematical formalism of the nowadays quantum mechanics and therefore it will be examined in the Part III of this work).

***3.*** The concept of probability-tree of an operation of generation of a microstate embodies and summarizes intuitively the whole complex and unexpected structure of the *genesis* of the form (9''') of the primordial transferred description of a microstate. When one progresses mentally bottom-up along the vertical of conceptualization, one can watch step by step on the probability-tree from the Fig.2 *how* a radical scission sets in between all the individual ***physical***-conceptual genetic *human actions* – that *do* involve space-*time* – and on the other hand the final global result (9''') of all these actions. On the graphic representation from the Fig.2, the purely numerical content of the final description (9''') appears displayed on the tops of the mutually disjoint purely spatial branch-zones of the tree. But this geometric tree-like disposition is only a globalized and residual purely spatial *trace* of the factual space-time emergence of the successive effects of the individual genetic operations that have generated the tree. The temporal aspects that, in their actuality, had individually and successively contributed to the globalized spatial splitting in distinct branches of the tree, have now evaporated in – literally – 'the air of time'. While the global final probabilistic description (9''') – considered by itself, separated from its genesis imposed by human aims and ways of thinking and by the human ways and technical possibilities of acting – *is radically devoid of any* ***own*** *space-time organization*.

The description (9''') is a purely numerical drop-off from the whole physically-operational, so *the space-time* ***factual realization***, of a conceptual-methodological design for reaching a cognitive purpose: the nature of this drop-off itself, a probabilistic-statistic, is abstract, and its own final global structure is a-spatial and a-temporal.

[52] In MMS [2002B]and [2006] it has been shown that the construction of the concept of material "object" in the classical sense involves precisely assignation of an own space-time support.

[53] The mentioned extensions of the classical concept of probability are intimately connected with basic extension of also the classical *logical* conceptualization. It has been shown in MMS [2002A], [2002B], [2006] that these extensions possess a general character and they admit *a unification of the logical and the probabilistic approaches* thereby dissolving obstacles that resisted since a long time.

[54] The concept of a primordial transferred description (9''') itself, in fact *founds universally the whole human conceptualization* (MMS [2002], [2006]), the macroscopic classical one as much as the conceptualization of the microstates. The only difference with microphysics is that inside the classical domain of conceptualization the direct perceptibility of the involved physical entities permits to economize an explicit *knowledge* of this foundational fact.

Of course, scissions of the same kind appear already in any human process of construction an abstract entity or even a material one; but nowhere with these radical characters. Nowhere erected upon a strict absence of *any* ever-perceived material instance of that what is studied, even inside only memories from human minds; and notwithstanding that what is to be studied is posited to be itself of physical nature, and to exist inside space and time. These radical descriptional specificities stem from the fact that inside *IQM*, for the case of microstates, we have been coercively led to ***entirely suppress*** any *specified* assignation to the specimens of the entity-to-be-studied, of any previously well-defined properties, and to ***replace*** these by exclusively a composition of a small number of basic *classes* of only posited and named concepts (cf. for instance *MD* and the definitions from (1.I).2) of only pre-formatted conceptual receptacles where to lodge later, in a prescribed way, a *knowledge* that is left for being generated and poured into these receptacles inside a *theory* of the microstates. This entails for the global structuration of the receptacles very purified inner links and general contours that induce a definite and detailed intelligibility. Such a result cannot emerge inside a mathematical theory of the microstates because there the general conceptual-methodological-operational imperatives get mixed from the start with the features of the mathematical tools and with the particular consequences of the particular model that is working – even if only implicitly – in order to specify appropriate measurement operations and coding-procedures.

Considered now as a whole, *IQM* illustrates two essential methodological facts, and it raises a major problem of the scientific conceptualization.

The methodological facts are the following ones.

* Taking systematically into account any involved descriptional relativity *restricts*, and thereby it specifies thus entailing *precision*. This is directly opposed to the meaning of the word 'relativ*ism*'. This huge confusion should be suppressed. Descriptional relativities are organically tied with reference, and reference installs methodological specifications instead of the vagueness governed by absolutes that usually are false absolutes. Descartes' concept of system of reference has organized our thought, our power of communication; it has enhanced to an unspeakable degree our material efficiency.

* The genesis of a description is the vehicle of the semantic contents poured into that description. So explicit geneses are precious to be known explicitly.

As for the announced problem of scientific conceptualization raised by *IMQ*, it is the following one:

* What, exactly, happens at each junction – inside a given theory of the microstates – between a factual effective realization of an output of the form (9'''), and a mathematical descriptor of it? *How, exactly, can radically singular and potentially so complex conceptual-**physical** contents carried by a description of type (9'''), be pertinently* ***loaded*** into an abstract construct like Schrödinger's differential equation, or a Hilbert vector-space?

This question, when one stops on it long enough, triggers a sort of stupefaction. I think that Wigner's famous considerations on *the* "unreasonable" power of mathematics concern very precisely *this* question. One senses a void of satisfactory analysis disguised in a feeling of miracle. This sort of void should be suppressed.

Finally let us recall that *IQM* is marked by construction by two related, big, *deliberate absences*. The absence of a general model of microstate and the absence of specified coding rules for assigning meaning to the observable result of a measurement succession *[G.MesA]* from (7). These two deliberate absences are conditions of the full generality of *IQM* because any manner of compensating them can stem only from *particularizing* postulations that can be introduced only inside a given theory of microstates.

By contrast and paradoxically, these absences are what imposes with full evidence a highly non-trivial assertion: Without a *model* of a microstate that permit to conceive 'appropriate' modalities for measuring a given quantity *A* on a given sort of a microstate, and without corresponding explicit *coding procedures* for translating the observable marks produced by one act of measurement *MesA*, into a meaning in terms of a definite value $a_j$ of *A*, the primordial transferred descriptions (9''') are just a heap of inert puppets. The necessary and sufficient strings that can bring these puppets to work in a controlled way and to create effective knowledge on microstates consist precisely of a general model of the concept of microstate and particular models drawn from this that permit to identify measurement-interactions of which the observable results can be *intelligibly* coded in terms of a definite value of the measured quantity.

Let us conclude.

Out of nearly a nothingness of explicit previously available knowledge on how consensual, predictive and verifiable knowledge on microstates can emerge, we have drawn an explicit methodological-conceptual-operational construct – the Infra-(Quantum Mechanics) *IQM* – where should be embeddable any acceptable theory of microstates. This construct has been endowed with a *formalized* though qualitative structure *tied* step by step with specifications of a *semantic* nature, in this sense that the whole construct '*IQM*' consists of a composition of void semantic moulds for lodging in each one of these a more specified content of the same semantic nature as itself.

While the Hilbert-Dirac formulation of the nowadays quantum mechanics raises problems of interpretation since decades, the Infra-(Quantum Mechanics) *IQM* symbolized above – ***by itself*** – seems already intelligible by construction, and even works. For instance, it elucidates already the endlessly discussed question of the 'primordial' or 'essential' statistical character of the modern microphysics, and "the locality problem" disappears. So when *IQM* will be compared with quantum mechanics, the confrontation will reveal differences, and thereby the comparison will act like a machine that produces guides for constructing a coherent and intelligible mathematical theory of microstates: A whole set of *referred criteri*a will be at work to help to reach this purpose.

Reference is a very powerful instrument, and *IQM* offers an organized recourse to reference. In its essence *IQM is* just organized reference, nothing more. However it is a whole coherent structure of elements of reference, namely a particular such structure that concerns specifically the generation of scientific knowledge on microstates.

But *IQM* has been organized inside the general Method of Relativized Conceptualization (MMS [2002A], [2002B], [2006]) and thereby it opens up a perspective that largely exceeds microphysics and traces on the horizon the path along which can be realized a deep-set methodological unification of the modern Physics, founded upon the distribution, inside this vast domain, of the various involved human cognitive situations. And furthermore – as it will appear at the end of this work – beneath the contours of the domain of modern Physics and far beyond them – one can discern lines that draw out a synthetic perception of the ***genetic*** *unity* of the whole of the human Science of what we call 'reality'.

**SYMBOLICALLY EXPRESSED SYNTHESIS OF *'IQM'*:**

***Operation of generation G of one factually defined microstate $ms_G$:***

$$\textbf{(1)}\ G \leftrightarrow ms_G, \quad ms_G \equiv \{\sigma(ms_G)$$

***One qualification of a microstate $ms_G$ by a value $a_j$ of a measured qualifying quantity A:***

**(2)** $[(G \rightarrow ms_G)$**.** $(MesA \rightarrow$ (group of observable marks $\{\mu\}_{kA}$ coding for one value $a_j$ of $A)]$, $k_A=1,2,...m_A$, $j=1,2..J$,

In short **(3)** $[G.MesA] \rightarrow (\{\mu\}_{kA} \approx a_j)$, $k_A=1,2,...m_A$, $j=1,2....J$, $A$ given

***The definitions from* (2.I).1 *of the main types of microstates $ms_G$***

***The factual predictive $(\varepsilon,\delta,N_0)$-probability law on the statistic of outcomes of A-measurements on the microstate $ms_G$, so inside the epistemic referential (G,A):***

$$\textbf{(5)}\ (\varepsilon,\delta,N_0)\text{-}\{(\pi(a_j)\}, \forall j)\}_G, \qquad (G,A)$$

**** The set of all the factual predictive $(\varepsilon,\delta,N_0)$-probability laws (5) for one given microstate $ms_G$, so 'description ofms$_G$ inside the epistemic referential $(G,V_M)$:***

$$\textbf{(5')}\ \{(\varepsilon,\delta,N_0)\text{-}\{(\pi(a_j)\}, \forall j)\}_G\}, \qquad (G,V_M)$$

**** The 'genetic triad' of one factual $(\varepsilon,\delta,N_0)$-probability law (5)***

$$\textbf{(6)}\ (G, ms_G, A), \quad (G,A)$$

**** The 'genesis' of one given law* (5)**

The set of successions of operations

$$\textbf{(7)}\ \{[G.MesA]\}), \qquad (G,A)$$

**** The brute observable output of the genesis {[G.MesA]} of one law (5)***

$$\{\{\mu_{kA}\}, \ k_A=1,2,...m_A, \quad (G,A)$$

**** The brute observable output of all the geneses {[G.MesA]} of all the laws (5'):***

The set of all the factual data produced by (5')

$$\textbf{(8)}\ \{\{\mu_{kA}\}, \ k_A=1,2,...m_A, \qquad (G,V_M)$$

**** Re-notation: 'primordial transferred description of $ms_G$ with respect to the mechanical qualification A, so inside the epistemic referential (G,A)'***

$$\textbf{(9)}\ \{(\varepsilon,\delta,N_0)\text{-}\pi(a_j)\}_G \approx D/A)(ms_G), \qquad (G,A)$$

**** Re-notation: the primordial transferred mechanical description of the microstate $ms_G$, so inside the epistemic referential $(G,V_M)$'***

$$\textbf{(9')}\ \{\{(\varepsilon,\delta,N_0)\text{-}\pi(a_j)\}_G \approx D_M(ms_G), \qquad (G,V_M)$$

**** Genetic symbolizations of the two sorts of primordial transferred descriptions:***

$$\textbf{(10)}\ (D/A)(G,ms_G, A), \ (G,A) \qquad \text{or} \qquad D_M(G,ms_G, V_M), \ (G,V_M)$$

**** The meta-probabilistic correlations $(M\pi c(G))_{AB}$ involved by (1) $G \leftrightarrow ms_G$ with respect to the pair (A,B) of qualifying quantities:***

$$\textbf{(11)}\ \pi(a_j)=\boldsymbol{F_{aj,B}}\{\pi(b_r), \forall r\}_G$$

$$\textbf{(11')}\ \boldsymbol{F_{AB}}(G)= \boldsymbol{\{F_{aj,B}}\{\pi(a_j), \forall j\}_G$$

where $\boldsymbol{F_{aj,B}}\{\pi(a_j), \forall j\}_G$ and $\boldsymbol{F_{AB}}(G)$ are two functionals that represent, respectively, the *individual* probability $\pi(a_j)$ in terms of the *whole* probability law $\{\pi(a_j), \forall j\}_G$ and the global correlation between the two whole laws $\{\pi(a_j), \forall j\}_G$ and $\{\pi(b_r), \forall r\}_G$.

**** The description (9') completed by (11), (11'):***

$$\textbf{(9'')}\ D_M(ms_G) \equiv \{[(\varepsilon,\delta,N_0)\text{-}\{\pi(a_j), \forall j\}_G, (M\pi c(G))_{AB}\,], \ \forall A, \forall AB \in V_M, \ (G,V_M)$$

**** Qualitative logical specification on the individual probabilistic predictions on one microstate with one micro-system and with composed operation of generation :***

$$\textbf{(12)}\ \pi_{12}(a_j)_{\boldsymbol{G}(G1,G2)} \neq \pi_1(a_j)_{G1} + \pi_2(a_j)_{G2}$$

where all the probabilities are *individual.*

**** Absorption in the operation of generation, of the evolution of a microstate $ms_G$ :***

$$\textbf{(13)}\ G_o, \ G_1=F(G,EC,(t_1\text{-}t_o)),..,G_k=F(G_o,EC,(t_k\text{-}t_{k-1})),..., G_f=F(G_o,EC,(t_f\text{-}t_{K-1})), \ k=1,2..K$$

$$\text{in short } \textbf{(13')}\ [G_o.(t\text{-}t_o)] \equiv G_t$$

**** Consequence of (13') on the transferred description (9''):$D_M(ms_{Gt})$:***

$$\textbf{(9''')}\ D_M(ms_{Gt}) \equiv \{ (D/A(ms_{Gt}), (M\pi c(G_t))_{XY} \}, \ (G_t,V_M)$$

# CONCLUSION ON THE PART I

When one watches the way in which *IQM* emerges, the naïvely realistic view that scientific knowledge is 'discovery' of pre-existing 'truth' collapses into dust. And in its place one sees, one *feels* in what a sense conceptual-operational *procedures* – involving physical operations or abstract ones – can progressively be assembled into *a method* born from the unlimited human curiosity and inventiveness, from the constraints imposed by the human ways of thinking and of acting upon what we call physical reality, and from explicit *purposes* chosen by men. What has been obtained here is such a particular piece of method. It is a global coherent piece of method for constructing a definite particular piece of procedural *scientific* knowledge directed by a definite project. It is not in the least a discovery of pre-existing 'intrinsic truths' about how physical reality *is*, absolutely, 'intrinsically', 'in itself'; not even is it – in the least – a way of 'approaching' such a discovery. Such discoveries, such asymptotic reaching, are mere illusion; just an emanation from the self-contradicting notion of *'scientific knowledge of reality-in itself'*; a genuine Fata Morgana, the original sin of scientific thought[55].

We are trapped in a cage where 'absolute intrinsic truth' is irrepressibly felt to pre-exist but constantly stays out of reach, change of direction, negate any definitive convergence, mark new starts, unpredictably, frustratingly, definitively hidden beyond a non-organized and changing swarm of lures toward ill-defined targets. The hope for final intrinsic scientific truths unavoidably entails assaults by a feeling of impotence, of inefficiency, of enslavement.

I perceive only one attitude that preserves from this sort of major fail: To realize fully that *a posit of **existence** of a physical reality, and **consensual knowledge** of 'how it truly is', are of **different essences***; that an absolute bare existence of 'reality' can be posited, but – as such, as exclusively a posit of existence – it is definitively imprisoned in metaphysics, inaccessible to *consensual* knowledge, notwithstanding that in the absence of this posit of pure existence "science" would seem to be just a game.

While inside science, with a blindfold deliberately fixed on our metaphysical eye and on the basis of entirely declared posits – metaphysical or not – and *data*, to *construct* consensual, predictive and verifiable knowledge, humbly, hypothetically, *relatively*, respecting step by step the unavoidable constraints as well as the deliberately chosen ones; and to construct *from the maximal possible depth, upward*. Thereby only restricted, finite and methodized knowledge can emerge; but a fully definite and consensual knowledge endowed with *an entirely exposed genesis* where the unending inflow of relative *meaning* can be watched and is constantly left open to return and to indefinite optimization, precisely because it is only hypothetical and finite and relative.

---

[55] MMS [2006], pp. 127-136. Human knowledge by itself is not a sin; but the posit that the scientific, consensual, predictive and verifiable knowledge is "discovered", not deliberately *constructed* accordingly to methodological-factual laws, is a huge sin because it imprisons the mind and exhausts it in vain goals.

# PART II

## CRITICAL-CONSTRUCTIVE PRELIMINARY GLOBAL EXAMINATION OF THE HILBERT-DIRAC QUANTUM MECHANICS, BY REFFERENCE TO *IQM*

# INTRODUCTION TO THE PART II

The second part of this work is devoted to a global preliminary examination of the Hilbert-Dirac formulation of Quantum Mechanics *$QM_{HD}$*, by reference to *IQM*.

*Throughout what follows $QM_{HD}$ is supposed to be well known.*

The main goals of the Part II are:

* By use of *IQM* to yield, from the outside of *$QM_{HD}$*, a critical perspective on the general structural features of *$QM_{HD}$*, and to *identify the model of a microstate that certainly does somehow work inside this theory* since in the absence of *any* model this theory would have been impossible.

* To establish explicitly all the *immediate* clarifications induced by reference to *IQM*.

* To identify the precise reason why the theory of measurements from *$QM_{HD}$* raises so stubbornly an unending variety of problems since soon a whole century.

The goals mentioned above are pursued inside two chapters, 5.II and 6.II.

- The chapter 5.II is devoted to a brief comparison between *$QM_{HD}$* and *IQM*.

- The chapter 6.II is devoted to critical and constructive clarifications, namely:

-- *Identification of the model of a microstate that works inside $QM_{HD}$*: an essential step.

-- Introduction of an explicit general use of the operations *G* of generation of a microstate in all the main mathematical writings from *$QM_{HD}$*.

-- Incorporation of the model of a microstate that works inside *$QM_{HD}$*, to the concept of operation *G* of generation;

-- An explicit *refusal* of the von Neumann representation of the quantum measurements.

-- Extraction of the essence of the *$QM_{HD}$* representation of measurements and a thorough critical identification of its implications.

Starting from the section (6.II)2 of 6.II – we shall begin by simply juxtaposing *IQM* to *$QM_{HD}$*. The simplistically enriched provisional framework obtained in this way will be denoted *[IMQ-$QM_{HD}$]*. The initial bare adjunction of *IQM* as just a reference-structure will already permit clarifications and modifications. These, progressively, will entail a fusion of *$QM_{HD}$* with *IQM*. So here we start an organic process of conceptual growth. At the end of the chapter 6.II this process will have brought us on the edge of the construction of a second quantum mechanics inside a new framework that will be re-noted *[IMQ↔$QM_{HD}$]* and that will then act and grow still more efficiently throughout the last Part III.

# 5.II [56]

# COMPARISON BETWEEN *$QM_{HD}$* AND *IQM* AND THE PLAN OF THE PART II

As already stated, $QM_{HD}$ is considered to be known. But for self-sufficiency and commodity we recall telegraphically the core-features of both representations to be compared.

## (5.II).1. THE $QM_{HD}$-REPRESENTATION

***The basic assumptions***

The $QM_{HD}$-formalism is founded on:

° Two basic *representational definitions*, namely a rule that defines the mathematical representation of the studied microstate and a rule (of 'quantification' [57]) that defines the mathematical quantum mechanical representation of the classical mechanical qualifying quantities.

° A *mathematical* principle.

° Three measurement postulates.

° A postulate of evolution.

* ***The rule of representation of a microstate)***. At any fixed time $t$ the studied microstate is defined by a state-vector (a *ket*) $|\psi(\mathbf{r},t)>$ from a Hilbert-space $\mathcal{H}$.

* ***The rule of representation of* the qualifying mechanical quantities.** Any classical measurable mechanical quantity $A(\mathbf{r,p})$ is represented by a corresponding operator $\boldsymbol{A}$ called an *observable* that acts on the elements from $\mathcal{H}$ ; this observable is constructed from $A(\mathbf{r,p})$ as follows:

- The classical quantity 'position' $\mathbf{r}(x,y,z)$ is represented by a position-observable $\boldsymbol{R(X,Y,Z)}$ where: the classical functional form that in $\mathbf{r}(x,y,z)$ relates the symbols $x,y,z$ is conserved; $\boldsymbol{X,Y,Z}$ represent, respectively, the classical space-coordinates entailed by the chosen Cartesian referential but each one of which is posited to act operationally by multiplying what follows it, which is expressed by re-writing it also as $X.$ , $Y.$ , $Z.$
- The classical quantity of momentum $\mathbf{p}(p_x,p_y,p_z)$ is represented by the momentum-observable $\boldsymbol{P(P_x,P_y,P_z)}$ where: $\boldsymbol{P_x}=i(h/2\pi)d/dx$, $\boldsymbol{P_y}=i(h/2\pi)d/dy$, $\boldsymbol{P_z}=i(h/2\pi)d/dz$; the classical functional form that in $\mathbf{p}(p_x,p_y,p_z)$ connects $p_x$, $p_y$ and $p_z$ being conserved for connecting also $\boldsymbol{P_x}$, $\boldsymbol{P_y}$ and $\boldsymbol{P_z}$.
- The classical mechanical quantity $A(\mathbf{r,p})$ is represented by the function $\boldsymbol{A(R,P)}$ of the operators $\boldsymbol{R}$ and $\boldsymbol{P}$ constructed first with the same functional form as in $A(\mathbf{r,p})$, and then symmetrized.

* ***Born's mathematical principle of spectral decomposability***. Any ket from $\mathcal{H}$ – state-ket or eigenket – can be decomposed on the basis of eigenket $\{|u(\mathbf{r},a_j)>\}$ introduced

---

[56] This notation is to be read: chapter 5 from Part II: The numbering of the chapters *continues* the numbering from the first part.

[57] Cohen-Tannnoudji, C., Diu, B. & Laloë, F., [1973],

in $\mathcal{H}$ by any observable $\boldsymbol{A}$, which yields the expansion of $|\psi(\mathbf{r},t)>/$ corresponding to $\boldsymbol{A}$, $|\psi(\mathbf{r},t)>/\boldsymbol{A}=\sum_j c(a_j,t)|u(\mathbf{r},a_j)>, \forall j$ of $|\psi(\mathbf{r},t)>$.

* ***The three measurement postulates***

- ***Measurement postulate 1.*** An act *MesA* of measurement of the observable $\boldsymbol{A}$ yields necessarily as result an eigenvalue $a_j$ of $\boldsymbol{A}$.

- ***Measurement postulate 2 (Born's probability postulate)***. When the observable $\boldsymbol{A}$ is measured at the time $t$ the probability $\pi(t,a_j)$ of outcome of the eigenvalue $a_j$ of $\boldsymbol{A}$ is calculated via the expression $|<u(\mathbf{r},a_j)|\psi(\mathbf{r},t)>|^2$ where $|u(\mathbf{r},a_j)>$ is the eigenket of $a_j$ and $|<u(\mathbf{r},a_j)|\psi(\mathbf{r},t)>|^2\equiv|c(a_j,t)|^2$ is the squared value of the coefficient $c(a_j,t)$ from the expansion $|\psi(\mathbf{r},t)>/\boldsymbol{A}=\sum_j c(a_j,t)|u(\mathbf{r},a_j)>, \forall j$ of $|\psi(\mathbf{r},t)>$.

- ***Measurement postulate 3 (of projection).*** If the act of measurement *MesA* has produced the result $a_j$ then immediately after the measurement the studied micro-*state* is represented by the re-normed projection

$$P_n|\psi(\mathbf{r},t)> / \sqrt{}|<\psi(\mathbf{r},t)|\psi(\mathbf{r},t)>|$$

of $|\psi(\mathbf{r},t)>$ on the direction in $\mathcal{H}$ of the eigenket $|u_{aj}(\mathbf{r})>$ of $a_j$.

* ***The postulate of evolution***: The evolution of the state-ket $|\psi(\mathbf{r},t)>$ is defined by the Schrödinger equation 'of the problem'

$$i(h/2\pi)d/dt\,|\psi(\mathbf{r},t)> = \boldsymbol{H}(t)$$

where $\boldsymbol{H}(t)$ is the hamiltonian observable that represents the total energy assigned to the studied microstate.

***The main algorithms***

The essence of the way of working of the $QM_{HD}$-formalism can be regarded to consist of four *purely formal* types of problems and of the correlative algorithmic procedures for obtaining the solution, and a fifth ***factual***-formal problem with its own solution:

***Problem 1***: Determine the state-ket $|\psi(\mathbf{r},t)>$ that represents the microstate to be studied inside the Hilbert space $\mathcal{H}$ assigned to this microstate.

*Solution to problem 1*: Write the Schrödinger equation of the problem; solve it; *introduce the limiting conditions* in order to identify the initial state-ket $|\psi(t_o)>$. Therefrom the Schrödinger equation is asserted to determine the state-ket $|\psi(t)>$ of the studied microstate for any time $t$.

***Problem 2***. For any state-ket $|\psi(\mathbf{r},t)>$ and any observable $\boldsymbol{A}$, determine the predictive probability law concerning the possible outcomes of measurements *MesA* performed on the studied microstate.

*Solution to the problem 2*:

Write the equation $\boldsymbol{A}|u(\mathbf{r},a_j)>=a_j|u(\mathbf{r},a_j)>, \forall j$, and calculate from it the basis of eigenket [58] $\{|u(\mathbf{r},a_j)>\}$ introduced in $\mathcal{H}$ by $\boldsymbol{A}$. Each eigenvalue $a_j$ of the quantum mechanical observable $\boldsymbol{A}$ is tied in this equation to a corresponding eigenket $|u(\mathbf{r},a_j)>$. According to the measurement-postulate 1, the result of any act of measurement *MesA* is necessarily an eigenvalue $a_j$.

In order to determine the probability of outcome of any *given* value $a_j$, form the set of squared absolute values $|c(a_j,t)|^2, \forall j$, drawn from the expansion $|\psi\, \mathbf{r},t)>/\boldsymbol{A}$ on the basis $\{|u(\mathbf{r},a_j)>\}$ of the eigenket of $\boldsymbol{A}$, and accordingly to the measurement-postulate 2 write Born's predictive probability law $\{\pi(a_j) \equiv |c_j|^2\}, \forall j$.

[58] As usual we write 'ket' without plural.

***Problem 3***. Specify the way in which you can transform the representation of the studied microstate in $\mathcal{H}$ relatively to $\boldsymbol{A}$, into the representation in $\mathcal{H}$ of the same microstate but relatively to another observable $\boldsymbol{B}$ with eigenvalues $b_k$ and eigenvectors $|v(\boldsymbol{r},b_k)>$, $\forall k$, $k=1,2,....K$.

*Solution to the problem 3*: Apply Dirac's 'theory of transformations':
For *any* given value of the index $k$ we have inside $QM_{HD}$ that

$$<v_k(\boldsymbol{r},b_k)\,|\,\psi_{G,\boldsymbol{H}}(\boldsymbol{r},t)> \;=\; e^{i\gamma(\boldsymbol{B},k)}\;|d_k(b_k,t)| = \textstyle\sum_j \tau_{kj}(\boldsymbol{A},\boldsymbol{B})\;c_j(t,a_j)\;,\quad \forall j,\ \forall t$$

where $\tau_{kj}(\boldsymbol{A},\boldsymbol{B})=<v_k|u_j>$, $\forall j$. So for any complex factor of given index $k$ there is a separate condition

$$e^{i\gamma(\boldsymbol{B},k)} = <v_k|\,\psi_G(t)>\,/\,|\,d_k(t,b_k)| \;=\; \textstyle\sum_j \tau_{kj}(\boldsymbol{A},\boldsymbol{B})\;c_j(t,a_j)\;/|\,d_k(t,b_k)|\,,\quad \forall \boldsymbol{A},\boldsymbol{B},\quad \forall t$$

(where ' / ' is to be read: *divided by*) so that

$d(b_k,t)=|v(\boldsymbol{r},b_k)>$ and $|\,\psi(\boldsymbol{r},t)>/\boldsymbol{B}=\sum_j \tau_{bk,aj}\;c(a_j,t)$ where $\tau_{bk,aj}=<v(\boldsymbol{r},b_k)|u(\boldsymbol{r},a_j)>$, $\forall j$, $\forall k$

***Problem 4***. Represent mathematically the measurement processes by which is verified the predictive probability law $\{\pi(t,a_j)\equiv|c(a_j,t)|^2\}$, $\forall j$, drawn from $|\,\psi(x,t)>/\boldsymbol{A}$, $\forall \boldsymbol{A}$.

*Solution to the problem 4*: Apply 'the quantum theory of measurement'.

***Problem 5***: Verify the statistical predictions of the formalism.

*Solution to the problem 5*: Accordingly to the quantum theory of measurements, '*prepare the measurement-evolution state-ket*' and operate the verification-measurements.

In the chapter 6.II it will appear that concerning this point 5 nothing is clearly specified.

- The term 'prepare' creates much confusion. Some authors seem to consider that the state-ket has to be prepared (*it* has to be replaced by eigenket of the measured observable, in connection with the measurement postulates 1 and 3); other authors, more numerous, seem to consider that the *microstate* has to be 'prepared' (or to be *also* 'prepared' in the sense that it has to be *physically* transformed in eigenstates, but it is not explained how, nor why); and still other authors consider that what is 'prepared' is the physical measurement operation (which – interestingly – would indicate a coding procedure), but once more this procedure is not analysed in general terms, nor referred to the sort of considered microstate. Anyhow:

- *The coding problem is not formulated* ***as such****, in general terms, nor, a fortiori, treated explicitly*; only two much discussed examples are given, for the cases of momentum-measurements or of spin-measurements, that seem to be considered to be valid quite generally, for any observable and any sort of microstate in the sense of (2.I).1, though when examined closely they raise questions and moreover they appear to possibly be valid only for microstates without quantum fields.

- The *factual and conceptual connections with the problem 4* are not worked out.

All this – as a whole – is what is called 'the measurement problem'. This problem is central. Here it is only recalled. In the chapter 6.II it will be thoroughly defined and examined.

## (5.II).2. THE *IQM* WAY OF REPRESENTING A MICROSTATE

This representation consists of the whole Part I of the present work where it stays available. It will act with all its features throughout what follows.

### (5.II).3. THE COMPARISON

When the two representations *IMQ* and $QM_{HD}$ are compared, the most striking conclusions are the following ones.

***Characterization of $QM_{HD}$.***

$QM_{HD}$ is expressed mathematically via continuous unlimited mathematical analysis or algebra that allow continuity and infinities. All that is explicitly ruled is defined in purely mathematical and algorithmic ways. The main descriptional element is the concept of a state-ket $|\psi>$ from the Hilbert-space assigned to the studied microstate. This state-ket is obtained exclusively via mathematical procedures and the statistical-probabilistic predictions from $QM_{HD}$ are generated *directly* and exclusively by $|\psi>$, so in an abstract mathematical way.

> With the unique exception of the measurement-postulate 1, *no* ***representations*** *of individual and factual entities or events or procedures do come in*, neither mathematically expressed ones, nor only qualitatively defined ones.

Even the concept of 'micro-*state*' – that is precisely what the whole formalism is implicitly asserted to 'represent' – is left devoid of a clearly stated definition, and even devoid of merely a specific symbolization. In the current way of speaking where the theory is imbedded this most basic concept is ambiguously indicated by both the words 'system' and 'particle'. A fortiori there is no defined concept of operation of generation of a microstate, and no model of a microstate is specified, nor legally accepted to be necessary. The coding problem, as stated above, is not declared and so it is not treated overtly.

> The representation of measurements remains an abstract problem that is dealt with via postulations and *statistical* mathematical algorithms, even though an act of measurement has quintessentially an individual nature.

This situation raises problems since decades. In such conditions there is no intelligibility. Not even only the statistical representation itself is intelligible. Indeed:

*The basic act of 'giving' the initial statistical state-vector $|\psi(t_o)>$, is not* ***generally*** *realizable.* This deserves being stressed: Let us suppose that it is possible to write down the Schrödinger equation for any factual situation and to calculate the general solution corresponding general solution (a very overrating assumption). The concept of 'initial limiting conditions' for 'determining' the initial state-ket $|\psi(t_o)>$ is used in a purely mathematical sense that should be clearly distinguished from a requirement to specify *all* the significant initial factual data. And a priori it seems very unlikely that these factual data be specifiable mathematically in *any* experimentally realizable situation (even if the basic sine qua non specification is reducible to only the statistical position-density at the initial moment at every point from the considered volume of space). Thereby already – and *in principle* – the domain of *rigorous* applicability of $QM_{HD}$-formalism is restricted a priori to the domain of physical problems that do permit to produce a mathematical expression of the initial state-ket. When the considered problem lies outside this domain the treatments are systematically marked by approximations (parcelling of the spatial support of the global phenomenon that is studied, more or less arbitrarily simplified mathematical translations of just imagined and posited micro-facts, etc.) of which the effects or absence of effects remain out of control.

Any predictive statistic, by definition, is referred both *mentally* and *factually* to *individual* things, operations, concepts, which are abstracted away from the statistical representation, but that – genetically – have determined it; and these – unavoidably – have to act again if one wants to verify the predictive statistic. In this sense a predictive statistic, and in particular the initial one, is not a self-sufficient concept, it cannot be represented *directly* and *exclusively* by itself, it has to be constructible factually from individual measurement operations. But this is not possible when a general and systematic treatment of any individual concept and any individual operation, is lacking.

Inside $QM_{HD}$ there remains a gap of possibility of a factually founded representation of the initial statistics involved by the initial state-ket $|\psi(t_o)>$, where arbitrary and approximation rush in uncontrollably.

***Characterization of IQM***

Inside *IQM* the descriptions (9), (9'), (9''), (9''') are *directly rooted into the factual microscopic a-conceptual physical reality* and are constructed out of this on the basis of individual definitions of concepts and of basic *physical* individual operations, via *factual*-conceptual procedures or conceptual-methodological posits declared as such.

Inside *IQM* the concept of an *individual* and *physical* operation *G* of generation of a microstate manifests a quite determining role, namely via (1) it *leads* to:

- The classification of the sorts of operation of generation (simple or composed, and actual or (in the case of bound states) revolved inside the past of a natural physical genesis.
- In consequence of the posit $ms_G \equiv \{\sigma(ms_G)\}$ the operation of generation *G* entails also the classification of the microstates defined in (2.I).1.
- Inside *IQM* the specification of a model of a microstate and of coding procedures for each sort of possible microstate and of possible qualifying quantity, have appeared as basic necessities for any theory of microstates.

*The whole statistical conceptualization from IQM* ***follows*** *from the individual conceptualization*, with distinct progressive stages. The basic tree-like structures from the figures 2, 3 and 4 that summarize graphically the whole *IQM*, stem all from *one* operation of generation, they continue with individual acts of measurement, these lead to factually constructed probability spaces of first order, and these spaces are mutually correlated on a meta-level of probabilistic qualifications of second order.

***The comparison***

$QM_{HD}$ contains no explicit representation of practically *none* of *all* the individual physical operations, concepts and entities that inside the reference-structure *IQM* have been shown to be basically necessary for an intelligible theory of microstates. The set of concepts

[*G*, $ms_G$, 'general model of a microstate', 'individual succession of operations *[G.MesA]*', coding procedures for translating the observable physical marks produced by *one* succession *[G.MesA]* in terms of *one* definite value $a_j$ of the measured quantity *A*],

*all* these individual descriptional elements that in our usual processes of *thought* irrepressibly come first, inside $QM_{HD}$ are devoid of any formal representation, of any only qualitative *definition*, even of any mere symbolization: The top-down historical approach did not reach the depths where they are placed, nor did it follow the right direction for reaching them, and we failed to notice this because we were advancing backwards toward the microscopic factuality, with our eyes fixed upon the classical level of conceptualization.

Whereas *IQM* draws bottom-up a statistical conceptualization *from* an individual one that is rooted into a-conceptual microphysical factuality, the $QM_{HD}$ top-down conceptualization has been formulated ***directly*** *in statistical terms* founded on conceptual extrapolations and on postulates and mathematical algorithms; and beneath this directly statistical conceptualization there remained a ***VOID*** of individual conceptualization. The Fig. 6 represents this situation.

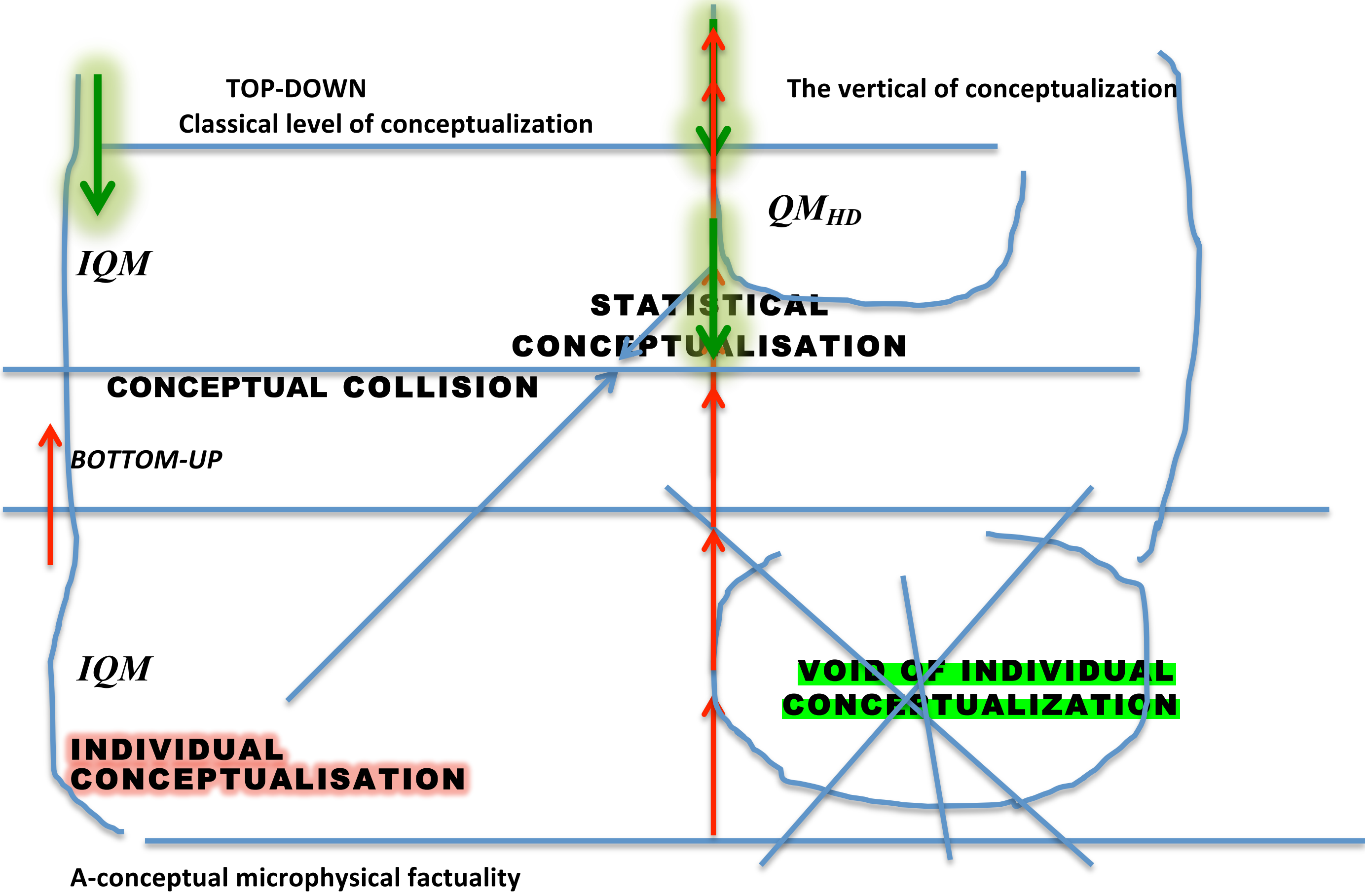

Fig 6. Comparison between $QM_{HD}$ and *IQM*

We also reproduce beneath the Fig. 5, for comparison:

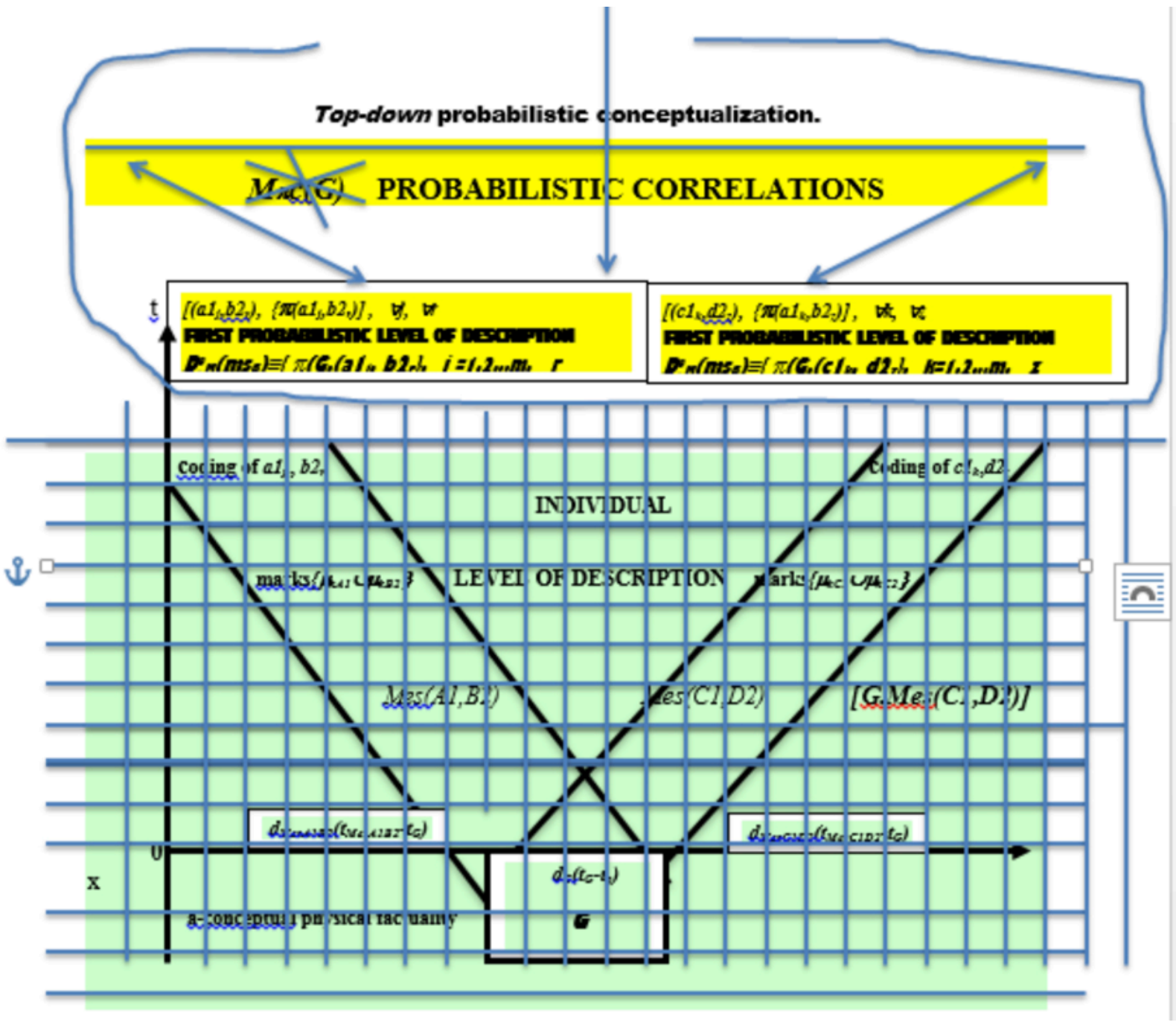

Fig. 5 (*A probability-tree T(G(2S),(A1,B2; C1,D2)) as encountered by a top-down approach that stops at the level of the probability spaces*).

This antagonism has settled in because in a top-down approach toward the development of a microphysics the statistical surface of representations of individual factual entities – micro-phenomena, operations – appeared *first*. Unavoidably it has constituted the first phase of the investigation. In this first phase Bohr's postulates acted as an efficient substitute for a general model of a microstate because at that time the bound states were at the core of all the attentions, and bound microstates – conceivable as spatially delimited and pre-existing entities – are much alike to classical 'objects'; moreover they permit measurements via interaction-'effects' with macroscopic fields (Stark, Zeeman), which eliminates major specificities of the quantum measurements. So all the main specificities of a factually founded representation of microscopic entities that are not directly perceivable – *a new cognitive situation* – remained hidden.

As for de Broglie's 'individual' model – that in fact has *not* been physically-operationally individualized by him (he individualized it only mentally, and in a non-stabilized way) – *it emerged some 25 years* ***after*** *Bohr's initial work*, in connection with also unbound microstates. And it emerged precisely at the time when Heisenberg was developing his new top-down offensive of mathematical representation by matrixes that was still taking the bound microstates into dominant consideration. So de Broglie's individual model, together with the corresponding Schrödinger equation, had to confront the already organized and solidly installed 'positivistic' attitude of the Copenhagen school, deliberately optimized for statistical predictions tied with bound microstates and measurements on these. The Copenhagen top-down approach opposed de Broglie's model under the protection of strong socio-psychological inertial forces. And so – up to this very day – *nowhere inside $QM_{HD}$ does one find a clear* ***distinction*** *between individual and statistical representations*.

So when *IQM* and $QM_{HD}$ are brought together like in the figures 5 and 6 there emerge anachronistic antagonisms and the Hilbert-Dirac statistical formulation $QM_{HD}$ appears as a conceptual bas-relief of which the surface is very finely crafted but of which the basic underlying forms are simply *undone*, mere potentialities still lost in an amorphous substratum.

# 6.II

# BASIC CLARIFICATIONS:
# * A GENERAL MODEL OF A MICROSTATE,
# * USEFULNESS OF '*G*',
# * REFUSAL OF:
# - VON NEUMANN'S REPRESENTATION OF QUANTUM MEASUREMENTS
# - THE WHOLE $QM_{HD}$ REPRESENTATION OF QUANTUM MEASUREMENTS

In what follows the local problems or insufficiencies from $QM_{HD}$ will constantly be referred to *IQM*. Therefore we introduce for our framework the symbol $[IQM\text{-}QM_{HD}]$. But to begin with, *IQM* and $QM_{HD}$ will act as two separate structures. The reference to *IQM*, however, will bring forth several basic clarifications that initiate a process of fusion.

## (6.II).1. THE $[IQM\text{-}QM_{HD}]$ MEANING OF AN EIGENFUNCTION OF AN OBSERVABLE AND CONSEQUENCES

***Digging out the detailed meaning of an eigenfunction***. Let us place ourselves inside $QM_{HD}$. Consider the equation $\boldsymbol{A}|u_j(\mathbf{r},a_j)\rangle = a_j|u_j(\mathbf{r},a_j)\rangle$, $j=1,2,...J$ $(\forall j)$, with $J$ finite[59] that determines the eigenfunctions $\{u_j(\mathbf{r},a_j)\}$ from the basis of eigenket introduced by $\boldsymbol{A}$ in the Hilbert space $\mathcal{H}$ of the studied microstate. In general such an eigenfunction is not square integrable. This is considered to be a 'problem', in the following sense. A state-function $\psi(\mathbf{r},t)$ from a state-ket $|\psi(\mathbf{r},t)\rangle$ is *required* to be square-integrable, since it represents a set of distributions of probability. But an eigenfunction in general is not square-integrable and furthermore it is *not required such*. Why, exactly, is that so*?* That is the 'problem'.

Bohm ((1954) p. 210-211) writes:

> « …We obtain $\psi = e^{ipx/\hbar}$…… Strictly speaking, the above eigenfunctions cannot, in general, be normalized to unity...Let us recall, however, ...that in any real problem the wave function must take the form of a packet, since the 'particle' is known to exist somewhere within a definite region, such as in the space surrounded by the apparatus. To obtain a bound and therefore normalizable packet, we can integrate over momenta with an appropriate weighing factor.»

So Bohm adopts an exclusively mathematical point of view. Not a moment does he focus upon the involved meaning. He does not even make use of a specific notation for distinguishing between eigenfunction and state-function. And in order to deal with the mathematical situation he accepts approximations without any hesitation, notwithstanding that the considered question seems to be a question of principle.

The same attitude is usually found in the textbooks.

Dirac (1958, p. 48), on the contrary, writes:

---

[59] From now on any index is posited finite and any finite spectrum is denoted by a notation of the type $\forall j$.

> « It may be that the infinite length of the ket-vectors corresponding to these eigenstates is connected with their unrealizability, and that all realizable states correspond to ket vectors that can be normalized so that they form a Hilbert space ».

(« connected with their 'unrealizability' » suggests that Dirac, in a certain subliminal way, perceived the absence – and the *utility* – of a representation of the way of generating a microstate, as well as the possibility of some specific significance of the non-integrability, in general, of the eigenfunctions).

As for the most outstanding didactic exposition of $QM_{HD}$, that by CTDL[60]*,* it proposes *«a physical solution to the difficulties»* (proposed also by Bohm (1954, p. 212), namely to replace the eigenfunction by a $\delta$-distribution centred upon the corresponding eigenvalue.

*Nobody – as far as I know – has conceived* ***that an eigenket might simply not represent a 'state'****.*

However recourse to history reveals that the 'problem' of non-integrability of an eigenfunction is a *false* problem because the concept of eigenfunction has a specific *meaning* that is radically different from that of a state-function. So the problem is not mathematical, it is conceptual. Indeed the meaning of an eigenket stems from Louis de Broglie's Thesis (1924, 1963). Louis de Broglie has derived his famous relation $p=h/\lambda$ from his well-known model of a microstate, (erroneously named the wave-*'particle'* model). The model itself stems from the usage made of Fourier decompositions inside classical electromagnetism. In a Fourier decomposition of an electromagnetic wave each constant value $\lambda$ of a monochromatic wavelength is associated with a corresponding plane wave. By analogy, to each value $p_{xj}$ of the classical mechanical fundamental quantity of momentum $p_x$ of an unbound electron, de Broglie has associated a plane wave with a *'corpuscular* ***phase****-function'* $\Phi(x,t)=ae^{(i/\hbar)\beta(x,t)}$ where '*a*' denotes an arbitrary and *constant* amplitude of vibration and the 'corpuscular phase' is written as $\beta(x,t)=(Wt-p_{xj}.x)$ where $W=m_oc^2/\sqrt{1-v^2/c^2}$ is the – relativistic – energy of the *'corpuscular-***like aspect** *of the corpuscular wave'* while $p_{xj}$ denotes the constant value posited for the momentum *of this 'corpuscular-like aspect'* (in one spatial dimension) **[61]**.

The 'corpuscular-like aspect of the corpuscular wave' remained devoid of representation inside the mathematical expressions that Louis de Broglie associated to his model. This has been a huge strategic error because in mathematical physics only what possesses a definite mathematical expression does subsist in the minds. The rest does not strike sufficiently numerous attentions and so at last it evaporates into the air of history. But verbally, de Broglie has clearly specified in his writings that he conceived the 'corpuscular aspect' to consist of a *singularity in the amplitude of the corpuscular wave*. Namely a very localized space-domain where this amplitude is so much bigger than its surrounding values that it concentrates in it practically the whole energy $W=mc^2/\sqrt{1-v^2/c^2}$ of vibration of the corpuscular-*like* singularity in the amplitude of the 'corpuscular ***wave***'. This singularity was posited to *glide* inside the wave "like a small classical mobile" that – in consequence of its strong spatial localization and its relatively very high energy – admits at any time the '*mechanical*' qualifications of position and momentum, from which in classical mechanics all the other mechanical qualifications can be constructed; whereas the rest of the wave, of course, does *not* accept mechanical qualifications.

---

[60] Cohen-Tannnoudji C., Diu B. and Laloë, F., 1973.

[61] De Broglie wrote in one dimension; here we introduce the notations '$\Phi$' and '$\beta$' in order to distinguish from the start the representation of a physical phase of a physical wave introduced by Louis de Broglie, from the phase $\varphi(x,t)$ of a mathematical 'state-function' $\psi(r,t)= ae^{(i/\hbar)\varphi(x,t)}$ introduced inside a $QM_{HD}$ state-ket $|\psi(x,t)>$ that represents a formal tool for statistical predictions on results of measurements on a microstate.

In short, *de Broglie's model does not introduce any 'particle' whatsoever*; it introduces a 'corpuscular-like wave', a wave inside which a very localized singularity of the amplitude admits mechanical qualifications that require a *wave-'mechanics'*.

In the course of the construction of the relation $p=h/\lambda$ de Broglie has proved the 'theorem of concordance of the phases' [62] according to which:

> The model of a microstate of an unbound electron can be *stable* if and only if the corpuscular-like singularity in the amplitude of its corpuscular wave glides inside the wave in a way *such* that the phase of the up-down-up vibration of the amplitude of the *localized* singularity – a *clock*-like phase in the sense of Einstein's special relativity, at any given location $\boldsymbol{r}$ – is at *any* moment $t$ identical to the phase-function $\beta(\boldsymbol{r},t)$ of the oscillation of the extended-in-space amplitude of the portion of wave that *surrounds* the singularity at that time $t$.

This condition of stability cannot be realized otherwise than via a convenient continuous displacement inside the wave, of the singularity from the amplitude of the wave, because from the standpoint of Einstein's special relativity $\beta(\boldsymbol{r},t)$ designates a *wave*-like phase and so, when one passes from one inertial observer to another one via the Lorentz-Einstein transformation of coordinates, it obeys another sort of variance then the localized clock-like phase of oscillation of the amplitude of the singularity **[63]**.

*This theorem is crucial for understanding the meaning of the $QM_{HD}$-concept of eigenket*. Indeed: Louis de Broglie's wave-function $\Phi(x,t)=a.exp^{(i/\hbar)\beta(x,t)}$ satisfies the equation $\boldsymbol{P_x}\Phi(x,t)=p_{xj}.\Phi(x,t)$ for eigenket and eigenvalues of the momentum observable from the Hilbert-Dirac formalism. And the $QM_{HD}$-equation $\boldsymbol{A}|u_j(\boldsymbol{r},a_j)>=a_j|u_j(\boldsymbol{r},a_j)>$, *($\forall j, \forall A$), generalizes* this particular mathematical fact to any quantum mechanical observable and introduces it in the bra-ket expressions of the Hilbert-Dirac formalism, *without any condition on integrability*. This leads immediately to the following identification of the general meaning of this equation:

> The eigen-function $u_j(\boldsymbol{r},a_j)$ from the eigenket $|u_j(\boldsymbol{r},a_j)>$ associated with the eigenvalue $a_j$ of the observable $\boldsymbol{A}$, plays the role of *a mathematical representation of a **sample** of a definite sort of wave-movement around the spatial location of the corpuscular-like singularity in the amplitude of the involved de Broglie 'corpuscular wave'*.

And if the wave-movement that surrounds the singularity is *constantly* represented by the eigenfunction $u_j(\boldsymbol{r},a_j)$, then – and only then – does stay constant the value $a_j$ of the mechanical quantity $A$ that qualifies in mechanical terms the displacement inside the wave, of the location of the corpuscular-like singularity from the amplitude of the wave.

As soon as this has been spelled out, it leaps to one's eyes that the form itself of the equation for eigenfunctions and eigenvalues of $\boldsymbol{A}$, simply cries it out on the roofs. So – no offense to Bohr – de Broglie's model of a microstate is quite basically present inside the whole formalism of $QM_{HD}$. It defines the physical-conceptual meaning of all the *bases* in the Hilbert-space of any microstate, as well as all the spectral decompositions of any state-ket. While furthermore these spectral decompositions are the core of the predictive formalism from $QM_{HD}$.

No more, no less.

---

[62] The conceptual content of this proof of only several lines is a jewel of human thought.

[63] We note that de Broglie's model is compatible with a relativistic model in the sense of Einstein's special relativity. Why, then, does $QM_{HD}$ as a whole oppose resistance to the realization of a general formal compatibility with Einstein's special relativity? This question should be elucidated before any attempt at unification of quantum mechanics with relativity (MMS [1994]) (the concept of 'mass' in de Broglie's sense (L. de Broglie [1956]) is *different* from Einstein's concept of mass).

The whole predictive $QM_{HD}$-algorithm is an undeclared infusion from de Broglie's model, wherefrom the physical significances are mutely drawn. Indeed in any spectral decomposition

$$|\psi(\mathbf{r},t)>/\mathbf{A}=\sum_j c(a_j,t)|u_j(\mathbf{r},a_j)>,\ \forall j$$

of a state-ket $|\psi(\mathbf{r},t)>$ with respect to the basis $\{|u_j(\mathbf{r},a_j)>\},\ \forall j$, introduced in $\mathcal{H}$ by an observable $\mathbf{A}$, the eigenket $|u_j(\mathbf{r},a_j)>$ from the term $c(a_j,t)|u_j(\mathbf{r},a_j)>$ symbolizes *the sample of that what is counted by the real squared modulus* $|c(a_j,t)|^2$ *of the complex coefficient* $c(a_j,t)$ (exactly as, in the expression *34m*, the symbol '*m*' means that the length that is measured is *34* times the length of the sample of a meter from the National Bureau of Standards of Weights and Measures). And *a sample of wave-movement – by the definition of the concept of 'sample' – has an arbitrary spatial extension*. So with respect to this significance of an eigenfunction, squared modulus integrability is simply senseless.

***Consequences of the identification of the meaning of an eigenfuction***. The preceding conclusion has noteworthy consequences.

- It evaporates the false 'problem' why an eigenfunction is in general [64] not required to be square-integrable: If it *were* required to always be square-integrable, *that* would be a real problem.

- In classical thinking a unique semantic dimension (for instance 'color') suffices for carrying *all* the 'values' ('red', green', etc.) that one wants to singularize on this dimension. But when a microstate has to be qualified it obviously is very useful – if not even necessary – to analyse the representation more, namely so as to compensate for the absence of any perception of a quale for assigning meaning to the brute result of each one act of measurement. The Hilbert-Dirac formalism realizes this analysis by a formal splitting: An observable $\mathbf{A}$ represents – separately – the considered semantic dimension – a qualifying quantity that qualifies from a mechanical point of view – and exclusively this ('a momentum' 'a total energy', etc.); it introduces a 'grid' for qualification, in the sense of (2). And on the other hand – like in a catalogue joined to $\mathbf{A}$ – inside the set of pairs $\{(|u_j(\mathbf{r},a_j)>,a_j)\},\ \forall j$, are represented separately each one of the 'values' singularized on the semantic dimension $\mathbf{A}$; and *each* 'value' is specified by a *pair* $(|u_j(\mathbf{r},a_j)>,a_j),\ \forall j,$ because the wave-movement of a corpuscular wave *and* a mechanical qualification of the 'corpuscular aspect' of that wave, are both involved, and are tied in a one-to-one connection [65]. This is marvellously expressive. And when it is discretized via an explicit adjunction of a unit for measuring the quantity *A* represented by the observable $\mathbf{A}$., it becomes also effective.

- This explains the high adequacy of the use of a Hilbert space $\mathcal{H}$ for representing mathematically the predictions on issues of measurements on a microstate: Each 'value' $a_j$ of $\mathbf{A}$, $\forall j,$ can be placed on a separate axis reserved to it, on which the state-ket $|\psi(\mathbf{r},t)>$, when projected onto that axis, determines the complex number $c(a_j,t)$, so also the probability $|c(a_j,t)|^2$ postulated by Born for the emergence of what is represented by the pair $(|u_j(\mathbf{r},a_j),\ a_j)$ if a measurement of $\mathbf{A}$ is performed upon the microstate with state-ket $|\psi(\mathbf{r},t)>$ (this mimics geometrically the expansion $|\psi(\mathbf{r},t)>/\mathbf{A}=\sum_j c(a_j,t)|u_j(\mathbf{r},a_j)>,\ \forall j$, of $|\psi(\mathbf{r},t)>$).

Thereby the representational roles of, respectively, a state-ket or an eigenket, are radically distinguished.

---

[64] In a bound state of a microsystem the eigenket of the total energy has the same mathematical expression as the state-ket; it is confounded with the state-ket and the eigenket is required to be square-integrable and it *is* such: from a conceptual point of view this is a mathematically 'degenerate' situation.

[65] Degenerate spectra are not considered here.

- The preceding remarks indicate that Dirac's 'theory of transformations' expresses in mathematical terms passages from a given *'semantic* space*'*, to another one: A considered semantic is defined formally by the corresponding pairs *(*$|u_j(\mathbf{r},a_j)>$,$a_j$*)*: *Dirac's calculus is potentially a calculus with semantic specifications* [66]. Thereby the Hilbert-Dirac formalism – in itself, *independently* of microstates and $QM_{HD}$ – can be useful in many disciplines. Precisely this is generally established by Gleason's theorem examined in (7.III).2.1.

## (6.II).2. FROM THE HIDDEN PRESENCE INSIDE $QM_{HD}$ OF de BROGLIE'S MODEL TO ITS EXPLICIT PHYSICAL-*OPERATIONAL* INCORPORATION INTO *[IQM-$QM_{HD}$]*

We have posited a framework denoted *[IQM-$QM_{HD}$]*. Inside this framework we shall now absorb de Broglie's ideal (non-operational) model of a microstate, into the physical-operational concept *G* of factual generation of the microstate-to-be-studied, as defined by the methodological decision *MD* from the Part I of this work.

What follows might be perceived as shocking after the cure of positivistic purity suffered by microphysics since nearly a century, according to which models were interdicted. But let us keep in mind that just above this interdiction has been shown to have never acted. And the announced absorption will soon appear to have been an essential step forward in the process of construction of a factually rooted second quantum mechanics. So let us vanquish the inhibitions. Everywhere in the sciences and the techniques the models are unavoidable and precious and their efficiency is increased when they are specified with detail.

***Association via G of the concept de Broglie's model, with a microstate $ms_G$***. The guiding ideas are the following ones:

***(a)*** We want to reconstruct a *mechanics* of microstates. According to de Broglie's model – that appeared to be quintessentially involved in $QM_{HD}$ – only the corpuscular aspects from a corpuscular wave do admit mechanical qualifications. Consider now the definitions from (2.I).1 of various sorts of microstates. It is clear that *what is called 'system' in (2.I).1 has to be identified with de Broglie's 'corpuscular-**like** singularity' in the amplitude of a corpuscular wave*. So we posit that:

> An operation of generation *G* of 'one micro-*state* of *one* micro-*system*' introduces ***one*** de Broglie singularity into the domain of what a human observer can qualify in a factual-operational and consensual way; whereas an operation of generation *G(ns)* (*ns*: *n* systems) of 'one micro-*state* of *n* micro-*systems*' introduces *n* de Broglie-singularities.

***(b)*** It seems obvious that it would be nonsense to conceive that an operation of generation *G* defined by the use of macroscopic apparatuses and macroscopically controlled parameters, cuts off radically the 'corpuscular-like' singularity[67] from the rest of the body of the involved specimen $\sigma(ms_G)$ of the studied microsystem $ms_G$ that has been generated by one realization of *G*; from the indefinitely extended 'corpuscular *wave*' that, before the action of *G*, incorporated the singularity and determined its displacement. Indeed de Broglie's corpuscular-like singularity is organically integrated to the 'corpuscular wave'. Outside this wave, by definition, it cannot be *conceived*. It *is* just a wave-aspect. So we *posit* that – in some sense and for some time – *G* just captures into the domain of what can be operated upon by human observers and what can produce observable marks, a portion of a corpuscular wave that carries *one* de Broglie singularity if *one* micro-state of *one* micro-system is generated, or carries *n* such singularities when *one* micro-state of *n* micro-systems is generated. While the main part of the wave-like phenomenon to which this portion of corpuscular wave *continues* to be incorporated after the action of *G*, remains in the physical substratum, even though via the microstate $ms_G$ generated by *G* it has been connected to the domain of the observable by men.

[66] In Dirac's mind might have worked implicit general criteria that he did not care to capture and communicate.
[67] We use singular expressions for the sake of simplicity.

Such a view violates the most current significance of the classical concept of 'object' *endowed with a definite spatial volume*, even if this 'object' is in a liquid or even a gaseous state. We are face to face with the modern view on the frontier between a representation of the Spinozian "Universal Substance" and a very *first* process of *individualized* human conceptualization. Indeed nowadays we are aware that any definite "entity", a chair, a living body, cannot be conceived to be cut in an absolute sense from the surrounding ‘physical waves’ that are posited to exist (electro-magnetic, gravitational, etc.). Any "object" in the classical sense is admitted to emit and to absorb waves of various natures and to be traversed by such waves.

But the *global* de Broglie model of the *whole* of the 'physical reality' (de Broglie [1956]) **[68]** does not incorporate the concept defined in *MD* of a *physical*-operational generation *G* of individualized, 'local' entities-to-be-studied. In de Broglie's global approach – like in Bohm's one – the physical-operational concept '*G*' from *IQM* is replaced by purely mental wandering focalizations of the attention, that do not entail consensual, physical-operational observable effects subjected to conditions of predictability and of verifiability. And in consequence of ***this*** it ***can*** be posited to be deterministic, like the classical disciplines of Physics **[69]**. Whereas in order to achieve *scientific* descriptions – consensually observable, predictive and verifiable – concerning consensually factually defined 'microstates', we inescapably have to make use of also local models; and these, via *MD*, introduce irrepressibly a primordial observational statistical-probabilistic character (MMS [2013]; [2017].

These considerations close the point *(b)*. We furthermore posit that:

***(c)*** The *location* of the de Broglie singularity inside the corpuscular wave of any specimen of a studied $ms_G$, in general *varies* arbitrarily from one individual specimen of $ms_G$ to another one (this is an essential element from de Broglie’s own view (de Broglie [1956]).

***(d)*** We do not try to specify other characters of the portion of a corpuscular-wave that is trapped into the domain of possibility of deliberate interaction with it, via by a repeatable operation of generation *G*. But we admit that, whatever these characters are, they constitute a class of similarities that justifies their common designation in *MD* and (1) as elements from the class $\{\sigma(ms_G)\}\equiv ms_G$ of emergences of ‘the *one* microstate $ms_G$ generated by *G*’: the language posited in this way has already shown its pertinence, so we keep it.

***(e)*** Finally, in agreement with de Broglie’s works and with those of the nowadays physicists from Bohm’s school (in particular Peter Holland 1993) *we also posit the famous guidance relation* according to which the phase of the corpuscular wave in the neighbourhood of the singularity ‘guides’ the singularity by determining its momentum.

***(f)*** Consider now a *composed* operation of generation $\mathbf{G}(G_1,G_2,...G_k)$. The methodological posit (1) requires a one-one relation between any operation of generation

---

[68] De Broglie has worked out the global De Broglie-Bohm model after Bohm's [1952]-work. This global model includes Bohm's work as well as the de Broglie's initial model of a microstate. As far as I know it is the unique global model of the 'physical substance'.

[69] Let us immediately and explicitly add that de Broglie representation of the Universal Substance does not in the least realize "knowledge of the physical reality such-as-it-is-in-itself"; it remains just a human model *relative* to the human way of perceiving and thinking from which it stems. This sort of imprisonment is definitively impossible to be transgressed.

$G$ and its result denoted $ms_G$. So – accordingly to the definitions from (2.I).2 – we have to posit that:

> An unbound *one* micro-state of *one* micro-*system* produced by a composed operation of generation involves only ***one*** singularity in the amplitude of the corpuscular wave of the generated microstate $ms_{\mathbf{G}(G1,G2,...Gk),cw}$.

But we also admit, accordingly to de Broglie [1956], that the other operations of generation $(G_1,G_2,...G_k)$ mentioned in connection with $\mathbf{G}(G_1,G_2,...G_k)$ – that *can* be realized separately but have *not* been separately realized when $\mathbf{G}(G_1,G_2,...G_k)$ has been realized – produce together some specific effects on the dynamic of this unique singularity such as it is asserted by the guidance law quoted in *(e)*. This will appear in the Part III to be a very important point in the process of construction of a satisfactory representation of the quantum measurements on any sort of unbound microstate.

On the basis of the assumptions *{(a),(b),(c),(d),(e),(f)}* we now finally take the following new step:

> ***The factual physical-operational modelling postulate*** $MP(\{\sigma(ms_{G,cw})\})$**.** In agreement with *MD*, (1) and (13'), it is posited that any one realization of $G_t$ generates one specimen of the studied microstate that is specified accordingly to the assumptions *{(a),(b),(c),(d),(e),(f)}*. Such a specimen will be denoted $\sigma(ms_{G,cw})$ (cw: corpuscular wave) and it will be called *the G-(corpuscular-wave)-model of a specimen of the studied microstate* (in short, a *G-cw* model). Correlatively the corresponding microstate can be re-noted $ms_{Gt,cw}$. So from now on – but only in so far that this is useful – the relation (1) $G \leftrightarrow ms_G$ from *MD* can be maximally specified as
>
> $$G_t \leftrightarrow (ms_{Gt,cw} \equiv \{\sigma(ms_{Gt,cw})\}) \qquad (1')$$

The *G-cw* model $MP(\{\sigma(ms_{G,cw})\})$ has been constructed under strict constraint of logical-semantic consistency with both *IQM* and the content of $QM_{HD}$. It introduces a crucial connection between *IQM*, $QM_{HD}$, and furthermore de Broglie's global *interpretational* approach constructed in [1956]), tied with Bohm's *interpretation* of $QM_{HD}$. But inside $[IQM\text{-}QM_{HD}]$, via the *G-cw* model, the concepts of microstate and of specimen of a microstate as defined in *MD cease to be imprisoned inside the category of 'interpretations' of* $QM_{HD}$: They become basic elements of the process attempted here of construction of a *new*, consensual, predictive and verifiable representation of the microstates.

Initially, in the relation (1) from *IQM*, the concepts $ms_G$ and $\sigma(ms_G)$ have been provisionally qualified exclusively by *the label 'G' of the way in which they are produced*; later in *IQM*, in (8), this initial definition has been enriched by a whole *set* of observable, brute, 'transferred' results $\{\mu\}_{kA}$, $k_A=1,2,...m_A$ of measurement-interactions. Both these definitions from *IQM* are already 'scientific' because they are *physical-operational, observable, communicable, consensual (so reproducible), and they are verifiable*. But they all *remain* ***exterior*** *to the concepts '*$ms_G$*' and '*$\sigma(ms_G)$*'* because *IQM* has been constructed as only a *general* structure of reference and insertion that rejects any specified model of a microstate, not as a definite theory of the microstates. While now, because we inside the new framework $[IQM\text{-}QM_{HD}]$ we research a new theory of the microstates, and in consequence of (1') $(ms_{G,cw} \equiv \{\sigma(ms_{G,cw})\})$, the results $\sigma(ms_{G,cw})$ of one operation $G$ and the result $ms_{G,cw}$ of a big set of repetitions of $G$ are posited to be endowed with characters that concern the *insides* of the specimens $\sigma(ms_{G,cw})$ *themselves*; they concern now the *own nature* assigned to the concepts $\sigma(ms_{G,cw})$ and $ms_{G,cw}$.

Thereby the present approach draws in, not only the possibility to investigate concerning factually individualized microstates, but *also the possibility to take into consideration data that are **interior to one studied microstate***: The fundamental inside-outside opposition (cf. Atmanspacher&Dalenoort [1994]) enters the mechanic of microstates that we are constructing here, and this can have many consequences because the main specificities of the notion of a factually defined microstate stem from its inner non-classical 'quantum-fields'.

So the *G-cw* local model enriches the initial concept (1) of a factually and consensually defined microstate. With respect to the absolute de Broglie-Bohm approach – let us denote it *dBB* – that has a basically deterministic character[70], the price of this gain is a primordially statistical-probabilistic observational character.

The results from (6II)2 constitute a noteworthy advance: They offer from now on clearly defined assumptions, concepts, and words for investigations that are placed – strictly – on the frontier between the still a-conceptual factuality and what is extracted therefrom *factually* for a radically first conceptualization able to lead to communicable, consensual and verifiable 'scientific' knowledge'. They also suggest subsequent developments[71]. And they stress explicitly the fact that *absolute* and intrinsic spatial delimitating contours of *individual* 'objects' in the classical sense are just a human construct that inside microphysics ceases to be useful and furthermore is likely to induce inner inconsistencies. *Individuality*, otherness, and also stable and homogeneous space-time *inner* structure are only pragmatic human assumptions of which the utility is relative to the local cognitive purposes and actions.

## (6.II).3. CLARIFICATIONS VIA THE CONCEPT OF OPERATION *G* OF GENERATION OF A MICROSTATE

It appeared above that the use of the concept *G* of generation of a microstate would have economized the false problem of why eigenkets in general are not square-integrable. Below we bring into evidence other three fundamental sorts of circumstances where the *[IQM-$QM_{HD}$]* concept '*G*' entails clarification of ambiguities or of latent problems that vitiate *$QM_{HD}$*.

Inside *$QM_{HD}$* works a mathematical principle of spectral decomposability of any state-ket $|\psi(\mathbf{r},t)>$, i.e. the posit that for any state-ket and any basis of eigenket $\{|u_j(\mathbf{r},a_j)>\}$, $j=1....,$ introduced by a *$QM_{HD}$*-observable ***A*** it is justified to assert the equality[72]:

$$|\psi(\mathbf{r},t)>/\mathbf{A}=\sum_j c_j(t)\ |u_j(\mathbf{r},a_j)>,\quad \forall j \tag{14}$$

[70] In the *dBB* approach the equation of evolution that yields solutions in terms of terms of statistical state-ket are just superposed to the deterministic basic representation, in absence of any *organically* incorporated explanation of the statistical-probabilistic character of the solutions.

[71] For instance: a microstate of one system (in the sense of the definitions from (2.I).1) with electric charge or magnetic moment can be drawn into the realm of the observable by use of classical macroscopic fields. But how could be *manipulated* the result of an operation *G* if this operation generates (for instance by a nuclear reaction) a microstate that is sensitive *exclusively* to a gravitational field? (Which probably means *a maximally 'simple' de Broglie singularity, a 'pure quantum of de Broglie-mass' (with non-null spin)? a 'graviton'*?). Such questions touch as much the most modern researches in gravitation, as the *dBB* representation of the sub-quantic substance: In de Broglie [1956], the chapter XI, pp. 119-131 is fascinating in relation with gravitation, teleportation, etc. And ***the pair ($MP(ms_{G,cw})$, (1')) offers legal scientific access to the dBB representation***.

[72] We recall that adaptation of the result of this work, to a coherently finite mathematical representation and to the correlative explicit finiteness of each domain of investigation – as required by our choice of effectiveness – will have to be introduced by conceptual-mathematical adjustments subsequent to the outline of a second quantum mechanics.

Furthermore, the general choice of a vector-space-representation *permits* to write the state-ket associated to a microstate $ms_{G(G1,G2,...Gn)}$ generated by a composed operation of generation *G(G1,G2,...Gk)* ((1.I).1), as a *mathematical* superposition

$$|\Psi_{12...k}(\mathbf{r},t)\rangle=\lambda_1|\Psi_1(\mathbf{r},t)\rangle+\lambda_2|\Psi_2(\mathbf{r},t)\rangle+......\lambda_k|\Psi_k(\mathbf{r},t)\rangle \qquad (15)$$

of the state-ket of the microstates $ms_{G1}$, $ms_{G2}$, …. $ms_{Gk}$ that *would* have been obtained *if* each one of the operations of generation *G1,G2,...Gk* that have been composed inside $G(G_1,G_2,...G_k)$ *would* have been realized *separately* [73].

In $QM_{HD}$ the Hilbert space $\mathcal{H}$ of a state-ket is extended into a 'generalized-Hilbert space' $\mathcal{H}$ where the eigenket are included as a limiting sort of vectors. This entails that from a strictly mathematical point of view both writings (14) and (15) are just superpositions of vectors inside $\mathcal{H}$, permitted by the mathematical *axiom* – included in the definition of the algebraic structure called a vector-space – that any two or more elements from a given vector-space admit an additive composition; which is expressed by saying that they can be 'superposed', i.e. added to one another. These formal features installed a purely mathematical language that calls *indistinctly* 'superposition' *any* additive combination of ket, whether only state-ket like in (15), or state-ket *and eigenket* like in (14), or only eigenket as in Dirac's theory of transformations.

One senses immediately what confusions can stem from such a mixing in-distinction between formal elements – state-ket and eigenket – that represent concepts that have been shown in (6.II).1 to be so deeply different from a semantic point of view. No physical criteria, nor conceptual ones, are made use of inside $QM_{HD}$ in order to make mutual specifications inside the general category of such additive compositions of ket, syntactically permitted in $\mathcal{H}$. In (6.II).2 we have seen an illustration of the consequences of precisely this sort of blindness with respect to conceptual-physical *meaning*, with respect to the semantic contents. A blindness of this sort is a major danger entailed by the intimate relation between physics and mathematics that grows inside mathematical physics. We have already seen how, under the protection of this intimacy, mathematics can simply chase the intelligibility out of physics.

In the case of the writing (14) the mathematical form induces into the minds the more or less explicit *semantic* interpretation that all the eigenket-terms $c_j(t)|u_j(\mathbf{r},a_j)\rangle$ from the second member are of the same *nature* as the state-ket $|\psi(x,t)\rangle$ from the first member, that all these formal elements carry the same sort of semantic contents. Which has been shown in (6.II).1 to be utterly false.

And in (15) this same sort of semantic blindness suggests that the state-ket $|\Psi_{12...k}\rangle$ points toward a *physical* 'superposition' of all the microstates $ms_{G1}$, $ms_{G2}$, …. $ms_{Gk}$, *themselves*, supposed to 'coexist inside $|\Psi_{12...k}\rangle$'. This formally induced suggestion is strongly privileged by the fact that inside Physics there exist '*principles* of superposition' – of physical superposition – that, when they are physically adequate, are mathematically expressed by an additive composition. In the chapter 7.II precisely this sort of confusion will play the main role.

Inside *IQM*, so now also inside the provisional framework *[IQM-$QM_{HD}$]*, any possibility of ambiguities of the sort specified above is avoided by construction. This however has to be stressed, both conceptually and via modified notations. So let us first detail the conceptual situation.

- According to *[IQM-$QM_{HD}$]*, in (14) only the state-ket $|\psi(\mathbf{r},t)\rangle$ from the first member corresponds – on the statistical level of conceptualization – to the studied microstate $ms_G$, while all the terms $c_j(t)|u_j(\mathbf{r},a_j)>$ from the right-hand expansion of $|\psi(\mathbf{r},t)>$ are symbols of a product of a *number* $c_j(t)$ and a *model* $|u_j(\mathbf{r},a_j)>$ of a possible corpuscular-wave-movement.

[73] The choice to make use of this formal possibility, inside a physical theory of microstates, will be criticized in the Part III, and replaced by another choice of representation.

- According to *[IQM-QM$_{HD}$]*, in (15) the resulting *one* microstate $ms_{G(G1,G2,...Gk)}$ that is effectively generated by the *unique* composed operation of generation $\mathbf{G}(G_1,G_2,...G_k)$, cannot be coherently conceived as a coexistence of all the microstates $ms_{G1}$, $ms_{G2}$, .... $ms_{Gk}$ that *would* have been obtained via the *separate* realizations of $(G_1,G_2,...G_k)$. This has been already much explained, first in the Part I and then in (6.II)2. So the requirement of self-consistency entails that all the state-ket from the second member of (15) have to be regarded as only *virtual* representational elements[74].

On the other hand inside $QM_{HD}$ these representational elements from (15) have been considered to be useful *precisely* ***in order to*** *represent mathematically the state-ket* $|\Psi_{12...k}(\mathbf{r},t)>$ *by the additive expression (15)*. Indeed this possibility permits to deal mathematically with the observable factual *in*-equality

$$\pi_{12...k}(\mathbf{G}(G_1,G_2,...G_k),a_j) \neq \pi_1(G_1,a_j)+\pi_2(G_2,a_j)+.........+pk(G_k,a_j) \qquad (16)$$

(that generalizes the relation (12)): via the spectral decomposition (14) of $|\Psi_{12...k}(\mathbf{r},t)>$ on the basis of eigenket of an observable $\boldsymbol{A}$ and application of Born's postulate of probability to the complex expansion coefficients $c_j(t)$, between the expansion coefficients there emerge *mathematical* 'interference'-terms that entail the inequality (16), which at a first sight seems very adequate for expressing the involvement of the operations of generation $(G_1,G_2,...G_k)$ in the unique effectively realized operation $\mathbf{G}(G_1,G_2,...G_k)$.

So $QM_{HD}$ and *IQM* involve distinct views concerning the semantic acceptability of writings of the form (15): This is possible because *IQM* is constructed to represent the involved meanings, while $QM_{HD}$ occults the involved meanings under purely mathematical requirements.

Such a situation is *doomed* to come to some factual confrontation. We are aware of this, and we stay attentive.

- Quite generally now: *When inside $QM_{HD}$ a state-ket seems to be 'absent', inside [IQM-QM$_{HD}$] this means that the microstate that corresponds to this state-ket* ***is not generated separately***. So it has not been brought into individualized factual existence. The most striking case of such an 'absence' of a state-ket is that of *one* micro-*state* of *two or several* micro-*systems*, tied with the problem of locality. The formalism of $QM_{HD}$ – rightly – represents such a micro-*state* by only *one* state-ket. But for each involved micro-*system* it introduces a distinct representation-space, and the mathematical relation between these representation spaces is specified in a way that is indicated by the now current words 'intrication' and 'non-separability'. Certain authors speak of « 'absence' of an 'own' state-ket for each 'system' » ; other authors speak of « absence of 'information' » (in what a sense, exactly?); as if a state-ket were a planet or a lake, something that 'is' somewhere outside, but nobody knows how to go and see where and how it 'is'. In the textbooks it is written that «often a micro-system 'is' represented by a state-ket and, if so, the state is 'pure'»; while if it 'is' not pure, then it 'is' a 'mixture', but in such a case (happily) one can nevertheless dispose of a statistical operator (how, exactly, one should factually act in order to construct this operator, is not clearly defined). But in the case of the 'problem of locality' not even a true statistical operator does 'exist', only 'a partial-trace' operator 'exists'; but this cannot change the *fact* that there is 'non-separability' because – curiously – a statistical correlation is observed even when the spatial distance between the involved systems is very big. All the mentioned ways of speaking suggest that the state-ket, the statistical operators, etc., are conceived to possess an existence quite independently of the representational choices, decisions, constructions, *of human beings, of physicists*. As soon as 'there *is*' a 'system', 'its' state-ket should also 'be', and nevertheless sometimes it is absent and we do not know why, nor where it is gone. The special case of one micro-state of *two* micro-systems in the sense of (2.I)1 is a particularly strong discloser of how the whole mathematical formalism of $QM_{HD}$ is currently conceived:

> *We are in presence of a huge reification of the mathematical formalism of $QM_{HD}$, considered to constitute the whole of $QM_{HD}$ by itself, by it alone.*

---

[74] We do not assert here any physical facts, only consequences of the requirement of inner consistency of the framework *[IQM-QM$_{HD}$]*. How this can be connected with physical facts will be examined later.

The fact that the whole $QM_{HD}$ is just a human construction achieved by men in order to represent what men can name 'factually definable microstates', has receded far out of the minds. How much more far, then, we still are from conceiving that not even the microstates to be studied do 'exist' out there, and sometimes not even the involved 'systems', in this sense that in general one has to generate them in order to study them!

This situation entails a sort of consternation. It even produces a sort of religious admiration for $QM_{HD}$, because the experiments on locality have confirmed the predictions of the formalism[75].

But inside *[IQM-$QM_{HD}$]* one understands that, and how, these attitudes stem from the following circumstances.

* Notwithstanding that in general the mathematical writings from $QM_{HD}$ are in agreement with the definitions from (2.I)1 *these definitions are not spelled out inside $QM_{HD}$*, and furthermore, in the case of one micro-*state* of two micro-*systems*, in the current language that accompanies the use of the formalism one speaks of two or several *'systems'* – never of one micro-*state* of two micro-*systems*. Therefore inside $QM_{HD}$ the one-one *indirect* and non-explicated connection

$$G\leftrightarrow|\psi_G(\mathbf{r},t)> \qquad (1'')$$

between $G$ and a state-ket that should be denoted $|\psi_G(\mathbf{r},t)>$ because the studied microstate $ms_G$ is generated by $G$, is simply out of perceptibility notwithstanding that it is generally accepted that always « a state-ket represents the studied *'system'* ». But inside *[IQM-$QM_{HD}$]* a connection (1'') is logically *entailed*, via (1) $G\leftrightarrow ms_G$.

* Going now to the roots, one finds that all the preceding examples illustrate how inside $QM_{HD}$ unintelligibility is entailed by the fact that no clear and systematic distinction is made between, on the one hand *individual* concepts ($ms_G$, specimens $\sigma(ms_G)$ of a microstate $ms_G$, or $|u_j(\mathbf{r},a_j)>$), or individual physical operations ($G$, acts of measurement *MesA*,), and on the other hand the *statistical* descriptors like $|\psi(\mathbf{r},t)>$.

---

[75] On the other hand it is true that it *does* seem amazing to find out to what a degree the mathematical formalism, in contradistinction to the physicists, is *observant* of (is compatible with) the involvement or not, in a given state-ket-symbol, of an operation of generation $G$ of a corresponding micro-state, and with the significance of the involved state-ket from the viewpoint of the definitions (2.I)1; *notwithstanding that inside the formalism the concept 'G' is neither defined nor represented, and the definitions (2.I)1 are not stated, while the specific meaning of an eigenket has not been recognized either.* Indeed: ***(a)*** a spectral decomposition (14) is usually conceived to involve an infinite number of terms, the coefficients from these terms are complex numbers dependent on time, while the eigenfunctions – models of wave-movement – are correctly written as independent on time; whereas ***(b)*** a superposition (15) of state-ket tied with a composed operation of generation is written as a finite number of terms, the coefficients are usually constant real numbers, and the ket from the superposition are dependent on time. Everything in the mathematical writings is fully concordant with the analyses made here inside *[IQM-$QM_{HD}$]*. This raises strongly a very interesting question concerning something that could be called *'the **semantic** expressivity of the mathematical internal syntactic coherence'*. The unique explanation seems to be that *the mathematicians, when they construct a mathematical structure, **imply the existence of operations of generation***.

In these conditions, inside the minds used to $QM_{HD}$ like a New-York boy is used to Manhattan, an explanation is badly needed indeed, *why* sometimes some of the two or several state-ket that would be so ‘necessary’, nevertheless are stubbornly ‘absent’.

We are now ready to close this point by the following convention:

***Notational convention 1***. Inside *[IQM-$QM_{HD}$]* any state-ket $|\psi(\mathbf{r},t)>$ that corresponds to a physically generated micro-*state* $ms_G$ will be re-noted as $|\psi_G(x,t)>$, and the sort of operation $G$ that indexes it will be explicitly stated, and when necessary its specific structure will be distinguished graphically.

So from now on (14) will be re-written as

$$|\psi(\mathbf{r},t)_G>/\mathbf{A} = \sum_j c_j(t)\ |u_j(\mathbf{r},a_j)>, j=1\ldots\ldots \qquad (14')$$

and (15) will be re-written as

$$|\Psi(\mathbf{r},t)_{\mathbf{G}(G1,G2,\ldots Gk)}> = \lambda_1|\Psi_{G1}(\mathbf{r},t)>+\lambda_2|\Psi_{G2}(\mathbf{r},t)>+\ldots\ldots\lambda_k|\Psi_{Gk}(\mathbf{r},t)> \qquad (15')$$

where, in the global symbol $\mathbf{G}(G_1,G_2,\ldots G_k)$, the first operation of generation $\mathbf{G}$ is written in bold font in order to express that it is the unique operation of generation that has been fully accomplished, while the only a priori possible but not *separately* realized operations of generation that in the second member are involved by the state-ket of virtual microstates-of-reference, will be written in non-bold font.

These specifications will entail clarification[76]**.** They also are an illustration of the gliding character of the level where takes place the collision between a *top-down* and a bottom-up approach and of the way in which the bottom-up approach can *incorporate* progressively the preceding top-down approach in a new, improving elaboration.

## (6.II).4. CRITICAL EXAMINATION OF THE *$QM_{HD}$* THEORY OF MEASUREMENTS

Here we definitely walk into Absurdland, so abruptly and totally that, by fear of being considered subjective and malevolent, I do not dare to immediately make use of my own voice. So I first offer an objective look at the conceptual situation by reproducing an extract from Wikipedia that consists of a collection of views.

« The **measurement problem** in quantum mechanics is the problem of how (or *whether*) wave-function collapse occurs. The inability to observe this process directly has given rise to different interpretations of quantum mechanics, and poses a key set of questions that each interpretation must answer. The wave-function in quantum mechanics evolves deterministically according to the Schrödinger equation as a linear superposition of different states, but actual measurements always find the physical system in a definite state. Any future evolution is based on the state the system was discovered to be in when the measurement was made, meaning that the measurement "did something" to the system that is not obviously a consequence of Schrödinger evolution.

To express matters differently (to paraphrase Steven Weinberg[1][2]), the Schrödinger wave equation determines the wavefunction at any later time. If observers and their measuring apparatus are themselves described by a deterministic wave function, why can we not predict precise results for measurements, but only probabilities? As a general question: How can one establish a correspondence between quantum and classical reality?[3].

**Schrödinger's cat**

The best known example is the "paradox" of the Schrödinger's cat. A mechanism is arranged to kill a cat if a quantum event, such as the decay of a radioactive atom, occurs. Thus the fate of a large scale object, the cat, is entangled with the fate of a quantum object, the atom. Prior to observation, according to the Schrödinger equation, the cat is apparently evolving into a linear combination of states that can be characterized as an "alive cat" and states that can be characterized as a "dead cat". Each of these possibilities is associated with a specific non-zero probability amplitude; the cat seems to be in some kind of "combination" state called a "quantum superposition". However, a *single, particular observation* of the cat does not measure the probabilities: it always finds either a living cat, or a dead cat. After the measurement the cat is definitively alive or dead. The question is: *How are the probabilities converted into an actual, sharply well-defined outcome?*

---

[76] The clarification will lead in the Part III to a ***suppression*** of the whole concept-and-writing (15').

**Interpretations** (*Main article: Interpretations of quantum mechanics*)

**Hugh Everett's** many-worlds interpretation attempts to solve the problem by suggesting there is only one wave-function, the superposition of the entire universe, and it never collapses—so there is no measurement problem. Instead, the act of measurement is simply an interaction between quantum entities, e.g. observer, measuring instrument, electron/positron etc., which entangle to form a single larger entity, for instance *living cat/happy scientist*. Everett also attempted to demonstrate the way that in measurements the probabilistic nature of quantum mechanics would appear; work later extended by Bryce DeWitt.

**De Broglie–Bohm** theory tries to solve the measurement problem very differently: this interpretation contains not only the wavefunction, but also the information about the position of the particle(s). The role of the wave-function is to generate the velocity field for the particles. These velocities are such that the probability distribution for the particle remains consistent with the predictions of the orthodox quantum mechanics. According to de Broglie–Bohm theory, interaction with the environment during a measurement procedure separates the wave packets in configuration space which is where apparent wave-function collapse comes from even though there is no actual collapse.

**Erich Joos and Heinz-Dieter Zeh** claim that the phenomenon of quantum decoherence, which was put on firm ground in the 1980s, resolves the problem.[4] The idea is that the environment causes the classical appearance of macroscopic objects. Zeh further claims that decoherence makes it possible to identify the fuzzy boundary between the quantum microworld and the world where the classical intuition is applicable.[5][6] Quantum decoherence was proposed in the context of the many-worlds interpretation[*citation needed*], but it has also become an important part of some modern updates of the Copenhagen interpretation based on consistent histories.[7][8] Quantum decoherence does not describe the actual process of the wavefunction collapse, but it explains the conversion of the quantum probabilities (that exhibit interference effects) to the ordinary classical probabilities. See, for example, Zurek,[3] Zeh[5] and Schlosshauer.[9]

**The present situation is slowly clarifying, as described in a recent paper by Schlosshaue**r as follows:[10]

Several decoherence-unrelated proposals have been put forward in the past to elucidate the meaning of probabilities and arrive at the Born rule ... It is fair to say that no decisive conclusion appears to have been reached as to the success of these derivations. ...

As it is well known, [many papers by **Bohr** insist upon] the fundamental role of classical concepts. The experimental evidence for superpositions of macroscopically distinct states on increasingly large length scales counters such a dictum. Superpositions appear to be novel and individually existing states, often without any classical counterparts. Only the physical interactions between systems then determine a particular decomposition into classical states from the view of each particular system. Thus classical concepts are to be understood as locally emergent in a relative-state sense and should no longer claim a fundamental role in the physical theory.

**A fourth approach is given by objective collapse models**. In such models, the Schrödinger equation is modified and obtains nonlinear terms. These nonlinear modifications are of stochastic nature and lead to a behaviour which for microscopic quantum objects, e.g. electrons or atoms, is unmeasurably close to that given by the usual Schrödinger equation. For macroscopic objects, however, the nonlinear modification becomes important and induces the collapse of the wavefunction. Objective collapse models are effective theories. The stochastic modification is thought of to stem from some external non-quantum field, but the nature of this field is unknown. One possible candidate is the gravitational interaction as in the models of Diósi and Penrose. The main difference of objective collapse models compared to the other approaches is that they make falsifiable predictions that differ from standard quantum mechanics. Experiments are already getting close to the parameter regime where these predictions can be tested.[11]

**An interesting solution** to the measurement problem is also provided by the hidden-measurements interpretation of quantum mechanics. The hypothesis at the basis of this approach is that in a typical quantum measurement there is a condition of lack of knowledge about which interaction between the measured entity and the measuring apparatus is actualized at each run of the experiment. One can then show that the Born rule can be derived by considering a uniform average over all these possible measurement-interactions. [12][13] ».

## (6.II).4.1. Refusal of von Neumann's representation of quantum measurements

I now dare to continue by my own summary of the situation. In what follows immediately I place myself inside $QM_{HD}$, not inside $[IQM\text{-}QM_{HD}]$. So I just express first the current nowadays language and reasoning about quantum measurements. The representation of measurements on microsystems is that one proposed by von Neumann in 1932:

The Schrödinger equation of the problem endows us with the state-ket of the problem, $|\psi(r,t)>$. So this state-ket is given mathematically, we dispose of it from the start in consequence of purely mathematical operations. We want now to *represent* the measurements. Therefore we have to write the state-ket for the measurement-interaction. For this we proceed as follows: Let $t=t_o$ be the initial moment given in $|\psi(x,t)>$. At a time $t_1 > t_o$ we want to measure the observable $\boldsymbol{A}$ on the 'system' represented by $|\psi(\boldsymbol{r},t)>$. We take now into account that for $t \geq t_1$ there is interaction between the studied system and the measurement-apparatus. "So", is it said :

For $t \geq t_1$ the measurement-evolution *must represent also the apparatus "because" the apparatus is also constituted of microsystems*.

So the measurement-evolution is to be represented by a state-ket of [(the studied system S)+(the apparatus for measuring $\boldsymbol{A}$)]. Let us then write, say, ($S$+*app(**A**)* and $|\psi_{S+app(A)}(x,t)$. Since we measure the observable $\boldsymbol{A}$, the expansion of $|\psi_{S+App}(\boldsymbol{r},t)>$ with respect to the basis of $\boldsymbol{A}$ comes in. Accordingly to the well-known quantum theory of a '*system* composed of two *systems*' we write the tensor-product expansion:

$$|\psi_{S+App}(\boldsymbol{r},t)>= \sum_k \sum_j c_j(t) d_k(t)\ |u_j(\boldsymbol{r},a_j)>|q_k(\boldsymbol{r},a_k>,\quad \forall j,\ \forall k \qquad (17)$$

where $|q_k(\boldsymbol{r},a_k>$, $k=1,2...$ are the eigenket of the observable called the 'needle-position of the *app(**A**)*, that can be denoted $\boldsymbol{\chi(A)}$, with eigenvalues, say ($v(a_k)$, $k=1,2...,$ that express, respectively, 'the needle-positions of *app(A)* that correspond to the eigenvalues $a_k$ of $\boldsymbol{A}$'. Furthermore – by the definition of the concept of 'apparatus for measuring $\boldsymbol{A}$' – the set $\{c_j(t)d_k(t)\}$ of the global, product-expansion coefficients $(c_j(t)d_k(t))$ from (17) reduces to a set $\{\alpha_{jj}(t)\}$ (with $\alpha_{jj}=c_jd_j$) of only the coefficients with non-crossed indexation, because the needle position ($v(a_j)$ of the *app(**A**)* *is what – **alone** – indicates the obtained eigenvalue $a_j$ of **A*** [77]. So in fact in this case we have only

$$|\psi_{S+App}(\boldsymbol{r},t)>= \sum_j \sum_j \alpha_{jj}(t)\ |u_j(\boldsymbol{r},a_j)>|v_j(\boldsymbol{r},a_j>,\quad \forall j \qquad (17')$$

The measurement evolution is produced accordingly to a measurement-Schrödinger equation *where the hamiltonian operator **H(A)** commutes with **A***. And it is *posited* that this evolution finishes with a definite needle position $\chi(a_j)$ that indicates *one* definite result $a_j$ [78].

Now, the above-mentioned representation is considered to raise two 'problems'.

- The *reduction problem*: what happened to all the terms from (17') with index $k \neq j$ that accordingly to a linear formalism should subsist? Where have they disappeared?

- The problem of 'decoherence': how can we *prove* that after the realization of the position $\chi(a_j)$ of the apparatus-needle that announces the result $a_j$ the measurement interaction really ceases?[79]

Here finishes my own summary of the general framework accepted for the representation of measurements. In what follows I go now back inside *[IQM-QM$_{HD}$]* and I speak again for myself and by use of the language introduced up to now in this work.

Bertrand Russell has written somewhere that aims are induced by temperament while the choice of a method when an aim is given is induced by intelligence. With respect to the aim to represent the measurements on microstates, von Neumann's choice of a method is stunning. If we followed his argument, in order to measure the position of a star by use of a telescope, given that the telescope and the star are both made of microsystems, we should represent [(the telescope)+(the star)+(the measurement interaction between these two entities)]; and we should prove in terms of the theory thus conceived, that the star and the telescope do really separate physically once the star's position has been established. Such an argument manifests luminously a total blindness with respect to the rather obvious fact that in science what decides the optimality of a representation is the *cognitive situation* of the observer-conceptualiser with respect to that on what he wants to obtain some knowledge, etc. The inner constitution of that what has to be qualified, or of the instruments that are made use of, has nothing to do with the criteria for generating the desired knowledge. Quite generally the functionality of a construct is not in a one-to-one relation with its material (or abstract) structure (those who have realized the aeroplanes have thoroughly understood that). Moreover, in the case of microstates, most often what *can* be registered is just marks on a sensitive registering device and/or durations determined by chronometers, not needle positions. From these data one has to *construct conceptually* the researched 'value' of the measured 'quantity' and this quantity in its turn is constructed beforehand on the basis of conceptual-mathematical operations. And finally, von Neumann's representation of measurements dodges the crucial coding problem. It simply makes it disappear behind an amorphous heap of words and symbolic writings void of definition, so of meaning. Indeed von Neumann's representation of measurements *transgresses* $QM_{HD}$:

> '*The* observable' $\boldsymbol{\chi(A)}$ called the 'needle-position of the *app(A)* is not a quantum mechanical observable, it cannot be constructed formally in a definite way from a definite classical mechanical quantity and so its eigenfunctions and eigenvalues cannot be *calculated*. The eigenvalues are just postulated to be the eigenvalues of the observable ***A***.

---

[77] So no coding problem arises according to this 'measurement-theory': One is protected from this problem, the apparatus will know where to settle its needle, since it is conceived for this aim.

[78] As far as I know, this has never been proved inside $QM_{HD}$ to be generally insured by the condition imposed upon the measurement evolution.

[79] The locality-problem incites to think that it might not do this, but so what?

All this simply is not acceptable. A measurement-apparatus that is made use of in a scientific description of something else than this apparatus itself has to be introduced as a *primary datum* that stays outside the representation of the act of measurement, if not, one enters indefinite regression[80]**.** This is a general interdiction of logical nature. Etc.

So I declare without shades that *I quite radically reject von Neumann's framework for representing quantum measurements*.

### (6.II).4.2. The implicit assumptions from the basic features of the $QM_{HD}$ theory of measurements

We shall now concentrate on the essential features of the representation of quantum measurements because: the core of the unintelligibility of $QM_{HD}$ is hidden there. The developments from 6.II and the refusal of von Neumann's 'theory' of quantum measurements leave us face to face with the real deep problems of intelligibility. These problems must now be stated independently of any superfluous representational clothing in order to have a chance to draw into light the prime source of the vices that obstruct the tortuous channels toward intelligibility. To guide toward this source I begin by a global sketch of the conceptual situation.

#### *(6.II).4.2.1. Preliminary questions and putative views*

One *constructs* statistics of numbers (or of other sorts of 'results') in order to be able to *predict statistically* on results of acts of examination indicated by the verbal label 'measurements'. Usually this construction itself is achieved by performing sets of individual measurements. When this is the case let us speak of *construction-measurements*. And then, in order to be certain that the construction of a predictive statistic has succeeded, one *verifies* the statistic before announcing it as useful for prediction. In this case let us speak of *verification-measurements*. It seems natural to assume that factually the results of construction-measurements and of verification-measurements identify when the measurements are the same, notwithstanding that they are performed with different aims. So knowledge of the way of factually performing the individual acts of measurement should be insured before trying to construct and to verify the statistics that involve such measurements.

[80] Wittgenstein has written somewhere: «There is one thing of which one can say neither that it is one metre long, nor that it is not one metre long, and that is the standard metre in Paris ». I dare to complete: At least one class of things cannot be absorbed into the quantum mechanical representation of measurements: The class of the measurement apparatuses.

But *where* can one find inside $QM_{HD}$ a thorough definition of the way of performing *factually* an individual act of measurement? *Anywhere,* strictly anywhere. Indeed:

In $QM_{HD}$ – via the Schrödinger equation – the representation of, directly, the predictive statistics is generated mathematically. This circumvents the necessity to define individual construction-measurements like in *IQM*.

But *the individual verification-measurements remain unavoidably necessary*.

So the absence of any definition of the individual acts of measurement is a striking gap, even *if* indeed such a definition were necessary exclusively for verification of the statistical predictions. (For bound states, collective acts of verification-measurement can often suffice (absorption or emission of radiation versus registration of the intensity of spectral lines of emission or of absorption of radiation)). But for unbound microstates, individually defined verification-measurements are not avoidable.

Now, inside $QM_{HD}$ the *process* of verification of the statistical predictions asserted by the state-ket of the studied microstate – considered globally – is represented mathematically by the Schrödinger *measurement-evolution of this state-ket, that is a* ***statistical*** *descriptor*. And it is admitted that such a measurement evolution somehow involves the factual *individual* acts of measurement that should verify the predictive statistic. Such a common formal location of two sorts of descriptional elements that belong to two different levels of conceptualization is immediately highly suspect. But furthermore, throughout this suspect representation, the explicit definition of an individual act of measurement is just supposed to be known and it is alluded to in mere words, without even being also *defined* at least in mere words: it is just pointed to with a verbal finger. And the *end* of such an undefined act of measurement that cohabits with the statistic to the verification of which it participates, is expressed formally via a sudden modification of the representation of the involved Schrödinger ket of measurement-evolution: a modification *that is* ***not*** *entailed by the mathematical formalism* but is just *posited separately* [81].

Some authors have remarked that in the classical theory of probabilities 'also' one considers a probabilistic distribution of the elementary events from a whole universe of possible elementary elements, while every individual act from the process of verification of this distribution produces just one definite result. But it seems worth noting that in the theory of probabilities the mathematical representation of the probabilistic predictive distribution is not used itself, and globally, for representing the individual acts from the process of verification of the predictive probabilistic distribution, and that the distribution itself is not provisionally *modified* by any individual act from the global process of verification of the probabilistic distribution, and re-asserted immediately afterward in connection with another act of verification. These coalescences between a statistical-probabilistic descriptional element, and the representation of individual operations for a progressive verification of a pre-existing probabilistic predictive distribution are specific of the quantum representation of measurements.

---

[81] Since already a century this situation seems to many physicists very strange or even scandalous; and it has led to a well-known very extravagant 'interpretation' by Everett that many physicists finally *do admit* "because it does not involve any *inconsistency*" (which is to be read, in fact, as '*mathematical* inconsistency', any semantic constraint being expurgated: This manifests the relations that work inside some minds of physicists concerning facts versus mathematics-and-logic inside a discipline of *physics*.

We finally add that – according to *IQM* – any individual act of measurement is accomplished on *one* specimen of a microstate and so it requires an immediately previous realization of an operation of generation of that specimen; and furthermore, *it also requires sine-qua-non an operational definition that involves a factually realizable coding procedure*; *which in its turn requires a general model of a microstate*. Inside *IQM* – that has been constructed as only a reference-and-imbedding structure of any theory of microstates – these requirements have been left blank, to be specified in any given theory of microstates. But here we try precisely to draw from *IQM* and $QM_{HD}$ a fully intelligible theory of the factually defined microstates $ms_G$ from (1) $G \leftrightarrow ms_G$ $ms_G \equiv \{\sigma(ms_G)\}$. Now, it appeared in (6II).2 that the formalism of $QM_{HD}$ involves surreptitiously but essentially de Broglie's model of a microstate, and we have re-defined there via (1') the concept of operation of generation of a specimen of the studied factually defined microstate. So in what follows it will have to be specified for each one of the various types of unbound microstates distinguished in the section (2.I)2, *what* explicit sort of definition of an individual quantum measurement is entailed via (1') by this de Broglie model, and what *corresponding coding-rule that definition can admit.*

The problems sketched out above can be summarized as follows: It is generally agreed that a statistical representation only mirrors the physical individual facts and operations that it involves genetically. But this does not exonerate from stipulating *of what consists that what is mirrored, when precisely that is studied.*

### *(6.II).4.2.2. A fundamental distinction:*
### ***Individual physical wave-function* versus *abstract statistically predictive 'state'-function***

Let us go back to the fact that the wave function initially introduced by Louis de Broglie was conceived to represent a *physical* 'corpuscular wave' $\Phi(\mathbf{r},t)=a(\mathbf{r},t)e^{(i/\hbar)\beta(\mathbf{r},t)}$ assigned to any micro-system; but very rapidly afterward this initial concept has transmuted into a mathematical representation of predictive statistics of results of measurements performed on some given sort of microstate. Thereby, *in de Broglie's mind*, the content of the initial descriptor $\Phi$ became just more 'complex'. The amplitude $a(\mathbf{r},t)$ of the function $\Phi$ pointed now toward a statistical-probabilistic prediction for results of repeated registrations of the position of the corpuscular-like singularity posited to be involved by the amplitude of any specimen of the considered microstate, while the phase $\beta(\mathbf{r},t)$ from $\Phi$ was conceived to continue to point toward the physical individual wave movement from the corpuscular wave of any one specimen of the studied microstate. But in fact this *violated* one of the numerous semantic rules that work inside our minds concerning the processes of conceptualization, which are still very ill-known. For instance: When one applies arithmetic to a factual problem it is explicitly interdicted – for semantic reasons – to add prunes with apples, notwithstanding that inside arithmetic the operation '+' is defined without restriction. Analogously, inside mathematical physics works implicitly an interdiction to represent by one same mathematical descriptor two features that qualify physical entities of different semantic natures, so two different physical entities: it is just a fact that if this interdiction is transgressed, sooner or later this leads to confusions. So quantum mechanics has *started* directly with a violation of a hidden semantic rule. The spontaneous public reasoning – that feels the semantic rules – has tended to compensate this violation by selecting *only* the statistical mathematical representation of a microstate. But this contributed to simply abandon in the non-spelled-out the individual physical descriptive elements.

Let us suppress the initial in-distinction mentioned above via the following second notational convention (to be added to the first one from (6II).3):

***Notational convention 2***. The physical individual wave-like phenomenon introduced in the domain of scientific conceptualization by one realization of the operation $G$ of generation of one specimen $\sigma(ms_{G,cw})$ of the studied microstate $ms_{G,mw}$ – *in the sense of (1')* – will be systematically denoted by an ***individual*** *wave*-function denoted $\Phi_{G,cw}(\mathbf{r},t)$ assigned to each *one* specimen $\sigma(ms_{G,cw})$ of the studied microstate $ms_{G,mw}$ [82]. This mathematical form is posited here to *include* the representation of also the 'corpuscular singularity(ies)' from the amplitude. While the *state*-function from the $QM_{HD}$-state-ket $|\Psi_G(\mathbf{r},t)>$ that is associated with $ms_{G,mw}$ will represent *exclusively* a mathematical tool for predictive statistics of results of measurements on individual specimens $\sigma(ms_{G,cw})$. With this convention in memory with its full expression, we usually shall go back to the simplified notations $\Phi_G$ and $|\psi_G>$.

The distinction introduced above does not in the least interdict any possible degree of similitude between the global mathematical forms of $\Phi_G$ and $|\psi_G>$ [83].

On the ground lighted by all the preliminary considerations from *(6.II).4.2.1* and the distinction from *(6.II).4.2.2* we shall now start a more detailed examination of the nowadays "theory of quantum measurements".

### (6.II).4.3. The coding rule implied by the $QM_{HD}$-formalism

Since we refuse von Neumann's representation of quantum measurements we go back to the initial representation of these, where the measuring-apparatus is *not* represented. But we conserve the Hilbert-Dirac representation. The other notations as well as the point of view remain those from the association *[IQM-QM$_{HD}$]* enriched with the *new* contents already gained in the chapters 5.II and 6.II; so we are no more inside the current $QM_{HD}$ alone. Armed in this way the main purpose in this section is the following one: Identify *how* – inside the current $QM_{HD}$ – the observable result of one *physical* measurement-evolution of the *one* specimen of the studied microstate that accordingly to *[IQM-QM$_{HD}$]* is involved in *one* act of measurement, is more or less explicitly supposed to be translatable in terms of *one* definite eigenvalue of the measured quantity.

We admit that via the Schrödinger equation of the problem acted by an initial evolution-hamiltonian operator $\mathbf{H}$ it has been possible to identify the state-ket $|\psi_{G,\mathbf{H}}(\mathbf{r},t)>$ of the microstate $ms_G$ to be studied, where $t \geq t_o$ and $t_o$ is the initial moment. If at a time $t_1 \geq t_o$ one wants to measure the observable $\mathbf{A}$ on a specimen of $ms_G$, the $QM_{HD}$-procedure is as follows ((5.II)1). Write the expansion of $|\psi_{G,\mathbf{H}}(\mathbf{r},t)>$ on the basis $\{|u(\mathbf{r},a_j,)>, \forall j\}$, of eigenket of $\mathbf{A}$ for the moment $t_1$ :

$$|\psi_{G,\mathbf{H}}(\mathbf{r},t_1)>/\mathbf{A} = \sum_j c_j(a_j,t_1)|u(\mathbf{r},a_j,)>, \quad \forall j \tag{18}$$

---

[82] We maintain the notation $\Phi$.

[83] For an unbound microstate there certainly exists a strong similitude between the mathematical function $\Phi_{G,cw}(\mathbf{r},t)$ appropriate for representing $\sigma(ms_{G,cw})$, and the state-function $\psi_G(\mathbf{r},t)=a(\mathbf{r},t),t)e^{(i/\hbar)\varphi(r,t)}$ from the state-ket of $ms_G$. But in consequence of the predictive task assigned to $\psi_G(\mathbf{r},t)=a(\mathbf{r},t)e^{(i/\hbar)\varphi(r,t)}$ – also certainly – there is no identity between these descriptors (this is now clear by the definition of the amplitude function $a(\mathbf{r},t)$; but even the phase $\varphi(\mathbf{r},t)$ from $\psi_G(\mathbf{r},t)$ might indicate only some sort of mean-phase with respect to the unknown individual phase functions $\beta(\mathbf{r},t)$ that are involved in the set of specimens $\{\sigma(ms_{G,cw})\}$ of the studied microstate $ms_{G,cw}$ (cf.(14). Anyhow, for now the fact is that in general we do not know what equation can yield as solutions the functional representations of the *individual specimens* $\sigma(ms_{G,cw})$; and even though de Broglie has asserted a common equation for $\psi_G$ and $\Phi_{G,cw}$ and has characterized it in detail (de Broglie [[1956],[1957]), nothing insures a priori the general validity of a common equation. During the years 1960-1980 neither de Broglie nor anybody else seems to have been aware of the essential role – for consensual verifiable knowledge – of the operation of generation of the entity-to-be-studied, *when the considered microstate is* ***put in this role***; nor, a fortiori, of the other related mathematical representations of individual operations. Correlatively, nobody seems to be aware that *'giving' the initial state-ket and the initial wave-function constitutes a key-action for quitting the realm of pure conceptuality and stepping over into the domain of factually predictive and verifiable consensual knowledge.*

The *statistical* prediction concerning the outcome of an ***A***-measurement on the microstate represented by $|\psi_{G,\boldsymbol{H}}(\boldsymbol{r},t_1)>$ is given by the set of numbers $\{|c_j(a_j,t_1)|^2, \forall j\}_{\boldsymbol{A}}$.

As for the *verification* of this prediction, one proceeds accordingly to the following algorithm: From $t_1$ on the action of the evolution-hamiltonian operator ***H*** that worked during the time-interval $(t_1\text{–}t_o)$ is stopped and the state-ket $|\psi_{G,\boldsymbol{H}}(\boldsymbol{r},t_1)>$ is subjected to a new sort of evolution represented by an ***A****-measurement-evolution-ket* $|\Psi_{G,\boldsymbol{H(A)}}(\boldsymbol{r},t_1\leq t\leq t_f)>$. This evolution is performed accordingly to the Schrödinger equation *acted by an* ***A****-measurement-hamiltonian* ***H(A)*** *that commutes with* ***A***. One ***A***-measurement-evolution *takes a time* $(t_1\leq t\leq t_f)$ where $t_f$ marks the end *of the considered* ***A****-measurement-evolution; thereby this evolution –is* ***individual***, it concerns *one* act of measurement. Nevertheless the measurement-evolution-ket $|\Psi_{G,\boldsymbol{H(A)}}(\boldsymbol{r},\ t_1\leq t\leq t_f)>$ is a ***statistical*** descriptor that is made use of exclusively via its expansion with respect to ***A*** where *all* the expansion coefficients are present.

Furthermore in

$$|\Psi_{G,\boldsymbol{H(A)}}(\boldsymbol{r},\ t_1\leq t\leq t_f)>/\boldsymbol{A} = \sum_j c_j(t_1))|u(\boldsymbol{r},a_j,)>, \quad \forall j \tag{19}$$

the expansion coefficients are the same as in (18). So the new evolution (19) of the initial expansion $|\psi_{G,\boldsymbol{H}}(\boldsymbol{r},t_1)>/\boldsymbol{A}$ – but with ***H(A)*** and $(t_1\leq t\leq t_f)$ – is supposed ***not*** to change the expansion coefficients from (18); while the eigenket $|u(\boldsymbol{r},a_j,)>$ are time-independent.

So one can ask oneself *what* changes – mathematically – during an "evolution" (19).

But this possible question, to be kept in mind, concerns mathematical changes; whereas we, in order to understand what coding rule is supposed to work, we have to first specify what is conceived to change *physically* during $(t_f\text{–}t_1)$, and how. This purpose is likely to lead us somehow from the representation with respect to ***A*** to the *space-time* representation of the measurement-evolution-ket $|\Psi_{G,\boldsymbol{H(A)}}(\boldsymbol{r},t_1\leq t\leq t_f)>/\boldsymbol{A}$.

Let us make a break at this point in order to bring in guiding data.

Remember Gottfiried's presentation of quantum measurements ([1966]) and de Broglie's analyses ([1957]); and also Bohm's analysis [(1951)] of the Stern-Gerlach method for spin measurements.

But quite especially, remember *the method time-of-flight for measuring the momentum observable* ***P***, which is ***basic*** since any observable ***A*** is defined as a function ***A=A(R,P)***, while the position-observable ***R*** has a particular character that can hide semantic specificities. This last method has been thoroughly studied by Park and Margenau ([1968]) and when it is examined closely it appears quite clearly that what is supposed to go on from a physical viewpoint during the process involved in (19) is *such* that:

For any given index *j* a measurement-evolution (19) represents globally the emergence via sufficiently many repetitions of one act of ***P***-measurement, of a statistical correlation between:

- On the one hand, a more or less extended space-time domain $(\Delta\boldsymbol{r}.\Delta t)_{j,A}$] that codes for *one* eigenvalue '$p_j$' of ***P***.

- And on the other hand, the (finite) number of times – posited to determine factually the probability for registering the result $p_j$ – in which the group of all the observable marks registered during one act of ***P***-measurement, is located *anywhere* inside $(\Delta\boldsymbol{r}.\Delta t)_{j,\boldsymbol{P}}$ (whereby the correlation is statistical).

*Which amounts to a coding-procedure.*

The method 'time of flight' has been formulated exhaustively and the case of the observable ***P*** is both basic and paradigmatic (the case of 'presence' ***R*** is trivial and degenerate while any observable ***A*** is a function ***A(R,P)***; whereas the case of spin-measurements is only paradigmatic). Therefore I summarize it explicitly below, in *[IQM-QM$_{HD}$]*-terms. We recal it beneath.

Let *δE(G)* be the space-domain covered by one realization of the operation *G* via an apparatus *A(G)* for generating specimens $\sigma(ms_{G,cw})_n$, $n=1,2,...N$, of the microstate $ms_G$ to be studied. Place a very extended detection-screen *S* sufficiently far from the space-domain *δE(G)* for permitting to assimilate *δE(G)* to only a point denoted *O*, *relatively* to the distance *OS* between *δE(G)* and *S* measured on an axis *Ox* that starts at *δE(G)≈O* and is perpendicular on the plane of *S*. We act as follows:

***(a)*** We effectively carry out with *A(G)* an operation $(G)_n$ and we denote by $t_{no}$ the time when $(G)_n$ *ends*, indicated by a system of two interconnected chronometers, one of which is connected to the generation-apparatus *A(G)* and the other one is connected to the screen *S* (the index '$_n$' individualizes the considered realization $(G)_n$ of *G*). The duration *δt(G)* of the operation of generation *G* does *not* come in, alone will matter the time elapsed between the moment $t_{no}$ when the operation of generation *G ends* and the time $t_n$ when an impact is recorded anywhere on *S*.

***(b)*** If between *δE(G)≈O* and the screen *S* there pre-exist macroscopic fields or material obstacles, at $t_{no}$ the fields are extinct and/or the 'obstacles' are removed by a convenient device.

***(c)*** After some time an impact is produced on a spot of the screen *S* that we indicate by $\mathcal{P}(x_n, y_n, z_n)$ where the coordinates are written with respect to a Cartesian system of reference that includes *Ox* and *(Oy, Oz)* are in the plane of *S*. When the impact-point $\mathcal{P}(x_n, y_n, z_n)$ emerges on *S* the system of chronometers indicates the time $t_n$ and *"the 'time of flight' $\Delta t_n=t_n-t_{no}$"* [84] (that has been automatically calculated in the system).

***(d)*** Let $\mathbf{d_n}$ designate the vector-value of the distance between *δE(G)≈O* and $\mathcal{P}(x_n, y_n, z_n)$. The square of the absolute value of this distance is $|\mathbf{d_n}|^2=d_{xn}^2+d_{yn}^2+d_{zn}^2$ where $d_{nx}$, $d_{ny}$, $d_{nz}$ are calculated with respect to the specified Cartesian referential.

***(e)*** The vector-eigenvalue $\mathbf{p_n}$ of the quantum mechanical momentum-operator $\mathbf{P}$ and its absolute value $|\mathbf{p_n}|$, are then calculated according to the formulas $\mathbf{p_n}=M(\mathbf{d_n}/\Delta t_o)=M\mathbf{v}$ *and* $|\mathbf{p_n}|=M\sqrt{(d_{xn}^2+d_{yn}^2+d_{zn}^2)}/\Delta t_o$, where *M* is the 'mass' associated with the involved specimen $\sigma(ms_{G,cw})_n$ of $ms_G$ such as this mass is defined in classical mechanics and in atomic physics[85], and $\mathbf{v}$ means 'velocity'. This completes the considered act of momentum-measurement.

Now we note what follows. The global apparatus is made up of: a generation apparatus *A(G)*; a system of two chronometers; the suppressor of external fields and/or obstacles; the screen *S*. The observable physical marks produced during the considered act of measurement, are: the point-like mark $\mathcal{P}$, the position of the needle of one chronometer at the beginning of the act of measurement, and the position of the needle of the other chronometer at the end of this act. The observable manifestations enumerated above are not themselves numerical values, nor do they 'possess' any quale of which the direct perception is necessarily associated in the observer's mind with the involved specimen of the studied microstate. They are only perceptible physical marks – say $\mu_{1n}$, $\mu_{2n}$, and $\mu_{3n}$, respectively – registered on 'recorders' of the utilized global apparatus.

---

[84] 'Flight' of what? Obviously a model is involved. Probably the model of a classical 'mobile'.

[85] So indeed here the ***classical*** *model of 'mobile' is involved*. For the mass *in de Broglie's sense* has a *different* definition, and this might come out one day to be very important. But probably the de Broglie-Bohm approach (Bohm [1952]), de Broglie [1956]) was not yet much known in 1968.

The *meanings* associated with the recorded marks as well as the numerical values associated with these are defined:

- With respect to permanent reference elements.

- By the way in which is *conceived* what is called 'one act of measurement of, specifically, the eigenvalue $p_n$ of the momentum-observable ***P***'. *This, in agreement with IQM, presupposes a general model of a microstate* with respect to which this way of conceiving the measurement procedure makes sense and can be conceptually integrated in the previously achieved structure of scientific knowledge. And it seems utterly clear that:

**-** The general concepts of 'momentum-value ***p***' and of 'mass *M*' that are involved in the time-of-flight method, such as it is *nowadays* formulated, are founded upon the ***classical*** model of a 'mobile'. This indicates that in the required conditions (in particular the extinction of all the exterior fields) the method works for a classical model even though in fact a specimen of a microstate is involved.

And the method as a whole seems 'reasonable' precisely *because, and* ***only*** *because* it is designed so as to change during one act of momentum-measurement certain features of the posited classical mobile – namely its position – but in a way that does not alter *that what has to be measured* for the involved specimen of such a mobile – namely the momentum-value $p_n$ of the momentum observable ***P*** at the *beginning* of the considered act of measurement. Indeed the procedure would be totally arbitrary in the absence of this assumption.

**-** The vector-value $d_n$ calculated from the observable marks $\mu_{1n}$, $\mu_{2n}$, and $\mu_{3n}$, has an origin that is permitted to *vary* inside some non-negligible space-domain $\Delta r$ (because the origin of the 'flight' of the posited 'mobile' simply cannot be defined strictly when the mobile indicates a physical entity that is a specimen $\sigma(ms_{G,cw})_n$ of a microstate $ms_G$).

The method time of flight is the only one that, accordingly to nowadays $QM_{HD}$, seems to be regarded as 'legal' for measuring the momentum observable. This is so because *it is implicitly supposed to realize the inner structure* ***of an eigenstate of the momentum observable*** *throughout the considered act of measurement*, namely the plane-wave structure required by the expansion (19) $|\Psi_{G,H(A)}(r, t_1 \leq t \leq t_f)>/P$ written for the observable ***P***.

In short, we conclude that the measurement-evolution (19) that is implied by the time-of-flight method, ***does*** *entail a coding rule*, via a statistical correlation. Indeed: Let us denote by $\mu_{pj} \equiv \{(\mu_k, k=1,2,3)_j\}_{pj}, \ j=1,2,...J_P,$ the group of *all* the observable marks produced by any one act of ***P***-measurement. On the basis of a reasoning drawn from the classical mechanics but that can be coherently related with the modelling postulate $MP(\{\sigma(ms_{G,cw})\})$and the relation (1') $G_t \leftrightarrow (ms_{Gt,cw} \equiv \{\sigma(ms_{Gt,cw})\})$, it is possible to specify for any eigenvalue $p_j$ of ***P*** a corresponding space-time domain $(\Delta r.\Delta t)_j$ such that if the whole group of marks $\{(\mu_k, k=1,2,3)_j\}_P$ is registered *anywhere* inside $(\Delta r.\Delta t)_j$) this means 'registration of the eigenvalue $p_j$'. So that one can write:

**[**(registration of the marks $\{(\mu_k, k=1,2,3)_j\}_P$ anywhere inside $(\Delta r.\Delta t)_j$ ) *means* the result ' $p_j$' **]** (20)

This, as announced, is a statistical correlation that acts like a coding rule for ***individual*** *acts of* ***P****-measurement*, in the sense defined in (1.I).2. So it seems likely that – on the basis of the method time-of-flight and comforted by the analyses of the Stern-Gerlach method for spin-measurements which mutatis mutandis leads to a similar conclusion – de Broglie, Bohm, Park and Margenau have admitted more or less implicitly that ***any*** act of quantum measurement involves a coding rule of the general form (20). We denote this view by *BBGPM*.

In the present context the conclusion (20) is very interesting because it immediately suggests the following new considerations.

- No general proof of the coding rule (20) has been worked out inside $QM_{HD}$ because *it concerns individual measurement-evolutions that find no formal place inside* $QM_{HD}$. But for our re-constructive purpose this is not a crucial circumstance. Indeed, more or less explicitly, Park and Margenau have succeeded to somehow 'prove' inside $QM_{HD}$ the correlation (20) for the particular case of the momentum observable $\boldsymbol{P}$. Moreover, as already remarked, for the basic position observable $\boldsymbol{R}$ a correlation of the same form is tautologically realized. And the $QM_{HD}$ 'postulate of representation' of an observable $\boldsymbol{A(R,P)}$ permits to form $\boldsymbol{A}$ by a simple calculus from the pair $\boldsymbol{(R,P)}$ of the two basic observables, so that also the eigenvalues $a_j$ of $\boldsymbol{A}$ can be constructed via definite algorithms from the eigenvalues of $\boldsymbol{R}$ and $\boldsymbol{P}$. So inside $QM_{HD}$ the coding procedure (20) can be applied to any observable $\boldsymbol{A}$, in some specifiable conditions.

- But the method time-of-flight involves the classical 'mobile'-model of a microstate; whereas inside $[IQM\text{-}QM_{HD}]$ the modelling postulate $MP(ms_{G,cw})$ from (6.II).2 introduces the *G-(corpuscular-wave)-model* of a microstate defined by (1') for any specimen $\sigma(ms_{G,cw})$ of the studied microstate $ms_G$. And according to this model, when the studied microstate involves inner quantum fields the condition of extinction of all the fields that act upon the corpuscular-like singularity from the wave of the involved specimen of the microstate cannot be realized any more.

> This means that inside $[IQM\text{-}QM_{HD}]$ the way of *conceiving* what can adequately be called 'an act of measurement of the observable $\boldsymbol{A}$', might in general *change* when the studied microstate involves inner quantum fields.

Nevertheless it remains quite conceivable a priori that for unbound microstates without inner quantum fields the form (20) of a coding rule might subsist and be coherently inserted in $[IQM\text{-}QM_{HD}]$ in agreement with the *G*-(corpuscular-wave)-model of a microstate. But in order to decide whether this is so or not, it seems necessary to better *understand intuitively* – just to thoroughly understand – on what grounds deeper than merely the above-mentioned algorithmic constructability of a one-to-one coding-rule with form (20) – *BBGPM* could have been led to admit implicitly a coding rule of this form for any observable $\boldsymbol{A}$. So we concentrate now upon discerning such grounds.

This brings us back to comparison with a classical mobile. The main obstacle in the way of such a comparison is that the formalism from $QM_{HD}$ does not distinguish between statistical descriptions and individual descriptions (the last ones are not even specified). Correlatively, in (19) the measurement-evolution-hamiltonian $\boldsymbol{H(A)}$ conserves only the *mean* value of the eigenvalues of $\boldsymbol{A}$. Whereas a coding relation of the form (20) for the result of one *individual* act of measurement on a microstate of one micro-system is quite essentially required to specify just *one* definite eigenvalue $a_j$ of $\boldsymbol{A}$ (accordingly also to the general concept of coding relation introduced in (1.I).2). So in order to be able to obtain the sharpest possible comparability of (19) with a measurement-evolution of a classical mobile, we should focus as strongly as possible the representation from (19) upon *one* eigenvalue $a_j$. Let us then make use of the legal *limit* of the concept '$|\Psi_{G,\boldsymbol{H(A)}}(t_1 \leq t \leq t_f)>/\boldsymbol{A}$' *by considering only one term from the sum (19)*:

$$c_{j'}(t_1)=0 \text{ pour } \forall j' \neq j, \quad |c_j(t_1)|=1, \quad c_j(t_1)=1.e^{i\alpha(x,\, t1)}, \qquad |\Psi_{G,\boldsymbol{H(A)}}(\boldsymbol{r},t_1 \leq t \leq t_f)>/\boldsymbol{A}=(1.e^{i\alpha(\boldsymbol{r},t1)}|u(\boldsymbol{r},a_j)>)/\boldsymbol{A} \tag{21}$$

where $\alpha(\boldsymbol{r},t)$ is an arbitrary phase-function. This – strictly speaking – does not lead outside the formalism of $QM_{HD}$; but it places just *upon* its frontier. In (21) the representation (19) becomes 'tangent' to $QM_{HD}$.

> According to $QM_{HD}$ the unique eigenvalue $a_j$ that is involved in (21) coincides with the mean value of the possible eigenvalues, so it is *itself* conserved by the measurement-evolution generated by $\boldsymbol{H(A)}$.

*Thereby we have finally extorted an individual qualification from the statistical formalism of $QM_{HD}$.*

However *what* ***changes*** in time in (19) – mathematically *and physically* – is not yet clear.

Then let us go over into the ‘physical’-space ***R***–representation of (21) where the considered measurement-evolution does play out inside the framework of our factual perceptions. In the physical space, the ket from (19) $|\Psi_{G,\mathbf{H(A)}}(\mathbf{r},t_1\leq t\leq t_f)>/\mathbf{A}$ is conceived as a 'wave-*packet*' endowed with one maximum to which it is possible to associate an 'individual' dynamic comparable with that of a classical 'mobile'. *But with the limit-form (21) of (19) the corresponding **R**-representation does not entail any more any evolution.* It simply stays unchanged; it ceases to serve the very aim for which the ***R***-representations have been defined inside $QM_{HD}$. So, while the limiting form (21) itself of the ***A***-representation (19) has permitted to obtain an *individual* insight by having started with a statistical descriptor, this same limiting form (21) cuts off quite radically any comparability between the statistical descriptions from $QM_{HD}$ and the individual description of a mobile in classical mechanics. This brings into evidence that – in general, and unsurprisingly – *it is impossible to pinpoint **individual** mechanical qualifications via a **statistical** description of mechanical movement.* It explodes into the attention that any quantum-mechanical state-ket (19) $|\Psi_{G,\mathbf{H(A)}}(\mathbf{r},t_1\leq t\leq t_f)>/\mathbf{A}$, even if it is represented in the 'physical' space, consists of just a set of classified *numbers* each one of which specifies a result – with respect to a referential – of a position-measurement on a specimen $\sigma(ms_G)$ of the studied microstate $ms_G$. The fact that the statistical-probabilistic distribution of these numbers possesses a maximum, in general, that changes its position in the 'physical' space, *is occulted in (19) by definition*; and even if it were not, it is a maximum of an abstract nature that cannot produce observable marks [86].

While according to *[IQM-$QM_{HD}$]* the ***R***-localization of that what *can* produce observable marks – namely the singularity from the amplitude of the corpuscular wave of the one specimen $\sigma(ms_{G,cw})$ of the involved studied microstate – is on the contrary very localized in the physical space, at any time, but it lies on the level of *individual* conceptualization and it remains exterior to the formalism of $QM_{HD}$.

Between the $QM_{HD}$ representation (19) and that what acts in the coding-procedure (20) there is a silenced semantic representational clash, a collision between abstract statistics of numbers and individual features of the physical individual entities that are involved, namely specimens $\sigma(ms_{G,cw})$ of the involved studied microstate.

This clash is silenced by the fact that the formalism of $QM_{HD}$ either absorbs any individual entity or feature in a conceptually inadequate way by placing it on the statistical level of conceptualization (think of the case of the eigenket in (II.6).1)), or it tolerates from them merely ghost-like verbal reflections on the statistical representations.

So here we have touched the limit of the investigation that can be made inside $QM_{HD}$ in order to found a general coding rule of the type (20). And notwithstanding the loss in the ***R***-representation of (21), of any mechanical significance, the hints entailed by the preceding considerations suffice already for having conveyed a direct, intuitive, analogical understanding of the *way* in which inside *[IQM-$QM_{HD}$]* it *will* be possible to connect intelligibly the $QM_{HD}$ representation (19) of the ***A***-measurement-evolution-ket

[86] This illustrates into what conceptual-descriptional impossibilities and inadequacies is cornered an exclusively statistical representation – realized ***before*** an individual representation because the approach has been developing ***top-down*** – when, later, this representation is confronted to the purpose to generate the individual physical entities and features on which any statistical description is necessarily ***founded***: a war of temporal order bursts out (We are in the domain of applicability of the theorem of Ehrenfest that 'connects' $QM_{HD}$ with classical mechanics; but the 'connection' is a clash (marked by the *general* assertion of Heisenberg's non-classical principle and theorem)).

$|\Psi_{G,\mathbf{H(A)}}(\mathbf{r},t_1\leq t\leq t_f)>/\mathbf{A}$, with the *G*-(corpuscular-wave)-model of a microstate, by going over into the *IQM*-domain of individual representations of the specimens of the studied microstate. For, going back in $QM_{HD}$, we can add that the space-time parameters that define a wave-packet are very adjustable. They permit to quite satisfactorily *approach* – via a statistical representation – the two essential features to be researched for the coding-purpose, namely:

***(a)*** Conservation of a mean value of the eigenvalue $a_j$ peaked around *one* value $a_j$ as strongly as one wants.

***(b)*** Choice of a definite direction and a convenient degree of stability for the dynamics of the maximum of the wave-packet throughout the time interval $(t_1\leq t\leq t_f)$, via the control of the 'external' fields involved by ***H(A)*** [87].

***(c)*** The possibilities *(a)* and *(b)* permit to identify the optimal choices for the *locations of the registering devices* (this is what permits the method 'time-of-flight'). (Inside *classical* mechanics it is obvious that a mechanical displacement of a mobile throughout which the value $a_j$ of a given mechanical quantity *A* keeps constant, leads the mobile into a *predictable* spatial domain $\Delta\mathbf{r}_j$ if the displacement lasts sufficiently; and that this domain can become as distant as one wants with respect to any other domain $\Delta\mathbf{r}_{j'}$ that corresponds to another value $a_{j'}$ of *A*, with $j'\neq j$ (the source-domain being the same). Which permits to mutually singularise these spatial domains).

So the $QM_{HD}$-parameters of the measurement-wave-packet (19) $|\Psi_{G,\mathbf{H(A)}}(\mathbf{r}, t_1\leq t\leq t_f)>/\mathbf{A}$ assigned to each measurement evolution of the studied microstate can be adapted to the goal of insuring that the initial – *individual* – value $a_j$ of ***A*** from the corresponding specimen of this microstate, stays constantly, with *controllable* approximations, on the direction of displacement that reaches a more or less spatially extended registering device where *any* impact means *'$a_j$'* [88].

Together, all the preceding considerations permit to understand inside *[IQM-$QM_{HD}$]* on what an extraordinary sort of mixture of concepts can have taken form the – implicit – supposition that *any* quantum measurement does admit a coding rule of the form (20); namely that the statistical $QM_{HD}$ representation of a microstate *can always* be brought to ***consist***, wholly, of exclusively an evolution of a form of the type (20)[89].

However in spite of the extraordinary character of this path, it leads us to *admit* here a priori – in agreement with *BBGPM* – the possibility of principle to transpose in *[IQM-$QM_{HD}$]* terms coding-measurement-evolutions of the formal type (20). But we know in advance that the reasons that ground this acceptance apply ***exclusively*** to *unbound* microstates of *one* microsystem and with *simple* operation of generation, i.e. to microstates that evolve ***in the absence of quantum-fields*** that can work on the involved 'singularity'. Which is a very severe restriction. So we know from the start that we shall

---

[87] 'External' in the sense that they can be manipulated (generated, suppressed) from our macroscopic level.

[88] What a prowess we have finally accomplished! The only way to find answer to *a basic* question of physics – how to know the meaning, in conceptualized terms, of the observable result of an act of measurement – inside *a basic theory of physics* called 'quantum mechanics', has been to do, ***what*** ?

***(a)*** To laboriously unmask individual intruders $\{|u(\mathbf{r},a_j,)>, \forall j\}$ surreptitiously injected into an abstract statistic of mere *numbers* $\{|c_j(a_j,t_1)|^2, \forall j\}_A$ drawn from the set of all such statistics of which a $QM_{HD}$-state-ket $\Psi_G$ – by itself – does *entirely* consist.

***(b)*** And therefrom, by use of faint reflections by these individual intruders $|u(\mathbf{r},a_j,)>$, of aspects of physical individual entities and features, to draw – at distance – on the level of the classical representation of the dynamics of another sort of individual real physical entities called 'mobiles', a virtual trace that is alike to a shadow of a 'mobile'. (The convolutions from this last verbal expression only translate the intrication of the reasoning that has permitted to knock out the conclusion formulated above). This conveys a feeling of the distance introduced between meaning and representation by the top-down constructed statistical formalism from $QM_{HD}$.

[89] And let us recall that – via the expansion-postulate – this concept is quite essentially tied with the reference-concept of an *eigenstate – an **a-temporal** sample of arbitrary spatial extension –* of the wave-movement in the neighbourhood of the singularity from a de Broglie corpuscular-like wave.

have to face the problem of defining also another coding procedure, valid for microstates that do involve quantum fields.

### (6.II).4.4. The major confusions from the $QM_{HD}$-representation of a measurement-evolution

We shall now concentrate upon the significances involved by the mathematical expression of the $QM_{HD}$-representation of quantum measurements. What follows might sometimes seem repetitive and trivial. But finally, we hope, this feeling will fade out and leave place to a feeling of full elucidation. Only a very detailed and rigorous analysis can dislodge sediments of thinking and expressing installed by more than a century of thought, of formal research, and of teaching.

#### *(6.II).4.4.1. Preliminary recall*

The evolving *state-ket* $|\psi_{G,H}(\mathbf{r},t)>$ of the studied microstate $ms_G$ is supposed for the moment to be always specifiable directly in mathematical terms by use of the Schrödinger equation of the problem.

- $|\psi_{G,H}(\mathbf{r},t)>$ works as a reservoir of all the *potentially* available predictive-numbers that $QM_{HD}$ can offer via algorithms concerning the studied microstate $ms_G$; namely the probabilities of results of measurements of any observable $\mathbf{A}$, performed on $ms_G$ at any time $t\geq t_o$ after the moment $t_o$ when the initial form $|\psi_{G,H}(\mathbf{r},t_o)>$ of $|\psi_{G,H}(\mathbf{r},t)>$ has been specified. The descriptor $|\psi_{G,H}(\mathbf{r},t)>$ has an only *potential* content that remains to be partially worked out according to particular definite predictive purposes. Its outputs are purely abstract, numerical and statistical.

- The expansion (18) $|\psi_{G,\mathbf{H}}(\mathbf{r},t_1)>/\mathbf{A}=\sum_j c_j(a_j,t_1)|u(\mathbf{r},a_j)>, \forall j, \forall t_1$ of $|\psi_{G,H}(\mathbf{r},t)>$ on the basis of eigenket $\{|u(\mathbf{r},a_j>, \forall j\}_A$ of a given observable $\mathbf{A}$, permits to explicate from $|\psi_{G,H}(\mathbf{r},t)>$ – via Born's postulate and for any moment $t_1\geq t_o$ and any observable $\mathbf{A}$ – the predictive probability law denoted $(\mathbf{A},\{|c_j(a_j,t_1)|^2, \forall j\})$ on the emergence by measurement of the eigenvalues $a_j$ of $\mathbf{A}$. This permits to *verify* this law via a subsequent sufficiently long succession of effective realizations of coding-measurement-evolutions for $\mathbf{A}$. So (18) permits to indicate the *total* predictive content of $|\psi_{G,\mathbf{H}}(\mathbf{r},t_1)>$ in the following more explicit form:

$$|\psi_{G,\mathbf{H}}(\mathbf{r},t_1)> \approx_{\text{pred.}} \{\forall \mathbf{A}, \forall t_1, |\psi_{G,\mathbf{H}}(\mathbf{r},t_1)>/\mathbf{A}\} \qquad (22)$$

- In (18) and so in (22) the particular amount of numerical predictive content from $|\psi_{G,H}(\mathbf{r},t)>$ that is tied with any *given* observable $\mathbf{A}$ and any given time $t_1$, ceases to be potential, it is explicated by already achieved calculi. Thereby, though they remain strictly abstract, the expressions (18) and (22) expresses now the whole essence of the $QM_{HD}$ predictive algorithm, namely: the concept of eigenstate of an observable $\mathbf{A}$ that plays the role of a sample of wave-movement around a singularity from a corpuscular wave (6.II)1); the equation $A|u_j(\mathbf{r},a_j)>=a_j|u_j(\mathbf{r},a_j)>, \forall j$ that determines the eigenket in connection with the algebra of observables defined inside $QM_{HD}$; so de Broglie's 'corpuscular-wave' model; and finally, Born's postulate $\pi(a_j)=|c_j(a_j,t_1)|^2$.

*The concept (22) is the explicit core of the $QM_{HD}$ predictive representation.*

- Consider now the second member of the coding-measurement-evolution ket (19) $|\Psi_{G,\mathbf{H(A)}}(\mathbf{r},t_1\leq t\leq t_f)>/\mathbf{A}=\sum_j c_j(t_1)|u(\mathbf{r},a_j)>$. It has the form of an expansion (18) of the state-ket $|\psi_{G,\mathbf{H}}(\mathbf{r},t_1)>$ with respect to $\mathbf{A}$ but that – from the time $t_1$ when one act of measurement-evolution begins and up to the time $t_f$ when this act of measurement evolution finishes – *continues* its evolution in new, '*coding*-conditions', defined by a measurement-hamiltonian $\mathbf{H(A)}$ that commutes with $\mathbf{A}$.

*The $QM_{HD}$ descriptor of any **one** coding-measurement-evolution is a **statistical** descriptor.*

But, as it is well-known, this same statistical descriptor is also posited to finish by a 'reduction' of its statistical character that reveals the *individual* result $a_j$ that the *one* considered act of ***A***-measurement-evolution has actualized out from the whole spectrum $\{a_j\}, \forall j$ of a priori possible results $a_j$.

*So from the beginning the descriptor (19) exhibits – formally – an ambiguous character in what concerns the level of conceptualization on which it is placed.*

- Let us finally notice also that the three descriptors $|\psi_{G,\mathbf{H}}(\mathbf{r},t)>$, $|\psi_{G,\mathbf{H}}(\mathbf{r},t_1)>/\mathbf{A}$ and $|\Psi_{G,\mathbf{H}(A)}(\mathbf{r},t_1\leq t\leq t_f)>/\mathbf{A}$ designate meanings of very different natures, in particular in what is tied with the involved time-parameters. So we stay attentive to the involved time parameters.

### *(6.II).4.4.2. Critical remarks and questions*

In what follows we adopt the point of view of *[IQM-$QM_{HD}$]*.

Consider the descriptor (19) $|\Psi_{G,\mathbf{H}(A)}(\mathbf{r},(t_1\leq t\leq t_f)>/\mathbf{A}=\sum_j c_j(t_1)|u(\mathbf{r},a_j,)>, \forall j$ of a coding measurement-evolution. It is often said that "at $t_1$ the 'system' is 'prepared' for measurement, and correspondingly is also 'prepared' the new ket $|\Psi_{G,\mathbf{H}(A)}(\mathbf{r},t_1\leq t\leq t_f)>/\mathbf{A}$ that represents the measurement-evolution". What happens *factually* during these 'preparations' is neither represented formally nor explicitly stated by consensually regulated ways of using words. In particular ***it is not said how the 'system' is obtained, physically, operationally, in order to (then?) 'prepare' it***. Nor is it explicitly stated how is to be carried out the factual coding-measurement-evolution *itself* for – specifically – a given observable ***A*** and a given sort of microstate. It is only specified that the hamiltonian must commute with ***A***, but on the physical coding-process there are no other indications than the directly *postulated* assertion that each act of ***A***-measurement produces, for *any* 'system' (i.e. microstate), an eigenvalue $a_j$ of ***A*** that is associated with the corresponding eigenket $|u(\mathbf{r},a_j,)>$ and is "indicated by the 'needles' of the registering devices" (just like in von Neumann's unacceptable representation). Everything hovers calmly in the mathematical-verbal spheres. No particular stress whatever is placed upon the fact that, in order for $a_j$ to become *known* and since obviously a statistic of abstract numbers cannot itself trigger physical marks by interaction with physical registering devices, one should be informed how the interaction has to be organized in order that the 'needles' inform us that they indicate one definite $a_j$ and not another $a_{j'}\neq a_j$, or even something else. While in fact the written expression of the descriptor (19) $|\Psi_{G,\mathbf{H}(A)}(\mathbf{r},t_1\leq t\leq t_f)>/\mathbf{A}=\sum_j c_j(t_1)|u(\mathbf{r},a_j>$ places us on the statistical level of conceptualization, as just a continuation of the expansion (18) in new external conditions expressed by the measurement-hamiltonian ***H(A)*** instead of the initially acting hamiltonian ***H***. And just like in the case of von Neumann's unreasonable representation of quantum measurements, here also is totally occulted the question of the procedure to follow in order to register observable physical marks $\{\mu_{kA}\}_j$, $k= 1,2,...n$, from which it be then possible for *us* to *construct the eigenvalue* $a_j$ that is to be regarded as the result of the considered act of coding-measurement-evolution. Even the concept itself of 'coding'-measurement-evolution – that in *IQM* is central – is devoid inside $QM_{HD}$ of a generally defined equivalent, notwithstanding that *precisely the requirement of coding the registered observable marks in terms of one definite eigenvalue $a_j$ of the observable **A** determines the whole definition of the physical content of 'one act of **A**-measurement'*, as it appeared clearly in (6.II).4.3. The method time-of-flight studied by Cohen and Margenau and Bohm's analysis of the Stern-Gerlach procedure have not been followed by an explicit conclusion formulated in general terms. In the text-books it is only added

sometimes that when $a_j$ 'is obtained' this 'fact' is accompanied by a ‘reduction’ of the measurement-ket (19) $|\Psi_{G,\mathbf{H(A)}}(\mathbf{r},t_1\le t\le t_f)>/\mathbf{A}=\sum_j c_j(t_1)|u(\mathbf{r},a_j,)>$ to only one of its terms.

> *But this 'fact' – the postulated obtainment of $a_j$ and the correlative reduction of (19) – is a **formal** 'fact'.*

The identification between mathematical writings and physical facts has become so deep in the minds and so perfect that *what is formal finally banishes what is factual*. Indeed strictly *nothing* is specified concerning what happens physically to the necessarily involved *specimen* of the studied microstate while the formal progression symbolized in (19) goes on. Namely that in general, while $a_j$ 'is obtained' – which in *IQM*-language means: while $a_j$ is *constructed* from physical marks, from the mathematical representation of $\mathbf{A}$, from a model posited for $\sigma(ms_G)$, and from the corresponding *way* of registering *relevant* physical marks $\{(\mu_{kA}\}_j$ – when all this has been done, then the physical *state* of the involved specimen is usually destroyed, even if the involved *system* subsists. (Which is one of the reasons that led us to speak inside *IQM* in terms of micro-*states* of micro-*systems*, *not* directly in terms of 'systems', if one wants to stay entirely clear). On the other hand, curiously, the concept of "successive measurements" has gained a solitary emergence and it has been variously represented in more or less fiction-forms that float high above genuine physical operability.

All this, though in a certain sense it is well known by many physicists, is not treated. Consequently, the necessity, in general, to generate *another* specimen of the studied microstate before entering upon a new measurement-evolution (19), does not in the least trouble the attention. So: *The necessarily repeated physical operation of generation G of a physical and individual specimen $\sigma(ms_G)$, remains more or less hidden in the void of factual meaning of the verbal expressions ‘preparation of the system' and 'preparation of the measurement-evolution-ket'. The basic concept of operation of generation is not consensually formed and stably asserted inside $QM_{HD}$.*

> In these conditions the unavoidable and obvious necessity to make use of a clearly individual representation of the measurement-evolutions for verifying a statistical description, remains obscure.

As for the *possibility* to make use of individual measurement evolutions for – *also* – constructing factually any statistical description, not only for *verifying* a mathematically defined statistical description as it is done in $QM_{HD}$, this does not even appear on the far horizon. (And why should it, when the basic descriptor $|\psi_{G,\mathbf{H}}(t)>$ is posited to be always available via exclusively mathematical means and to involve already any expansion (18) $|\psi_{G,\mathbf{H}}(\mathbf{r},t_1)>/\mathbf{A}$, which, directly, generates mathematically the probability-law to be verified?).

In short, all the individual descriptors *(G, $G_t$, $ms_G$, $\sigma(ms_G)$, MesA, [G.MesA], [$G_t$.MesA], etc.)* that inside *IQM* are singularized and mutually distinguished, not only are not represented in the mathematical formalism of $QM_{HD}$, but moreover, in the verbal expressions that accompany the mathematical $QM_{HD}$-representations they act *without being defined*, in a lacunar and chaotic way, intermittent and uncontrolled. And so a thick conceptual mud has banked up. The Fig. 6 is strictly valid.

### *(6.II).4.4.3. The reduction problem*

Consider now the postulated ‘reduction’ of the descriptor (19) $|\Psi_{G,\mathbf{H(A)}}(\mathbf{r},t_1\le t\le t_f)>/\mathbf{A}=\sum_j c_j(t_1)|u(\mathbf{r},a_j,)>$, $\forall j$, at the final moment $t_f$. Let us recall, for self-sufficiency of the present argument, that some authors have asserted that from a general conceptual viewpoint this ‘reduction’ does not seem unacceptable when one thinks of the general formal representation of the calculus of probabilities. Each one realization of the

involved 'experiment' that generates by repetition the whole universe $U=\{(e_j, j=1,2,...J)\}$ of possible elementary outcomes $e_j$, actualizes only one outcome from $U$. This can be verbally indicated in a loose way by saying that "each realization of this experiment 'reduces' the permanent and a priori global potentiality $U$ to only one elementary event $e_j$"; why not? Well, let us repeat that in the case of the $QM_{HD}$-descriptor $|\Psi_{G,\mathbf{H(A)}}(\mathbf{r},t_1\leq t\leq t_f)>/\mathbf{A}$ it *remains* that the 'reduction' of the $QM_{HD}$-descriptor $|\Psi_{G,\mathbf{H(A)}}(\mathbf{r},t_1\leq t\leq t_f)>/\mathbf{A}$ is not compatible with the mathematical rules imposed by a linear formalism, whereas in the case of general probabilistic writings an incompatibility of this sort is not imposed [90]. But one can also go a little further and suggest that it might be found to be *conceptually* inadequate to express, inside *one* same descriptor, one actual individual outcome of one given eigenvalue $a_j$, *and* also the representation of the whole a priori set of potential individual outcomes.

But beyond all this, what remains most mysterious is:

> *Why* should one *desire* that an individual procedure of measurement performed on an individual physical specimen of a physical entity-to-be-studied in order to verify via repetitions of this procedure a statistical prediction concerning this entity, be *itself* represented statistically? *Why should one want to get entangled in such a circle?*

When one stops a sufficiently long moment to focus genuine attention upon this strange conceptual situation that has been brought into evidence so repetitively that it vanishes by trivialization, all of a sudden, like in certain optical illusions, a surprising sort of summarizing explanation leaps to one's eyes as an obviousness: Just because historically only the statistical descriptor became first available and then it stayed for a long time the unique available conceptual resource:

> The $QM_{HD}$-descriptor (19) $|\Psi_{G,\mathbf{H(A)}}(\mathbf{r},t_1\leq t\leq t_f)>/\mathbf{A}=\sum_j c_j(t_1)|u(\mathbf{r},a_j)>, \forall j,$ of a measurement-evolution, with the 'reductions' that it requires, is just a desperate aborted attempt at crowding inside a unique mathematical statistical descriptor all that is mentioned below.

***(a)*** On the one hand, the representation of ***all*** the *individual*, physical, actual, and *successively* realized coding-measurement-evolutions $[G_t.MesA]$ defined in (3.I).4, each one of which ends with the individual factual, actual registration of its *own* result that consists of a group $\{\mu_{kA}\}_j$, $k=1,2,...n$, of physical marks that code in terms of *one* eigenvalue '$a_j$'.

And on the other hand *also*

***(b)*** The *globalized*, unique, statistical, a-temporal, abstract assertion of the $QM_{HD}$-predictive distribution $\{|c_j(t_1)|^2\}, \forall j$ of pure numbers, each one of which is the cardinal and also the posited probability of a corresponding class of outcomes of one a priori possible result '$a_j$'.

All this has been crowded inside one – statistical – descriptor.

But such an extraordinary attempt quite certainly involves gross confusions, and these do certainly entail various uncontrollable vicious consequences. One such consequence concerns the time-parameter that comes in, and it is identified below.

---

[90] Moreover, such a mathematical representation might even suggest in certain minds that it is a 'fact' that the statistic *itself* achieves the individual experiments, so that one only has to find out what goes on factually when this happens? Who knows? Concerning $QM_{HD}$ anything succeeds to seem conceivable. Think of the current face-value way of understanding Schrödinger's ironical cat-example, or much better, of Everett's infinity of parallel universes that – without any irony in this case – are asserted to be 'really' generated by each 'reduction' of a mathematical writing on a sheet of paper: In such views the mixture between formal descriptors written on paper or screens, and physical facts, reaches not only perfection, but also greatness.

Consider the chain of statistical descriptors that leads to a measurement-evolution (19) and then includes it. Let us denote this chain as (ch):

$$[\,|\psi_G(\mathbf{r},t_o)>-|\psi_{G,\mathbf{H}}(\mathbf{r},(t_1-t_o))>-|\psi_{G,\mathbf{H}}(\mathbf{r},t_1)>/\mathbf{A}-|\Psi_{G,\mathbf{H}(A)}(\mathbf{r},t_1\le t\le t_f)>/\mathbf{A}\,] \qquad (ch)$$

The measurement-evolution (19) is connected to this chain via the *statistical* time-parameter $t_1$. Verbally, the statistical time-value $t_1$ from (ch) is indicated as the time when "a measurement-evolution begins", and the statistical time-value $t_f$ from (ch) is indicated as the time when "the measurement-evolution finishes". As for the statistical time-value $t_o$, it has not a very definite significance because the whole concept of operation of generation is lacking inside $QM_{HD}$ and so also the notion (13') $[G_{t1}.MesA]$ is lacking (*with* $G_{t1}\equiv[G_o.(t_1-t_o)]$ and also $G_{t1}=F(G_o,EC,(t_1-t_o))$ where $G_o\leftrightarrow[ms_G\equiv\{\sigma(ms_G)]$ according to (1)). Nevertheless it seems clear that *the whole **statistical** chain (ch) is subtended by any **one** individual coding-measurement-succession* $[G_{t1}.MesA]$ (13') with $G_{t1}\equiv[G_o.(t_1-t_o)]$ and also $G_{t1}=F(G_o,EC,(t_1-t_o))$ where $G_o\leftrightarrow[ms_G\equiv\{\sigma(ms_G)]$ according to (1)).

But inside $QM_{HD}$ there is no indication whatever of where and how a physical specimen $\sigma(ms_G)$ of the studied microstate is generated and loaded factually inside the one individual thread of *factual* coding-measurement-succession $[G_{t1}.MesA]$.

In these conditions the two concepts of a $QM_{HD}$ statistical chain (ch) and an individual *factual* coding-measurement-succession $[G_{t1}.MesA]$ are radically disconnected from one another. But there are correspondences of meaning that can be established as it is shown in the following figure 7. And for once we shall go to an extravagant limit of explicit statements of obvious trivialities, in order to radically expurgate the unacceptable $QM_{HD}$ representation of the measurements of its entire absurdity.
What follows applies to both sorts of representation of quantum measurements, either of generation of the statistical predictions, or verification-measurements of already constituted statistical predictions.

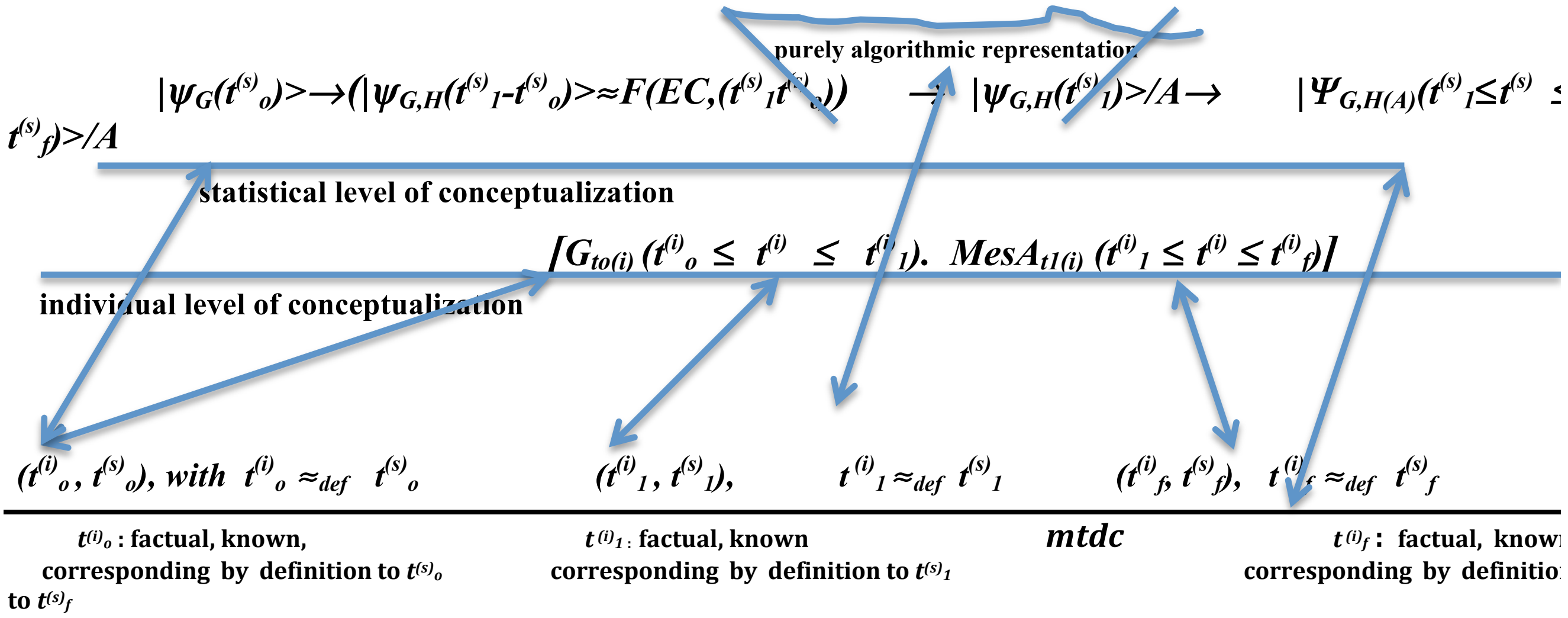

**Fig.7.** The correspondences on a meta-temporal dimension between statistical $QM_{HD}$ times
and *IQM* individual times

The formal $QM_{HD}$ chain (ch) concerning a given problem of measurement is entirely placed on a level of statistical conceptualization.

*Each* thread of an individual, physical coding-measurement-succession $[G_{t1}.MesA]$ is represented on an individual level of conceptualization of the microstates and it is achieved in order to contribute – by many repetitions of it – to the verification of the predictions from (ch). Each *one* realization of a succession $[G_{t1}.MesA]$ runs beneath the *whole* statistical chain (ch) and along it, in its own specific, factual and individual temporal universe, carrying each time *one* physical specimen $\sigma(ms_G)$ of the studied microstate, from the moment $t^{(i)}_o$ when this specimen has been factually generated by the corresponding realization of the operation $G_o$ from $G_{t1(i)}=F(G_o,EC,(t^{(i)}_1-t^{(i)}_o))$, to the final moment moment $t^{(i)}_f$ when it triggers its *own* contribution of observable marks that codes for one given eigenvalue $a_j$. For indeed, quite obviously:

> It is the one specimen $\sigma(ms_G)$ carried by each one individual coding-measurement-evolution $[G_{t1}.MesA]$ that triggers upon physical devices one group of physical observable marks $\{\mu_{kAj}\}_j$, $k=1,2,...n$, that can be translated into one eigenvalue $a_j$ of which the realization is immediately counted as a contribution of one unity to the future final estimation of the degree of factual verification of the result '$a_j$', to the statistical prediction asserted before by the term $|c_j(t_1)|^2$ from the expansion $|\psi_{G,\mathbf{H}}(\mathbf{r},t_1)>/\mathbf{A}$ from (ch). This is 'the needle of the apparatus', nothing else.

But from the point of view of $QM_{HD}$ the presence *inside the chain (ch)* of the considered one specimen $\sigma(ms_G)$ of the studied microstate is a *Deus ex machina*, since the formalism ignores it.

Now, it is obvious that one *common* succession of the statistical time-parameters from (ch) and the individual time-parameters involved by a coding-measurement-succession $[G_{t1(i)}.MesA]$ *cannot be conceived*: The operational-conceptual process exposed in *IQM* by which a statistical representation of the studied microstate emerges by repeated realizations of a very long sequence of repetitions of factual successions $[G_{t1}.MesA]$ – either for generating a statistical prediction or for verifying a pre-established one – is such that the individual physical content from *one* realization of the succession $[G_{t1}.MesA]$ is entirely eliminated as soon as its own input in the statistical representation (ch) is achieved and inscribed: Thereby it has been transformed in just that humble contribution of one unity to the emerging so the changing 'verification-cardinal', say $ver.|c_j(t_1)|^2$, that has been mentioned above and that leads to the stable cardinal $|c_j(t_1)|^2$ asserted in the element $|\psi_{G,\mathbf{H}}(\mathbf{r},t_1)>/\mathbf{A}$ from (ch) for the class of outcomes of the considered eigenvalue $a_j$ of $\mathbf{A}$. If at the end of a sequence of a very big number $N$ of repetitions of a succession $[G_{t1}.MesA]$, all the final 'verification-cardinals' $ver.|c_j(t_1)|^2$ – with any $j$ – are sufficiently equal to those from the element $|\psi_{G,\mathbf{H}}(t^{(s)}_1)>/\mathbf{A}$ from (ch), then the verification of the $QM_{HD}$ prediction concerning $\mathbf{A}$ has succeeded. But this can happen only long after the individual and repeated times $t^{(i)}_o$, $t^{(i)}_1$ and $t^{(i)}_f$. These individual times are all essentially disconnected from the statistical times $t^{(s)}_o$, $t^{(s)}_1$ and $t^{si)}_f$, respectively.

But let us introduce a meta-temporal *dimension of comparison* between (13') and the chain *(ch)*. Let us denote it *mtdc*. Let us project upon *mtdc* both the statistical times and the individual times from the Fig.4. This brings into evidence the possibility to define the *semantic correspondences* $t^{(i)}_o \approx_{def} t^{(s)}_1$, $t^{(i)}_1 \approx_{def} t^{(s)}_1$, and $t^{(i)}_f \approx_{def} t^{(s)}_f$. So on *mtdc*, in the sense just specified, it is finally possible to achieve ***one** common **meta**-succession of 'corresponding meanings'* of the statistical times and the individual times from the Fig.4. This will now permit to genuinely understand the mental sources of the 'reduction problem.

The degree of specification that marks the correspondences from *mtdc* is not uniform. *$QM_{HD}$ does* contain the non-formalized, the only spoken concepts of '*the time when a measurement begins*' and '*the time when the measurement finishes*' of which the statistical or individual status is left unspecified. But the concept of 'the time when the involved *individual specimen* of the studied microstate begins to exist' is not even only verbally contained by $QM_{HD}$. So *in the correspondence denoted in the Fig.4 by '*$t^{(i)}_o \approx_{def} t^{(s)}_o$*'* ***the second time involved,*** $\boldsymbol{t^{(s)}_o}$***, is entirely absent from*** $\boldsymbol{QM_{HD}}$. So we stress that the denotation '$t^{(s)}_o$' – insofar as in the figure 7 it stems from the statistical chain (ch) – is a ***purely*** conventional insertion for our present analytic aim, because its factual significance in connection with the operation of generation (1) *G* (renoted '$G_o$' in (13')) in fact stems exclusively from *IQM* and it exists exclusively on the individual level of conceptualization.

More: In fact in the statistical chain (ch) the time-parameter – re-noted in the Fig.7 with an upper index '$^{s}$' – is only a *global* and *conventional label* of the epoch when this statistic has been achieved. Any temporal regulation of the physically and conceptually meaningful temporal features of a coding-measurement-evolution is strictly individual; *it concerns a corresponding individual succession [*$G_o$*.MesA]*, nothing else; the $QM_{HD}$ values of the time-parameter from the statistical chain (ch) are just a reflexion from this succession, like the light of the moon that stems from the sun. Their insertion in the elements fro (ch) is misleading.

But when – in the botched statistical common representation of all these various individual and statistical times that are implicitly involved by the mixed $QM_{HD}$ representation of the quantum measurements – the *last* moment $t^{(s)}_f$ is reached and an individual outcome $a_j$ is explicitly asserted, this, I dare think, brings forth in the minds at least a vague conceptual uneasiness; possibly even a weak disagreement concerning the in-distinction between statistical features and individual ones; and, may be, a glimmer of wonder on *where* some representation of an agent of interaction with the apparatus has entered the chain (ch) $[|\psi_G(\mathbf{r},t_o)>–|\psi_{G,\mathbf{H}}(\mathbf{r},(t_1-t_o))>–|\psi_{G,\mathbf{H}}(\mathbf{r},t_1)>/\mathbf{A}–|\Psi_{G,\mathbf{H(A)}}(\mathbf{r},t_1\leq t\leq t_f)>/\mathbf{A}]$ of purely numerical abstract structures from the formalism of $QM_{HD}$; a wonder associated with some resurgence of the entirely forgotten common-sense *necessity* of some *individual* and *material* entity able to bring into *physical being* the asserted observable effect. And so, to face at least this last necessity when exclusively a purely statistical formalism was available, the 'reduction' of the measurement-ket $|\Psi_{G,H(A)}(\mathbf{r},t^{(s)}_1 \leq t^{(s)}\leq t_f)>/\mathbf{A}$ has been postulated.

Which then has been felt to be scandalous *from a mathematical point of view !!!!*

While the physical point of view seems to have never been declaredly required, nor a conceptual one, notwithstanding that this whole saga happens inside a theory of *physical* entities.

So a fortiori no physical-and-conceptual explanation has been identified.

That is how the general mixture between vague and lacunar individual concepts and statistical mathematical representations that flaws the whole $QM_{HD}$-formalism, finally came to an outburst in the last link from (ch). This happened because *there – inside one same descriptor –* while the time-parameters $t^{(s)}_1$ and $t^{(s)}_f$ do possess a verbal significance that focuses attention upon meanings, furthermore, the formal representation *has* to be dashed explicitly onto the individual level in order to generate some minimal intelligibility. And this has been done by a verbal diktat that violates the rules of the utilized *mathematical* language, which – consensually – is sacrilege.

From a purely conceptual viewpoint, now, the preceding scan brings forth strikingly that:

> A unique absolute well *ordered* temporal succession of time parameters some of which characterize a statistic to be verified (or generated), and others characterize individual measurement-evolutions tied with the chosen purpose, is just *an impossible concept*[91].

[91] As soon as it is stated this seems trivial; but in fact it certainly is *not* for if it were the reduction problem would not have subsisted so long and steadily.

The 'reduction' problem, when it is transposed in conceptual terms, brings into full light this gross, utterly trivial fact that at the public time when one given individual succession *[$G_t$.MesA]* in the sense of *IQM* is repeated for verifying a statistic, *that* statistic has already been entirely established before, both mathematically and factually, while in order to verify it, one has to construct *another* statistic that is not yet there, and to construct it factually. So conceiving *in one ordered succession*, on one same temporal dimension regarded as a temporal dimension *for a physically realized order of physical facts* – not as an only imagined methodologically introduced meta-dimension –, time parameters that label a representation of physical individual entities and operations, and time parameters that label merely abstract predictive numbers that qualify these entities and operations in globalized statistical-probabilistic terms, is just non-sense.

On the basis of the preceding examination *we refuse, not only von Neumann's representation of quantum measurements, but also the essence itself of the $QM_{HD}$-representation of quantum-measurements*.

### (6.II).4.5. Conclusion on (6.II).4

The analyses from (6.II).4 entail that:

*$QM_{HD}$ is devoid of an acceptable representation of measurements*.

How has it been possible for such a situation to establish itself in the most fundamental among the nowadays theories of physical domains of reality? How has it been possible that so well known and trivial considerations as those brought forth above, have stayed inactive such a very long time?

The answer might lie in the circumstance that the nowadays representation of quantum measurements has emerged under the pressure of the fact that the $QM_{HD}$-formalism, such as it has been formulated in its first phase by a purely mathematical top-down approach, offered mathematically pre-organized conceptual moulds for lodging in them exclusively what seemed immediately useful for prediction. The content in terms of individual physical operations repeated upon individual physical entities, though it is by definition and obviously involved by the concept of a 'statistic', remained entirely non-described by this purely mathematical top-down construction of the predictive statistics from $QM_{HD}$ because it was not directly necessary for prediction. So the question of the physical individual *constructability* of the predictive statistics from $QM_{HD}$ did not even arise. Correlatively, the very distinction between the concept of a predictive statistic of abstract numbers, and its individual, physical genesis, that is involved by definition, faded out from the minds: these two fundamentally different concepts – one with a conceptual status of 'cause' and the other one with a conceptual status of 'effect' tended to identify on a fictitious common *statistical* level of conceptualization. The void of a bottom-up organization of also the individual level of representation of the microstates *confined* the expressivity – ***and the thought*** – on exclusively the statistical level, the first one that had been encountered in the historical development of the quantum theory (and mainly for bound microstates). The minds remained imprisoned in a statistical conceptual fortress that floated on a morass of vague and chaotic individual concepts, as it still does nowadays. There was no organized individual conceptual level of conceptualization on which to leap down and break free. What was lying deeper than the constructed statistics was not visible in the undone beneath them.

It is very noteworthy that such a process has been possible: This draws attention upon the control that must be kept active, in Physics, upon the connections between mathematical expressions and physical-conceptual contents.

*IQM* with its bottom-up approach that starts from local zeros of knowledge and generates intelligibility while it constructs its representations, was not even conceivable at that initial time. This, probably, was why de Broglie's physical and individual concept of a corpuscular-like wave has been hauled upside into the emergent statistical formalism, disguised in two *ill*-understood mathematical concepts, the concept of eigenstate of an observable and the concept of a physical wave-*function*. Under the protection of the void of any conceptual control, the concept of wave-function *moved* immediately into the statistical descriptor called a wave-packet, and then into the still more abstract concept of a state-ket; while the concept of eigenstate was assimilated in its essence to that of an unintelligible limiting sort of state-ket. Whereby the hegemony of *a mathematic of pure statistics* stepped in, and the meanings were silenced. And so the individual features – that irrepressibly do impose themselves to the minds when one deals with measurements – have been stuffed together with statistical features into one common statistical descriptor devoid of inner semantic consistency, the coding-measurement-evolution-state-ket $|\Psi_{G,H(A)}(\mathbf{r},t^{(s)}{}_1 \leq t^{(s)} \leq t_f)>/\mathbf{A}$. This offered at least a way of speaking. And therefore, in spite of all the confusions and inadequacies that somehow – from the very start – *must* have acted in the minds as a resistance, this incoherent descriptor (19) continues to be *taught* up to this very day, even though it has been criticized so persistently and so variously[92]. (Think of Schrödinger's cat that has been written in order to criticize the emerging representation of the quantum measurements and nevertheless it entered the minds as an illustration of 'the strange *behaviour* of the microstates).

On the other hand *the mathematical formalism of $QM_{HD}$ itself*, precisely by the reduction problem, *rejects* the in-distinction between an individual level of conceptualization and the statistical one. The formalism rejects the inclusion of *both* these mutually distinct levels of conceptualization, in one same descriptor; it succeeds to express that *these should be formally represented in some other well-constructed way*: The inner consistency of the mathematical structures – as well as of the logical ones – is sensitive to the semantic contents.

This also is very noteworthy.

This whole history deserves being kept alive in the minds and its implications deserve being made use of.

Notwithstanding the situation brought forth above, $QM_{HD}$ has worked and it continues to work. This theory has achieved remarkable successes and it still will achieve other successes even if it is left just such as it now stands. Indeed the curious omnipresent genius of human mind invents local and individual more or less implicit understandings that permit to act adequately there and when one actually does want to *act*. General methods for thinking well are massively left to the theorists. And it seems that for the experimenters it suffices to believe that a quantum theory of measurements exists in order to measure adequately and to constantly make progress. This teaches humility to those who try to construct theories.

This also *proves that a fully satisfactory theory of quantum measurements is possible* since no doubt it is quite often ‘applied’ without being known.

So, practically, there is no urgency.

But conceptually there is urgency. Indeed, what value of principle – as a *theory* – does a representation of *non*-perceptible microstates possess, if it predicts via purely mathematically constructed predictions and does not state in a clear and generally valid way how to conceive-and-perform measurements for *verifying* the predictions?

[92] Somebody has asserted that "only a new construction can ruin a previously installed construction".

# CONCLUSION ON THE PART II

We have first identified in 5.II the void, inside $QM_{HD}$, of an individual conceptualization of the microstates, which has led to a first clear perception of the concept of *(top-down)(bottom-up) anachronistic collision* between the approach from $QM_{HD}$ and the approach from *IQM*.

Then, inside 6.II we have brought forth that an eigenket $|u(\mathbf{r},a_j,)>$ of a quantum mechanical observable ***A*** has the meaning of a definite mathematically expressed model of wave-movement around the singularity in the amplitude of de Broglie's general corpuscular-wave model of a microstate; namely, a wave-movement that keeps constant the corresponding eigenvalue $a_j$ while the singularity "glides inside its wave".

This has triggered inside *[IQM-*$QM_{HD}$*]* a first – important – constructive step toward the new theory of microstates that is researched here. Namely, we have defined a '***G****-corpuscular-wave model'* of a microstate, denoted $ms_{G,cw}$, and to this we have associated a *modelling postulate* $MP(ms_{G,cw})$. Thereby de Broglie's general model of an 'individual' specimen of a microstate – an *ideal* purely mental model – is translated in physical-operational terms that permit to incorporate it to the approach develop here, marked by a general physically-operational character, consensually predictive and verifiable. Correlatively, via the relation (1') $G_t \leftrightarrow (ms_{Gt,cw} \equiv \{\sigma(ms_{Gt,cw})\})$ the initial relation (1) $G \leftrightarrow (ms_G)$ with $(ms_G) \equiv \{\sigma(ms_G)\}$, that inside *IQM* defines already a new sort of *factually* generated concept of a microstate '$ms_G$' but still possesses an only ***general,*** *purely methodological character*, is enriched with features that concern the *inner* structure assigned to the *IQM*-concept '$ms_G$' on the basis of a particular *model*, de Broglie's 'corpuscular-like wave' model, that is specific of the new theory of the microstates that we try to construct.

We have then brought into evidence the *general* power of clarification entailed by a systematic specification of the existence – *or not* – of a connection between a *ket* from a mathematical $QM_{HD}$-expression, with *a physical operation of generation G of a microstate*, *and a fortiori with the character of this operation of generation (simple or composed).* This led to a useful new denotation of the two sorts of ket from the formalism of $QM_{HD}$.

Finally we have examined the quantum theory of measurements from $QM_{HD}$. We have refused von Neumann's representation on the basis of general conceptual reasons. Then we have identified the coding rule that is implicitly assumed inside $QM_{HD}$ and we have explicated that it is likely to be *devoid of a* ***general*** *validity.*

Then we have examined the essence itself of the $QM_{HD}$-representation of quantum measurements and we have brought into evidence that – and why – it is not acceptable either.

Thereby the preliminary global critical examination of $QM_{HD}$ by reference to *IQM* has come to its end. We can now enter upon an attempt at constructing a second quantum mechanics.

# PART III

# THE PRINCIPLES OF A SECOND QUANTUM MECHANICS

## *rooted into the microphysical factuality in a physical-operational way*

# INTRODUCTION TO PART III

« It would seem that we have followed as far as possible the path of logical development of the ideas of quantum mechanics as they are at present understood. The difficulties, being of a profound character, can be removed only by some drastic change in the foundations of the theory, probably a change as drastic as the passage from Bohr's orbit theory to the present quantum mechanics. »

P.A.M. Dirac, *The Principles of Quantum Mechanics*, Oxford at the Clarendon Press, 4th edition 1958 (1st edition 1930).

The third part of this work is resolutely constructive. On the new basis offered by the clarifications, improvements and purges from the Part II we shall now delineate the essential features – *only* these – of a new sort of mathematical Hilbert-Dirac representation of the quantum predictions and of the verification of these that is operationally rooted directly into the physical factuality and therefrom it is constructed bottom-up.

The whole representation is rigorously inserted in *IQM* that – accordingly to the aim for which it has been constructed – acts as a structure of insertion and reference.

The framework *[IQM-$QM_{HD}$]* that has been introduced in the chapter (6.II).2 by simply juxtaposing *IQM* and *$QM_{HD}$* is conserved. The conjugate use of both these structures has already initiated in 6.II the emergence of a common language and a common system of notations that manifest the growth of a new whole. This growth will continue throughout the Part III. And so at the end of the Part III it will have become clear that the framework *[IQM-$QM_{HD}$]* has never possessed the nature of a scaffold, that, from the start, it has acted and evolved like an organic embryo that has been incorporated to the growth of the conceptual substance and form of a second quantum mechanics, *QM2*.

> *QM2 is not conceived as a new "interpretation" of the nowadays quantum mechanics*. Nor as an achieved new theory of microstates; nor as a didactic itemization of something that already exists. It is a first outline of a fundamentally new representation of microstates required to be general, scientific, and *fully intelligible*.

*In the chapter 7.III* we construct the contours and the main lines of a new representation of the quantum-measurements for unbound microstates (the case of bound microstates is absorbable). This representation has a *factual*-formal character and distinguishes explicitly and constantly between the individual level of conceptualization, and the statistical one

*In the chapter 8.III*, around the core constituted by the new representation of the quantum measurements constructed in the chapter 7.III are sketched out very succinctly the main lines of the whole 'second quantum mechanics', *QM2*. The de Broglie-Bohm approach is *included*, as *a general model of what can be called 'the Universal Physical Substance'*.

> But this model is radically *distinguished* from the operational, consensual, predictive-verifiable Hilbert-space representation developed here.

The juncture between the de Broglie-Bohm basic model and our approach is defined on the basis of the modelling postulate $MP(ms_{G,cw})$ and the relation (1') $G_t \leftrightarrow (ms_{Gt,cw} \equiv \{\sigma(ms_{Gt,cw})\})$, and de Broglie's guiding relation (33) $\mathbf{p}(\mathbf{r},t) = -\boldsymbol{\nabla}.\beta(\mathbf{r},t)$.

*In the chapter 9.III* the second quantum mechanics *QM2* is briefly examined from its outside. From its genesis and its structure is first drawn a general interrogation concerning the relations between facts and mathematical structures inside the modern theoretical physics.

Then a major conclusion is formulated on the nature of scientific knowledge, and a path is specified toward a *methodological* unification of modern physics and even of the whole of modern science.

# 7.III

# A NEW REPRESENTATION OF THE QUANTUM-MEASUREMENTS FOR UNBOUND MICROSTATES

## (7.III).1. THE SEMANTIC SELF-CONSISTENCY OF *[IQM-$QM_{HD}$]* AND ORGANIZATION OF A NEW PERSPECTIVE ON THE REPRESENTATION OF QUANTUM-MEASUREMENTS

This following short section is devoted to the global inner consistency of what will be constructed in the third part of this work.

### (7.III).1.1. Apparent absence of unity inside *[IQM-$QM_{HD}$]* on the statistical predictions and their verification

Consider $QM_{HD.}$

Inside $QM_{HD}$ the predictions on results of quantum measurements are obtained exclusively by mathematical operations. This is not disturbing for didactical idealizations. But when real physical situations are considered that are not heavily a priori restricted, in general it is much more difficult to work out mathematically verifiable predictions than it is asserted in textbooks (think of Schrödinger's Memoire for solving the 'simplest' real case of the one electron from an atom of hydrogen). *In order to dispose of the state-ket* $|\psi_{G,\mathbf{H}}(\mathbf{r},t)>$ *of the problem – that is the source of the whole predictive $QM_{HD}$ algorithm – one has to **'give'** the initial conditions via the initial state-ket* $|\psi_{G,H}(r,t_o)>$, which often might be impossible without admitting *basic* approximations (as for instance the Laplace principle of an initial uniform distribution of the probabilities of the elementary events, *so in particuklar of the initial distribution of 'presence' in space*, etc.). And when the acting hamiltonian cannot be considered to be stationary, or when it simply *is entirely unknown* because unspecified quantum fields are acting, then even the writing down of the Schrödinger equation of the problem itself is impossible. While when this equation *can* be written, nearly always the mathematical generation of a general solution involves already various approximations of which the factual effects cannot be imagined a priori, so they cannot be controlled mathematically.

So the general formal *constructability* of the predictions is far from being generally insured inside $QM_{HD}$.

Furthermore, as it appeared in (6.II).4, inside $QM_{HD}$ when it has been possible to acceptably establish the predictions in a purely mathematical way, the *verification* of these predictions is treated in a way that is dramatically deficient as much from a conceptual point of view as from a mathematical point of view. This deficiency is tied with the fact that the individual elements, entities, concepts, physical operations – which $QM_{HD}$ *does involve* quite fundamentally – are not represented formally, nor are they defined with mere current words, even though in the $QM_{HD}$ representation of the quantum measurements one is systematically brought *by postulation* on the individual level of conceptualization when a result '$a_j$' of one act of measurement has to be taken into account, which result itself also is considered to be individual *by mere postulation*.

Consider now *IQM*.

Inside *IQM* the predictions are constructed *only factually* and then they can be verified *only* by repeating the factual construction (cf. 3.I).5).

So when the fundamental question of prediction and of verification of the predictions is considered, it seems at a first view that inside the framework *[IQM-$QM_{HD}$]* even the slightest degree of unity is lacking between the *IQM* representation and the representation form *$QM_{HD}$*.

But this is only an appearance, and a very misleading one. Just below will appear a deep unity inside *[IQM-$QM_{HD}$]* between predictive probabilities and the verification of these that – remarkably – has silently emerged.

### (7.III).1.2. The conditions of inner semantic self-consistency of the global framework *[IQM-$QM_{HD}$]*

*IQM* has been constructed like a reference-and-immersion-structure for understanding any given theory of the microstates, for estimating its adequacy, and for improving it. As such *IQM* has been deliberately endowed with the maximal generality compatible with its required status. This entailed leaving *undefined* the model of a microstate. And in consequence of the absence of a definite model, the content of a measurement operation '*MesA*' remained equally unspecified inside *IQM*, as well as the 'external conditions' *EC* from the generalized definition (13') $G_t=F(G_o,EC,(t-t_o))$ of an operation of generation of a specimen of the studied microstate. This entailed that throughout the Part I and the Part II of this work the conditions of a full comparability between the *semantic* contents of *$QM_{HD}$* and *IQM* have remained un-defined.

But in the chapter 6.II we have realized several basic *semantic* elucidations and this has changed the conceptual situation. For self-sufficiency of this constructive chapter 7.III we reall these conditions below:

- In (6.II).1 we have identified the meaning of the concept of eigenket of a quantum mechanical observable and its intimate connection with the model of a microstate that acts quite systematically inside *$QM_{HD}$*.
- In (6.II).2 we have incorporated this model to the physical-operational approach practised in this work, via the generalizing redefinition (1') $G_t \leftrightarrow (ms_{Gt,cw} \equiv \{\sigma(ms_{Gt,cw})\})$ of an operation of generation, and the *modelling postulate* $MP(\{\sigma(ms_{G,cw})\})$.
- In (6.II).3 we have stressed the clarifying role inside the *$QM_{HD}$* writings, of the existence – or not – of a direct connection between, respectively, a state-ket or an eigenket, and the operation of generation *G* of the involved specimens of the studied microstate, and we have introduced specifying notations.
- In *(6.II).4.3* we have brought into evidence the implicit existence, inside *$QM_{HD}$*, of a general type (20) of concept of *coding-measurement-evolution* that seems to be valid indeed for an *unbound microstate without quantum fields*.
- In *(6.II).4.4.3* we have shown that the *$QM_{HD}$* representation of one act of coding-measurement-evolution, namely (19) $|\Psi_{G,\mathbf{H(A)}}(\mathbf{r},t_1)>/\mathbf{A}=\sum_j c_j(t_1)|u(\mathbf{r},a_j,)>, \forall j$, attempts to express the second factor *MesA* from an *IQM individual* coding-measurement-succession $[G_t.MesA]$ by use of a *statistical* descriptor that *occults the originating first factor* $G_t$ (cf. the Fig.7 and the corresponding comments) and asserts a final fall on the individual level of conceptualization; which is unacceptable both conceptually and mathematically.
- In *(6.II).4.4* we have reached the general negative conclusion that *$QM_{HD}$* is devoid of an acceptable representation of quantum measurements.

The semantic progresses listed above entail also a massive positive conclusion that has taken form silently and that concerns specifically the general question of prediction and verification of results of quantum measurements. We explicate it below.

To begin with, suppose optimistically that we are in a physical situation that has permitted to write the Schrödinger equation of the considered problem, to solve it, to write down the initial state-ket $|\psi_{G,\mathbf{H}}(t_o)>$, and so, to identify the state-ket of the studied microstate, $|\psi_{G,\mathbf{H}}(\mathbf{r},t)>$ for any moment $t \geq t_o$. Consider now an expansion (18) $|\psi_{G,\boldsymbol{H}}(\boldsymbol{r},t_1)\rangle/\boldsymbol{A}=\sum_j c_j(a_j,t_1)|u(\boldsymbol{r},a_j)\rangle, \forall j$, of this state-ket $|\psi_{G,\boldsymbol{H}}(\boldsymbol{r},t_1)\rangle$ at a given moment $t_1$, and the predictive probability law $(\boldsymbol{A},\{|c_j(a_j,t_1)|^2, \forall j\})$ defined by this expansion inside $QM_{HD}$. The examination from *(6.II).4.4.3* of the $QM_{HD}$ representation of the quantum measurements induces the following remark:

> In general the statistical prediction $(\boldsymbol{A},\{|c_j(a_j,t_1)|^2, \forall j\})$ from an expansion (18) *can* ***not*** be *verified* experimentally otherwise than via a very big number of repetitions of whole individual coding-measurement-succession $[G_{t1}.Mes\boldsymbol{A}]$, *in the sense defined in IQM.*

And in order for the verification to be expressible inside $[IQM\text{-}QM_{HD}]$, the merely general structural definitions imposed by *IQM* and the clarifications from 6.II have to be now completed by specifying in an organized way the *conditions of operational semantic compatibility* between *IQM* and $QM_{HD}$. These are the following ones:

***(a)*** We make use of the model $ms_{G,cw}$ of a microstate, such as this model is expressed by the modelling postulate $MP(ms_{G,cw})$ and by (1') $G\leftrightarrow ms_{G,cw}$, with $ms_{G,cw}\equiv\{\sigma(ms_{G,cw})\}$.

***(b)*** We posit that in (13') $G_t=F(G_o,EC,(t-t_o))$ with $G_o\leftrightarrow ms_{Go}$ and $[G_o.(t-t_o)]\equiv G_t$, the external conditions '*EC*' have to be those expressed inside $QM_{HD}$ by the hamiltonian operator $\boldsymbol{H}$ that, in the Schrödinger equation of evolution of the problem, acted abstractly upon the state-ket $|\psi_{G,\boldsymbol{H}}(t)\rangle$, $(t_o\leq t\leq t_1)$ thus determining it, *when this equation can be written and solved*. This permits then to write inside $[IQM\text{-}QM_{HD}]$

$$G_{t1}=F(G_o, \boldsymbol{H}, (t_1-t_o)) \qquad (13'')$$

where '*F*' means ' a functional of ' and $\boldsymbol{H}$ includes what is called 'obstacles' (walls, barriers, wells).

***(c)*** When inside the succession of operations $[G_{t1}(t_1-t_o).Mes\boldsymbol{A}(t_f-t_1)]$ the act $Mes\boldsymbol{A}(t_f-t_1)$ of an $\boldsymbol{A}$-coding-measurement operation begins at a time $t_1$, $\boldsymbol{H}$ is replaced by a measurement-hamiltonian $\boldsymbol{H}(\boldsymbol{A})$ that commutes with $\boldsymbol{A}$; and then, *throughout the duration* $(t_{f1}-t_1)$ *of the act* $Mes\boldsymbol{A}(t_f-t_1)$, *the measurement hamiltonian* $\boldsymbol{H}(\boldsymbol{A})$ *acts upon the (unknown)* ***individual*** *physical wave-function* $\Phi(\boldsymbol{r},t)=ae^{(i/\hbar)\beta(r,t)}$ *of the specimen* $\sigma(ms_G)$ *of the studied microstate* $ms_G$ *that is involved, accordingly to the distinction introduced in (6.II).4.2.2 between a state ket and a wave-function*[93].

***(d)*** The act of coding-measurement-evolution $Mes\boldsymbol{A}(t_f-t_1)$ from the succession $[G_{t1}.Mes\boldsymbol{A}(t_f-t_1)]$, and its formal representation, must be explicitly defined in a manner consistent with: the modelling postulate $MP(ms_{G,cw})$ accordingly to *(a)*; the type of microstate that is considered (in the sense of the definitions from (2.I).1); the mathematical language chosen for constructing the representations.

The specifications *(a),(b),(c),(d)* define the a priori conditions of inner semantic-operational self-consistency of the framework $[IQM\text{-}QM_{HD}]$. *Inside* $[IQM\text{-}QM_{HD}]$ *they apply a posteriori to IQM also.* All these specifications must be added to those made in *(6.II)4.4.3* (that are illustrated in the Fig.4) in what concerns time parameters versus level of conceptualization. On the basis of requirements obtained in this way we shall now bring into evidence the *nature* of the unity that can be realized inside $[IQM\text{-}QM_{HD}]$ with respect to statistical predictions on results of quantum measurements and verification of these. As long as this unity will be realizable we shall stay inside the framework *[IQM-$QM_{HD}$]*. When this will cease to be possible – which will happen – the framework *[IQM-$QM_{HD}$]* will have to be improved explicitly.

---

[93] This condition brings forth the general ambiguity, inside $QM_{HD}$, of the significance of a solution of the Schrödinger equation, namely whether this solution points toward a physical wave-phenomenon or a statistical descriptor. Indeed the condition (13'') holds also for the indirect acts of measurement on a bound microstate from an atomic or molecular structure where it is clear that *one specimen* $\sigma(ms_G)$ of the studied microstate, with its ***physical*** wave, subsists for an arbitrarily long time and meanwhile interacts from time to time for measurements with test-particles or other devices (Zeeman or Stark effects, etc.); so that in this case both $\boldsymbol{H}$ and $\boldsymbol{H}(\boldsymbol{A})$ act on a physical wave, while the solution $|\psi_{G,\boldsymbol{H}}(t_1)\rangle$ of the equation permits to calculate ***statistical*** predictions. This situation has to be thoroughly understood.

In this way at the end of this chapter we shall be endowed with a general representation of the quantum measurements endowed with intelligibility.

### (7.III).1.3. Basic assertion on the prediction-verification unity inside *[IQM-$QM_{HD}$]*

Consider the factually constructed *IQM*-descriptor (9) $(D/A)(ms_{Gt1})\equiv\{(\varepsilon,\delta,N_0)\text{-}\pi(a_j,t_1)\}_{Gt1}$, $\forall j$. We formulate the following 'assertion' *Ass.1* supported by a corresponding 'argument' *Arg(Ass.1)*[94] :

> ***Ass.1.*** If the *IQM* description (9) $(D/A)(ms_{Gt1})\equiv \{(\varepsilon,\delta,N_0)\text{-}\pi(a_j,t_1)\}_{Gt1}$ has been *constructed* by use of coding-measurement-successions $[G_t.MesA(t_f\text{-}t_1)]$ of which the content has been specified accordingly to the conditions *(a),(b),(c),(d)* of semantic-operational inner self-consistency of *[IQM-$QM_{HD}$]*, then the statistical predictive $QM_{HD}$-law $(A,\{|c_j(a_j,t_1)|^2, \forall j\})$ defined by the expansion (18) can be found to be *verified **if and only if*** it identifies in content – inside the limits permitted by the parameters $(\varepsilon,\delta,N_0)$ – with the statistical assertions of which the factual description (9) consists.

***Arg(Ass.1).*** Obvious: Since inside *IQM* the description (9) $(D/A)(ms_{Gt1})$ is constructed *factually*, in order to verify this description inside *IQM* one is obliged to just repeat its construction. No other way is conceivable inside *IQM*. This means that inside *IQM the experimental verification of $(D/A)(ms_{Gt1})$ is certain a priori, by construction* (3.I).5)[95]. So if a very big number of repetitions of the succession of operations $[G_{t1}.MesA]$ accomplished for *verifying* the $QM_{HD}$ prediction $(A,\{c_j(a_j,t_1)^2\}),\forall j$, are realized in the same way permitted by the conditions *(a),(b),(c),(d)* as the successions of operations $[G_{t1}.MesA]$ by which the *IQM* description (9) $(D/A)(ms_{Gt1})$ has been constructed, then these successions of operations, while they verify $(A,\{|c_j(a_j,t_1)|^2, \forall j\})$, they also reconstruct $(D/A)(ms_{Gt1})$ ■

In other words, the conditions *(a),(b),(c),(d)* of semantic-operational inner self-consistency of *[IQM-$QM_{HD}$]* entail the following unifying identifications:

- Under the constraint of these conditions the predictive content of the *IQM* description (9) $(D/A)(ms_{Gt1})\equiv\{(\varepsilon,\delta,N_0)\text{-}\pi(a_j,t_1)\}_{Gt1}, \forall t_1, \forall j$ – a 'factual probability law' – identifies with the predictive content denoted $\{A,|c_j(a_j,t_1)|^2,\forall j\}$ of the $QM_{HD}$ expansion (18) $|\psi_{G,\mathbf{H}}(\mathbf{r},t_1)>/A\}$; only the notations differ.
- Under the constraint of these conditions the *IQM* 'complete' description (9'') $D_M(ms_G)\equiv\{[(\varepsilon,\delta,N_0)\text{-}\pi_{t1}(G,a_j)\}, (M\pi c(G))_{AB}], \forall A, \forall AB\}$, $\forall j$ has the same *global* predictive content as the $QM_{HD}$ representation (22) $|\psi_{G,\mathbf{H}}(\mathbf{r},t_1)> \approx_{\text{pred.}} \{|\psi_{G,\mathbf{H}}(\mathbf{r},t_1)>/A\}$, $\forall A$, $\forall t_1$ of the state-ket $|\psi_{G,\mathbf{H}}(t_1)>$.

At a first sight it might seem that the pair (*Ass.1, Arg(Ass.1)*) expresses a circularity or at least a triviality. But in fact this is not at all the case because the identifications stated above under the sole constraint of the conditions *(a),(b),(c),(d)*, stem from the specification, in rigorously defined and structured terms, of ***the factual-operational source** – on the **individual** physical-operational level of conceptualization from IQM –* of the statistical-probabilistic ***exclusively*** numerical contents of the two basic $QM_{HD}$ descriptors $|\psi_{G,\mathbf{H}}(\mathbf{r},t_1)>$ and $\{|\psi_{G,\mathbf{H}}(\mathbf{r},t_1)>/\mathbf{A}$. So:

> The enlarged and *common* framework *[IQM-QM_HD]* that acts in the pair (*Ass.1, Arg(Ass.1)*) offers quite non-trivial complete *'vertical'* specifications – from zeros of local individual knowledge up to statistical predictions constructed from these – that inside $QM_{HD}$ alone are entirely lacking (cf. Fig.6 and the comments, and (6.II).1, *(6.II).4.4.3*).

---

[94] Throughout what follows we speak in terms of 'assertions' and 'arguments' because we are not yet inside a formally closed structure where can be given 'proofs' in the strict sense. From now on, for brevity, we drop the specification $(t_f\text{-}t_1)$.

[95] In so far, of course, that $(D/A)(ms_{Gt1})$ has been considered to have been accomplished only when a convenient choice in (9) of the set of parameters $(\varepsilon,\delta,N_0)$ has stabilized the quasi-identical recurrence of $(D/A)(ms_{Gt1})$ when one reconstructs it inside the correspondingly admitted fluctuations.

Via the conditions *(a),(b),(c),(d)* the pair (*Ass.1, Arg(Ass.1)*) *knits together the physical, factual* ***genesis*** *of a* $QM_{HD}$ *predictive statistical law* $(\mathbf{A},\{|c_j(a_j,t_1)|^2,\forall j\})$, *with this law itself, so with the Hilbert-Dirac mathematical formalism.* Thereby any mathematical law appears now as the result of thoroughly defined and structured individual physical operations[96].

Inside the framework *[IQM-QM_HD]* the pair (*Ass.1, Arg(Ass.1)*) fills now entirely the void of formalized organization of the individual-factual level of conceptualization that flaws $QM_{HD}$ (Fig.6) and it does this by beginning to specify the semantic contents of the conceptual moulds deliberately left void inside *IQM* – model of a 'microstate', 'coding-measurement-successions' – that from now on will permit to construct defined ways of ***controlling*** any $QM_{HD}$ predictive statistical law $(\mathbf{A},\{|c_j(a_j,t_1)|^2,\forall j\}$ via actions from the individual, physical-operational level of conceptualization. The force and fertility of this pair will appear below. It is a first massive manifestation of a process of intimate fusion of *IQM* and $QM_{HD}$.

### (7.III).1.4. An immediate consequence of the assertion *Ass.1*: Possibility of principle to circumvent the Schrödinger equation or to complete its performance

As soon as the first impression of triviality is dominated and the assertion *Ass.1* is really understood, it immediately points further toward a very surprising and remarkable possibility.

The $QM_{HD}$-predictive mathematical representations should be, not only *verifiable* by a long set of repetition of coding-measurement-successions $[G_t.MesA]$ that are specified accordingly to the conditions *(a),(b),(c),(d)*, but also ***constructible*** in this same way, radically, just like inside *IQM*.

Indeed, since the results that are obtained by a long set of repetitions of coding-measurement-successions $[G_t.MesA]$ realized accordingly to the conditions *(a),(b),(c),(d)*, do verify the prediction $(\mathbf{A},\{|c_j(a_j,t_1)|^2,\forall j\})$ entailed by the $QM_{HD}$ expansion (18) $|\psi_{G,\mathbf{H}}(\mathbf{r},t_1)>/\mathbf{A}=\sum_j c_j(a_j,t_1)|u(\mathbf{r},a_j)>$ only if they *re-construct* its predictive content, these results also *construct* the content of (18) *exactly in so far that this content is factually true*. So:

> Whenever this is convenient, it should be possible to *circumvent the use of the Schrödinger equation of evolution for generating the state-ket of the studied microstate.*

---

[96] Inside *IQM* the statistical writings (9) $(D/A)(G_t,ms_G,A)$ and (10) $D_M(G_t,ms_G,V_M)$ stress precisely the organic *unity*, in the case of microstates, between the statistical predictive knowledge that, once established, can be considered and made use of separately, and on the other hand the conceptual-physical-operational genesis of this knowledge – respectively the individual genetic triads $(G_t,ms_G,A)$ and $(G_t,ms_G,V_M)$ – via repeated actions $[G_t.MesA]$, $\forall\mathbf{A}$, of the human observer-conceptor, wherefrom the *intelligibility* stems. But there this is done in only general and qualitative terms, whereas inside the framework these terms are particularized via the modelling postulate $MP(\{\sigma(ms_{G,cw})\}$ and they will be worked out quantitatively.

This possibility is the consequence of the change introduced by *IQM*, of origin on the vertical of conceptualization, and so the order of conceptualization, that now progresses bottom-up.

The problem to be solved for attaining this purpose of liberation of the hegemony of the *directly* statistical outputs of the Schrödinger equation is purely representational. Namely one has to find the means, inside *[IQM-$QM_{HD}$]*, to express the *content* of (18) in the same mathematical Hilbert-space form as that from (18), in order to conserve access to Hilbert-space calculi, like inside $QM_{HD}$. The solution might consist of just dropping the results of a long set of factually realized repetitions of coding-measurement-successions *[$G_t$.Mes**A**]*, into the *pre*-imposed Hilbert-Dirac mathematical form of a corresponding expansion (18). Indeed the Schrödinger equation of evolution itself is not a necessary element of a Hilbert-Dirac representation of the predictions on results of quantum measurements. And furthermore, a factual generation of a Hilbert-space representation of these predictions would simply suppress all the restrictions of a purely mathematical nature that the Schrödinger equation carries with it and imposes not only upon the very possibility to obtain these predictions directly by calculus, but also upon the contents of the predictions, in consequence of the so often necessary idealizing approximations. So if a factual generation of (18) were realizable this would entirely free (18) of any purely syntactical restriction that does not concern specifically the physical and conceptual nature of the knowledge that is researched; this knowledge would be obtained just as free of restrictions as it is inside *IQM*.

But of course one would have to survey the possibility of specific consequences of a factual bottom-up approach for generating the Hilbert-space representation.

On the other hand, when it has been possible to write down the Schrödinger equation of the problem, once the predictions have been factually generated for a time-value $t_o$ *and* expressed in the mathematical form of an expansion (18) of a state-ket, this would permit to make then use of the equation for calculating by its use the predictions for subsequent times $t_1>t_o$. Which would *simplify* notably the mathematical tasks; *extend* the domain of effective utility of the Schrödinger equation; and clean the results of any effect of a priori mathematical approximations.

In short, the two distinct representations *IQM* and $QM_{HD}$, far from being unrelated inside *[IQM-$QM_{HD}$]* concerning prediction and verification, quite on the contrary appear now all of a sudden to be soldered to one another in this respect, and under a vast new horizon.

### (7.III).1.5. Possibility of a 'normal' relation between a predictive statistics and the individual measurements that verify it

In (6.II).4 we have analysed why and how the reduction problem has emerged, and it appeared that the main cause has been the absence inside $QM_{HD}$ of an organized level of individual conceptualization where to lodge the representations of individual physical entities and operations. But inside the framework *[IQM-$QM_{HD}$]* the infra-quantum mechanics offers now a strongly organized level of individual conceptualization, while the assertion *Ass.1* specifies the 'vertical' connection between the individual level of conceptualization, and the statistical level. So nothing withstands any more a 'normal' relation between a predictive statistics and the measurements that verify this statistic. Of course, specifying such a normal relation more than this is already very well known by everybody can only consist of sheer triviality from *A* to *Z*. Nevertheless, given the so astonishingly long-lasting more or less passive acceptance of the reduction-problem, it seems necessary to state these trivialities explicitly. So:

- We require that any statistical prediction on results of measurements on microstates be available *before* the *verification*-measurements *begin*. We state this common-sense requirement no matter how the predictive statistic has been constructed – mathematically or factually – and how it is expressed – as a formal law $(\boldsymbol{A},\{|c_j(a_j,t_1)|^2,\forall j\})$, or as a factually-probabilistic description (9) $(D/A)(ms_{Gt1})\equiv\{(\varepsilon,\delta,N_0)\text{-}\pi(a_j,t_1)\}_{Gt1}$.

If the statistical-probabilistic prediction is constructed factually as in the case of (9) $(D/A)(ms_{Gt1})\equiv\{(\varepsilon,\delta,N_0)\text{-}\pi(a_j,t_1)\}_{Gt1}$, the requirement formulated above means that before any action of verification begins, a sequence of many repetitions of an *individual* coding-measurement-succession *[$G_{t1}$.MesA] that – itself – is* ***not*** *expressed equally in statistical terms*, has been repeated by adjusting the parameters $(\varepsilon,\delta,N_0)$ from (9) until the global result exhibits the desired degree of stability of the normed cardinals of the various classes of mutually distinct registered results $a_j$ (the $\pi(a_j,t_1),\ \forall j$), i.e. *they stay constant up to* $(\varepsilon,\delta)$ *as long as* $N_0$ *is kept invariant*; the involved predictive statistic is available only once this has been done[97]. Now:

- A factually constructed probabilistic prediction (9) is verified by its re-construction (*IQM* and *Ass.1*).

- An only mathematically calculated probabilistic prediction $(\boldsymbol{A},\{|c_j(a_j,t_1)|^2,\forall j\})$ is verified *iff* the corresponding stable factually generated $(\varepsilon,\delta,N_0)$-probabilistic distribution obtained as required in *IQM* for (9) *and* accordingly to the conditions of compatibility *(a),(b),(c),(d)* from *(7.III).1.2* – is $(\varepsilon,\delta,N_0)$-identical to this mathematically established prediction $(\boldsymbol{A},\{|c_j(a_j,t_1)|^2,\forall j\})$.

Any experimentalist, no doubt, conceives and works precisely in this way.

While on the individual level of conceptualization, during an *individual* time-interval $(t_f\text{–}t_o)$, *each one* individual coding-measurement succession re-written more explicitly as $[G_{t1}.MesA\rightarrow a_j$, creates *outside* the stable formal statistical chain (ch'), *one* whole thread of operational evolution that begins with an operation of generation (13") and is closed by the registration of its own result $a_j$.

In this way no mixture is made between individual time parameters and statistical time parameters: When the individual successions $[G_{t1}.MesA]\rightarrow a_j$ are repeatedly performed for verification of a previously fully accomplished prediction, the expression of the expansion $|\psi_{G,\boldsymbol{H}}(t_1)>/\boldsymbol{A}$ and of the corresponding predictive law $(\boldsymbol{A},\{|c_j(a_j,t_1)|^2,\forall j\})$ are already *just history, just a pre-constructed reference that waits to be made use of later*[98]. And when finally – *a posteriori* – the verifying confrontation has to be done *globally,* nothing withstands it any more, because at *that* time the mathematically (or factually) established statistical-probabilistic prediction $(\boldsymbol{A},\{|c_j(a_j,t_1)|^2,\forall j\})$ and the factually constructed verifying statistical-probabilistic distribution (9) of all the results $a_j$ produced by the repeated coding measuring successions $([G_{t1}.MesA]\rightarrow a_j)$, are conceptually homogeneous, they are both already fully accomplished statistics.

*No scandalous reduction is necessary any more*. Everything is plainly 'normal', and is intelligible.

[97] In the top-down approach from $QM_{HD}$ this constructive phase is absorbed in the abstract construction of the state-ket $|\Psi_{G,\boldsymbol{H}(A)}(\boldsymbol{r},t>$ via calculus, i.e. in the construction of the Schrödinger equation of the problem, of its solution, and of the determination of the initial state-ket $|\Psi_{G,\boldsymbol{H}(A)}(\boldsymbol{r},t_o>$.

[98] It is not any more an on-going *process*, like in (19), that, while it is statistical i.e. *abstract* and carries *in* it also all the seeds '$c_j(t_1)$' of the statistical predictive law $(\boldsymbol{A},\{|c_j(a_j,t_1)|^2,\forall j\})$, nevertheless is *also* posited to generate *actual* verifying individual and physical data (as it is supposed in *(19)*.

### (7.III).1.6. Global formulation of the purpose formed in (7.III).1

In consequence of the inclusion of the general reference-and-embedding structure *IQM*, the framework *[IQM-QM$_{HD}$]*, such as it has been specified in (7.III).1, offers the possibility of a radically bottom-up conceptualization of the microstates.

*Such an approach, while it generates representation, generates also the corresponding intelligibility.*

The imprisonment into an exclusively top-down approach that initially permits only a *directly* mathematical definition of the statistical-probabilistic predictions conceived under the mental hegemony of the classical thinking, is overcome.

Indeed, the preliminary results obtained above suggest that finally it should be possible to construct a very monolithic and intelligible new Hilbert-space representation of the microstates. A representation where the expansions (18) $|\psi_{G,\mathbf{H}}(\mathbf{r},t_1)>/\mathbf{A}$, $\forall \mathbf{A}$, can be determined *factually* for any time $t_1$.

When the Schrödinger equation of the problem *can* be written, this would permit to instil in it a factually established initial state-ket $|\psi_{G,\mathbf{H}}(\mathbf{r},t_o)>/\mathbf{A}$ that carries in it *indubitably reliable* factual truth, precisely because it *consists* of factual data instead of an priori mathematical representation of factual data that is affected by basic restrictions and approximations; while for any time $t_1>t_o$ the equation would be very useful for *calculating* the expansion (18) $|\psi_{G,\mathbf{H}}(\mathbf{r},t_1)>/\mathbf{A}$ that is *entailed* by this certainly true factually constructed initial expansion $|\psi_{G,\mathbf{H}}(\mathbf{r},t_o)>/\mathbf{A}$; which would be economical, synergetic. Moreover the equation would also be useful in *a new way*, namely as a generator of elements of reference or comparison between mathematically established assertions and factually established ones.

While when the Schrödinger equation of the problem *cannot* be written or when it is difficult to be solved, the factual generation of the needed expansions (18) for *any* time could suffice for nevertheless coming in possession of predictive and verifiable knowledge. We are at the time of big data and of a vertiginous progress of the computing tools. The limitations that stem from the nowadays constraint to make use exclusively of the Schrödinger equation for generating predictive and verifiable consensual knowledge knowledge on microstates, can be dissolved. Only the conceptual-semantic restrictions from *IQM* would remain active.

But *such 'restrictions' are knowledge themselves.*

Our further purpose is to achieve the possibility outlined above.

## (7.III).2. CONSTRUCTION OF A *FACTUAL*-MATHEMATICAL *[IQM-$QM_{HD}$]*-REPRESENTATION OF MEASUREMENTS ON UNBOUND MICROSTATES

Initially in Dirac's Hilbert-space reformulation of the wave-mechanics the projections $(Pr._j|\psi_G)=c(a_j,t_1)$ of the vector state-ket $|\psi_G(\mathbf{r},t_1)>$ of the studied microstate on a basis defined in the Hilbert space of this vector were more or less explicitly considered to be a descriptive feature *specifically* tied with microstates. But since 1954 Gleason's theorem contradicts this view.

This is at the same time a very important and a very seldom-understood fact.

Indeed Gleason's theorem establishes a fundamental and *general* connection between the Hilbert-space mathematical structure, and the *mathematical* representation of the basic concept of a 'measure', in the mathematical sense. So, in particular, *this theorem establishes also a fundamental and general connection between the Hilbert-space mathematical structure and* ***the possible mathematical representation*** *of the omnipresent measures of probability*. Thereby Gleason's theorem concerns also Quantum Mechanics, but in particular.

In short, Gleason's theorem dissolves the belief that the Hilbert-space formalism is specific of the $QM_{HD}$ representation of microstates.

### (7.III).2.1. Gleason's theorem on a Hilbert-space representation of mathematical measures, *versus* Born's postulateand Hilbert-space representation of a probability laws,

***Gleason.*** Gleason's own formulation of his theorem – let us denote it *Gth* – is as follows:

***Gth.*** « Let $\mu$ be a measure on the closed subspaces of a separable (real or complex) Hilbert-space $\mathcal{H}$ of dimension at least three. There exists a positive semi-definite operator $T$ of the trace class such that for all closed sub-spaces $A$ of $\mathcal{H}$

$$\mu(A) = \text{trace}\ (TP_A)$$

where $P_A$ is the orthogonal projection of $\mathcal{H}$ onto $A$. »

This formulation includes the case of mathematical representation of probability measures, of *any* probability measure. So *Gth* applies also to the *factually* established probability measures (5) $(\varepsilon,\delta,N_0)\text{-}\{\pi(a_j),\ \forall j\}_G$ concerning outcomes of measurements on microstates, defined inside *IQM* and associated there with Kolmogorov's mathematical representation. Thereby *Gth* ***offers a choice between Kolmogorov's mathematical representation of probability laws, and Gleason's Hilbert-space representation***.

But *Gleason's theorem itself is quite independent of the concept of 'microstate'. It is independent of even any theory of the microstates.*

Let us briefly examine, in relation with *Gth*, the historical evolution of the mathematical representation of the probabilities that concern outcomes of quantum measurements.

***The Schrödinger-deBroglie Wave-Mechanics versus Born's postulate.*** Consider a problem concerning a microstate $ms_G$. Inside the Schrödinger-deBroglie '*Wave*-Mechanics' – denote it *WM* – let $\psi(\mathbf{r},t)$ be the solution of the Schrödinger equation of that problem; and let $c(a_j,\psi)$ denote at any time the coefficient from the term of index '*j*' in the expansion $\psi(\mathbf{r},t)/\boldsymbol{A}=\sum_j c_j(a_j,t)(u(\mathbf{r},a_j))$ of $\psi(\mathbf{r},t)$ with respect to the observable $\boldsymbol{A}$. Born's postulate – let us denote it ***Bp*** – asserts, about *individual* probability values, the *calculated, predictive*, numerical equalities

$$|c(a_j,\psi)|^2 =_{Bp} \quad \pi(a_j,\psi) \quad \forall j \qquad \textbf{(Bp)}$$

where: the symbol '$=_{Bp}$' is to be read 'equal according to Born's postulate' [99]; $\pi(a_j, \psi)$ is the *abstract*, *predictive*, *individual* probability for the outcome of the eigenvalue $a_j$ if $\boldsymbol{A}$ is measured on the microstate represented by the wave-*function* $\psi$ that has been identified by Schrödinger-calculi, and via the expansion $\psi(\mathbf{r},t)/\boldsymbol{A}$. So:

> *Born's postulate is about the* ***mathematical representation*** *of probabilistic* ***prediction*** *inside WM.*

This prediction *remains to be verified*, which can be realized only via individual *factually* realized coding-measurement-evolutions from repeated successions of operations *[G.MesA]*. And in (6.II).4.4 it has been shown that the 'quantum theory of measurement' does not offer an acceptable conceptual and *mathematical* representation of the verification of the Born-predictions. Not even the distinction between probabilistic predictions and verification of these, is fully expressed.

***Dirac's formalization $QM_{HD}$* versus *Bp and Gth.*** Let us recall that: In the first edition ([1930]) of his famous book Dirac has had the very useful idea to represent the solution $\psi(\mathbf{r},t)$ of Schrödinger's equation of a problem by a *vector* in a Hilbert vector-space. He called such a vector 'the state-ket' associated to the studied microstate and denoted it '$|\psi(\mathbf{r},t)>$'. The set of eigen-*functions* $\{u_j(\mathbf{r},a_j)), \forall j\}$ of any observable $\boldsymbol{A}$ from the wave-mechanics became a basis of eigen-*'ket'* $\{|u_j(\mathbf{r},a_j)>), \forall j\}$ regarded as vectors that are tangent to the Hilbert-space of $|\psi(\mathbf{r},t)>$, which leads to a generalized Hilbert vector-space denoted $\mathcal{H}$.

Thereby Dirac constructed the 'modern' formulation $QM_{HD}$ inside which he gained the possibility of the well-known and very expressive vector-representation of Born's postulate from the *WM*:

Consider a microstate to be studied. Let $|\psi>$ be its state-ket[100] with expansions (18) $|\psi(\mathbf{r},t)>/\boldsymbol{A}=\sum_j c_j(a_j,t)|(u(\mathbf{r},a_j,)>, \forall \boldsymbol{A}$. Let $\{|u_j(\mathbf{r},a_j)>, \forall j\}$ be the basis of eigen-*ket* in $\mathcal{H}$ defined by the observable $\boldsymbol{A}$ and let $\{\pi(a_j), \forall j\}$ be the probability measure on the Universe of elementary events $\{a_j, \forall j\}$ from the Kolmogorov probability-space tied with $\boldsymbol{A}$. In this new mathematical vector-context, Born's postulate acquires the representation

$$\pi(a_j, |\psi>) \quad =_{Bp} \quad |c(a_j, |\psi>)|^2 \; =_{\mathcal{H}} \; |Pr._j|\psi>|^2 \qquad \forall j$$

where: the symbol '$=_{\mathcal{H}}$' is to be read 'equal inside $\mathcal{H}$'; and $Pr._j|\psi>$ is the projection of $|\psi>$ on the $\mathcal{H}$-direction of the eigenket $|u_j>$ from the basis of eigenket $\{|u_j(\mathbf{r},a_j)>, \forall j)\}$ of $\boldsymbol{A}$ and the numerical real value of this projection is $|c(a_j, |\psi>)|^2$. This last geometrical feature is *new* and it is essentially associated with the fact that upon the direction introduced in $\mathcal{H}$ by an eigenket $|u_j(\mathbf{r},a_j)>$ one can explicitly associate the real number $|Pr._j|\psi>|^2 =_{Bp} \pi(a_j, |\psi>)$ with an explicit writing of, also, the eigenket $|(u(\mathbf{r},a_j,)>$ itself, which brings forth the meaning identified in (6.II).1 for the eigenket $|u_j(\mathbf{r},a_j)>$ from the expansion-term $c(a_j,t)|u_j(\mathbf{r},a_j)>$[101].

---

[99] This is not at all an assertion a priory obvious. It is only suggested by the significance of a Fourier coefficient in the classical electromagnetism (de Broglie has made use of the suggestion) and it is active in the association between the eigenket $\{|u(\mathbf{r},a_j>$ and the eigenvalue $a_j$ in the equation of eigenket and eigenvalues of $\boldsymbol{A}$.

[100] For grasping the essence it suffices to consider exclusively pure states.

[101] I quote: "*the sample of that what is counted by the real squared modulus* $|c(a_j,t)|^2$ *of the complex coefficient* $c(a_j,t)$ (exactly as, in the expression *34m*, the symbol '*m*' means that the length that is measured is *34* times the length of the sample of a meter from the National Bureau of Standards of Weights and Measures)". (The mathematical structures, by their axiomatic construction, incorporate meanings that remain often hidden in their applications; this one of the sources of Wigner's wonderment about the power of mathematics.

Now, according to Dirac's initial conception the equality $|c(a_j,|\psi>)|^2 =_{\mathcal{H}} |Pr._j|\psi>|^2$ was just a geometrical feature entailed by a vector-space, that entailed no particular relation with probabilities. So Born's postulate seemed to express a *necessary* and independent supplementary datum. But as soon as Gleason's theorem is known it *entails* for this case that the whole probability measure $\pi_A=\{\pi(a_j),\forall j\}$ can be represented by a trace as defined in *Gth*, while for the *individual* outcomes $a_j$ one can write

$$\pi(a_j,|\psi>) \quad =_{Gth} \quad [\ |Pr._j|\psi>|^2 \quad =_{def} \quad |c(a_j,|\psi>)|^2\ ]\ , \qquad \forall j \qquad (23)$$

where : The symbol '$=_{Gth}$' and '$=_{def}$' is to be read 'equal according to Gleason's theorem'. And the symbol '$=_{def}$' is to be read 'equal by definition', because $|c(a_j,\psi|^2$ *just denotes by definition* the numerical value of the projection $|Pr._j|\psi>|^2$ that is assigned by *Gth* to $\pi(a_j,|\psi>)$. The new information introduced by *Gth* consists precisely of a mathematical relation between a Hilbert-space structural feature – traces (in the mathematical sense) – and a whole probability measure, again in the mathematical sense; which entails a mathematical relation between another Hilbert-space structural feature – $|Pr._j|\psi>|^2$ – and an individual abstract, pre-calculated, predictive probability value $\pi(a_j,|\psi>)$ [102]. ***There is nothing factual in (23)***. This is noteworthy.

***Conclusion on Gth versus Born's postulate***. Born's postulate and Gleason's theorem have different semantic natures; there is no conceptual identity between them, even though they make use of a same mathematical descriptor, namely the representation by $|c(a_j)|^2$ of the real number that measures the individual probability $\pi(a_j)$ of an outcome $a_j$ if ***A*** is measured on a given microstate. Indeed:

- Born's postulate concerns explicitly and exclusively microstates. Like any 'postulate' from Physics it involves assertions of specified facts that concern a definite category of physical entities (there exist probability laws on outcomes of measurements on microstates, the state-ket $|\psi_G>$ of the studied microstate is known, etc.).

- Whereas Gleason's theorem is strictly void of any assertion of this sort; it has the status of a purely logical implication 'if-then' *that concerns a general relation between a mathematical 'measure' and the mathematical structure called Hilbert-space*. When a Hilbert space is made use of for the mathematical representation of microstates, this general relation entails that [***if*** *a probability-measure* $\{\pi(a_j),\forall j\}$ *is somehow mathematically* ***given as such*** *(for instance like in the case of Bp),* ***then*** for the individual abstract predictive probability values $\pi(a_j)$ the writing $\pi(a_j,|\psi>) =_{Gth} |Pr._j|\psi>|^2$ from (23) that involves a Hilbert-space representation, is by itself mathematically self-consistent, and the descriptor $|c(a_j,|\psi>)|^2$ drawn from an expansion (18) $|\psi>/$***A*** permits to effectively calculate the value $|Pr._j|\psi>|^2$; which entails that in such a case Born's postulate is not necessary any more.

> Throughout the history evoked above, the distinction between *predictive* probabilistic assertions and *verification* of these has been very vague and fluctuating, or radically absent. This is intimately tied with the fact that the theoretical microphysics does not represent the individual physical entities or operations, only statistics tied with these while verification is necessarily via individual measurements.

---

[102] One can also say, more clearly perhaps, that the abstract, predictive, individual probabilities $\pi(a_j,|\psi>)$ calculated from $|\psi>$ can be asserted to have the numericalvalue $|c(a_j,\psi)|^2$ that, by geometric definition of $|\psi>$ in $\mathcal{H}$, denotes the value $|Pr._j|\psi>|^2$.

***Gth versus [IQM-QM$_{HD}$].*** Let us now consider *Gth* relatively to *[IQM-QM$_{HD}$]*. Gleason's theorem permits to regard (23) as just a *generally* valid translation from the mathematical language of the general Kolmogorov-theory of probabilities into the mathematical language of Hilbert-space representations[103].

Inside the Dirac formulation *QM$_{HD}$*, Schrödinger's equation is made use of for defining an abstract, calculated global representation of a given probability law, by a Hilbert-vector $\psi>$. But *Gth* does not *require* any equation. It involves a radical *separation* between, on the one hand the nature, the semantic features of the facts and entities to which the considered probabilistic concepts are applied, and on the other hand the mathematical representation of these probabilistic concepts themselves. In particular, ***Gth does not exclude factually constructed mathematical representations***. So:

> *Gth* can be applied *directly* to the *factually* constructed probability laws (5') $\{(\varepsilon,\delta,N_0)\text{-}\{(\pi(a_j)\}, \forall j)\}_G\}, \forall A$, from *IQM*, as well as to the more complex concept of probability law that is involved in the transferred description (9'') where are explicitly included also the meta-correlations $M\pi c(G_t))_{\mathbf{AB}}$ between distinct probability laws that are involved in (5').

Inside *[IQM-QM$_{HD}$]* one can connect each elementary event from the universe $U=\{a_j,\forall j\}$ of possible individual outcomes $a_j$ of a coding-measurement-succession *[G.MesA]* on a studied microstate $ms_G$, with a Hilbert-space-vector – say $V_{\mathcal{H}}$ – *such* that the semantic specificities of each $a_j$ inside the set $U=\{a_j,\forall j\}$ be represented in $\mathcal{H}$ by a corresponding 'semantic direction' $(\sigma\delta)_j$. Then according to *Gth* the set $\{(\sigma\delta)_j,\forall j\}$ of semantic directions defines in $\mathcal{H}$ a basis of eigenket $\{|\sigma\delta_j>,\forall j\}$ and the numerical value of the projection $|Pr._j(V_{\mathcal{H}})|$ of $V_{\mathcal{H}}$ on the direction of a given eigenket $|\sigma\delta_j>$ from this basis can be utilized to represent the absolute numerical value of the individual ***factual*** probability $\pi(a_j,V_{\mathcal{H}})$ from a *factual* probability law (5) $\{(\varepsilon,\delta,N_0)\text{-}\{(\pi(a_j)\}, \forall j)\}_G]$. Then the whole law itself can be represented by the use of an expansion of $V_{\mathcal{H}}$ on the basis of eigenket $\{|\sigma\delta_j>,\forall j\}$.

> The possibility of such an expansion, instead of being insured by postulation as it is done in *QM$_{HD}$* (cf. 5.II).1), *can be realized by a factual construction, thereby charging it with* ***a priori certain factual truth***, like in the case of the *IQM* description (9'') that emerges "verified", i.e. the process of verification just repeats the process of factual construction of the predictive statistics by repeated individual coding-measurement-successions *[G.MesA]*.

Indeed one can construct the Hilbert-vector $V_{\mathcal{H}}$ progressively, via individual coding-measurement-succession *[$G_{t1}$.MesA]*, under a priori organized conditions of *formal* mutual consistency with Dirac's top-down statistical ket-representation of the expansions $|\psi>/\mathbf{A}\ \forall \mathbf{A}$, *from QM$_{HD}$*. Such a factually and progressively generated equivalence $V_{\mathcal{H}}\approx|\psi>$ between $V_{\mathcal{H}}$ and a *QM$_{HD}$* state-ket $|\psi>$ would realize a *connective encounter* between the historical top-down approach that has generated *QM$_{HD}$*, and the bottom-up approach from *IQM*. In this way would be accomplished a new representation of the microstates that incorporates the individual entities and operations with which the predictive statistics are organically tied.

[103] Pitowsky (2008) has drawn this essence in connection with 'quantum logic'. He asserts that for him "Hilbert spaces are a new theory of probabilities". For me the probability-trees of a transferred description in the sense of *IQM* are indeed a new theory of probabilities – of probabilities of *facts*, not of mathematical elements, and that introduces fundamentally new and basic semantic probabilistic features – whereas Gleason's theorem offers just a new and very efficient mathematical language for representing the nowadays classical Kolmogorov-concept of probability.

The final result of such an approach would admit a conceptual-mathematical representation of the quantum-probability laws by *a very synthetic and expressive fusion* between:
- the tree-like representation defined in *IQM* of the *factual emergence* of Kolmogorov-spaces and of correlations between these, via repeated *individual* coding-measurement-successions *[G.MesA]*,

and on the other hand
- the Dirac-Gleason Hilbert-*vector* representation of the probability laws from these spaces.

We shall now try to realize this inside *[IQM-QM$_{HD}$]*. We begin by constructing the announced sort of representation for unbound microstates without quantum fields.

### (7.III).2.2. Factual-formal construction of a Hilbert-space representation of quantum measurements on unbound microstates without quantum fields

The condition of semantic consistency *(d)* from *(7.III).1.2* requires to specify consistently the *IQM*-concept of a *coding*-measurement-evolution *MesA* from an individual measurement-succession *[G.MesA]* performed on a microstate of any sort (in the sense of the definitions from (2.I).1). For the basic case of measurements on unbound microstates with non-composed operation of generation we are already prepared to achieve this. The case of the other types of microstates will be considered later.

#### *(7.III).2.2.1. The coding-postulate for unbound microstates ms(unbound,1)$_{G(n-c)}$ with non-composed operation of generation*

The analyses from (7.III).1 have suggested that the sort of measurement-evolution that is presupposed inside *QM$_{HD}$* by the *BBGPM* approach can – in coherence with the Hilbert-Dirac representation – be conceived to perform implicitly a coding-measurement-evolution of the general form (20). But it also has appeared there that the mentioned supposition has a restricted validity, in this sense that it can be 'understood' only in the absence of quantum fields. (Why this is so will appear later). So in (6.II).4.3 – founded upon the analysis of the method 'time-of-flight' for measuring the basic momentum observable, and by reference also to the Stern-Gerlach procedure for spin-measurements – we have *admitted* the efficiency of a coding-measurement-evolution of type (20) in the absence of quantum fields, so for unbound microstates with non-composed operation of generation. Such microstates will be denoted *ms(unbound,1)$_{G(n-c)}$*. For these we shall now postulate explicitly, in strict and detailed *[IQM-QM$_{HD}$]*-terms, the coding-measurement-evolution of the general type (20) required by the condition *(d)* of inner consistency and that is compatible with the other conditions, *(a),(b),(c)*.

***Recall.*** In (6.II).1 it has been found that inside *QM$_{HD}$* works implicitly de Broglie's general 'corpuscular-wave' model. In (6.II).2, via the modelling postulate *MP(ms$_{G,cw}$)*, this general model has been brought to consistency with the general *IQM* concept (1) of an operation of generation *G* of a specimen $\sigma(ms_{Gt})$ of the studied microstate. So here the coding-measurement-evolution admitted in *[IQM-QM$_{HD}$]* for the particular sort of microstate denoted *ms(unbound,1)$_{G(n-c)}$* will have to be specified in consistency with the modelling *MP(ms$_{G,cw}$)* and with the observable *A* to be measured; *and* furthermore it must obey all the conditions of semantic consistency formulated in *(7.III).1.2.*

Consider a coding-measurement-succession *[G$_{t1}$.MesA]*.

According to the conditions of semantic consistency we have that in (13'') $[G_o.(t-t_o)]\equiv G_t$, with $[G_t=F(G_o,EC,(t-t_o)), G_o\leftrightarrow ms_{Go}$, and the external conditions '*EC*' during $(t_1-t_o)$ are those expressed in $QM_{HD}$ by the hamiltonian operator ***H*** from the Schrödinger equation of the problem. So $G_{t1}$ has to be written as $G_{t1}=F(G_o,\mathbf{H},(t_1-t_o))$ ('*F* 'means: 'a functional of').

According to the modelling postulate $MP(ms_{G,cw})$ the operation $G_{t1}$ introduces via $G\equiv G_o$ a specimen $\sigma(ms_{Go})$ of the initial state of the studied micro-state $ms_{Gt1}$ that is represented by an unknown physical individual wave-*function* $\Phi_{Go}(\mathbf{r},t=a(\mathbf{r},t)e^{(i/\hbar)\beta(\mathbf{r},t)}$ (cf.*(6.II).4.2.2*). So until the moment $t_1$ when the act ***MesA*** of measurement-evolution begins, the evolution of $\sigma(ms_{Go})$ is represented by the action upon the individual *physical* wave-function $\Phi_{Go}(\mathbf{r},t)=a(\mathbf{r},t)e^{(i/\hbar)\beta(\mathbf{r},t)}$, of the exterior fields from the hamiltonian ***H*** that determines also the evolution of the statistical $QM_{HD}$ descriptor $|\psi_{G,\mathbf{H}}(t_1-t_o)>$.

In *(6.II).4.3* we have admitted the *BBGPM* implication that a measurement hamiltonian operator ***H(A)*** that commutes with the measured observable ***A*** – if it works on $\sigma(ms_{Gt1})$ ***after t₁*** and *in the absence of quantum fields* – installs for the wave-function $\Phi_{Gt1}(\mathbf{r},t)=a(\mathbf{r},t)e^{(i/\hbar)\beta(\mathbf{r},t))}$ of $\sigma(ms_{Gt1})$ with $t\geq t_1$, a structure of wave-movement represented by an eigenket of ***A***. While correlatively, for the corpuscular-like singularity in the amplitude of $\Phi_{Gt1}(\mathbf{r},t)=a(\mathbf{r},t)e^{(i/\hbar)\beta(\mathbf{r},t)}$ it generates a dynamic that leads it into a space-domain $\Delta\mathbf{r}_j$ (or a space-time domain $(\Delta\mathbf{r}\Delta t)_j$) that is in a one-one relation with a given eigenvalue $a_j$ of ***A***.

***Formulation***. The above recall leads to the following $[IQM\text{-}QM_{HD}]$ version of a coding-postulate of the general form (20):

> ***P(cod)$_{G(n-c)}$.*** A coding-measurement-evolution ***MesA*** from a succession $[G_{t1}.\mathbf{MesA}]$ performed upon a microstate $ms(unbound,1)_{G(n-c)}$, admits the general representation:
>
> $[(G_{t1}\rightarrow\sigma_\Phi).\mathbf{MesA}(\sigma_\Phi)] \quad \rightarrow_{\mathbf{H(A)}} \quad$ *(marks registered in* $(\Delta\mathbf{r}\Delta t)_j \simeq$ *'$a_j$')* (20')
>
> ($\sigma_\Phi$: an abbreviation for $\sigma(ms_{Gt1})$ and $G_{t1}\equiv[G_o.(t_1-t_o)]$: a functional $F(G_o,\mathbf{H(A)},(t_1-t_o))$.

If in particular it is supposed that the coding-measurement-evolution ***MesA*** is performed by starting it at the time $t_o$ when the initial operation of generation $G_o$ finishes, then we make use of the corresponding particular form of $P(cod)_{G(n-c)}$ :

$$[(G_o\rightarrow\sigma_\Phi).\mathbf{MesA}(\sigma_\Phi)] \quad \rightarrow_{\mathbf{H(A)}} \quad \text{(marks registered in in } (\Delta\mathbf{r}\Delta t)_j \simeq \text{'}a_j\text{')} \qquad (20'')$$

So, like (20), the postulate $P(cod)_{G(n-c)}$ concerns an *individual specimen of the studied microstate*: Inside $[IQM\text{-}QM_{HD}]$ this postulate 'explains' the non-analysed $QM_{HD}$-postulation of 'emergence' of an eigenvalue $a_j$ of the measured observable ***A***[104].

> And it is noteworthy to remark that the $QM_{HD}$ hamiltonian ***H(A)*** works on the individual level of conceptualization, while the $QM_{HD}$ hamiltonian ***H*** works on the statistical level of conceptualization.

---

[104] We are so deeply used to the purely mathematical and purely statistical representations from $QM_{HD}$, that the content of the whole point *(7.III).2.2.1* might seem queer inside a work of theoretical physics. But the reader is asked to remember that we want to root quantum mechanics in factuality, and in a non-perceivable and as yet a-conceptual physical factuality. This requires suppression of the inertial psychological refusals induced by feebly intelligible, purely algorithmic top-down representations, supported by philosophical interdictions, that vitiate the modern microphysics since more than 100 years. *Human minds act on the basis of models* and this fact has to be incorporated explicitly and fully to the processes of generation of scientific knowledge.

This is coherent with our critique of the reduction problem in *(6.II).4.4.3*.

The coding postulate (20') is presupposed throughout the constructive action from the following point.

### *(7.III).2.2.2. Factual-formal construction of a Hilbert-Dirac representation for the results of coding-measurement-evolutions (20') on microstates $ms(unbound,1)_{G(n\text{-}c)}$*

We shall now proceed in a radically constructive way.

We go back to the $QM_{HD}$-concept (22) $|\psi_{Gt1,\mathbf{H}}(\mathbf{r},t_1)>$ $\approx_{\text{pred.}}$ $\{\forall \mathbf{A},\forall t_1, |\psi_{Gt1,\mathbf{H}}(\mathbf{r},t_1)>/\mathbf{A}\}$ that has been mentioned to be the core of the $QM_{HD}$-representation of prediction. We make the following assertion *Ass.2* :

> ***Ass.2.*** Inside *[IQM-QM$_{HD}$]* – via repeated individual coding-measurement-successions *[G.MesA]* that obey the coding-procedure (20') and the consequence (23) of Gleason's theorem – it is possible to generate for a microstate of the type $ms(unbound,1)_{G(n\text{-}c)}$ a formal-*factual* equivalent of the Hilbert-Dirac expression (22) of a state-ket as the set $\{\forall \mathbf{A},\forall t_1, |\psi_{Gt1,\mathbf{H}}(\mathbf{r},t_1)>/\mathbf{A}\}$ of all its expansions, where the involved Hilbert-Dirac state-ket $|\psi_{Gt1,\mathbf{H}}(\mathbf{r},t_1)>$ :
> ***(a)*** *Is independent of the Schrödinger equation of the problem.*
> ***(b)*** Involves an *extension* of its semantic contents from (22), in this sense that – by construction – it offers for any observable $\mathbf{A}$ and any factual initial situation, predictive probability laws and correlations between these that are a priori *certainly* endowed with factual truth inside the limits of the a priori chosen and arbitrarily small $(\varepsilon,\delta,N_0)$ imprecisions from the *IQM*-description (9'') that corresponds to the constructed state-ket in the sense of the assertion *Ass.1*.

***Arg(Ass.2).*** Consider the $QM_{HD}$-state-ket $|\psi_{G1,\mathbf{H}}(\mathbf{r},t)>$ of a microstate $ms_{Gt1}$ of the type $ms(unbound,1)_{G(n\text{-}c)}$.

- The assertion *Ass.1* and the corresponding argument have established that, *if* the probabilistic predictions (in the sense defined in (3.I).1) of any given state-ket $|\psi_{Gt,\mathbf{H}}(\mathbf{r},t)>$ are available and are *verified* by the repeated realization of coding-measurement-successions $[G_t.Mes\mathbf{A}]$, $\forall \mathbf{A}$, $\forall t$ that satisfy the conditions of inner consistency *(a),(b),(c),(d)* defined in *(7.III).1.2*, *then* these predictions are necessarily $(\varepsilon,\delta,N_o)$-identical to those asserted by the factual $(\varepsilon,\delta,N_0)$-probability laws (5') from the *IQM*-description

$$(D_M(ms_G))_t \equiv [(\varepsilon,\delta,N_o)\text{-}\{\pi(G_t,a_j)\}, (M\pi c(G_t))_{\mathbf{AB}}], \ \forall \mathbf{A}, \forall \mathbf{AB}, \forall j, \forall t \qquad (9'')$$

that is constructed by use of the same set of coding-measurement-successions. We select one *given* description of this kind in order to work with it and we denote by $D_M$ this *selected* description.

- So, ***if*** – inside the new framework *[IQM-QM$_{HD}$]* – a *factually constructed* state-ket of the Hilbert-Dirac formal type – let us denote it $|\psi_{Gt1,\mathbf{H}}(\mathbf{r},t_1)>_{fact(DM)}$ – can be associated to the selected *IQM*-description of type (9'') (i.e. to the *factual* probability laws (5') from *it* and to the corresponding correlations $(M\pi c(G_t))_{\mathbf{AB}}$]), ***then*** the *Ass.1* and *Gth* entail that up to imprecisions $(\varepsilon,\delta,N_0)$ we must have the equalities:

$$|c_j(a_j,t)_{\mathbf{fact(DM)}}|^2 \ =_{(\varepsilon,\delta,No)\ (Ass.1,\ Gth)} \ \pi(G_t,a_j)_{fact(DM)}, \ \forall j, \forall \mathbf{A}, \forall t \qquad (23')$$

where: $|c_j(a_j,t)_{\mathbf{fact(DM)}}|^2$ denotes a factually defined representation of the absolute numerical value of the expansion coefficient of index $j$ in the expansion of $|\psi_{Gt,H}(r,t)>_{fact(DM)}$ with respect to the observable $\mathbf{A}$;

- the symbol '$=_{(\varepsilon,\delta,No)\ (Ass.1,\ Gth)}$' is to be read '$(\varepsilon,\delta,N_0)$-equal according to the *Ass.1*' and to the consequence (23) of Gleason's theorem;

- $\pi(G_t,a_j)_{fact(DM)}$ is the $(\varepsilon,\delta,N_o)$-probability of $a_j$ according to the factually constructed description (9'') from *IQM*.

We shall now actually construct factually the state-ket $|\psi_{Gt,H}(r,t)\rangle_{fact(DM)}$, by use of (23').

***1.*** We first prepare the necessary 'materials'. Namely:

***1a.*** The *form* of an as yet unknown state-ket $|\psi_{Gt,H}(r,t)\rangle_{fact(DM)}$ that is conceived a priori as a Hilbert-space vector of the type (22) $|\psi_{G,\mathbf{H}}(\mathbf{r},t_1)\rangle \approx_{\text{pred.}} \{\forall \boldsymbol{A}, \forall t_1, |\psi_{G,\mathbf{H}}(\mathbf{r},t_1)\rangle/\boldsymbol{A}\}$ where each $QM_{HD}$-observable $\boldsymbol{A}$ introduces a basis of eigenket $\{|u_j(\mathbf{r},a_j)\rangle, \forall j,\}$ so that we lodge the factually generated data directly in a Hilbert space $\mathcal{H}$ associated to the microstate to be studied. Therefore we write for each observable $\boldsymbol{A}$ the expansion-form

$$|\psi_{Gt,H}(r,t)\rangle_{fact(DM)}/\boldsymbol{A} = \sum_j e^{i\alpha(A,j)}\, |c_j(a_j,t)_{\mathbf{fact(DM)}}|\, |u_j(\mathbf{r},a_j)\rangle, \ \forall j, \forall t_1 \qquad (24)$$

where the expansion coefficients are deliberately written as a product in order to be able to treat separately the question of the numerical values of the two factors $e^{i\alpha(A,j)}$ and $|c_j(a_j,t)_{\mathbf{fact(DM)}}|$.

***1b.*** We *establish factually* for the studied microstate $ms_{Gt1}$ the $(\varepsilon,\delta,N_o)$-probability laws $\{\pi(G_t,a_j)\}, \forall \boldsymbol{A}, \forall t$, from the description $D_M$ of type (9'') of the microstate to be studied $(ms_G))_t$. We act accordingly to *IQM*, by use of (13'') and of coding-measurement-successions *[G<sub></sub>t.MesA]* that obey the coding postulate (20') $P(cod_{\mathbf{G(n\text{-}c)}})$, and by respecting all the conditions *(a),(b),(c),(d)* of semantic compatibility defined in *(7.III).1.2*).

This exhausts the first purely factual phase of the construction.

***2.*** We shall now establish the points *(a)* and *(b)* from the assertion *Ass2*, denoted respectively *(2a)* and *(2b)*.

***(2a).*** Consider now the *general graphic **form*** (24) – still void of any numerical specification – of the $QM_{HD}$-expansion of $|\psi_{Gt,H}(r,t)\rangle_{fact(DM)}$ with respect to $\boldsymbol{A}$. For each coefficient from (24) we have:

$$c_j(a_j,t)_{\mathbf{fact(DM)}} = e^{i\alpha(A,j)}\, |c_j(a_j,t)_{\mathbf{fact(DM)}}| \ \forall j, \ \forall t_1 \qquad (25)$$

We concentrate upon the factors $|c_j(a_j,t)_{\mathbf{fact(DM)}}|$ (the factors $e^{i\alpha(A,j)}$ will be examined later). So far in (25) the expansion coefficients $c_j(a_j,t)_{\mathbf{fact(DM)}}$ are endowed with, exclusively, the general geometrical meaning that is assigned to them by definition, namely that of the projections onto the directions introduced by the eigenket $\{|u_j(\mathbf{r},a_j)\rangle\}$ of $\boldsymbol{A}$ in the Hilbert-space $\mathcal{H}$ of the factual state-ket $|\psi_{Gt,H}(r,t)\rangle_{fact(DM)}$ that *is to be* constructed. The expressions (24), (25) are still *void* mathematical molds. We shall now endow them with defined contents.

Let us *imagine* for a moment that we are inside $QM_{HD}$ and that it has been possible to write the Schrödinger equation of the problem and to solve it, so that the researched state-ket $|\psi_{Gt,\mathbf{H}}(\mathbf{r},t)\rangle$ and its expansion (18) $|\psi_{Gt,\mathbf{H}}(\mathbf{r},t_1)\rangle/\boldsymbol{A}=\sum_j e^{i\alpha(A,j)}|c_j(a_j,t_1)||u_j(\mathbf{r},a_j)\rangle$ are known. However – because of quasi ubiquitous mathematical idealizations or approximations, in general we would be unable to know also to what a degree Born's postulate is indeed ***factually** true* in this case; concerning the factual validity of the expansion coefficients from (18) there would subsist doubts.

Whereas in our case we are no more inside $QM_{HD}$, we are in the framework $[IQM\text{-}QM_{HD}]$ where according to the achieved point 1 the probabilities $\pi(G_t,a_j)_{fact(DM)}, \forall j, \forall \boldsymbol{A}, \forall t_1$, have been established factually, not by calculi via the Schrödinger equation. And now – by just a *constructive* ***definition*** – we *posit*, up to an $(\varepsilon,\delta,N_0)$-imprecision, the *new* formal-factual equality

$$|c_j(a_j,t)_{\mathbf{fact(DM)}}|^2 \quad =_{;constr,def,,\,(\varepsilon,\delta,No)} \quad \pi(G_t,a_j)_{fact(DM)}\,, \quad \forall j \qquad (26)$$

where the symbol '$=_{constr,\ def.\ (\varepsilon,\delta,No)}$' is to be read 'is equal by definition and with $(\varepsilon,\delta,N_o)$-precision'; the index '$_{fact(DM)}$' is to be read ' factual, involved by *DM* '.

Let us stop here to note immediately that the genesis and the final numerical content of the first member from (26) – $|c_j(a_j,t)_{\mathbf{fact(DM)}}|^2$ – are ***essentially*** *different in nature* from those of the $QM_{HD}$ descriptors from (23) $\pi(a_j) =_{Gth} \; [\,|Pr._j|\psi>|^2 =_{def} |c(a_j,\psi)\,|^2\,]$, $\forall j$, tied with Gleason's theorem (and via ***(Bp)*** with Born's postulate). This is so because of the different sorts of geneses of these two expansion coefficients.

- In (23) that concerns the Hilbert-Dirac $QM_{HD}$ the numerical value of the descriptor $|c(a_j,\psi|^2$ – and also that of the *predictive* probability $\pi(a_j,\psi)$ – have been only calculated from the solution $|\psi>$ of the Schrödinger equation of the problem (hypothetically considered to have been available), as the inserted symbol '$\psi$' reminds explicitly. So this predictive value ***remains*** *to be also verified*, necessarily by repeated individual coding-measurement-successions *[G.MesA]* that incorporate the coding procedure (20').

- Whereas in (26) the descriptor $|c_j(a_j,t)_{\mathbf{fact(DM)}}|^2$ has first been just written in the graphic form of a coefficient from the expansion of a Hilbert-Dirac state-ket, and its numerical value $\pi(G_t,a_j)_{fact(DM)}$ is ***known*** to possess the nature of a 'probability' and to be factually true within an 'imprecision $(\varepsilon,\delta,N_o)$' because it has been constructed such via individual coding-measurement successions *[G.MesA]* and accordingly to the definition of $D_M$ in (9'') from *IQM*; but that now inside $[IQM\text{-}QM_{HD}]$ incorporates also the coding procedure (20'). So *no verification is necessary any more, the verification is insured by construction*[105].

We can now re-write (24) with the following '*factual*-mathematical' form of the right-hand member where instead of the form $|c_j(a_j,t)_{\mathbf{fact(DM)}}|^2$ is inserted the number $\sqrt{\pi(G_t,a_j)_{fact(DM)}}$ :

$$|\psi_{Gt,H}(r,t_1)>_{fact(DM)}\,/\boldsymbol{A} = \textstyle\sum_j e^{i\alpha(A,j)}.\ \sqrt{\pi(G_t,a_j)_{fact(DM)}}.\ |u_j(\boldsymbol{r},a_j)>),\ \ \forall j,\ \forall t_1, \qquad (27)$$

The same procedure is valid for the spectral decomposition of $|\psi_{Gt,H}(r,t_1)>_{fact(DM)}$ with respect to any other dynamical observable different fro $\boldsymbol{A}$. So this settles for any observable the question of the absolute numerical values of the expansion coefficients from (24), (25).

Consider now the imaginary factors $e^{i\alpha(A,j)}$ from (24), (25).

In the expression $e^{i\alpha(A,j)}$ the observable $\boldsymbol{A}$ is a variable. When one passes from $\boldsymbol{A}$ to another *given* observable $\boldsymbol{B}$ the $QM_{HD}$ concept of a state-ket $|\psi_{Gt1,\boldsymbol{H}}(\boldsymbol{r},t)>$ involves conditions of mutual consistency between the expansion-coefficients of $|\psi_{Gt1,\boldsymbol{H}}(\boldsymbol{r},t)>$ with respect to $\boldsymbol{A}$ and the expansion-coefficients of $|\psi_{Gt1,\boldsymbol{H}}(\boldsymbol{r},t)>$ with respect to $\boldsymbol{B}$. These conditions are taken into account in Dirac's theory of transformations and they can be specified via a trivial lemma *L(Ass.2)* made explicit inside $QM_{HD}$ for any state-ket $|\psi_{Gt1,\boldsymbol{H}}(\boldsymbol{r},t)>$:

---

[105] This is nearly a procedural and semantic *opposition* with respect to those assigned to the descriptor $|c(a_j,\ \psi|^2$ from (23); this sort of opposition reflects the general distinction between the top-down abstract and directly statistical approach that marks the formalism of $QM_{HD}$, and the bottom-up factual approach practised here, where the descriptors $|c_j(a_j,t)_{\mathbf{fact(DM)}}|^2$ are constructed by factual individual coding-measurement-successions $[G_{t1}.MesA]$. (This distinction must be somehow tied with the change of the 'good' order of reading, and the choices, in Physics, of a direction of rotation; indeed (26) is 'better' read from right to left, so with the motion of the pointer of a clock on its projection on a line).

***L(Ass.2).*** If in (25) an *arbitrary* set $\{e^{i\alpha(A,j)}\}$ of complex factors is introduced for the observable ***A***, then Dirac's theory of transformations determines *consistently* with this initial choice, all the complex factors to be introduced in all the other expansions of $|\psi_{Gt1,\mathbf{H}}(\mathbf{r},t)>$ corresponding to any other given dynamical observable ***B***, that does not commute with ***A**, so [A,B]≠0.*

***Proof of L(Ass.2).*** Consider the expansion

$$|\psi_{Gt1,\mathbf{H}}(\mathbf{r},t)>/\mathbf{B} = \sum_k e^{i\gamma(\mathbf{B},k)}\,|\,d_k(t,b_k)\,|\,|\,v_k(\mathbf{r},b_k)>, \quad k=1,2\ldots K, \quad \forall t$$
(18')

of $|\psi_{Gt,\mathbf{H}}(\mathbf{r},t)>$ on the basis $\{|\,v_k(\mathbf{r},b_k)>\}$ of eigenket introduced in $\mathcal{H}$ by a given observable ***B***, $[A,B]\neq 0$ that does not commute with ***A***, where the numbers $e^{i\gamma(\mathbf{B},k)}$ $|\,d_k(t,b_k)|$ are the expansion coefficients. For *any* given value of the index *k* we have inside $QM_{HD}$ that

$$<v_k(\mathbf{r},b_k)\,|\,\psi_{G,\mathbf{H}}(\mathbf{r},t)> = e^{i\gamma(\mathbf{B},k)}\,|d_k(t,b_k)| = \sum_j \tau_{kj}(\mathbf{A},\mathbf{B})\, c_j(t,a_j)\,, \quad \forall j,\ \forall t \qquad (28)$$

where $\tau_{kj}(\mathbf{A},\mathbf{B})=<v_k|u_j>$, $\forall j$. So for any complex factor of given index *k* we have a separate condition

$$e^{i\gamma(\mathbf{B},k)} = <v_k|\,\psi_G(t)> / |\,d_k(t,b_k)| = \sum_j \tau_{kj}(\mathbf{A},\mathbf{B})\, c_j(t,a_j) /|\,d_k(t,b_k)|\,, \quad \forall \mathbf{A},\mathbf{B}, \quad \forall t$$
(29)

(where ' / ' is to be read: *divided by*).

So the lemma is proved for a $QM_{HD}$ state-ket $|\psi_{Gt1,\mathbf{H}}(\mathbf{r},t)>$ if the condition (29) is fulfilled.

We impose the condition (29) for also the formal-factual state-ket (27) $|\psi_{Gt,H}(r,t_1)>_{fact(DM)}$.

Thereby the Hilbert-Dirac representation of any expansion of the factual-formal state-ket $|\psi_{Gt,H}(r,t_1)>_{fact(DM)}$ is now achieved: all the contents of the representational elements are fully specified, semantically as well as structurally and numerically. So we finally write:

$$|\psi_{Gt,H}(r,t_1)>_{fact(DM)}/A \equiv \sum_j e^{i\alpha(A,j)}.\sqrt{\pi}\,(G_{t1},a_j).\,|u_j(\mathbf{r},a_j)>, \quad \forall j,\ \forall t_1,\ \forall A \qquad (27')$$

And this in its turn permits to write for the *global* factual-formal construct $|\psi_{Gt,H}(r,t_1)>_{fact(DM)}$ the synthetic integrated form:

$$|\psi_{Gt,H}(r,t_1)>_{fact(DM)} \approx_{\mathbf{pred.}} \{\forall \mathbf{A}, \forall t,\ (|\,\psi_{Gt,H}(r,t_1)>)_{DM,ff}/\mathbf{A}\} \qquad (30)$$

like in (22) for a $QM_{HD}$ state-ket.

The representation of $|\psi_{Gt,H}(r,t_1)>_{fact(DM)}$ from (27') specifies explicitly – and directly – all the first order probability laws involved in (30), from all the probability spaces that crown the branches of the *IQM*-probability-tree $T(G,\forall A)$ of the studied microstate (Fig.2). Furthermore –the imaginary factors $e^{i\alpha(A,\,j)}$ from the expansion-coefficients and the lemma *L(Ass.2)* – (27') are likely to determine also the $[IQM\text{-}QM_{HD}]$ mathematical representation of the general syntactic structures (11) and (11') of the meta-probabilistic correlations $(M\pi c(G_{t1}))_{\mathbf{AB}}$, $\forall \mathbf{AB}$ that have been defined in *IQM*. Indeed – insofar that also the $QM_{HD}$ expression of the *IQM* meta-correlations $(M\pi c(G_t))_{\mathbf{AB}}$ have been specified in $|\psi_{Gt,H}(r,t_1)>_{fact(DM)}$ via (9") and by the condition (29) – the formal-factual relation (30) has necessarily emerged as a *full* correspondent of the $QM_{HD}$ equivalence (22) $|\psi_{Gt,\boldsymbol{H}}(\boldsymbol{r},t)>\approx\{|\psi_{Gt,\boldsymbol{H}}(\boldsymbol{r},t)>/\boldsymbol{A}\}, \forall\boldsymbol{A},\forall t$ [106]. And this correspondent has been constructed without making use of the Schrödinger equation of the problem.

So the point *(2a)* from the *Ass.2* is established.

Let us add that in the equivalence (30), in consequence of the fact that no Schrödinger equation has been made use of, the global factual-formal state-ket $|\psi_{Gt,H}(r,t_1)>_{fact(DM)}$ is *not* endowed with an own explicit integrated functional expression: such as it has emerged, it has to be regarded as just *defined* by the second member of (30).

***(2b)***. We have repeatedly stressed that a $QM_{HD}$ state-ket $|\psi_{Gt,\boldsymbol{H}}(\boldsymbol{r},t)>$ obtained via the Schrödinger equation of the problem is in general marked by purely mathematical constraints – idealizations, approximations, or just guesses, etc.; which – in the absence of any defined individual genesis of the calculated statistical predictions – can induce into the consequences of the mathematical constraints uncontrollable deviations from the unknown factual truth. Whereas in (30) all the involved numerical values $\sqrt{(\pi(G_t,a_j))}$ from the predictive probability laws imported from (27') have been defined via factual individual successions of operations $[G_t.MesA]$ and accordingly to the definitions (5') and (9'') from *IQM*. So, with chosen and arbitrarily small $(\varepsilon,\delta,N_0)$-uncertainties, the semantic contents involved in (30) are by construction in conformity with the factual truth of the asserted predictions ***and*** *with the* ***verification*** *of thes*e, *realized together*: In what concerns the factual truth of the predictions the argument *Arg(Ass.2)* insures it a priori by construction; and in what concerns the verification of the predictions, it follows from (3.I).5 and from the assertion *Ass.1*.

So, in consequence of their independence of an equation of evolution, the predictions drawn from the formal-factual state-ket $|\psi_{Gt,H}(r,t_1)>_{fact(DM)}$ are in general ***different*** from those that are drawn from the corresponding $QM_{HD}$ state-ket $|\psi_{Gt1,\boldsymbol{H}}(\boldsymbol{r},t)>$ and they are certainly true factually, by construction. In this sense (30) *extends* the $QM_{HD}$-domain (22) of possibility of factually true predictions on microstates.

Which establishes also the point *(2b)* from the *Ass.2*.

The whole assertion *Ass.2* is now fully established. ■

---

[106] We believe that the imaginary factors $e^{i\alpha(A,j)}$ determine the $QM_{HD}$ mathematical expression of the *IQM* meta-probabilistic correlations $(M\pi c(G_{t1}))_{\boldsymbol{AB}}$ from a probability-tree and from (9''). Consider the *IQM* descriptor (11) $p(b_k)=\boldsymbol{F}_{\boldsymbol{bk,A}}\{p(G_t,a_j)\}, \forall k, \forall j, \forall(\boldsymbol{A,B})$ that denotes meta-probabilistic correlations $(Mpc(G_t))_{\boldsymbol{AB}}$ between whole probability laws that crown two distinct branches of a given probability-tree. Inside $QM_{HD}$ the Dirac transformation from the Hilbert-space representation of the state-ket $|\psi_G(t)>$ of the studied microstate with respect to the eigenvalues $a_j$ of an observable $\boldsymbol{A}$, to the representation of $|\psi_{Gt}(t)>$ with respect to the eigenvalues $b_k$ of another observable $\boldsymbol{B}$ that does not commute with $\boldsymbol{A}$, is defined by $d_k(t,b_k)=\sum_j \tau_{kj}(\boldsymbol{A,B})\ c_j(t,\boldsymbol{A}), \forall j, \forall k, \forall t$. Of course, the aim of Dirac's $QM_{HD}$ calculus of transformations is entirely ignorant of the *IQM* operational-semantic categorization of the set of all the considered pairs of observable events $\{(a_j,b_k)\},\forall(\boldsymbol{A,B}),\forall t$ tied with the studied microstates, *inside a tree-like probabilistic whole founded upon the operation of generation G or* $G_t$ *that corresponds to the state-ket* $|\psi_{Gt}>$ *of the studied microstate*. This is so because the individual operations $G$ or $G_t$ – like also any corresponding coding-measurement-evolution – are not represented inside $QM_{HD}$. So inside $QM_{HD}$ Dirac's calculus of transformation from one 'representation' of $|\psi_G(t)>$ with respect to an observable $\boldsymbol{A}$, to its resentation with respect to another observable $\boldsymbol{B}$, is asserted as just a mathematical algorithm devoid of any more general meaning. Nevertheless the isomorphism between the writings (11), (11') and those from Dirac's theory of transformations suggests that these formulas point toward the *possibility of a much more general Dirac-like 'calculus of* ***semantic proximities****'*, that remains to be made explicit and exploited: For instance, the scalar product of two distinct state-ket of two different microstates but expressed inside one *same* representation, might be used as a *measure of a concept of 'degree of angular proximity' inside this representation,* so relatively to qualifications by the observable that determines the representation (MMS [1993]); while a Dirac transformation alone leads from one semantic universe to another one that is disjoint from the first. It seems rather clear that inside $QM_{HD}$ the imaginary factors from the expansion coefficients determine phase-relations between different terms in one expansion and furthermore *they define meta-probabilistic correlations* that are likely to coincide with those denoted in *IQM* by the sign $(Mpc(G_t))_{\boldsymbol{AB}}$.

We close this argument by adding explicitly to the formal writing (30) a graphic indication of the remarkable fact that *the predictive content from (30) emerges already verified*:

$$|\psi_{Gt,H}(r,t_1)>_{fact(DM)} \neq_{\mathbf{pred.}} |\psi_{Gt1,\mathbf{H}}(\mathbf{r},t)>, \qquad (31)$$
$$[\ |\psi_{Gt,H}(r,t_1)>_{fact(DM)} \approx_{\mathbf{pred.}} \{|\psi_{Gt,H}(r,t_1)>_{fact(DM)}/\mathbf{A}\}, \forall \mathbf{A}, \forall t_1\}\ ]_{\mathbf{verif}}$$

***Important remark.*** Throughout what follows the equivalence from (31) is just *posited* to entail the possibility to construct for the first member $(|\psi_{Gt,H}(r,t_1)>)_{DM,ff}$ an *integrated* mathematical functional expression of the form $\psi_{Gt1,\mathbf{H}}(\mathbf{r},t)>_{ff}=a(\mathbf{r},t)e^{(i/\hbar)\varphi(\mathbf{r},t)}$ (there should even exist an infinite family of such possibilities, since in the lemma *L(Ass.2)* the initial choice of imaginary phase-factors has been arbitrary). Inside $QM_{HD}$ this integrated expression is obtained by calculus from the Schrödinger equation of the problem (*iff* this equation can be written and the solution can be found on the basis of acceptable approximations [107]). But we stress that *an integrated solution is not more useful than (31) itself* because any prediction-verification process is founded on expansions (22) and the Hilbert-space representation (23) of these, which (31) permits directly. Moreover, since (31) emerges verified, the Hilbert-space representation – in contradistinction to how it is used inside $QM_{HD}$ – has here mainly a role of consensual communicability of probability laws, un-separated in terms of prediction and verification.

We summarize: Via the assertion *Ass2* we have constructed a new representation of the quantum measurements for the quasi-classical case of the microstates $ms(unbound,1)_{G(n-c)}$, no matter whether the aim of these is to construct a prediction or to verify it. This new representation is rooted directly into the a-conceptual physical factuality, it is constructed bottom-up, and – by construction – it *emerges* both predictive and verified. Since furthermore, while it constructs a representation of the results of quantum measurements on microstates $ms(unbound,1)_{G(n-c)}$, it also constructs factually a Hilbert-Dirac mathematical representation (31) $|\psi_{Gt,H}(r,t_1)>_{fact(DM)}$ of the studied microstate itself, the approach brought forth by the argument that proves the assertion *Ass.1* appears to be remarkably synthetic.

[107] **We quote as it stands the following extract from the French Wikipedia: Rareté d'une résolution analytique exacte**

La recherche des états propres de l'hamiltonien est en général complexe. Même le cas analytiquement soluble de l'atome d'hydrogène ne l'est rigoureusement sous forme simple que si l'on néglige le couplage avec le champ électromagnétique qui va permettre le passage des états excités, solutions de l'équation de Schrödinger de l'atome, vers le fondamental.

Certains modèles simples, bien que non tout à fait conformes à la réalité, peuvent être résolus analytiquement et s'avèrent très utiles :

- particule libre (potentiel nul) ;
- oscillateur harmonique (potentiel quadratique) ;
- particule se déplaçant sur un anneau ;
- particule dans un puits de potentiel rectangulaire ;
- particule dans un guide d'onde annulaire ;
- particule dans un potentiel à symétrie sphérique ;
- particule dans un réseau unidimensionnel (potentiel périodique).

Dans les autres cas, il faut faire appel aux diverses techniques d'approximation :

- la théorie des perturbations fournit des expressions analytiques sous la forme de développements asymptotiques autour d'un problème non perturbé exactement soluble.
- l'analyse numérique permet d'explorer des situations inaccessibles par la théorie de perturbation.

### (7.III).2.3. Hilbert-space representation of quantum measurements for unbound microstates with quantum fields

We consider now microstates of one microsystem but *with a composed operation of generation* (cf. the definitions from (2.I).1). Such a microstate incorporates a phenomenon of 'interference of corpuscular-waves', so a non-null quantum potential wherefrom quantum fields can emerge. *Such a microstate is denoted* $ms(unbound,1)_{cG(qf)}$ (the index '$_{cG(qf)}$' is to be understood as: 'composed operation *G* of generation involving possibility of quantum fields'). Thereby we quit now the first, superficial stratum of $QM_{HD}$ that deals with the quasi-classical dynamics of the corpuscular-like singularity from the specimens of a microstate $ms(unbound,1)_{G(n\text{-}c)}$.

The microstates $ms(unbound,1)_{cG(qf)}$ constitute the key-category in the endeavour toward an intelligible consensual predictive and verifiable conceptualization of microstates and micro-phenomena. Though they are still unbound states and so, essentially endowed with unstable characters, the *inside* of the specimens of a microstate $ms(unbound,1)_{cG(qf)}$ cannot any more be conceived to be transparent to the *exterior* context, as it is presupposed to be by the coding postulate (20'). With respect to microstates $ms(unbound,1)_{cG(qf)}$ the concepts of 'outside' and 'inside' (cf. Atmanspacher&Dalenoort [1994]) acquire full power, and this, in the processes of conceptualization, plays an outstanding role[108]. The opposition outside-inside displaces us from upon the very frontier between classical physics and quantum physics, into the depth of the 'quantic' stratum of physical substance. It draws us deep into the unlimited and a-conceptual 'Universal Physical Substance' in the sense of Spinoza (*Natura Naturans*), as this substance has been specifically conceptualized by de Broglie and Bohm. Only here, *beneath* the frontier with the classical mechanics, does acquire *general* and *full* contour and density the fundamental question ***how*** something of which the concept and the name have been formed in the minds, but that is as yet un-*known* in its strict specificity, can be 'separated' from the whole where it belongs in a sense that permit to assert that it has become 'the entity to be studied' *consensually*, and how *that* can be effectively 'studied', which means *qualified*.

In the Part I of this work we have dealt already precisely with this question, but only in principle, without specifying a model of a microstate; in the Part II we have specified the model that works implicitly inside $QM_{HD}$; and in the chapter 9.III, when we shall integrate the structure of "the Second Quantum mechanics *QM2*" we shall come back to the general question mentioned above, and the answer will place us on the path toward an answer to the problem of 'Unification'.

While here we shall try for the moment to continue the construction of a new representation of the quantum measurements researched inside this chapter 7.III. for microstates $ms(unbound,1)_{cG(qf)}$, the constructive approach achieved in (7.III).2.3 *via* the Assertion *Ass2* for the case of microstates $ms(unbound,1)_{G(n\text{-}c)}$.

#### *(7.III).2.3.1. Critique of the* $QM_{HD}$ *representation of microstates* $ms(unbound,1)_{cG(qf)}$ *and of momentum-measurements on such microstates*

For *microstates* $ms(unbound,1)_{cG(qf)}$ the dynamic of the singularity from the physical wave of the specimens of the studied microstate cannot be conceived any more to be influenced exclusively by the classical macroscopic fields from a Schrödinger Hamiltonia; this Hamiltonian becomes insufficiently rich [109]. Correlatively, the $QM_{HD}$

---

[108] Intimately tied with the opposition 'open'-'closed' that lies on the ground of the concept of 'formal system'.

[109] In $QM_{HD}$ this is compensated by the help of the concepts like "obstacle", "potential-wall", approximations, etc I reproduce from: file:///Users/mms/Desktop/Équation%20de%20Schrödinger%20—%20Wikipédia.webarchive, 14 March 2018 (without translating):

Hilbert-space representation of the microstates *microstates* $ms(unbound,1)_{cG(qf)}$ becomes itself inadequate, which brings forth a limit of the framework *[IQM-QM$_{HD}$]*.

In order to show this we consider for simplicity a microstate $ms(unbound,1)_{cG(qf)}$ of type (15') with only two components in the operation of generation, for instance a Young-interference $ms_{G12}$ generated by an operation of generation $\boldsymbol{G(G_1,G_2)}$. The $QM_{HD}$ Hilbert space-representation of the state-ket of such a microstate (and with the revised notations from *[IQM-QM$_{HD}$]*) is:

$$|\Psi_{G(G1,G2)}(\mathbf{r},t)\rangle = \lambda_1|\Psi_{G1}(\mathbf{r},t)\rangle + \lambda_2|\Psi_{G2}(\mathbf{r},t)\rangle \qquad (15'')$$

Consider the momentum-observable $\boldsymbol{P}$. This observable plays a basic role inside $QM_{HD}$ (the position-observable $\boldsymbol{R}$ is degenerate from various points of view, while the eigenvalues of any other observable $\boldsymbol{A}$ can be represented as a function $\boldsymbol{A(R,P)}$ that permits to calculate the eigenvalues of $\boldsymbol{A}$ from those of $\boldsymbol{R}$ and $\boldsymbol{P}$ (cf. 5.II).1)**)**. The eigenfunctions of the momentum-observable $\boldsymbol{P}$ are plane wave-functions $a.exp(i/\hbar)\mathbf{p_j.r}$ where $\mathbf{p_j}$ designate the eigenvalue with index $j$ of $\boldsymbol{P}$ that is tied with the *plane*-wave eigenfunction $a.exp(i/\hbar)\mathbf{p_j.r}$. The expansion $|\Psi_{G(G1,G2)}(\mathbf{r},t)\rangle/\boldsymbol{P}$ where $|\Psi_{G(G1,G2)}(\mathbf{r},t)\rangle$ is defined by (15'') yields for the predictive probability of the outcome $\mathbf{p_j}$ – let us re-note it $\mathbf{p_{j,12}}$ – the well-known representation of form

$$\pi_{12}(\mathbf{p_{j,12}}) = \pi_1(\mathbf{p_{j,1}})+\pi_2(\mathbf{p_{j,2}})+F(\pi_1(\mathbf{p_{j,1}}),\pi_2(\mathbf{p_{j,2}})) \qquad (32)$$

with $\mathbf{p_{j,1}}$ and $\mathbf{p_{j,2}}$ the eigenvalues introduced by the expansions $|\Psi_{G1}(\mathbf{r},t)\rangle/\boldsymbol{P}$ and $|\Psi_{G2}(\mathbf{r},t)\rangle/\boldsymbol{P}$, respectively.

We assert that inside *[IQM-QM$_{HD}$]* the prediction (32), and correlatively also the representation (15'') are inadequate from conceptual points of view. But furthermore the $QM_{HD}$ (32) raise questions concerning the applicability of the coding-postulate (20'), as well as questions of effectiveness and of inner logical consistency. We are in presence of a sort of block of questions that manifests *basic* insufficiencies. These have to be explicitly formulated and solved.

***Conceptual inadequacy of (15'') with respect to [IQM-QM$_{HD}$].*** As it has been already stressed several times, the additive representation (15'') of the state-ket $|\Psi_{G(G1,G2)}(\mathbf{r},t)\rangle/$ of a microstate $ms_{G12}$ – though mathematically it is *permitted* in a vector-space – suggests semantic features that are misleading from the standpoint of *[IQM-QM$_{HD}$]*: In a certain sense the state-ket $|\Psi_{G1}(\mathbf{r},t)\rangle$ and $|\Psi_{G2}(\mathbf{r},t)\rangle$ from the second member from (15'') do indeed concern the microstates $ms_{G1}$ and $ms_{G2}$. But according to *[IQM-QM$_{HD}$]* these microstates are not physically realized, and so they do not 'exist': By the basic methodological decision *MDI* from I.1 we have posited a one-to-one relation (1) $G\leftrightarrow ms_G$ between any given operation of generation $G$ and its result denoted $ms_G$, and in the case considered in (15'') the unique operation of generation that is conceived to have been physically *realized* is $\boldsymbol{G(G_1,G_2)}$. So exclusively the microstate $ms_{G12}$ represented by the state-ket $|\Psi_{G(G1,G2)}(\mathbf{r},t)\rangle$ is realized. Full stop. The microstates '$ms_{G1}$' and '$ms_{G2}$' have *not* been *individualized* by $\boldsymbol{G(G_1,G_2)}$[110] because $G_1$ and $G_2$ have not been fully and separately realized. These two possible but unrealized microstates have only contributed genetically to the global – and *abstract* – mathematical expression of the limiting conditions to be asserted for the solution of the Schrödinger equation of the problem, that is tied with the inner 'interference' structure of the microstate $ms_{G12}$ generated by $\boldsymbol{G(G_1,G_2)}$.

However nothing interdicts to suppose that in $QM_{HD}$ the contents of the second member from (15'') have been defined with a purely algorithmic purpose, namely for calculating the probabilities of outcomes of measurements on the unique generated and studied microstate $ms_{G12}$. If this view is adopted then, in so far that the algorithm is found to lead indeed to factually true predictions, this can be considered to suffice for choosing the representation (15'') of the state-ket $|\Psi_{G(G1,G2)}>$. So let us focus now on verifiability.

---

[110] This is not visible inside $QM_{HD}$ where the whole notion of an individual operation of generation of the studied microstates remains hidden.

***Inadequacy with respect to (15'') of a coding postulate of type (20)-(20').*** A coding postulate (20') of the type (20) appears to be inadequate with respect to momentum measurements on microstates *ms(unbound,1)*$_{cG(qf)}$, no matter whether one places oneself inside $QM_{HD}$ or inside *[IQM-*$QM_{HD}$*]*. Indeed:

*Inadequacy inside* $QM_{HD}$

**1.** The eigenstates of the momentum-observable are plane waves and one of the measuring-postulates of $QM_{HD}$ asserts that the result of a measurement on a microstate is always an eigenvalue of the measured observable that emerges ***with*** the corresponding eigenstate (and that immediately after the end of the measurement the 'system' *remains in this eigenstate*) [111].

On the other hand:

**2**. Let us first remark that *in a Young interference, a plane wave-structure is* ***nowhere*** *spontaneously realized*. Strictly speaking, even at an infinite distance from the holed screen there still is superposition of contributions from both holes.

What about the coding-measurement-evolution? Inside $QM_{HD}$ it is explicitly admitted that for the momentum-measurements one has to apply the method time-of-flight. This method requires basically and explicitly the suppression of all the 'external' fields that are working (no doubt precisely in order to start a coding-measurement-evolution that shall *generate* at finite distance-and-time an eigenstate of the momentum, starting from the moment that the measurement-evolution begins). But the quantum fields that work in a Young interference are not classical 'exterior' fields and the human observer cannot suppress them without suppressing the microstate-to-be-studied. In these conditions the coding-postulate (20)-(20') *cannot bring forth an eigenstate of the momentum-observable*.

So the predicted eigenvalues – with *their* eigenstates and *their* probabilities to emerge – in fact cannot emerge at all; the $QM_{HD}$ prediction (32) cannot be verified.

One might then suggest that – since the structure of the superposition wave-function from a Young interference becomes spontaneously as *near* to that of a plane wave as one wants when the distance-and-time increase toward infinity – it suffices to end each coding-measurement-succession *[G.Mes**P**]* by a position-registration on a very distant sensitive-screen, which will constitute a satisfactory measurement of the momentum eigenvalue.

However it seems clear that it is not acceptable to admit a definition of a coding-measurement evolution that *in its very principle* involves either approximations, or a non-effective distance-and-time of evolution. A measurement operation has to be a priori required rigorous and effective; in its principle.

**3**. In short, the situation just described in this point 2 is in direct contradiction with the $QM_{HD}$ *view* recalled in 1, and the prediction (32) involved by it cannot be verified. (So the contradiction can last undisturbed).

*Inadequacy inside [IQM-*$QM_{HD}$*]*

**1'**. In *[IQM-*$QM_{HD}$*]*-terms now, the descriptive structure of an interference microstate with state-ket $|\boldsymbol{\Psi}_{G(G1,G2)}(\mathbf{r},t)>$ is tied with the two operations of generation $G_1$ and $G_2$ involved in the unique realized operation $\mathbf{G(G_1,G_2)}$, in the sense of the last definition from (1.I)1; and that what is considered to have a 'trajectory' along a maximum of presence-probability (the 'particle') is the singularity from the amplitude of the

[111] Cf. (5.II).1 (measurement postulates); and also Cohen-Tannnoudji, C., Diu, B. et Laloë, F. [1973]).

individual physical wave represented by the *individual* de Broglie wave-function $\Phi_{G(G1,G2)}(\mathbf{r},t)=a(\mathbf{r},t)e^{(i/\hbar)\beta(\mathbf{r},t)}$ of the specimen $\sigma_{\Phi(G(G1,G2))}$ of $ms_{G12}$ that is involved in any one coding-measurement-succession *[**G**(G1,G2).Mes.**P**]* (cf. *(6.II).4.2.2*). Now, *each* one of the *two* operations of generation $G_1$ and $G_2$ involved in the unique realized operation of generation ***G**(G1,G2)* can be considered to induce into the movement of the *unique* singularity from one given $\Phi_{G(G1,G2)}(\mathbf{r},t)$ a directional *trend*, and accordingly to de Broglie's guiding law these two distinct directional trends are conceived to *combine* so as to determine for the one involved singularity a trajectory along a maximum of the presence-probability determined by the state-ket $|\Psi_{G(G1,G2)}>$ that corresponds to $\Phi_{G(G1,G2)}(\mathbf{r},t)$. Insofar that this is correct:

> The neighbourhood of the singularity from $\Phi_{G(G1,G2)}(\mathbf{r},t)$ can ***never*** be conceived to be populated by a form of wave-movement representable by only ***one*** *plane wave*, not even approximately and at ideal infinity; for in the case of a microstate $ms(unbound,1)_{cG(qf)}$ ***precisely*** *this small neighbourhood is conceived to be quite* ***essentially*** *populated by a* ***physical*** *superposition of* ***two*** *wave-movements of distinct directions that –* ***together*** *– determine the trajectory.*

And conceiving to wait that 'near infinity' the obstacles disappear amounts in fact to waiting (non-effectively) that what you want to study, disappear.

**2'**. Since $QM_{HD}$ is part of the framework *[IQM-$QM_{HD}$]*, the $QM_{HD}$ measuring-postulates are still acting.

**3'.** So again the points 1' and 2' are not mutually consistent, just like above according to $QM_{HD}$.

So we conclude that:

In fact the 'prediction' (32) consists of just a mathematical *definition* of the predictive probabilities $\pi_{12}(\mathbf{p}_{j,12})$ that is postulated to be factually true but that cannot be verified to be such; it follows that for the category of microstates $ms(unbound,1)_{\mathbf{c}G(qf)}$ a new coding postulate has to be formulated and that, correlatively, new definitions are needed for the concepts of eigenstates and eigenvalues of the momentum-observable, as well as for the Hilbert-space representation of the involved state-ket. Nothing interdicts such a search.

The mathematical linear representations of the type (15") of microstates $ms(unbound,1)_{\mathbf{c}G(qf)}$ is just a hurried choice that is *permitted* by the Hilbert-vector-space structure but is devoid of any necessity of semantic or of logical nature. On the contrary, from the standpoint of inner logical-semantic coherence it appears to be inadequate.

***Comment on the critique.*** One might believe that this whole problem is a false problem that stems from *IQM*, while nothing imposes to accept *IQM*. One can hold that this problem can be evacuated by just refusing the concept of a 'composed operation of generation' defined in (2.I)1 and by just accepting the direct postulation from $QM_{HD}$ of the adequacy of additive representations of state-ket like in (15") and its consequence (32'). But this is *not* in the least the case. A rejection of the concept of composed operations of generation would not change the fact that the predictions (32) raise questions *with respect to basic postulates of $QM_{HD}$*; nor would it change the fact that the coding-postulate (20') cannot be used for verifying the prediction (32) of (15"). The fact that this unacceptable situation has been brought forth by the concept of composed operation of generation is a strong manifestation of the *relevance* of this concept.

So if one wants to construct an effective and intelligible representation of the microstates, it is imperative to deal overtly with the problems raised by measurements of the momentum observable on microstates $ms(unbound,1)_{\mathbf{c}G(qf)}$.

This leads us to the works of Broglie and Bohm, the only ones that have taken fully into explicit consideration the quantum potentials and quantum fields involved by most microstates, and which are the core of the specificity of quantum physics.

### *(7.III).2.3.2. Louis de Broglie, David Bohm, and de Broglie's Double-Solution interpretation of $QM_{HD}$*

What now is called the 'de Broglie-Bohm *interpretation* of quantum mechanics' has been much developed in works by P. Holland, B. Hiley, D. Dürr, S. Goldstein, N. Zanghi, M. Towler, W. Struyve, and many others [112]**.** But this vast and complex evolving domain of research is *exterior* to the purpose of this work; in this work we make direct use of, exclusively:

- Louis de Broglie's initial model of a micro-'system', as introduced in his Thesis and that has been identified in (6.II).1, (6.II).2, *(6.II).4.2.2* to be basically incorporated in the $QM_{HD}$-formalism via the concepts of eigenstates and eigenvalues of an observable;

- The general features of Louis de Broglie's 'double-solution causal interpretation of quantum mechanics' ([1956], [1957]) that developed the work of Bohm [1952] wherefrom de Broglie has drawn basic features as well as moral support (this has to be acknowledged).

Only these two essential features – and only in very succinct terms and very selected respects – will be connected to the present approach in what follows from this point and up to the end. Louis de Broglie's double solution interpretation will be denoted $dB_{DS}(B)$ (to be read: ' de Broglie's double-solution interpretation as conceived in relation with Bohm's interpretation).

Via the 'theorem of concordance of the phases' cited in (6.II).1, Louis de Broglie's model is tied with the well-known 'guiding law' of the momentum by the phase-function of the *physical* corpuscular-like wave assigned to any micro-system[113]. So the guiding law also – alike to the corpuscular-wave model of a specimen of a microstate and via this model is equally omnipresent in the $QM_{HD}$-formalism, implicitly, and *carried by a* ***restricting*** *equation for eigenfunctions and eigenvalues of a quantum mechanical observable* (cf. (6.II).1 and what follows); which is organically tied with the critique of the representation (15")-(32) from *(7.III).2.3.1)*. And along this way the guiding-law is implicitly but organically available inside the framework *[IQM-$QM_{HD}$]*, in a still potential way that has to be now actualized in suitable *generalized* terms.

We begin below by trying to understand what – exactly – withstands application of the coding-procedure (20') in the case of a microstate $ms(unbound,1)_{cG(qf)}$ of the type of that from (15"). This draws attention upon the so widely accepted distinction between 'observables' and 'beables' introduced by Bell.

### *(7.III).2.3.3. On 'observables' and 'beables'*

It has become current to distinguish between 'beable' qualifications and $QM_{HD}$ 'observables'. The position-vector observable ***R*** is considered more or less implicitly to behave like a beable, in this sense that the registration of perceivable marks produced by a specimen of the studied microstate can be considered to trivially show where 'the system' was 'really' placed when the mark emerged. But the case of ***R*** is regarded as a degenerate case. In general a $QM_{HD}$ observable – the momentum observable included – is currently conceived to manifest eigenvalues that are *created* by the measurement-interaction: What is observed and coded by an eigenvalue of the measured quantity is conceived to be in general different from the initial beable value, i.e. to have been created by the measurement-evolution *out* of this beable value[114].

---

[112] Cf. Wikipedia file:///Users/mms/Desktop/De%20Broglie–Bohm%20theory%20-%20Wikipedia.webarchive.

[113] Let us notice that in contradistinction to de Broglie, *Bohm does not explicitly distinguish* between the physical wave $\Phi(r,t)=a(r,t)e^{(i/h)\beta(r,t)}$ of a specimen of the microstate-to-be-studied, and the statistical state-ket associated to this microstate *(cf.(6.II).4.2.2)*.

But the analyses from (6.II).3 and (6.II).4 – and particularly of the method 'time of flight' for measuring momentum observable ***P*** – bring forth a view that is *directly opposed to that recalled above*:

A coding-measurement-evolution of the general form (20') – that, implicitly, is conceived inside $QM_{HD}$ to be applicable to any sort of microstate – in fact is *expressly* constructed such as to ***conserve unchanged** the initial 'beable' value of the measured quantity and to draw it into the realm of consensual observability – so knowledge –*, via perceivable marks that permit to identify it on the basis of theoretical arguments and calculi. Indeed this is the very essence of the criterion that acts in (20) and in (20') for asserting that the procedure *can* be considered to act as as 'a *coding*-measurement-evolution'. And precisely *this* ceases to be realizable in a controlled way for momentum measurements on microstates $ms(unbound,1)_{cG(q\text{-}f)}$.

But why, exactly, does it cease?

Certainly one of the causes is that in the case of a microstate $ms(unbound,1)_{cG(qf)}$ – in contradistinction to what happens with microstates $ms(unbound,1)_{G(n\text{-}c)}$ – the dynamics of that what, in a specimen of the studied microstate, does admit *'mechanical'* qualifications [115], can *never* be brought under the controllable dependence of ***exclusively*** the 'exterior' context *of that*. According to *[IQM-$QM_{HD}$]* this is so by the definition of the operation that generates a microstate $ms(unbound,1)_{cG(qf)}$, namely a composed operation of generation $\boldsymbol{G}(G_1,G_2,...G_m,...G_M)$. The singularity of *any* specimen of a microstate $ms(unbound,1)_{cG(qf)}$ is genetically immersed in a quantum potential entailed by the physical superposition of the mutually distinct directions of wave-movement instilled by the components $(G_1,G_2,...G_m,...G_M)$ of ***G***. The mechanical displacements of this singularity are, both, determined by this inner quantum potential, and shielded by it from the outside of the whole specimen, continuously, irrepressibly, and in a way that cannot be controlled because any inner fluctuation that propagates from the space-time support assigned to the operation of generation $\boldsymbol{G}(G_1,G_2,...G_m,...G_M)$, can generate unpredictable quantum forces that in general add unknowable changes to the initial value of the momentum that characterizes the mechanical displacement to be qualified. So the inside of the specimens of a microstate $ms(unbound,1)_{cG(qf)}$ is out of the human observer's control. No act of measurement tied with 'extinction' and control of the *inner* fields can be conceived, that would – at least only theoretically – preserve from changes the value of the mechanical quantity to be measured, throughout the measurement-evolution. All that seems possible to be done in this respect is to maintain as stable as possible the *global* immediate outside of any whole specimen of the studied microstate [116].

But certainly this is not the fundamental explanation of the resistance opposed by $QM_{HD}$-postulates to the application of the coding-postulate (20') for verifying the predictions (32) on the results of momentum-measurements drawn from the $QM_{HD}$ representation (15'') of a microstate $ms(unbound,1)_{cG(qf)}$. *This* resistance stems from a still more basic level that manifests itself in the various critical remarks from *(7.III).2.3.1*, some of which concern specifically the *contradiction* between the $QM_{HD}$ measurement-postulates 1 and 2 and the implicit but *clear* assumption of the unrestricted validity of the coding-postulate (20'): this contradiction points toward a too restricted form of the $QM_{HD}$ equation for eigenstates and eigenfunctions of the momentum-observable.

---

[114] I have myself held this view, strongly and for a long time.

[115] In Bohmian mechanics, for instance, the spin quantity is considered *not* to be a 'mechanical' quantity; it is regarded as a qualification of the wave movement.

[116] Here, at this point of the present inquiry, one can realize to what a point the physical reality opposes obstacles to factually gained *knowledge* – not to *mental* modelling, nor to mental separation from the 'rest' of the Universal Physical Substance, but to ***factual*** generation in the role of entity-to-be-studied; which, like precisely in the case of a specimen of a microstate, can even not insure a global spatial delimitation (cf. (6.II).2), so that the very basic separation 'inside-outside' becomes vague. And nevertheless it remains important because it has *observational* consequences, like precisely in the case of a *factual* operation of generation and the inside of the generates entity-to-be-studied.

Let us conclude on the concepts of 'beable' and of 'observable'. These concepts do *not* reach the basic feature that opposes resistance when one wants to apply the coding postulate (20') to the case of momentum- measurement upon microstates $ms(unbound,1)_{cG(qf)}$.

> The belief carried by these concepts, that in general a $QM_{HD}$ measurement-evolution necessarily changes the value possessed by the measured quantity when the measurement-evolution begins, might simply be *false*.

The essential difference – intuitively targeted but not reached by the concepts of 'observable' or 'beable' – *does not concern the* ***numerical value*** *of the considered quantity*; it concerns the ***way*** *in which it is possible to bring this value into consensual and verifiable* ***knowledge***. And *how* it concerns this, though not fully clear yet, is likely to be connected somehow with the measurement-postulates 1 and 2 from $QM_{HD}$, so with the equation for eigenstates and eigenvalues of the momentum observable ***P***, because in a $ms(unbound,1)_{cG(qf)}$ it is not possible to generate by a coding-measurement-evolution that lasts a finite time, an plane-wave-eigenstate of the momentum observable, as it is asserted by these postulates and as it is supposed by the coding-postulate (20), admitted implicitly inside $QM_{HD}$ for *any* sort of unbound microstate.

This is the problem on which we have to focus.

This problem does not incriminate the representation $dB_{DS}(B)$ since this is a priori introduced as only an *interpretation* of $QM_{HD}$. But it does flaw $QM_{HD}$ itself that claims true and verifiable predictions for any sort of microstate, while for the microstates $ms(unbound,1)_{cG(qf)}$ – that constitute the most specific core of nowadays fundamental microphysics – the predictions (32) are not verifiable [117]. And it quite essentially incriminates the framework $[IQM\text{-}QM_{HD}]$ where this flaw emerges.

> The conditions for satisfying the fundamental assertion *Ass.1* cease to be realizable inside the framework $[IQM\text{-}QM_{HD}]$. In order to treat the case of the microstates $ms(unbound,1)_{cG(qf)}$ the framework $[IQM\text{-}QM_{HD}]$ must be modified so as to include an adequate coding-procedure for momentum-measurements.

Here we suspend the general considerations on 'beables' and 'observables', in order to focus specifically on the formulation of such a coding-procedure.

### *(7.III).2.3.4. Doubt on the presumed non-measurability of the de Broglie instantaneous guided value of the momentum*

Since in the case of a microstate $ms(unbound,1)_{cG(qf)}$ the stable conservation of the initial value of the momentum throughout a lasting momentum-measurement-evolution cannot be certainly insured until a final interaction that produces observable coding-marks [118], one is led to try to identify a sort of coding-measurement-evolution for this case that *consumes* the instantaneous value possessed by the momentum at the time when the measurement *begins*, in a process that – *itself* – codes for this value: a 'value-consuming' process of coding-measurement-evolution. No solution of another sort comes to our mind.

---

[117] This, I think, is intimately connected with the very fundamental exchange between Englert & alt. [1992], Dürr & alt. [1993], Finkelstein [1995] and Hiley&alt. [2000], on which we come back at the end of this chapter.

[118] Which usually *destroys* the measured momentum-value, so also the corresponding eigenstate, contrarily to the $QM_{HD}$ postulate that any act of measurement leaves the 'system' in the eigenstate of the obtained eigenvalue. But it might be possible to realize conditions in which this final destruction is avoided.

This brings us to the $dB_{DS}(B)$ approach because it is the only one that penetrates explicitly into the 'inside' of the microstates $ms(unbound,1)_{cG(qf)}$. This approach is the source of the concepts of quantum potentials and quantum fields.

Inside *[IQM-QM$_{HD}$]* the microstates $ms(unbound,1)_{cG(qf)}$ are characterized by composed operations of generation $\mathbf{G}(G_1,G_2,...G_m,...G_M)$, $m=1,2,..M$ [119] and these introduce a wave-function $\Phi_{G(G1,G2)}(\mathbf{r},t)=a(\mathbf{r},t)e^{(i/\hbar)\beta(\mathbf{r},t)}$ of which the phase-function $\beta(\mathbf{r},t)$ defines a sort of momentum-value that inside $QM_{HD}$ is not defined, and which, in the context of the problem formulated in *(7.III).2.3.1* draws attention upon it: Consider de Broglie's well-known concept of 'guiding relation'

$$\mathbf{p}(\mathbf{r},t) = - \nabla.\beta(\mathbf{r},t) \qquad (33)$$

where $\mathbf{p}(\mathbf{r},t)$ is the 'guided' momentum of the corpuscular-like singularity at the time $t$ and $\beta(\mathbf{r},t)$ is the *phase*-function at $t$ from the physical individual wave-function $\Phi_G(\mathbf{r},t)=a(\mathbf{r},t)e^{(i/\hbar)\beta(\mathbf{r},t)}$ that represents the wave of each specimen $\sigma_{\Phi(G)}$ of the studied microstate (cf. (6.II).2 and (7.III).1.2). This is the very definition of the momentum of the singularity from a physical 'corpuscular-like' wave. And inside $dB_{DS}(B)$ :

> The guidance law (33) is asserted deductively and with full generality, both in the presence and the absence of quantum fields.
> *But this law is quasi-unanimously considered to be **un**-observable.*

Even de Broglie himself adhered to this view. It is believed that as soon as one would try to register the guidance-trajectory in a specimen $\sigma_{\Phi(G)}$ of the studied microstate, the beginning of the interaction would immediately destroy the inner structure of the phase represented by the phase-function $\beta(\mathbf{r},t)$ and this would compromise any global relevance of the data drawn from the interaction. This idea however seems to have been admitted on the sole basis of the powers of a priori submission to the formalism of $QM_{HD}$. And notwithstanding these powers:

- Trace-registrations are currently used in Wilson-chambers, since a long time.
- For *photonic* interference states a guided trace has already been experimentally registered (cf. A. Steinberg [2011]), *which is a very strong indication that an experiment with **heavy** microsystems could equally succeed.*

In fact, I think that nobody as yet has genuinely analysed and tested whether yes or not it is possible to choose the values of the parameters involved in a trace-registration on a *heavy* microstate so that to be able to compute out of the registered data the value of the momentum at the moment when the trace registration *began*, in full agreement with the theoretical assumptions from $dB_{DS}(B)$ [120]. So this analysis remains to be done. But it has to be done *without* any use of the formalism of $QM_{HD}$ such as it now stands *because any such use is **circular** in consequence of the fact that $QM_{HD}$ involves a **general** validity of the coding–postulate (20)-(20') which is incompatible with measurability of guided trajectories (33)* [121, 122].

---

[119] The $dB_{DS}(B)$ interpretation is devoid of any feature connectable with the *[IQM-QM$_{HD}$]* concept of operation of generation. But in the nowadays 'Bohmian mechanics' the concept of 'conditional wave-function' plays the role of a sort of **imagined** operation of generation that separates mentally a definite specimen of a microstate inside the universal state-function (cf.IOP Science, IP address 195.132.213.223.). This is similar to procedures from the cosmologic theory of gravitation.

[120] It even seems that Louis de Broglie's works [1924], [1956], [1957] are not yet available in English, which simply is aberration.

[121] I once accomplished a theoretical examination of the measurability of the guided momentum (33) in an interference state (MMS [1963]) and it led to a 'proof' that for a globally *stable* interference microstate (with non-null quantum potential but with null permanent quantum fields) it *is* possible – *in full compatibility with $dB_{DS}(B)$* – to register data

The time of idolatry with respect to the $QM_{HD}$ formalism and its explicit or implicit diktats seem to have revolved. History, once more, brings face-to-face with the as yet undone[123]:

> In an interference-microstate $ms(unbound,1)_{cG(qf)}$, the momentum-value $\boldsymbol{p_j}$ of the *unique* corpuscular-like singularity from any *specimen* of that microstate generated by one realization of the corresponding operation of generation $\boldsymbol{G}(G_1,G_2,...G_m,...G_M)$, $m=1,2,..M$, is determined according to (33) by the phase of a wave-movement represented by a wave-function $\Phi_{G(G1,G2)}(\mathbf{r},t)=a(\mathbf{r},t)e^{(i/\hbar)\beta(\mathbf{r},t)}$ *where the phase-function* $\beta(r,t)$ *has* ***never*** *the form of one plane wave*.

One immediately perceives that this is fully consistent with the critique from *(7.III).2.3.1* of the $QM_{HD}$-representation (15") of the state-ket of the considered microstate via a linear superposition of other state-ket. Furthermore, surreptitiously, the notion begins to gain form that this critique might touch all the $QM_{HD}$-observables $\boldsymbol{A}$, since any such observable is a function $\boldsymbol{A(R,P)}$ where the eigenket of the observable $\boldsymbol{P}$ might appear to be *arbitrarily* restricted by the equation for eigenket and eigenvalues of $\boldsymbol{P}$; while the Schrödinger equation of evolution does not involve the quantum-potentials though these are present in nearly any paradigmatic microstate, like those with "barriers" or "walls" or "wells", as well as in any bound microstate.

### *(7.III).2.3.5. Proposal of an experiment conceived inside [IQM-QM$_{HD}$]*

Let us imagine the experiment represented in the figure 8 – denoted *EXP* – that involves a very simplified interference-state generated by an operation of generation $\boldsymbol{G}(G_1,G_2)$.

---

that *do* permit to calculate from them the corresponding momentum-value from (33) for the time $t$ when these registrations have *begun*. But in that proof the representation of the state-ket had the form (15"). So the cited work, more or less implicitly, is likely to involve elements that are not coherent with the present development. Nevertheless it still *does* show that nothing of experimental nature withstood the idea of principle that the de Broglie momentum-value (33) for an unbounded interference-microstate can be measured, while this idea was – and still is – unanimously banished.

[122] There has been a debate on "Surrealistic Bohmian Trajectories". One of the involved works (Hiley&alt.) considers Bohm's approach alone, while the others *mix* Bohmian formalism with $QM_{HD}$ formalism. But *all these works make full abstraction of de Broglie's* ***individual physical model as well as of de Broglie's physical wave-function*** $\Phi(\mathbf{r},t)=a(\mathbf{r},t)e^{(i/\hbar)\beta 12(\mathbf{r},t)}$ ***of one specimen of the microstate to be studied.*** In short, the whole own approach of de Broglie is ignored; exclusively the statistical state-ket is made use of. All the involved authors accept $QM_{HD}$ just as it now stands – with "collapsing" measurements ('reduction') included; and they follow the purpose to show full agreement between Bohmian Mechanics and $QM_{HD}$. Thereby these works – though very interesting and likely to cooperate volens-nolens toward a new microphysics – are all quite fundamentally different from the construction attempted here inside *[IQM-QM$_{HD}$]*.

[123] This conclusion, I think, is backed by the Steinberg experiment quoted above.

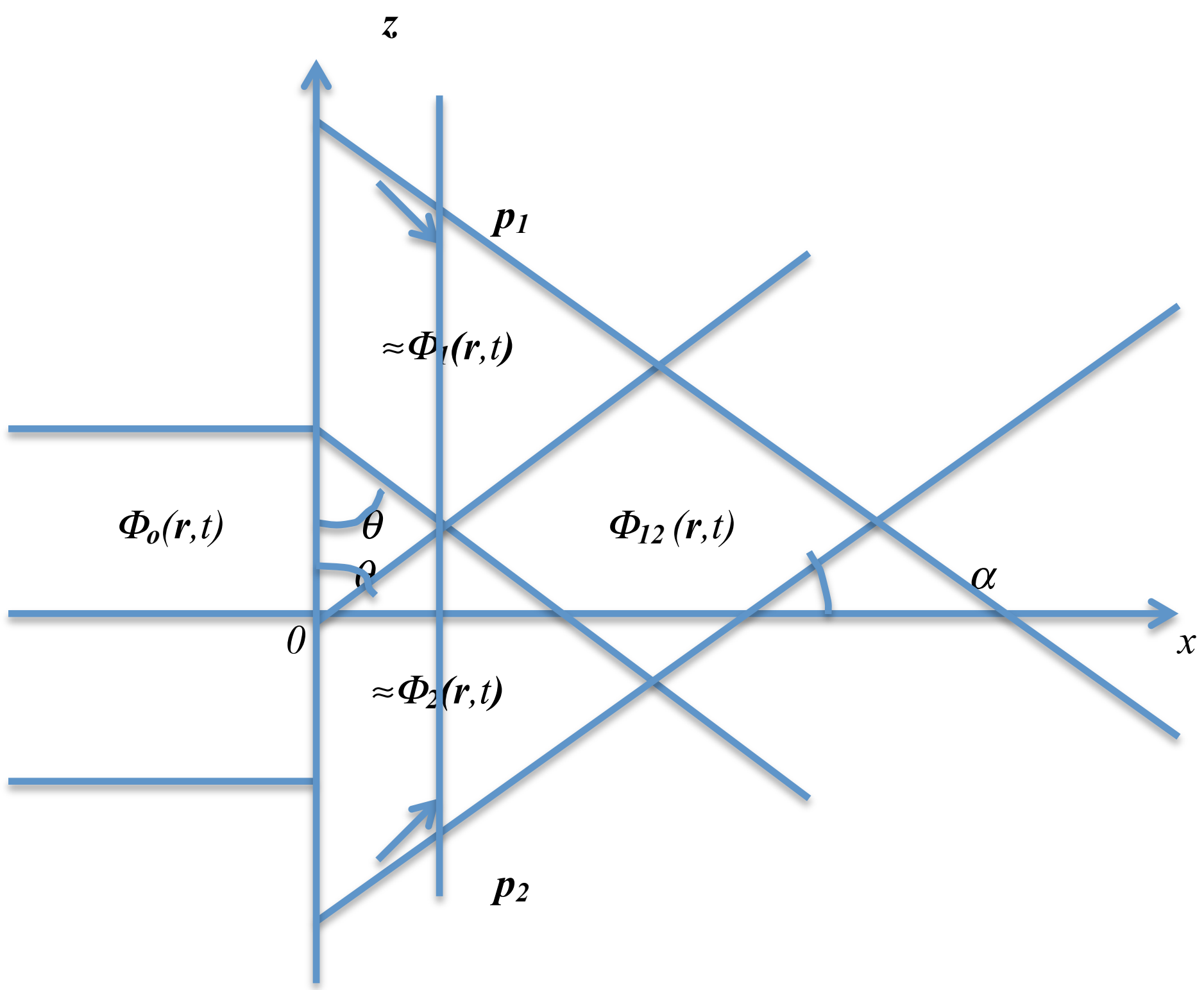

Fig.8. A microstate $ms(unbound,1)_{cG(q\text{-}f)}$ with operation of generation $\mathbf{G}(G_1,G_2)$.

The role of the figure 8 and of the corresponding experiment described below, is only preliminary: We want to analyse the main features of the physical situation that is involved and of the physical interactions entailed by a 'value-consuming' coding-measurement-evolution that would consist of a registration of a trace left in a sensitive medium by $\mathbf{p}(\mathbf{r},t)$ from (33) in conditions that permit to *calculate* the value of $\mathbf{p}(\mathbf{r},t)$ from characters of this registration. If this analysis leads to a global conclusion of technical possibility, the actual experiment can then be realised with a Young interference where any a priori approximation or/and restrictive feature in the structure of the studied microstate, is eliminated.

On the Figure 8 the symbol $\Phi_o$ denotes the wave-function of a physical individual specimen of a 'preliminary' microstate $ms_{Go}$. Out of $\Phi_o$ a front-wave-divisor splits $\Phi_o$ in two parts that are *approximately* described by, respectively, two other waves '$\approx\Phi_1$' and '$\approx\Phi_2$' ($\approx$: approximately) – more exactly, trends of wave-movement – tied in the sense defined at the end of (1.I).1.2 with the composing operations of generation $G_1$ and $G_2$. These *combine* their effects inside the unique fully realized operation $\mathbf{G}(G_1,G_2)$. The directions of propagation of $\approx\Phi_1$ and $\approx\Phi_2$ make a mutual angle $\alpha$, while with the axis $0z$ they make angles $\theta$ of the same absolute value. In this simplified preliminary representation the considered specimen of the interference-microstate to be studied $ms_{\mathbf{G}(G1,G2)}$ is realized only inside the space-time domain where the individual physical wave-movements denoted '$\approx\Phi_1$' and '$\approx\Phi_2$' superpose physically into one individual physical interference-wave-movement denoted by the wave-function $\Phi_{12}$ that 'takes place' only on a restricted and definite domain of what is called the 'physical' space-time and ceases outside this domain, while inside this domain and on the individual level of conceptualization, the phenomenon evolves.

For each specimen $\sigma_{\Phi(\mathbf{G}(G1,G2))}$ of $ms_{G(G1,G2)}$ and with respect to the introduced referential, the $dB_{DS}(B)$ guidance relation (33) asserts that the corpuscular-like singularity in the amplitude of the respective physical wave with wave-function $\boldsymbol{\Phi}_{12}(\mathbf{r},t)=a(\mathbf{r},t)e^{(i/\hbar)\varphi(\mathbf{r},t)}$ has a velocity with components $v_x=v_o \sin\theta=const,\ v_y=v_z=0$. So the momentum-components are

$$p_x=Mv_x=Mv_o \sin\theta, \qquad p_y=p_z=0 \qquad (34)$$

where *M denotes the time-dependent 'quantum mass' introduced by the involved specimen of the studied microsystem in the sense of* $dB_{DS}(B)$ ([1956]). For indeed the singularity is not a 'particle', it is a localized aspect from the amplitude of the individual physical wave represented by the wave-function $\boldsymbol{\Phi}_{12}(\mathbf{r},t)=a(\mathbf{r},t)e^{(i/\hbar)\varphi(\mathbf{r},t)}$.

Imagine now a factual context that is stabilized as much as possible. Imagine a big number of repetitions of the operation of generation $\mathbf{G}(G_1,G_2)$ defined above. Each one of these repetitions involves its own specimens $\sigma_{\Phi o}$ and $\sigma_{\Phi(\mathbf{G}(G1,G2))}$, so its own individual physical wave-movements described by corresponding wave-functions $\boldsymbol{\Phi}_o$, $\approx\boldsymbol{\Phi}_1$, $\approx\boldsymbol{\Phi}_2$ and $\boldsymbol{\Phi}_{12}(\mathbf{r},t)=a_{12}(\mathbf{r},t)e^{(i/\hbar)\beta 12(\mathbf{r},t)}$. That what inside *[IQM-QM$_{HD}$]* is represented by a state-ket is the global result of all these repetitions placed on the statistical level of conceptualization and considered mentally and retroactively as a whole. We just denote this state-ket by $|\boldsymbol{\Psi}_{\mathbf{G}(G1,G2)}(\mathbf{r},t)>e^{(i/\hbar)\varphi 12(\mathbf{r},t)}>$ but we do not know yet its Hilbert-space-representation because its $QM_{HD}$ representation (15") has been examined and found to be deficient from the viewpoint of *[IQM-QM$_{HD}$]*). However it is generally accepted on the basis of *factual* data that the presence-probability inside de space-time support of $|\boldsymbol{\Psi}_{\mathbf{G}(G1,G2)}(\mathbf{r},t)>$ consists of a pattern of fringes of high presence-probability ('brilliant' fringes), all parallel to the *0x* axis and mutually separated by fringes of quasi-zero presence-probability ('dark fringes'). (We note that there is no way to directly *observe* these 'fringes', since they are only a statistical concept totalized outside any individual observation-time, while the individual realizations of a de Broglie guided trajectory do not leave spontaneously a perceivable trace. Only cumulated final impacts on a sensitive screen can be perceived).

We now start describing the proposed experiment *EXP*.

The presupposed trajectory of the corpuscular-like singularity from any one physical individual $dB_{DS}(B)$ wave-function $\boldsymbol{\Phi}_{12}$ is posited to be parallel to *0x*. Now, any experimental intrusion in the inside of a given specimen $\sigma_{\Phi(\mathbf{G}(G1,G2))}$ of $ms_{G(G1,G2)}$ is currently posited to entail quantum-fields that destroy the physical phase relation from the physical wave of this specimen, represented by the phase-function $\beta_{12}(\mathbf{r},t)$ from $\boldsymbol{\Phi}_{12}(\mathbf{r},t)=a_{12}(\mathbf{r},t)e^{(i/\hbar)\beta 12(\mathbf{r},t)}$. One of the main purposes of *EXP* will be to figure out a system of choices of the parameters that – by trial and error – permit to *prevent* this to happen and to control whether yes or not this purpose has been realized. The most important parameter is likely to be the kinetic energy of the specimens of $ms_{G(G1,G2)}$ that are made use of; this should be sufficiently high with respect to:

- the medium value of the energy of ionization of a molecule from the sensitive medium that is made use of;
- the medium value of the spontaneous fluctuations of the $dB_{DS}(B)$ quantum potential (anyhow the forces entailed by such fluctuations act only via the *0z* component so that – at most – their effect consists of a displacement of the involved singularity on another 'brilliant fringe', without suppression of the phase relation that determines the momentum-value $p_x$ from (34));

This might already suffice for insuring stability of, both, the phase relations from $\boldsymbol{\Phi}_{12}(\mathbf{r},t)$ and the supposed direction of displacement of the singularity from the amplitude of $\boldsymbol{\Phi}_{12}(\mathbf{r},t)$.

The *EXP* can be structured as a sequence of distinct tests:

At a distance $0x_1$, near the entry into the zone of interference, is placed on *0x* a very thin layer *L1* of sensitive substance permitting with maximal probability at most *2* successive initial acts of ionization. At a second distance $0x_2$ placed near the end of the interference domain is placed a thick layer *L2* of photographic emulsion with high density of molecules. When the first ionization occurs in *L1* at a time $t_1$ a conveniently connected chronometer registers this time. As soon as the corpuscular singularity reaches the second layer it produces there nearly certainly and practically on the edge of its entry in this layer, a third ionization that is recorded at a time $t_2$. Then other ionizations follow until the energy of the corpuscular-like energy is consumed.

These ionization constitute the operation of measurement, let us denote it *Mes.**p*** (where: ***p*** denotes the de Broglie guided momentum value from (33), not the quantum mechanical momentum-observable ***P***).

A big number of coding-measurement successions *[**G**(G₁,G₂).Mes.**p**]* is realized. We keep all the cases in which either one or two initial ionizations have been registered.

* When two first ionizations are available they permit to establish via the direction of the small segment of line that unites them whether the perturbing quantum-force has effectively displaced the corpuscular singularity on another fringe of high presence-probability, or not. This permits to be aware of the existence of perturbing quantum fields and the strength of their effects, and to keep for use in the final calculus only the cases without displacement on another fringe-direction.

* The two ionizations at the two times $t_1$, $t_2$ registered respectively in *L1* and *L2* can furthermore be regarded to define explicitly the direction of the momentum (34) and they yield a first estimation of its value (a sort of time-of-flight method 'internal' to the involved specimen of the studied microstate).

* The ending set of ionizations inside *L2* permits to calculate the absolute value of the momentum out of the energy consumed by one ionization and the number of ionizations.

* The statistic of the positions at the time $t_2$ permits to know whether the position distribution after the first one or two ionizations is still organized in maxima and minima indicating interference fringes; so *it verifies the conservation of the initial phase-relation*.

* Since the first impact defines also the initial position ***r*** with respect to the referential, *considered globally the set of registrations specified above **would violate the Heisenberg principle*** [124]. This would establish that the validity of Heisenberg's non-mathematical 'principle' in fact is relative to the experimental procedure. It would also establish that:

> The concept of incompatible observables is tied with coding-measurement-*evolutions* that ***freeze** the eigenvalue to be identified, as required in (20')*. While for microstates $ms(unbound,1)_{cG(qf)}$ this concept escapes human control.

In consequence of this the domain of validity of the mathematical uncertainty *theorem* from $QM_{HD}$ might appear to be devoid of general validity, which, again, points toward *the* adequacy of a non-linear general equation of evolution for a concept of state-ket that is not implicitly and arbitrarily restricted a priori.

* The statistic of the registered momentum-values would permit now confrontation with the $QM_{HD}$-prediction (32).

These considerations establish *the very particular stake* of an experiment EXP.

The preceding indications are generalizable to *any* microstate $ms(unbound,1)_{cG(qf)}$.

Finally, let us stress that *stricto-sensu* $QM_{HD}$ concerns exclusively *heavy* microsystems. So only a realization of *EXP* for systems with non-null rest-mass would possess a full significance of principle (furthermore this mass should be conceived to be a variable de Broglie-mass from (34) (cf. de Broglie [1956], chapter X). The best choice would be to work with a neutron-Young-interference (two holes) that would introduce relatively high kinetic energies even for moderate velocities and would involve exclusively quantum potentials and fields, thus avoiding any electromagnetic effect during the ionizations.

---

[124] Such a violation – of which the possibility has been very explicitly asserted for heavy microstates in MMS [1964] – has been recently *proven* experimentally for photons (cf. Piacentini & altera [2015]).

### *(7.III).2.3.6. On a debate related with the experiment EXP*

Just a few words on the debate Englert&alt. [1992], Dürr&alt [1993], Finkelstein [1995], Hiley&alt [2000] cited in a preceding note.

It is already clear at this point, I think, that the semantic differences that separate from one another the approach practised inside $QM_{HD}$ from that practised in $dB_{DS}(B)$ are so numerous and fundamental that any argument on the *theoretical* possibility or not to register a 'trace' in the sense of $dB_{DS}(B)$ is simply *devoid of significance if it is formulated by the use of the* $QM_{HD}$ *formalism*.

On the other hand the stake of a valid answer to this problem – centred upon the more definite question of the *experimental* possibility of trace-registrations of the value of the $dB_{DS}(B)$ 'guided momentum' (33) – appears to be very high. Indeed this answer can decide between:

♦ *Either* the nowadays $QM_{HD}$ representation of the microstates which in fact – as it now stands – offers reliable consensual and verifiable predictions only on the microstates $ms(unbound,1)_{G(n-c)}$ that are devoid of any inner interference phenomenon, *or* a new representation of the microstates that offers verifiable predictions on any sort of unbound microstates, with or without inner interferences.

And it will also decide between:

♦ *Either* a de Broglie-Bohm approach $dB_{DS}(B)$ that – if it is organically incorporated into a new intelligible quantum mechanics might play there a basic conceptual, operational and representational role in which $QM_{HD}$ fails, *or* a de Broglie-Bohm approach $dB_{DS}(B)$ that conserves its present status of only a mathematically expressed metaphysics of microphysics that keeps fighting for the title of "interpretation" of an un-intelligible $QM_{HD}$.

In these conditions the performance of the experiment *EXP* proposed in *(7.III).2.3.5* deserves fully the effort for being realized.

### *(7.III).2.3.7. Extended framework [IQM-$QM_{HD}$-$dB_{DS}(B)$] and a new coding-postulate for the momentum-values of* **any** *unbound microstate*

In the present work I shall not await the verdict of *EXP* for achieving the started construction of a global outline of an acceptable representation of the microstates. I shall just admit by hypothesis that *EXP* has been performed and has established the possibility to observe experimentally instantaneous guided momentum-values (33). This, I think, is very likely to happen soon, in our era of nanotechnologies. This likelihood can only be increased by an already pre-existing global outline of a view where such an investigation possesses its own and central conceptual place.

In order to achieve the outline of a fully intelligible second quantum mechanics, from now on the framework *[IQM-$QM_{HD}$]* is enlarged into a new framework that includes $dB_{DS}(B)$ and is denoted *[IQM-$QM_{HD}$-$dB_{DS}(B)$]*.

We first recall that the guidance law (33) is asserted deductively inside $dB_{DS}(B)$, and with *full generality*, both in the presence and in the absence of quantum fields. So inside *[IQM-$QM_{HD}$-$dB_{DS}(B)$]* we formulate the following coding-postulate for momentum-measurements on *any* sort of unbound microstate, *without* or *with* inner quantum potential, as well as with *one* or with *several* micro-*systems* (in the sense of the definitions from (2.I)1).

We posit that:

***P(cod)***$_{\forall unb.ms}$. The instantaneous momentum-value of *any* unbound microstate – whether a microstate $ms(unbound,1)_{cG(qf)}$ or $ms(unbound,1)_{G(n\text{-}c)}$ – can be determined by a coding-measurement-succession that obeys the representation

$$[G_t.Mes(\mathbf{r},\mathbf{p})] \quad \rightarrow_{\mathbf{trace}} \quad (\mathbf{r_k},\mathbf{p_j})_t, \quad k=1,2,..K; \quad j=1,2,..J; \quad \forall ms_{Gt} \qquad (35)$$

where: the sign "$_{\forall unb.ms}$" is to be read "valid for *any* unbound microstate", ; $G_t$ is posited to generate at the time *t – in the sense of (13")* – one unbound individual physical specimen $\sigma_{(Gt)}$ of the studied microstate $ms_{Gt}$, initially represented by an unknown individual wave-function $\Phi_{Gt}$ ; $Mes(\mathbf{r},\ \mathbf{p})$ denotes one full act of measurement of the *pair* $(\mathbf{r},\ \mathbf{p})_t$ formed *at the time t when the measurement begins*, by [the instantaneous values at *t* of the position $\mathbf{r}$ and the momentum $\mathbf{p}$ from (33), of the singularity from $\sigma_{(Gt)}$] ; the sign '$\rightarrow_{trace}$' is to be read ' identified via a process of registration in the sense of *EXP* ' of the trace of the $dB_{DS}(B)$ guidance-trajectory of the singularity from the specimen $\sigma_{\Phi(Gt)}$ during $Mes(\mathbf{r},\mathbf{p})$; $(\mathbf{r_k},\mathbf{p_j})_t$ is the pair of *values* registered for the pair $(\mathbf{r},\mathbf{p})_t$ of qualifying quantities.

Like the coding postulate (20'), the coding-postulate $P(cod)_{\forall unb.ms}$ also acts exclusively on the *individual* level of conceptualization. Furthermore, the symbol "$_{\forall unb.ms}$" can indicate *exclusively* a guiding interference effect (33) in absence of any classically definable field, and in such a case *it transgresses the capacity of representation of a* $QM_{HD}$ *measurement-Hamiltonian* ***H(A)***. Then (35) falls entirely outside the domain of facts that can be represented by a $QM_{HD}$-Hamiltonian. Which means that the Schrödinger equation of evolution simply ceases to be applicable.

And finally, let us note that:

*The coding postulate (35) violates overtly Heisenberg's principle, the principle of complementarity, as well as the Heisenberg theorem from* $QM_{HD}$ [125].

The postulate $P(cod)_{\forall unb.ms}$ is a new descriptional feature from the $dB_{DS}(B)$ approach that, inside the present approach, is injected into the domain of 'scientific' i.e. communicable, consensual, predictive and verifiable knowledge. The channel of adduction into 'scientificity' of descriptional prime-matter from the purely 'interpretative' $dB_{DS}(B)$ theory becomes now:

$$[MP(ms_{G,cw}),\ \ (1')\ G_{cw} \leftrightarrow ms_{G,cw},\ \ (14)\ ms_{G,cw} \equiv \{\sigma(ms_{G,cw})\},\ \ (35)\ P(cod)_{\forall msG})] \qquad (36)$$

Thus enriched, the channel (36) might entail a radical transmutation of the mutual status of the $dB_{DS}(B)$-conceptualization and that from $QM_{HD}$; a genuine *inversion* of their relative conceptual position. Indeed, while $QM_{HD}$ is found to be devoid of verifiability for its predictions on momentum-measurements on unbound interference microstates – which is a major gap – the present approach might suppress this flaw by use of elements from the works of de Broglie and Bohm that are arbitrarily considered to express 'purely' interpretive approaches.

---

[125] These 'principles' and this theorem – formulated in absolute terms – constitute together a knot of confusions between individual temporal characters and statistical characters. This knot deserves a detailed analysis of the type of that from (*6.II.4.4.3)* for the $QM_{HD}$ representation of quantum-measurements: Heisenberg's 'principle' mixes present and future individual data, while the 'principle' of complementarity is in fact a consequence of the $QM_{HD}$ Hilbert-space definitions of ket-states and of observable-operators.

### *(7.III).2.3.8. Identification of an adequate Hilbert-space representation of an interference microstate ms(unbound,1)$_{cG(q\text{-}f)}$*

**Rule seventeen**
The proposed difficulty must be gone through be making abstraction of the fact that some of the involved terms are known and other ones are unknown, by following in a genuine walk the mutual dependences.
**Rule nineteen**
It is by this method that one must research all the dimensions (physical quantities) that are expressed in two different ways, in terms either known or unknown, in order to browse directly through the difficulty; for by these means we shall obtain just as many comparisons between equal things.
**Rule twenty.**
*It is **after** having obtained the equations that we must achieve the omitted operations, without ever making use of multiplication when division is necessary*[126].

Descartes, The Rules for the direction of mind
(Regulae ad directionem ingenii), toward 1628 – 1629,
Letters to Elisabeth, Wikipedia (our translation from French)

Inside *[IQM-QM$_{HD}$-dB$_{DS}$(B)]* we are still devoid of an acceptable Hilbert-space formal representation for interference-microstates *ms(unbound,1)$_{cG(qf)}$*. So we cannot as yet extend to the microstates *ms(unbound,1)$_{cG(qf)}$* a given Hilbert-space form of state-ket to be reproduced factually and that permit to obtain an equivalent of the relation (31) established for the microstates *ms(unbound,1)$_{G(n-c)}$*. Hence for interference-microstates *ms(unbound,1)$_{cG(qf)}$* we cannot as yet *represent* in Hilbert-space terms factually generated probability laws that permit to calculate predictions via expansions of such a state-ket.

We shall now compensate this major lacuna. This – from a methodological viewpoint – will be a peculiar exploration.

***On the relation between $\Phi(r,t)$ and $|\Psi(r,t)>$.*** Since *(6.II).4.2.2* we have constantly distinguished between the individual wave-function denoted $\Phi_{G,}(\mathbf{r},t)$ assigned to each one specimen $\sigma(ms_G)$ of the studied microstate $ms_G$, and the $QM_{HD}$ state-ket $|\Psi_G(\mathbf{r},t)>$ associated with $ms_G$. What is the relation between these two sorts of descriptors?

Consider the coding-postulate (35) that is defined for any sort of unbound microstate. According to the completed framework *[IQM-QM$_{HD}$-dB$_{DS}$(B)]* each one given realization of the operation of generation *G* introduces into the domain of the observable one physical individual specimen $\sigma_{\Phi(G)}$ of the studied microstate $ms_G$, represented by a wave-function $\Phi(\mathbf{r},t)$. But all this is still only *model* conceived by human conceptor-observers, and *operations* realized by these. *Observation* can arise only when a whole sequence $[G_t.Mes(\mathbf{r}, \mathbf{p})]$ is realized. And *prediction-verification* can arise only if many repetitions of this sequence are realized, the respective results are observed and noted, and then – on the statistical level of conceptualization – the whole set of obtained results is considered globally, mentally and *retroactively*, and is represented by a well-defined state-ket $|\Psi_G(\mathbf{r},t)>$ that permits a calculus of probabilistic predictions that can be followed by verification of these. So – in basic agreement with the relations (1), (1') – the physical and individual descriptor $\Phi(\mathbf{r},t)$ has no access to the level of prediction-verification. Therefore its mathematical structure can be constrained only a posteriori and statistically, via the $QM_{HD}$ use of this tool for predictive calculi that consist of a Hilbert-space state-ket $|\Psi_G(\mathbf{r},t)>$. Nothing more can be done for defining the form of the wave-function $\Phi(\mathbf{r},t)$.

---

[126] My italics and bold fonts.

In this situation, in order to capture criteria for forming a notion on the relation between the mathematical forms of the state-ket of a microstate-to-be-studied and the wave functions of its specimens, we first notice that they are both written a priori with the *same* usual general functional form of a "wave-like' representation: $|\Psi(\mathbf{r},t)\rangle=|a_\Psi(\mathbf{r},t)e^{(i/\hbar)\varphi(\mathbf{r},t)}\rangle$ and $\Phi(\mathbf{r},t)=a_\Phi(\mathbf{r},t)e^{(i/\hbar)\beta(\mathbf{r},t)}$. Then we consider a set of many repetitions of a succession $[G_t.Mes(\mathbf{r_t}, \mathbf{p_t})]$ from (35) concerning the chosen microstate and we focus upon the obtained set of pairs of results $\{(\mathbf{r_k},\mathbf{p_j})\}_t$. In connection with these we introduce the following two rather straightforward *posits* P1 and P2.

**(P1)**. Any probability of an observable outcome is operationally definable – so also verifiable – only by *individual* acts of measurement. And the square root of the 'presence'-probability $\pi(\mathbf{r_{k,t}})$ (assimilated for the sake of effectiveness to $(\varepsilon,\delta,N_0)$-relative frequency) found inside a whole rich set of individual results $\{(\mathbf{r_k},\mathbf{p_j})\}_t$ from (35) for the outcome $\mathbf{r_k}$, is identical to Born's $QM_{HD}$ postulate of presence-probability asserted by $|\Psi(\mathbf{r},t)\rangle$, so at the time $t$ we have $\pi_\Psi(\mathbf{r_k},t) =(a_\Psi(\mathbf{r_k},t))^2$. On this basis and in agreement with history and with nowadays practice, for the amplitudes we posit here – like in $dB_{DS}(B)$ – that $(a_\Psi(\mathbf{r},t))^2=(a_\Phi(\mathbf{r},t))^2$ (normed to 1 and both members being expressed numerically in terms of a same convenient unity of length).

**(P2)**. Consider now the phase-function $\varphi(\mathbf{r_k},t)$ from the state-ket $|\Psi(\mathbf{r},t)\rangle=|a_\Psi(\mathbf{r},t)e^{(i/\hbar)\varphi(\mathbf{r},t)}\rangle$ of a microstate $ms_G$, and the phase-function $\beta(\mathbf{r},t)$ from the individual wave-function $\Phi(\mathbf{r},t)=a_\Phi(\mathbf{r},t)e^{(i/\hbar)\beta(\mathbf{r},t)}$ of a *specimen* $\sigma_{\Phi(G)}$ of $ms_G$: These phase-functions also are assigned a priori the same general functional structure. On this basis we posit here that at the space-time point $(\mathbf{r_k},t)$, the numerical value of the phase function $\varphi(\mathbf{r_k},t)$ is equal to the *mean* of the numerical values of the physical individual phase-functions $\beta(\mathbf{r},t)$ from the wave-functions $\Phi(\mathbf{r},t)=a_\Phi(\mathbf{r},t)e^{(i/\hbar)\beta(\mathbf{r},t)}$ of the set of the *specimens* $\sigma_{\Phi(G)}$ of the studied microstate $ms_G$ that are generated successively by the set of repeated coding-measurement-successions $[G_t.Mes(\mathbf{r_t}, \mathbf{p_t})]$ from (35) (in $dB_{DS}(B)$ the concept of coding-measurement-succession is absent) [127, 128].

***Back to the notions of observables and beables discussed in (7.III).2.3.3.*** Consider again the microstates $ms(unbound,1)_{cG(qf)}$ (for simplicity we make us of the basic case of an operation of generation $\mathbf{G}(G_1,G_2)$ with only two components $G_1$ and $G_2$ and we write in short $\Phi_{G(G1,G2)}(\mathbf{r},t)\equiv\Phi_{12}(\mathbf{r},t)$ and $|\Psi_{G(G1,G2)}(\mathbf{r},t)\rangle\equiv|\Psi_{12}(\mathbf{r},t)\rangle$). The examination of the microstate $ms_{G12}$ represented in the Fig.8, associated with the tentative posit P2 advanced above, focus attention upon the general fact that features of the phase-function $\beta_{12}(\mathbf{r},t)$ from the wave-functions $\Phi_{12}(\mathbf{r},t)=a_{12(\Phi)}(\mathbf{r},t)e^{(i/\hbar)\beta 12(\mathbf{r},t)}$ that represent the individual physical specimens $\sigma_{\Phi(G(G1,G2))}$ of a microstate $ms_{G(G1,G2)}$ involved in ***any*** measurement on $ms_{G(G1,G2)}$ – and that play a determining role in each individual outcome – might have remained simply *non-represented* inside $QM_{HD}$ because there – from the start and deliberately – only some sort of means have been researched. For momentum measurements on microstates $ms(unbound,1)_{cG(qf)}$ this has even certainly happened, along the following way. The $QM_{HD}$ representation of quantum measurements

[127] In consequence of this absence, inside $dB_{DS}(B)$ the phase functions $\varphi(\mathbf{r_k},t)$ and $\beta(\mathbf{r},t)$ are ***not** mutually distinguished*.

[128] The two posits P1 and P2 might be found later to be insufficient, or even inadequate (for instance, it might appear preferable to posit two different equations of evolution for $|\Psi\rangle$ and $\Phi$). Nevertheless we do formulate these posits in order to convey what sort of problems and possibilities emerge when one wants to create a consistent connection between $dB_{DS}(B)$ and $QM_{HD}$. This permits also to better understand why the descriptors $\Phi$ and $\Psi$ are currently confounded, and even in de Broglie's famous Thesis (cf. (6.II).1) as well as ***throughout the whole Bohmian Mechanics***. Only in de Broglie's final approach $dB_{DS}(B)$ [1956] are they distinguished from one another (but the notion of repeated coding-measurement-successions *[G.MesA]* is absent and so the idea of *mean* value in P2 did not emerge and $\varphi$ and $\beta$ have just been posited to be equal). And, like in the case of the $QM_{HD}$ representation of quantum measurements, we are often in presence of the tendency to wipe away the distinction (statistical, so abstract)-(individual and physical), via a vertical projection onto one common fictitious level of conceptualization.

has been constructed without any explicit recourse to de Broglie's Thesis. So, under the opaque cover of a total in-distinction between the individual and the statistical (cf. *(6.II).4.4.3*), it remained unnoticed that:

**(a)** According to the present approach *any* act *Mes.**P*** of momentum-measurement is performed on only *one **specimen** $\sigma_{\Phi(G(G1,G2))}$ of the **factually** generated microstate-to-be-studied $ms_{G(G1,G2)}$*, never on this microstate itself that is just *an abstract classical concept* (cf. (1.I).1.3 and the relations (1), (1')) ; while inside $QM_{HD}$ the concept of a factually generated microstate is not even defined, only the shadow of the abstract concept of microstate floats there in a basic mist, tied with the word 'system'.

**(b)** In *each* coding-measurement-succession $[G_t .Mes(\mathbf{r_t} ,\mathbf{p_t})]$ from (35), the structure of the phase $\beta_{12}(\mathbf{r},t)$ from the individual physical wave with wave-function $\Phi_{12}(\mathbf{r},t)=a_{12(\Phi)}(\mathbf{r},t)e^{(i/\hbar)\beta 12(r,t)}$ of the involved individual physical specimen $\sigma_{\Phi(G(G1,G2))}$ of the studied microstate $ms_{G(G1,G2)}$, plays a determining role in the emergence of the registered vector-value $\mathbf{p_j}$.

**(c)** For the particular case of a microstate $ms(unbound,1)_{cG(qf)}$, the structure of the wave-movement of the singularity from the wave of $\sigma_{\Phi(G(G1,G2))}$ can ***never*** be conceived to be that of a plane wave.

These lacunae explain how it has been possible to formulate the $QM_{HD}$ measurement-postulates 1 and 2 according to which the result of *any* act of momentum measurement performed on *any* sort of microstate would yield an eigenvalue $\mathbf{p_j}$ of the momentum-observable ***P*** that:

♦ Is 'created' by the measurement process, i.e. is *different* from the beable value at the moment when the act of measurement begins.

♦ Emerges necessarily tied to ***one*** corresponding ***plane***-wave eigenstate $a.e^{(i/\hbar)\mathbf{pj.r}}$ of the $QM_{HD}$-concept of momentum observable ***P*** (the measurement postulate 1.

♦ Subsists even after the closure of the act of measurement, *together* with its corresponding one-plane-wave eigenstate (the measurement-postulate 2).

When in fact, as already asserted, when one analyses carefully the situations that led to (20') and to (35) it appears that :

★ The momentum-value $\mathbf{p_j}$ that is observed for a microstate is *never "created"* by the measurement process, *neither* for a microstate $ms(unbound,1)_{G(n-c)}$ subjected to coding-measurement-evolutions that obey (20'), *nor* for a microstate $ms(unbound,1)_{cG(qf)}$ subjected to coding-measurement-evolutions that obey (35): i*n both these cases the momentum-value at the time t when an act of momentum-measurement **begins** is also the final value $\mathbf{p_j}$ that is **observed** at the end of the coding-measurement-evolution*; only the coding-process is different for these two sorts of microstates.

★ In the case of a microstate $ms(unbound,1)_{G(n-c)}$ subjected to momentum-measurements that accept the coding-postulate (20'), a convenient coding-measurement-evolution acts as follows: at the time *t* when a measurement-evolution begins, the structure of the wave-movement of the singularity from the wave of $\sigma_{\Phi(G(G1,G2))}$ – in general – might not be that of a plane wave; but when the exterior fields are all suppressed it can coherently be posited that it *immediately* acquires a plane-wave structure – at least in the quasi point-like neighbourhood of the singularity from the wave of $\sigma_{\Phi(G)}$ – just in consequence of the immediate lack of reasons that settles in for admitting any further influence on the subsequent dynamics of the singularity from $\sigma_{\Phi(G)}$ (The Occam-razor argument). And for the same sort of reason one has to conceive that the wave-movement from the whole wave of $\sigma_{\Phi(G)}$ acquires 'rapidly' a plane wave structure throughout the process of measurement. So *the momentum-value is **conserved** such as it was at the moment t when the considered act of measurement has begun*; this is the very aim of the coding postulate (20'), and for a momentum-measurement (20')

amounts to the method time-of-flight with coding procedure (20) from which it stems) [129]. So *the observed value is the beable value itself, not a created value*.

★ In the case of a microstate *ms(unbound,1)*$_{cG(qf)}$ subjected to momentum-measurements, at the moment *t* when a coding-measurement-evolution (35) begins, the singularity from the wave of the involved specimen $\sigma_{\Phi(G(G1,G2))}$ of the studied microstate $ms_{G(G1,G2))}$ is already genetically immersed inside the quantum-potential produced by the corresponding operation of generation ***G****(G1,G2))* (cf. *(7.III).2.3.3*). And even though the coding-measurement-evolution imposed upon $\sigma_{\Phi(G(G1,G2))}$ by (35) has to be conceived to destroy – to consume – the momentum-value $p_j$ possessed by the singularity from the wave of $\sigma_{\Phi(G(G1,G2))}$ at the time *t* when the coding-measurement-evolution begins, nevertheless, even though it is consumed, ***this*** initial value $p_j$ is what is finally translated from all the data gathered via the posited coding-measurement-evolution. So *in this case also the observed momentum-value* $p_j$ *is the beable-value itself, not a value created by the process of observation*[130]**.**

In short, we can now firmly assert that, as already sensed in *(7.III).2.3.3*:

> The essential difference between the two sorts of coding-measurement-evolutions considered above does not concern the *value* that is observed; it lies exclusively in *the way in which the observed momentum-value* $p_j$ *is drawn into consensual knowledge*.

And *this* difference depends on whether the coding-measurement-evolution is of the type (20') or of the type (35); which in its turn depends on the sort of microstate that is studied, *ms(unbound,1)*$_{G(n-c)}$ or *ms(unbound,1)*$_{cG(qf)}$. But in *both* cases the registered momentum-value $p_j$ is that one that is possessed by the singularity from the involved specimen of the studied microstate at the time *t* when the considered act of measurement *begin*s: in both cases it is a beable value, it is the value of *an 'observable beable'*.

The distinction between beables and observable' evaporates.

The preceding considerations finally do fully specify in what a sense the concepts of 'beable' and 'observable' are very confusing indeed: they pend stagnant *above* the genuinely basic difference that has intuitively suggested them, without reaching it, and mirroring it falsely. This leads us to a modification of language: From now on inside the framework *[IQM-QM*$_{HD}$*-dB*$_{DS}$*(B)]*:

> The concept of 'observables' distinct from 'beables' is banished.
> Instead, for the momentum quantity we shall speak of *value-conserving* coding-measurement-evolutions (20') that bring into the domain of the observable the beable momentum-value of a microstate *ms(unbound,1)*$_{G(n-c)}$, and of *value-destructive* coding-measurement-evolutions (35) that bring into the domain of the observable the beable momentum-value of a microstate *ms(unbound,1)*$_{G(qf)}$.

All the critiques made in *(7.III).2.3.1* and *(7.III).2.3.3* on the representation (15") and its consequences emanated from this hidden but precise and basic conceptual source finally specified just above [131].

---

[129] So in the case of a microstate *ms(unbound,1)*$_{G(n-c)}$ it is at most the structure of the global wave-movement from the whole wave of a specimen $\sigma_{\Phi(G)}$ of the studied microstate $ms_G$ – not the registered momentum-value $p_j$ – that is created by the act of measurement. But this does not qualify a mechanical observable, it qualifies a conceptual feature of the model of a microstate *ms(unbound,1)*$_{G(n-c)}$.

[130] And let us add that in both cases, at the end of the considered measurement-evolution the involved specimen of the studied microstate has ceased to belong (in the sense of (1)) to the studied micro-*state*.

[131] This is remarkable. It is a strong retroactive confirmation of *IQM* with its basic specificities with respect to $QM_{HD}$: The definition (1) that introduces the concept of operation of generation *G*; a relation between each *G* and *one* 'factually generated microstate' that introduces a whole set of distinct physical specimens; the concept of *composed operation of generation*; the concept of coding-measurement-evolution; etc. All this is radically absent in $QM_{HD}$. While here, once more, it permits quite basic elucidations, like those obtained in (6.II) for the significance of an eigenket' and the $QM_{HD}$ representation of quantum-measurement and the consequences entailed by this significance. *I do not speak here of*

As for what happens once a coding-measurement-evolution is finished, it does not seem useful to try to state this in general terms. This can be examined from case to case accordingly to deliberate future operational purposes of the conceptors-observers [132].

Finally we are ready now to enter upon the main problem, that of an acceptable Hilbert-space of the microstates $ms(unbound,1)_{cG(q\text{-}f)}$.

***Hilbert-space representation of the microstates*** $\boldsymbol{ms(unbound,1)_{cG(qf)}}$**.** We go back to the restriction to a ***one***-*plane-wave* structure $a.exp((i/\hbar)\mathbf{p_j}.\mathbf{r})$ imposed by the $QM_{HD}$ definition of the momentum-observable $\boldsymbol{P}$ upon the solutions of the $QM_{HD}$-equation $\boldsymbol{P|u(p_j,r)>=p_j|u(p_j,r)>}$ for eingenket and eigenvalues of $\boldsymbol{P}$. This restriction entails that a basis of eigenfunctions of $\boldsymbol{P}$ consists exclusively of *one*-plane-waves. We have exposed above for what complex reasons this requirement is not consistent with the factual situation that is realized for microstates $ms(unbound,1)_{cG(qf)}$. These reasons converge in suggesting that the a priori restriction of the $QM_{HD}$-concepts of momentum-observable $\boldsymbol{P}$ and of the corresponding $QM_{HD}$-equation $\boldsymbol{P|u(p_j,r)>=p_j|u(p_j,r)>}$ are source of the resistance that withstands a uniform applicability of a fully intelligible Hilbert-space representation of predictive probabilities for *any* sort of unbound microstates, on the basis of an association of $\big[(22)\ |\psi_{Gt1,\mathbf{H}}(\mathbf{r},t_1)>\approx_{\text{pred.}}\{\forall \boldsymbol{A},\forall t_1,\ |\psi_{Gt1,\mathbf{H}}(\mathbf{r},t_1)>/\boldsymbol{A}\}$ with Gleason's theorem (23)$\big]$, but transposed in terms of *factually* generated state-ket, as it has been realized for the case of microstates $ms(unbound,1)_{G(n\text{-}c)}$ via the relation

$$(31)\ |\psi_{Gt,H}(r,t_1)>_{fact(DM)}\ \neq_{\textbf{pred.}}\ |\psi_{Gt1,\mathbf{H}}(\mathbf{r},t)>,$$
$$\big[|\psi_{Gt,H}(r,t_1)>_{fact(DM)}\ \approx_{\textbf{pred.}}\ \{|\psi_{Gt,H}(r,t_1)>_{fact(DM)}/\boldsymbol{A}\},\forall\boldsymbol{A},\forall t_1\}\big]_{\textbf{verif.}}$$

In what follows we obey this suggestion. We proceed in 4 steps.

***1.*** *Summarizing recall.* According to *[IQM-QM$_{HD}$-dB$_{DS}$(B)]* any *one* act of momentum-measurement that finishes by a registration of one eigenvalue – whether it involves a microstate $ms(unbound,1)_{G(n\text{-}c)}$ or a microstate $ms(unbound,1)_{cG(qf)}$ – is *always* performed upon *one physical and individual* specimen $\sigma_{\Phi(G)}$ of the studied factually generated microstate $ms_G$, never upon this microstate itself that, by the basic definitions (1), (1'), consists of the abstract set of all such specimens and is 'represented' by a state-ket, in the sense (22). So an adequate equation for eigenvalues and eigen-functions of an observable $\boldsymbol{A}$ must concern the *individual physical* wave-functions $\boldsymbol{\Phi_G(r,}t)=a(\mathbf{r},t)e^{(i/\hbar)\beta(\mathbf{r},t)}$ that describe the specimens $\sigma_{\Phi(G)}$ of the studied microstate $ms_G$[133].

As for the probabilistic predictions concerning a microstate, these are represented in $QM_{HD}$ by expansions of the form (22). Inside the framework *[IQM-QM$_{HD}$]* and for microstates $ms(unbound,1)_{G(n\text{-}c)}$ the expansions from (22) have been transposed in factual terms via individual coding-measurement-successions, and *this* involves a fully intelligible access to Gleason's Hilbert-space representations (23) of the probabilistic predictions.

But our critiques from *(7.III).2.3.1* have shown that for microstates $ms(unbound,1)_{cG(qf)}$ and the basic momentum observable $\boldsymbol{P}$ the path recalled above for passing from the individual level of conceptualization to the statistical one, is blocked.

---

*'truth' in an absolute sense, which is mere illusion, but of elucidations in the sense of construction of intelligibility by inner semantic-formal consistency of a structure of representation.*

132 All the innumerable $QM_{HD}$ considerations on 'successive measurements' are certainly devoid of general validity. Moreover they are much more speculative than factually realizable. In general the effects of the final registration of the result of an act of measurement contradict the $QM_{HD}$ postulate 2 (while the postulate 1 is already incompatible with the critique of (15") and (32) and with the consequence (35) of this critique).

133 Inside $QM_{HD}$ the general occultation of the individual physical features might weaken the perception of this fact.

This is so because a microstate $ms(unbound,1)_{cG(qf)}$ is generated by a composed operation of generation $\boldsymbol{G}(G_1,G_2,...G_m,...G_M)$ where the composing operations $G_1,G_2,...G_m,...G_M$ from the unique realized global operation $\boldsymbol{G}$ entail always *more* than only one trend of wave-movement at *any* possible space-time location of the singularity from the physical individual wave $\boldsymbol{\Phi}_{\boldsymbol{G}t(G1,G2,...Gm,...GM)}$ of which *consists* any specimen $\sigma_{\Phi(G)}$ of the studied microstate; which in general entails for the singularity from $\boldsymbol{\Phi}_{\boldsymbol{G}t(G1,G2,...Gm,...GM)}$ momentum vector-values $\boldsymbol{p_j}$ of the 'guided' form (33) that are never tied to only *one*-plane-wave, and so are not compatible with the value-*conserving* coding-measurement-evolutions (20)-(20') implicitly presupposed inside $QM_{HD}$ to be universally valid for unbound microstates: Guided momentum-values like in (33) require the coding-postulate (35).

At this point our recall has reached its goal. Namely it has brought forth clearly that: In order to construct now inside the framework $[IQM\text{-}QM_{HD}\text{-}dB_{DS}(B)]$ a predictive algorithm of the form [(22),(23),(31)] that be applicable to also the microstates $ms(unbound,1)_{cG(qf)}$, the basic concept of momentum-observable $\boldsymbol{P}$ has to be generalized in a way *such* as to make it compatible with also the value-*destructive* coding-measurement-evolutions (35). Which requires a corresponding generalization of also the $QM_{HD}$-equation $\boldsymbol{P}|u(\boldsymbol{p_j},\boldsymbol{r})\rangle=\boldsymbol{p_j}|u(\boldsymbol{p_j},\boldsymbol{r})\rangle, \forall j$, for eigenket and eigenvalues of the $QM_{HD}$-momentum-observable.

In what follows we want to realize this [134].

***2.** A generalized concept of momentum-observable and its equation for eigenket and eigenvalues*. Let us call the researched generalized momentum-operator *a 'value-conserving or value-destructive momentum-observable*, and let us denote it $\boldsymbol{P_{vc\text{-}vd}}$.

Consider a microstate $ms_{\boldsymbol{G}t(G1,G2,...Gm,...GM)}$ of type $ms(unbound,1)_{cG(qf)}$. A given succession $[\boldsymbol{G}_{(G1,G2,...Gm,...GM)t}.Mes\boldsymbol{P_{vc\text{-}vd}}]$ from (35) is related via (33) with the phase-function $\beta(\boldsymbol{r},t)$ from the wave-function $\boldsymbol{\Phi}_{\boldsymbol{G}t(G1,G2,...Gm,...GM)}(\boldsymbol{r},t)=a(\boldsymbol{r},t)e^{(i/\hbar)\beta(\boldsymbol{r},t)}$ of the involved specimen $\sigma_{\Phi \boldsymbol{G}t(G1,G2,...Gm,...GM)}$ (in short $\boldsymbol{\Phi}_{\boldsymbol{G}t}$ and $\sigma_{\Phi \boldsymbol{G}t}$).The phase-function $\beta(\boldsymbol{r},t)$ from $\boldsymbol{\Phi}_{\boldsymbol{G}t}$ can be conceived to stem from an inner structure of the global wave-movement expressed by $\boldsymbol{\Phi}_{\boldsymbol{G}t}$ that involves *more* then only one plane-wave, in the following sense: It seems necessary to conceive that (in general at least) at any space-time point $(\boldsymbol{r},t)$ where $\beta(\boldsymbol{r},t)$ is defined, each component operation $G_m$ from the operation of generation $\boldsymbol{G_t}$ $(G_1,G_2,...G_m,...G_M)$ induces its own directional trend of wave-movement. Let us denote by $\boldsymbol{k_m}(\boldsymbol{r},t)$ the unit-wave-vector of the wave-contribution stemming from the component $G_m$ from $\boldsymbol{G}_t$. According to (1) $\boldsymbol{G}_t$ is posited to come out 'the same' – with respect to the set of parameters that define it – each time that it is re-produced. So the local contribution from $G_m$, with its own wave-vector $\boldsymbol{k_m}(\boldsymbol{r},t)$, is conceived by definition to be invariant *with respect to the repetitions of* $\boldsymbol{G}_t$. But in each *one* given succession $[\boldsymbol{G_t}.Mes\boldsymbol{P_{vc\text{-}vd}}]$, inside the one involved wave-function $\Phi_{Gt}$, the wave-vector $\boldsymbol{k_m(r,t)}$ is in general a variable *with respect to passage from one space-time point (r,t) to another one*, for any given $m$ [135].

[134] This is not a minor purpose. For in our view the case $ms(unbound,1)_{cG(qf)}$ plays the role of the *rule* while the case of a simple operation of generation $G$ is a rare exception (as already remarked, even in textbook examples with potential-'barriers', 'walls', 'wells', harmonic oscillators, etc. in fact we dwell with microstates $ms(unbound,1)_{cG(qf)}$ that involve interference and non-null quantum potentials). This remains non-singularized by the $QM_{HD}$ formalism where the concept of a physical-conceptual operation of generation is absent and its physical-conceptual basic role is replaced by the exclusively mathematical notion of 'limiting conditions' imposed upon the Schrödinger solution 'of the problem'. In such circumstances one is led to wonder *whether the $QM_{HD}$ predictions on the results of momentum measurements have ever been seriously subjected to verification*; and if this has been done, *how the measurements have been realized*, given that also – or only – non-classical quantum-potentials do act, that are not accessible to human manipulation, so that the method time-of flight cannot be applied; while inside $QM_{HD}$ trace-registrations, in principle, are not 'legal' procedures for measuring the momentum observable. This whole – enormous – set of questions goes lost in just language-solution ("quantum-tunneling", etc.); or, for *bound* states, in global "effects" (Zeemann, Stark", etc.) that are made use of as measurement-operations, after many approximations for defining the state-ket.

We *posit* that, at the time $t$ when the act of measurement *begins* (that indexes $G_t$), the momentum vector-eigenvalue of $\boldsymbol{P_{vc\text{-}vd}}$ registered by *any* momentum-measurement-succession $[G_t.Mes\boldsymbol{P_{vc\text{-}vd}}]$ – whether making use of the coding postulate (20') or (35) [136] – is :

$$p_j(\boldsymbol{r},t) = -\ \boldsymbol{\nabla}.\beta(\boldsymbol{r})_t\ =\ (\textstyle\sum_m (p_{jm}\,\boldsymbol{k_m(r)}))_t\ , \qquad m=1,2,..M \qquad (33')$$

where:

- $p_{jm}$ is the projection of $p_j(\boldsymbol{r},t)$ on $\boldsymbol{k_m(r)}$ at the time $t$ ;
- $j=1,2,.....J$ and $J$ is an arbitrarily big *finite* integer (the discrete character assumed for the succession of the values $p_j(\boldsymbol{r},t)$ being entailed and determined by, and relative to, the choices of the involved units of measurement (we write in short $\forall j$)).

The relation (33') amounts to an explicit and *general* connection between the de Broglie guidance law (33), and the basic *IQM*-concept of operation of generation (via its re-definition (1')). As such it *completes* the modelling postulate (35) $MP(\{\sigma(ms_{G,cw})\})$ from (6.II).2 and the channel (36) of adduction into consensual predictive-verifiable 'scientificity' of descriptional prime-matter from the purely 'interpretative' $dB_{DS}(B)$ theory; this channel now becomes

$$\Big[MP(ms_{G,cw}),(1')G_{cw}\leftrightarrow ms_{G,cw},(14)ms_{G,cw}\equiv\{\sigma(ms_{G,cw})\},(35)P(cod)_{\forall msG}),(33')p_j(\boldsymbol{r},t)=\boldsymbol{\nabla}.\beta(\boldsymbol{r})_t=\textstyle\sum_m(p_{jm}\boldsymbol{k_m(r)})_t\Big] \qquad (36')$$

According to (33') the 'mechanical' momentum vector-value $p_j(\boldsymbol{r},t)$ of the singularity from each one *wave*-function $\Phi_{Gt}(\sigma_{\Phi(Gt)})$ of the one specimen $\sigma_{\Phi(Gt)}$ of the studied microstate $ms_{Gt}$ involved in one measurement-succession $[G_t.Mes\boldsymbol{P_{vc\text{-}vd}}]$ – at the space-point $\boldsymbol{r}$ where it happens to be – *depends essentially on time as soon as m>1.*

The relation (33') is just a posited definition that amounts to assert that the eigen-*function* – an eigen-***wave**-function* – that corresponds to $p_j(\boldsymbol{r},t)$ is:

$$\Pi_m\Big(a.exp((i/\hbar)\,(p_{jm}.\boldsymbol{k_m(r)})_t.\boldsymbol{r}\Big) = a.exp\,\Big((i/\hbar)\,\big(\textstyle\sum_m p_{jm}\,\boldsymbol{k_m(r)}\big)_t\,.\boldsymbol{r}\Big), \qquad \forall j,\ m=1,2,..M \qquad (37)$$

where $M$ is a *small* integer.

The calculus that *totalizes* the coefficient $\sum_m(p_{jm}\boldsymbol{k_m(r)})_t$ of the position-vector $\boldsymbol{r}$ is deliberately left non-effectuated in (37) because the effectuated sum would lead to the first member from (33') that has been assigned the *mechanical* meaning of a momentum-vector; while – in order to strictly respect rules of conceptual homogeneity that work more or les non-explicitly throughout the human conceptualization [137] – *we want to keep mutually distinct the two meanings of, on the one hand a wave-movement, and on the other hand a mechanical qualification of the mechanical displacement of the singularity from the amplitude of this wave-movement.* So from a purely mathematical point of view, inside (37) the signs $p_{jm}$, $m=1,2,..M$ work as just numerical coefficients, one can made *total* abstraction of their connection (33') with a concept of 'momentum', so (37) defines *pure **wave**-functions*, there is nothing specifically mechanical in these as long as the definition (33') is not added. On the other hand, the writing (37) expresses explicitly in mathematical terms the inner structure assigned by (33') to the phase $\beta(\mathbf{r},t)$ of the considered physical wave $\Phi_{Gt}$ – namely the structure of a *physical* superposition of $M$ plane-waves imparted to this wave at any given point $(\mathbf{r},t)$ by its operation of generation $\mathbf{G_t}(G_1,G_2,...G_m,...G_M))$ (with $m\geq 1$); concerning the eigenvalue of the mechanical concept of momentum it asserts strictly nothing.

---

[135] Moreover $\boldsymbol{k_m(r,t)}$ can be symbolically tied with a corresponding whole plane-wave $a.exp[(i/\hbar)\,(p_{jm}\boldsymbol{k_m}).\boldsymbol{r}$ – where $p_{jm}$ is the projection on $\boldsymbol{k_m}$ of $\boldsymbol{p_j}$ at the time $t$ – that acts like just a *sample* of arbitrary extension of the direction of wave-movement instilled at $(\boldsymbol{r},t)$ by $G_m$ (cf. (6.II).1).

[136] We recall that inside $[IQM\text{-}QM_{HD}\text{-}dB_{DS}(B)]$ the guiding relation (33) can be conceived to be valid consistently with the whole $QM_{HD}$ formalism, in consequence of the identification in (6.II).1 of the significance of any eigenstate and of the fact that in $dB_{DS}(B)$ the guiding theorem is obtained for any sort of microstate, with or without an inner quantum potential.

[137] The fact that Louis de Broglie has written "corpuscular phases" has introduced much confusion, and in particular, the arbitrary restriction that just now we are trying to suppress.

So finally inside *[IQM-QM$_{HD}$-dB$_{DS}$(B)]* the equation for the eigen-***waves*** and eigen-values of the *mechanical* observable $\mathbf{P_{vc\text{-}vd}} = (\hbar/i)\,\nabla$ has to be written as:

$$\mathbf{P_{vc\text{-}vd}} \,.\, \Pi_m \Big(a.\exp((i/\hbar)\,(p_{jm}.\mathbf{k_m(r)})_t.\mathbf{r}\Big) = \big(\textstyle\sum_m p_{jm}.\mathbf{k_m(r)}\,\big)_t \,.\, \Pi_m \Big(a.\exp((i/\hbar)\,(p_{jm}.\mathbf{k_m(r)})_t.\mathbf{r}\Big) =$$

$$=\big(\textstyle\sum_m p_{jm}.\mathbf{k_m(r)}\big)_t \,.\, a.\exp\Big((i/\hbar)\,\big(\textstyle\sum_m p_{jm}\,\mathbf{k_m(r)}\big)_t.\mathbf{r}\Big), \qquad \forall j,\ m=1,2,..M, \qquad (38)$$

The calculus from (38) isolates now the sum $(\sum_m p_{jm}.\mathbf{k_m(r)})_t$ from the phase of the eigenfunction and it expulses this sum in front of the expression of the wave-function. And since now this sum is available in a position that separates it from the expression of the eigen-wave, we finally can, without violating semantic homogeneities, make use of "both equations" required by Descartes in the motto from the beginning of *(7.III).2.3.7* (in our case these two equations are (33') and (38)). So accordingly to the twentieth rule quoted in this motto we can *perform* the addition from (33') and write a last unifying equation for eigenstates – eigen-*waves* in fact – and eigenvalues $\mathbf{p_j(r},t)$ of the generalized *mechanical* momentum-observable $\mathbf{P_{vc\text{-}vd}}$; we can write this *safely from a semantic viewpoint*, i.e. without mixing the meanings. Which finally yields the researched generalized equation

$$\mathbf{P_{vc\text{-}vd}} \,.\, \Pi_m \Big(a.\exp((i/\hbar)\,(p_{jm}.\mathbf{k_m(r)})_t.\mathbf{r}\Big) = \mathbf{p_j(r},t).\ a.\exp\Big((i/\hbar)\,\big(\textstyle\sum_m p_{jm}\,\mathbf{k_m(r)}\big)_t.\mathbf{r}\Big), \quad \forall j,\ m=1,2,..M, \qquad (38')$$

that – quite remarkably – *distinguishes* efficiently between the eigen-*values* $\mathbf{p_j(r,t)}$ of the *mechanical* momentum that appear only in the coefficient placed in front of the second member, and the eigen-wave-functions $a.\exp\Big((i/\hbar)\,\big(\sum_m p_{jm}\,\mathbf{k_m(r)}\big)_t.\mathbf{r}\Big)$, where the ***wave***-vector-values

$$\big(\textstyle\sum_m p_{jm}\,\mathbf{k_m(r)}\big)_t.\mathbf{r}$$

from the phase-functions are now 'free' to define accordingly to (33') the eigenvalues of the ***mechanical***-momentum-observable $\mathbf{P_{vc\text{-}vd}}$ with the whole generality imposed by the structure of the acting operation of generation $\mathbf{G_t}(G_1,G_2,...G_m,...G_M))$ (with $m\geq 1$. Indeed the equation 38') respects now explicitly that the two quantities represented by the symbols $\mathbf{p_j(r},t)$ and $\big(\sum_m p_{jm}\mathbf{k_m(r)}\big)_t.\mathbf{r}$ have *different conceptual-physical natures* and can be combined numerically only because the corresponding mathematical concepts have from the start the same *dimensional* definition [138]. This procedure avoids the confusions

---

[138] This lesson delivered by Descartes over centuries seems to me so noteworthy that I take the liberty to confess the process that led to (38). After more than a whole year of incapacity to formulate a wholly convenient generalization of the *QM$_{HD}$*-equation for eigenvalues and eigenket of the momentum observable, I had an accidental insight that clarified the conceptual situation. I told this to a very learned friend, Carlos Lobo, and he showed me the Descartes rules. These brought into luminous evidence why I had stagnated such a long time: *Mathematically* there is no imposed distinction between the non-effectuated form of the sum $\sum_m (p_{jm}\mathbf{k_m})_t.\mathbf{r}$ from the second member of (38), and its effectuated form $\mathbf{p_j.r}$, and in consequence of this one glides nearly irrepressibly into the effectuated form. But on the other hand the

entailed throughout $QM_{HD}$ by the heritage of de Broglie's initial notations that had mixed wave-qualifications and mechanical qualifications (deliberately no doubt, in order to draw attention upon his 'corpuscular-like' model (then usually called the wave-'particle' model))[139].

For a microstate $ms(unbound,1)_{G(n-c)}$ we have $m=1$ ; so $\boldsymbol{G_t}(G1,G2,....Gm,.....GM) \equiv \boldsymbol{G_t}$ and the index $t$ can be dropped; everything becomes time-*in*dependent and $\boldsymbol{P_{vc\text{-}vd}}$ reduces to $\boldsymbol{P}$. So (38') becomes the $QM_{HD}$ equation for eigenfunctions and eigenvalues of the momentum observable $\boldsymbol{P}$ and the coding postulate (20') also can be utilized. While for $m\geq 2$ exclusively the coding-postulate (35) is valid.

The researched generalization of the $QM_{HD}$ equation is now fully achieved.

Which finally closes a rather complex loop.

***3.*** *The state-ket of a microstate* $ms(unbound,1)_{cG(qf)}$. Instead of the inadequate additive $QM_{HD}$-representation of type (15''), inside $[IQM\text{-}QM_{HD}\text{-}dB_{DS}(B)]$ we now can assign to a microstate $ms(unbound,1)_{cG(qf)}$ a *one*-term representation

$$|\Psi(\boldsymbol{r},t)_{\boldsymbol{G}t(G1,G2,...Gm,...GM)}(\boldsymbol{r},t)> \qquad (39)$$

The physical content of $|\Psi(\boldsymbol{r},t)_{\boldsymbol{G}t(G1,G2,...Gm,...GM)}(\boldsymbol{r},t)>$ is determined by:
- the operation of generation $\boldsymbol{G_t}(G_1,G_2,...G_m,...G_M))$;
- the model $MP(ms_{Gt,cw})$ of a microstate;
- the individual wave-function $\Phi_{\boldsymbol{G}t(G1,G2,...Gm,...GM)}(\boldsymbol{r},t)=a(\boldsymbol{r},t)e^{(i/\hbar)\beta(\boldsymbol{r},t)}$ of any specimen $\sigma_{\Phi(\boldsymbol{G}t)}$ of the studied microstate $ms_{\boldsymbol{G}t(G1,G2,...Gm,...GM)}$;
- the coding-postulate (35);
- the connection between $\Phi_G$ and $|\Psi_G>$ posited by P1 and P2.

Together, the sources enumerated above constitute a rather complex constraint.

***4.*** *Hilbert-space representation of the predictions.* Inside $[IQM\text{-}QM_{HD}\text{-}dB_{DS}(B)]$ the expansion with respect to $\boldsymbol{P_{vc\text{-}vd}}$ of a state-ket (39), $|\Psi(\boldsymbol{r},t)_{\boldsymbol{G}t(G1,G2,...Gm,...GM)}(\boldsymbol{r},t)>/\boldsymbol{P_{vc\text{-}vd}}$, is to be written by use of an ortho-normalized basis of *wave*-eigenket

$$\{\ |\Pi_m\left(a.exp((i/\hbar)\ (p_{jm}.\boldsymbol{k_m}(\boldsymbol{r}))_t.\boldsymbol{r}\ >\right\} = \{\ |a.exp\left((i/\hbar)\ (\textstyle\sum_m p_{jm}\ \boldsymbol{k_m}(\boldsymbol{r}))_t\ .\boldsymbol{r}\right)\ >\},\ \ \forall j,\ m=1,2,..M$$

(in short $\{\,|\,u(\boldsymbol{r}, \boldsymbol{p_j}, M)>\}_t$ )[140] (40)

---

effectuated form $a.exp[(i/\hbar)\ \boldsymbol{p_j}.\boldsymbol{r}]$ describes ***one*** plane wave. Whereas *physically*, at the moment *t* when *begins* the act of measurement $Mes\boldsymbol{P_{vc\text{-}vd}}$ from the succession $[\boldsymbol{G_t}.Mes\boldsymbol{P_{vc\text{-}vd}}]$, the wave-function of a specimen of a microstate $ms(unbound,1)_{cG(q\text{-}f)}$ has, at any space point $\boldsymbol{r}$, the inner structure of an *evolving* superposition of *M* plane waves, *not* the *stable* and *restrictive* form of one plane-wave function, and *this is why in general the vector-value* $\boldsymbol{p_j(t)}=\sum_m(p_{jm}\boldsymbol{k_m})_t$ *cannot be conserved during a measurement evolution as it is required for applicability of the coding-postulate (20')*. And precisely this has led in this work to the necessity of recourse to the value-destructive trace-registrations (33) and to the coding-postulate (35). Which shows that the choice of the adequate coding-measurement-evolution depends quintessentially on the inner structure imparted to a physical individual wave $\Phi_{Gt}$ of a physical individual specimen $\sigma_{\Phi(Gt)}$ of the studied microstate $ms_{Gt}$, by its operation of generation. So it was crucial indeed not to glide too early into the effectuated form of the sum $\sum_m(p_{jm}\boldsymbol{k_m})_t$ (when one wants to make the sum of several apples with several prunes, one also has to *first* have become able to speak of 'fruit'). But such a warning cannot be *conceived* inside the statistical algorithmic representations from $QM_{HD}$ where even the basic significance of the mathematical concept of eigenfunction is entirely ignored. This illustrates the unpredictable specificities that, inside Mathematical Physics, can – and must – distinguish a representation constantly guided by *semantic*-mathematical criteria, from a representation guided by purely mathematical criteria. *Mathematical Physics cannot be reduced to Mathematics*. And it is impressing to learn how explicitly Descartes was aware of this.

[139] In its turn, the distinction between (38) and (38') constitutes an explicit connection with our critique in (7.III).2.3.1 of non-verifiability of the prediction from (15)-(32).

[140] In de Broglie ([1956], pp. 119-133, a fascinating science-fiction-like chapter) in a microstate obtained by reflection on a mirror of an incident state there are places where the corpuscular-like singularity – endowed with a ***'quantum***

So finally:

Gleason's theorem (23) permits to place now the numbers

$$\pi(\mathbf{p_j}) = \left|c_j(t)\right|^2 = \left|Pr._j \,|\, \boldsymbol{\Psi}(\mathbf{r},t)_{\mathbf{G}t(G1,G2,...Gm,...GM)}(\mathbf{r},t) > \right|^2$$

from the expansion $|\Psi(\mathbf{r},t)_{\mathbf{G}t(G1,G2,...Gm,...GM)}(\mathbf{r},t)>/\mathbf{P_{vc\text{-}vd}}$ on the axes of the basis (40) introduced in the Hilbert-space of state-ket $|\Psi(\mathbf{r},t)_{\mathbf{G}t(G1,G2,...Gm,...GM)}(\mathbf{r},t)>$, each one of these axes being labelled by an eigenket from (40).

The mathematical representation of *the physical effects entailed by the co-presence of* $(G_1,G_2,...G_m,...G_M))$ *inside the unique fully realized operation of generation* $\mathbf{G_t}(G_1,G_2,...G_m,...G_M))$, upon the prediction-verification probability-distributions of registered momentum-values, is now entirely taken in charge by the enriched basis of *wave*-eigenket and by the corresponding enriched equation for eigenket and eigenvalues (38') and the basis of eigenket (40). The $QM_{HD}$ mathematical interaction between expansion-coefficients from the additive combination of the two state-ket from (15") – that leads to the non-verifiable prediction (32) – is finally replaced by a conceptually grounded mathematical representation that inside $[IQM\text{-}QM_{HD}\text{-}dB_{DS}(B)]$ is logically coherent with all the three involved sources, *IQM* and $QM_{HD}$ and $dB_{DS}(B)$.

Since any quantum-mechanical observable is by construction a symmetrized function $\boldsymbol{A(R,P)}$ of the position-observable $\boldsymbol{R}$ and the momentum-observable $\boldsymbol{P}$, the results obtained above for the momentum observable can be extended to any observable $\boldsymbol{A}$.

So the relation (31) can now be extended step by step – ***factually*** – to also the microstates $ms(unbound,1)_{cG(qf)}$.

*This leads to a fully intelligibe,* ***consensual, observational, predictive and verifiable*** *representation of the quantum measurements on unbound microstates of any kind (in the sense of the definitions from* (2.I).1*).*

### *(7.III).2.3.9. Conclusion on the* $[IQM\text{-}QM_{HD}\text{-}dB_{DS}(B)]$-representation of the microstates $ms(unbound,1)_{cG(qf)}$

We summarize.

The main lines of a Hilbert-space representation inside $[IQM\text{-}QM_{HD}\text{-}dB_{DS}(B)]$, of the microstates $ms(unbound,1)_{cG(qf)}$, researched in (7.III).2.3, are now sketched out. Of course many points remain to be worked out (for instance the orthogonalization of (40), the Dirac-transformations, etc.). But the essence of the specific problems and of the corresponding solutions is clarified. And, as asserted above, on this new foundation and by use of:

- the coding postulate (35);
- the definition (33');
- the equation (38');
- the definition (39) of the state-ket of a microstate $ms(unbound,1)_{cG(qf)}$ ;
- the expansions (22) with respect to the generalized momentum-observable $\mathbf{P_{vc\text{-}vd}}$, so on the basis (40);
- Gleason's theorem (23)

---

**mass'** – is at rest, or in other circumstances it becomes imaginary; so *one should systematically work with the concept of 'quantum mass' and admit a priori also eigenvalues* $\mathbf{p_j}(r,t)=0$ (and also 'imaginary' eigenvalues ?). This is likely to be tied with teleportation.

The proof of the assertion *Ass.2* can now obviously be extended point by point to the microstates *ms(unbound,1)*$_{cG(qf)}$. Which leads to a representation (31) of the state-ket (39). The desired *general* factually constructed equivalent of the central predictive $QM_{HD}$ algorithm of form ((22)-(23)), is gained for also the microstates *ms(unbound,1)*$_{cG(qf)}$.

We can now proceed to close our action of Hilbert-space representation of the category of unbound microstates inside the framework *[IQM-*$QM_{HD}$*-*$dB_{DS}$*(B)]* by a succinct examination of also the last case defined in (2.I).1, of one unbound microstate of two or more micro-systems.

## (7.III).3. THE CASE OF *ONE* UNBOUND MICRO-*STATE* OF *TWO* OR MORE MICRO-*SYSTEMS*

Consider now the case of *one* micro-state of *n* micro-systems, which we denote by $ms(unbound,n)_{cG(qf)}$. In what follows we restrict ourselves to the case $ms(unbound,2)_{cG(qf)}$ with only *two* micro-systems that is involved in Bell's theorem on non-locality (cf. (3.I).1) (the generalization of the considerations to cases $n\geq 2$ is obvious).

I have already examined Bell's theorem in other works, variously and thoroughly [141]. So here I shall add only the following very rapid considerations specifically connected with the context from the Part II of this work.

Consider the probability tree from the part I of this work that is tied with a Bell-experiment (fig 4, (3.I).2). Since the two involved micro-systems *S1* and *S2* belong by definition to one *common* micro-state $ms_{G(2S)}$ (in the sense of (2.I).1) there is no reason whatever to posit a priori that the spin-values registered for *S1* and *S2* by one measurement-succession $[G(2S).Mes_{12}(spin1.spin2)]$ should come out to be *non*-correlated (the sign *Mes(spin1.spin2)* applies the general notation $Mes_{12}(A1,B2)$ from (3.I).2); quite on the contrary, and all the more so as in this case the considered sort of correlation is to be awaited to be *stronger* than the generally present 'meta'-correlations (11), (11') from the probability-tree of *any* sort of unbound microstate, because by conceptual-factual construction *it stems from the* ***interior*** *of each one* ***individual*** *event brought forth by each one complete measurement-succession from one branch of the corresponding probability-tree*, whereas the correlations (11), (11') involve globally considered whole probability laws. This is a *very* unusual conceptual situation, as much with respect to the classical mechanical characterizations of 'mobiles', as with respect to the classical calculus of probabilities [142]. The *mechanics* of microstates brings in – *both* – the individual specimens of the studied microstates considered globally *and* their inner structure, because that what admits mechanical qualifications is inside the wave-like 'whole' called 'one specimen of a microstate' (in the sense of (1), (1') and represented by individual wave-functions $\Phi_{Gt}(\mathbf{r},t)=a(\mathbf{r},t)e^{(i/\hbar)\beta(\mathbf{r},t)}$ [143].

In *such* conditions, *what is the point* in hasting for changing the orientations of the spin-measurement devices just at the last moment before the registration? A predictive calculus can determine quite calmly all the possible observable spin-correlations generated by the various orientations of the registering devices. These potentialities do not depend on time. A high-pressure last-moment choice of the orientations of the devices that register spin-values is expected to do, *what*? To trick a mathematical implication of the formalism of $QM_{HD}$ that violates Einstein's principle of macroscopic locality by forcing the observational effects of micro-phenomena to dominate this mathematical violation? Such a procedure possesses meaning only with respect to:

- *An a priori refusal* of the possibility of existence of any correlation.

---

[141] It has the merit to have released a revolution inside the Bohr-orthodoxy. It has acted as a very active ferment in the scientific conceptualization of the microphysical reality. In order to fix Psycho-social reference-elements on the time-dimension, I have reproduced in the Prologue to the first part of this work an 1979 lecture on Bell's theorem; and in (MMS [2013] **v.3** and in [2017], French texts) can be found what I call a 'conceptual' invalidation of Bell's proof (I have shown that the conclusion – such as it is expressed verbally by Bell – ***does not follow*** *from the mathematical proof*, though this – considered independently – is *valid*. I have also succeeded to construct a counterexample to Bell's formulation of his only asserted conclusion (MMS [1987]) that has been confirmed by Bordley [1989]) as 'factually possible'. But the most relevant *new* data concerning the problems raised by Bell consist of the content from the sub-section (3.I).2 from this work (completely established only since 2012).

[142] In order to bring into explicit evidence the general peculiarities of such a situation, a systematic preorganization of the involved concepts and language – like that from *IQM* – is *a sine qua non pre-condition*; if only concepts and words from the current languages are made use of, in this case one gets lost for speaking and reasoning in a precise way.

[143] That *corresponds to* but is *distinct from* the involved state-ket $|\Psi_{Gt}>$ that – via its expansions (22) $|\Psi_{Gt}>/\mathbf{A}$ – describes abstract statistics of results of individual coding-measurement-successions *[G.MesA]* measurements where are involved the individual wave-movements described by the wave-function $\Phi_{Gt}(\mathbf{r},t)$. In order to genuinely understand the physical and the conceptual situation one should have a clear view on all this.

- The research of a factual estimation of the value of the velocity of a 'transmission of influences' between the two involved micro-systems, that *according to Bell's theorem would necessarily exist*.

But *the second assumption recalled above is **false***: Bell's mathematical calculus and logical reasoning prove *exclusively* the existence of a correlation accordingly to the $QM_{HD}$-formalism, which is known ; they do *not* prove also the necessity of an *'influence'* that entail the correlation (MMS [2013] [144]).

> The notions of 'influence' and so also of a velocity of transmission of it, are exterior to Bell's logical-mathematical proof. They are just asserted – *directly and exclusively* – in Bell's verbal formulation of *his* conclusion.

And these notions are founded upon a furtive supplementary and *independent* assumption, namely that: The "two systems" *must* have 'separated' when they have come 'far enough' from one another [145]**.** Which amounts to wipe away the inside-outside specificities of the mechanics of a micro-'system' (a de Broglie corpuscular-like singularity in the wave of a specimen of a microstate) and to treat it like a classical mobile [146].

The preceding remarks brings us now to the following central question: What is the *ground* for imposing so dramatically a *universal* Einstein condition of locality, when this condition has been endowed with a well-defined significance and role for – specifically – macroscopic mobiles that are directly perceived with definite global contours that individualize them mutually, by human observers that are imprisoned in differently moving inertial frames of reference wherefrom *they communicate via light signals*?

When two or more *such* observers do *all* survey *such* a mobile, and in *such* conditions, a scientific representation does indeed require some rules of consensus concerning the identification of the mobile and of its dynamic. Some invariants are indeed necessary for generating, for instance, a consensual significance for the basic assertion that these observers are all perceiving the 'same' mobile.

But for microstates, each human observer gathers a knowledge of the kind structured in *IQM*; a knowledge constructed very indirectly and in a solitary way, by each researcher isolated alone in his own Laboratory, without perceiving anything else than cryptic marks on devices out of which he draws some significance only by use of previously elaborated conceptual-mathematical rules and treatments and of operational actions (physical or abstract) decided on the basis of a model that involves unlimited waves, etc., etc.. In *such* circumstances, the a priori importation into fundamental microphysics of all the requirements of the macroscopic relativistic mechanics is very far from being an obvious necessity. It even is a highly *arbitrary* constraint that manifests the impressive blinding force of the inertial urge to assert the absolute general validity of anything that has been strikingly efficient inside some particular context.

Einstein's requirement of locality has a *methodological* nature. It is not an absolute factual truth. A fortiori there is no reason whatever to require it universally. Not any group of invariance has to be uniformly asserted in any cognitive situation.

We conclude that the structure of Bell's theorem seems to have been entailed by a model of two solid balls that are receding from one another[147].

---

[144] The proof of this last assertion is what I have called *'conceptual'* invalidation of Bell's theorem.

[145] In the text-books one finds systematically the assertion that when they are still close to one another "the systems can interact" but "when the systems are sufficiently far from one another they must have separated".

[146] This illustrates the major role of the model in a consensual predictive and verifiable description of the microstates, that Bohr wanted to 'free' from any model, by a philosophical diktat.

[147] Of course all the preceding considerations also are founded upon models. But by now, I think, it has become clear that without any model one cannot try to construct a theory of microstates; one even cannot reason, prove, *conceive*, speak and write. And indeed the whole non-locality problem concerns a model, a more or less hidden one that involves

The *cognitive situation* of the human conceptor-observer with respect to this or that domain of physical entities is what decides the relevant conditions of consensus to be required.

A unification of Physics cannot be realized directly by representations of assertions of absolute factual truths. It can be realized only indirectly, only methodologically, inside a general and basically relativized common methodological framework that, by convenient particularizations, leads to a relativized representation of this or that particular domain of facts.

We now quit the problem of non-locality and, from the specific and humble viewpoint of the purpose of this chapter 7.III, we consider specifically the case of one microstate of two or more microsystems. From this point of view the significant conclusion is the following one:

Nothing withstands the extension of the assertion *Ass.2* and of the equivalence (31), to the microstates of one unbound micro-state of two or more micro-systems. The proofs worked out in (7.III).2.3 for microstates $ms(unbound,1)_{G(qf)}$ can be transposed point by point to also this category of microstates. Louis de Broglie's model of a specimen of the considered microstate suggests that inside any microstate of two or more micro-systems, non-null *quantum* potentials act, tied with the distinct dynamics of the two or more involved corpuscular-like singularities from the amplitude of the common physical wave $\Phi_G$: The global operation of generation $G$ that has to be realized for generating one unbound micro-state of two or more micro-systems, somehow involves corresponding two or more composing operations of generation, i.e. it is of the form $\boldsymbol{G}(G_1,G_2,...G_m,...G_M)$, which involves a quantum potential and possible quantum fields. So the Schrödinger equation of evolution is not appropriated in the strict sense of this word, because the $QM_{HD}$-hamiltonian does not include quantum potentials.

Thereby the exploration begun in (7.III).2.1 is closed and we can formulate the following global conclusion on the whole sub-section (7.III).2.

---

itself huge questions like the nature of space and time ('physical' nature, basically? primarily an only psychical nature, for time (Bergson)?); and what about the indefinitely multiform and variable separations in relative wholes that the human minds instil in what we call 'physical reality', each one of which entails its own pair of an inside and an outside?

## (7.III).4. THE RELATION (31) FOR UNBOUND MICROSTATES *VERSUS* THE EQUATION OF EVOLUTION

When also classical macroscopic fields are active in the considered global "external situation" situation, one can still write a $QM_{HD}$-Hamiltonian where only these classical fields are explicitly present, while *the quantum-fields are only implicitly introduced by the limiting conditions*; which then obliges to research a solution of the equation by fragmenting the spatial domain in separate zones with mutually different limiting-conditions and by producing an expression of the interference effects via additive superpositions of local solutions (like in the paradigmatic cases of "potential barriers", "walls", etc.) that inside the framework *[IQM-$QM_{HD}$-$dB_{DS}$(B)]* are rejected. But when any classical field is absent, it appears nakedly that *the Schrödinger equation of evolution is not sufficiently comprehensive*. The interference effects should be representable basically and independently inside the equation *itself*, not only a posteriori, in the solution, where they are introduced independently, furtively and approximately, by ad hoc mathematical procedures and correlative ad hoc considerations and denominations ("quantum tunneling", etc.). Limiting conditions imposed by composed operations of generation of type $\boldsymbol{G_t}(G_1,G_2,...G_m,...G_M)$ and the re-formulations (38') and (40) concerning the equation for eigenket and eigenvalues of the momentum-observable, cannot compensate fully the absence of any correlative representation of the quantum potentials inside the equation of evolution itself. It seems obvious that such an absence must entail some specific deleterious descriptional consequences that cannot be avoided by idealizations and approximations.

Thereby we enter the domain of de Broglie's "double solution" representation [1956], actively developed now by the new Portuguese School of Broca and Araujo [2010] (cf. Gatta, Rica da Silva, Araujo, Croca and Silva, Cordovil,Moreira, Magalhäes, Alves, Santos) into a very interesting and fundamental new representation of what here we call 'the Universal Substance'. We come back upon this in the chapter (8.III).

But inside the framework *[IQM-$QM_{HD}$-$dB_{DS}$(B)]* the face-à-face between the bottom-up construction of the representation (31), and the Schrödinger predictions (22) constructed top-down, entails already noteworthy new possibilities of direct optimization of the predictive outputs concerning quantum measurement.

Consider the Schrödinger equation $\sigma_{\Phi(Gt)})$

$$i\hbar(d/dt)\,|\psi_{G,\boldsymbol{H}}(\boldsymbol{r},t)\rangle = \boldsymbol{H}|\psi_{G,\boldsymbol{H}}(\boldsymbol{r},t)\rangle$$

Schrödinger, when he his equation, imposed a priori a general wave-like form of the solutions – $\Psi(\boldsymbol{r},t)=a(\boldsymbol{r},t)e^{(i/\hbar)\varphi(\boldsymbol{r},t)}$ – *that, at that time, he conceived to represent* ***physical*** *waves*. Therefrom stems, and stubbornly persists up to this day, a total confusion between *individual* entities – specimens $\sigma_{\Phi(Gt)})$ of the studied microstate, wave-*functions* $\Phi_{Gt}(\boldsymbol{r},t)$, individual coding-measurement-successions *[$\boldsymbol{G_t}$.MesA]*) that are involved in coding-postulate (20') where operates a *measurement*-hamiltonian ***H(A)*** – and on the other hand state-ket $|\psi_{G,\boldsymbol{H}}(\boldsymbol{r},t)\rangle$ that represent *abstract* statistics. This permitted to start, to connect the mathematical techniques to those from Maxwell's mathematical representation of electro-magnetic waves, and to have rapid and spectacular successes. But very soon the ambiguity between 'statistical waves' and physical waves as well as the unintelligible dimensionality and structure of the 'propagation- space' [148] , began to deteriorate the conceptual situation. The dimensionality and the lack of conceptual homogeneity of the 'propagation-space' of the equation involve an inextricable mixture of representation of individual features (tensor-products of spaces of individual "systems") and representation of statistical features, that prolong and amplify the multi-face mixture that flaws the $QM_{HD}$-representation of quantum measurements. The present approach reacts to this situation. Now, what is the possible role of the Schrödinger equation, inside the present approach?

To begin with, consider a Hamiltonian situation. It is often said that the initial state-ket $|\psi_{G,\mathbf{H}}(\mathbf{r},t_o)>$ is determined by imposing the limiting conditions upon the general solution of the Schrödinger equation of the problem. But if *on the whole spatial domain globally delimited by the limiting conditions* the physical initial situation varies in space in a way that is stable but nevertheless escapes mathematically specifiable knowledge (as nearly always is the case in a non-idealized factual situation), the limiting conditions alone i.e. not *exhaustively* the initial state-ket $|\psi_{G,\mathbf{H}}(\mathbf{r},t_o)>$), might not suffice for a satisfactory degree of specification of the statistical contents of the initial state-ket. Moreover when the problem involves a non-stationary hamiltonian it is *generally* questionable whether the corresponding Schrödinger equation can determine with an acceptable accuracy the time-dependent probabilistic predictions. In short, when the equation exists, the predictions, in general, emerge flawed by ignorance and approximation.

> *And **how** they are flawed, and to what a degree, can be known only a posteriori, by **verification**-measurements.*

But consider the expression (31). The involved prediction involves already verification, by construction. So if the predictions calculated via the Schrödinger equation of the problem are founded upon factually constructed initial predictions from (31) (the initial state-ket from (31), for $t_o$), the conclusions calculated via the equation for subsequent time $t>t_o$ will also be endowed with factual truth, insofar that the equation itself is well-constructed.

Consider now a physical situation that is not Hamiltonian. Then the Schrödinger equation cannot be written.

> *In such conditions the equivalence (31) entails the possibility of principle to **fully replace** the equation.*

So inside *[IQM-$QM_{HD}$-$dB_{DS}$(B)]* the optimal use of the Schrödinger equation is to compose it with the use of the equivalence (31), as follows:

- One can *systematically begin* by constructing factually the involved expansion $|\psi_{Gt,H}(r,t_1)>_{fact(DM)}/\boldsymbol{A}$ from (31) for the considered observable $\boldsymbol{A}$, via individual measurements (like in the description (9") $D_M$ from *IQM*) by following the factual-formal constructive procedure from *(7.III).2.2.2*. The expansions from the second member of (31) express *the whole and certainly true initial statistical situation* that – factually – generates for any subsequent time $t>t_o$ a corresponding statistical distribution that, factually, is certainly true. The involved $G_o$ or $G_t$ embodies the limiting conditions that act factually, while the *factual*-formal character of the whole constructive procedure entails maximal sensitivity with respect to the specificities of the *whole* factual initial ground encompassed by the limiting conditions, which specificities can (in principle) acquire a formal expression only if the limiting conditions are furthermore completed by

[148] The dimensionality and the lack of conceptual homogeneity of the 'propagation-space' involve an inextricable mixture of individual physical features and statistical-abstract ones that prolong and amplify the mixture that flaws the $QM_{HD}$-representation of the quantum measurements.

the construction for $t=t_o$ of the whole considered factual-formal expansion with respect to ***A*** of the state-ket $|\psi_{Gt,H}(r,t_1)>_{fact(DM)}$ from (31).

Once the initial factually generated expansions (31) is thus determined:

- If the Schrödinger equation of the problem exists, then one can continue by determining mathematically by its use the state-ket entailed by it for any moment $t>t_o$, when this seems economical.

- One can also generate factually the state-ket (31) just *in order to compare it* to the corresponding expansions (22) determined by the equation.

In short, inside *[IQM-QM$_{HD}$-dB$_{DS}$(B)]* the assertion *Ass2* and the formal-factual construct (31) that expresses it, endow the representation of prediction-verification for unbound microstates with the possibility:

- To replace radically this equation when it is not available, or too difficultly available, or too marked by unavoidable idealizations and approximations.
- To associate the use of the equation, if it is available, with that of the factually generated representation (31), in order to realize various sorts of optimization (extended domain of applicability, control, economies).

In a phase when the computational powers have become so high and continue to increase so rapidly, and when the techniques for making use of big data are in vertiginous progress, the conclusion stated above seems natural.

## (7.III).5. BOUND MICROSTATES

We finally consider also the case of bound microstates. In *[IQM-QM$_{HD}$-dB$_{DS}$(B)]*-terms these are:

*stable* micro-structures of specimens of several sorts of microstates, all the corpuscular-like singularities of which circulate inside a very limited domain of the 'physical' space.

Thereby we come back to the general introduction to this work. Let us stabilize this loop by expressing explicitly the connexion.

Historically the representation of this sort of micro-structures has begun inside the classical physics via a top-down approach that associated classical 'planetary' models to entities from chemistry and molecular and atomic physics. Around 1900 this extrapolating progression came short of logical consistency with the classical physics, and so to stagnation, and consequently the first 'quantum' postulates were formulated (Bohr, Plank, Einstein's explanation of the photoelectric effect). These postulates have acted like close precursors of the dramatic full *reversion*, by Louis de Broglie, of the direction and the nature of the construction of scientific knowledge on micro-phenomena; de Broglie's corpuscular-wave model initiated a decidedly bottom up approach along the vertical of conceptualization, that ties the classical level of conceptualization to the as yet never represented before 'Universal Substance' (in the Spinozian sense). And Schrödinger's equation, by its first applications, has yielded so striking results that a new phase of construction instilled itself furtively into the scientific thought.

But it began there where Schrödinger's equation placed the new start. Namely with mathematical representations of predictive statistics of results of quantum measurements, of which the processes of verification by long series of *necessarily individual* acts of measurement, *that were not yet represented.*

Only now, after nearly a century, *IQM* represents an explicitly structured representation of this lacking foundation.

But with respect to the essential specificities of the descriptions of microstates such as these have been organized in this work, first inside *IQM* and then inside the two successive frameworks *[IQM-QM$_{HD}$]* and *[IQM-QM$_{HD}$-dB$_{DS}$(B)]*, the case of bound microstates is now placed far up under the classical roof, under an accumulation of a mixture of hyphens of various natures between specifically quantum features and classical ones. Under the cover this transitional texture the strongest conceptual and formal specificities of the microstates remain devoid of a clean contour:

- For a bound microstate the human observer did not himself achieve deliberately the involved operation of generation $G_t$ . This operation has been achieved naturally before the beginning of the human investigation. So from the point of view of its availability for being studied, a bound microstate is like a classical “object”, *it just pre-exists ‘there outside'.* In the light spread by *IQM* one can understand what loss of specificity is entailed by this absence:

- In the absence of a deliberate human operation of generation of the microstate, the measurements have to be conceived in the classical manner, without successions *[$G_t$.MesA]* each one of which involves first a realization of $G_t$. So, explicitly at least, the measurement-operations *MesA* are achieved *directly*, by use of test-microstates (for which the measurement evolution can obey the coding-postulate (20') (for instance Compton collisions), or they are incorporated to field-effects (Stark, Zeeman). Furthermore the measurement-interaction *itself* is often realized *statistically*, by the action, from the start, of a big set of simultaneous 'test-operations' upon a big set of replicas of the studied microstate (monochromatic radiation incident on a collection of atoms of a definite kind, etc.) so that the result is expressed as a *mean* drawn factually from the involved eigenvalues, that remain unknown individually.

- Because a bound microstate (its specimens) is permanently stably captured inside a small global space-time support assigned mentally to the physical entities that are involved, the only mental nature of this spatial delimitation has been abstracted away; which entails that the spatially unlimited wave-aspects that are devoid of a *separable* observable significance, have been posited to be altogether devoid of any sort of significance. Which brings back to the remarks from (7.III).2.4 on the configuration spaces introduced via the Schrödinger equation; indeed this artificially delimited physical space-time support is what is associated with an abstract representation space where are lodged all the values considered on the mechanical dimensions of qualification of all the involved micro-systems. In a classical configuration space this happens currently and nobody wonders why more than only four dimensions of 'physical' representation are considered. If in the case of a bound microstate one begins to wonder about this, then much place for confusion is opened up. But precisely this happens because the general absence of intelligibility of the formalism entails a general attitude of suspicion.

- For a bound microstate – like in the case of an unbound one – the *formal* distinction is very feeble between the state-function from the statistical state-ket of a considered microstate $|\psi_{Gt}>$ and the corresponding de Broglie wave-function $\Phi_{Gt}$. But for a bound state even the conceptual-*physical* distinction – so progressively and painfully specified in this work – is subjected to a new sort of confusion. Namely, for a bound structure of microstates *the physical* ***specimens*** *of the involved microstates are all* ***stably*** *co-present* inside a very small domain of physical and they concentrate the attention on *them*, while the abstract statistical features represented by the corresponding state-ket are *perceived as such*, like in the classical physics; whereas for an unbound microstate the strict reverse happened, the existence and the role of the physical specimens have been occulted and the abstract representation by a state-ket has been reified because the formalism works exclusively with this. Moreover *the involved state-ket is from the start conceived to be also one* ***actually*** *realized eigenket of the total energy observable* $\boldsymbol{H}$, which is represented by *one* ***wave***-function$\Phi_{Gt}$ (in this case this is formally possible by 'degeneration' *and* factually true at any given time. But potentially there are many such wave-functions, since the equation is linear. So, conceptually, a huge stratified confusion steps in between: the descriptor of a given actual physical wave-movement; the abstract set of possible mutually distinct such descriptors; and the descriptor of a set of sets of predictive counts on results of measurement on the one involved sort of physical entity.

Etc.

All this creates many conceptual ghost-problems that here we want to avoid. All the more so as a theory of, specifically, quantum measurements, is not imperatively necessary for measurements on bound states.

In these circumstances and since in the present work the aim is to bring forth the *specific* principles of a fully intelligible representation of the microstates, the bound microstates have a very marginal conceptual role, which permits to postpone their treatment.

## (7.III).6. SUMMARIZING CONCLUSION ON THE CHAPTER 7.III

The global results from the chapter 7.III can be stated as follows. First in the framework *[IQM-$QM_{HD}$]* and then in the completed framework *[IQM-$QM_{HD}$-$dB_{DS}$(B)]*, we have outlined constructively a new representation of the quantum measurements.

This representation has been centred upon:

- The necessary performance of an operation of generation *G* that necessarily has an individual and factual character.

- The necessarily individual and physical-operational character of the coding-measurement-successions *[G.MesA]* that unavoidably must be performed.

- The deliberate requirement of a general and fully intelligible Hilbert-space representation of the factual probability laws constructed for the outcomes of long series of coding-measurement-successions *[G.MesA]*, because – ***quite independently of the concept of a microstate*** – a Hilbert-space representation of these is indeed a very expressive and efficient representation of *any* probability law, accordingly to Gleason's theorem.

In more detail now:

In the section (7.III).1 – via a fundamental first assertion *Ass.1* and the argument *Arg(Ass.1)* that founds it – we have brought into evidence a conceptual constraint that permits to organize a basic operational and logical coherence between *IQM* and *$QM_{HD}$* with respect to the basic requirement of statistical-probabilistic predictions and the verifiability of these, that inside *IQM* have been insured *factually and bottom-up* while inside *$QM_{HD}$* the predictions have been claimed to be established via *top-down and mathematically* and the representation of the process of verification raises problems since nearly a century. At a first sight the assertion *Ass.1* seems to be trivial. But a closer examination has brought forth an essential character tied with major consequences:

- As soon as a bottom-up approach is practised the operations of quantum measurement can play a major constructive role for also the *elaboration* of predictions, not only a role of verification of pre-calculated predictions.

- By a strict use of a small number of definite conditions of compatibility, *IQM* can strongly guide the elaboration of a new, fully intelligible Quantum Mechanics that *incorporate the whole stratum of individual conceptualization of the microstates*, thus compensating a huge lacuna that distorts *$QM_{HD}$*.

In the section (7.III).2, by systematic reference to *IQM*, we have proved a second assertion *Ass.2*, and this – for the particular case of microstates *ms(unbound,1)$_{G(n-c)}$* – produced the *formal-factual* equivalence

$$(31)\ [|\psi_{Gt,H}(r,t_1)>_{fact(DM)}\ \approx_{\textbf{pred.}}\ \{|\psi_{Gt,H}(r,t_1)>_{fact(DM)}/A\}, \forall A, \forall t_1\}]_{\textbf{verif}}$$

where the factually constructed state-ket $[|\psi_{Gt,H}(r,t_1)>_{fact(DM)}$ and all its expansions $\{|\psi_{Gt,H}(r,t_1)>_{fact(DM)}\ /A\}$ *emerge by construction already verified, so endowed with certain factual truth*; thereby these factual-formal descriptors entail predictions that – in general – are different from those drawn the corresponding *$QM_{HD}$*-state-ket:

$$|\psi_{Gt,H}(r,t_1)>_{fact(DM)}\ \neq_{\text{pred.}}\ |\psi_{Gt1,\boldsymbol{H}}(\boldsymbol{r},t)>$$

Just like to the expansions (22) from *$QM_{HD}$*, the expansions $\{|\psi_{Gt,H}(r,t_1)>_{fact(DM)}\ /A\}, \forall A, \forall t_1\}$ from (31) insure, for microstates *ms(unbound,1)$_{G(n-c)}$*, a ***direct*** access to a Hilbert-space representation (23) of the predictions on the results of measurements. They do this without the use of the Schrödinger equation of the problem, via the 'factual-formal' procedure that generates them. This feature is fully consonant with the very new and specific computing possibilities of our present time.

In the section *(7.III).2* we have tried to extend the result (31) to also the microstates *ms(unbound,1)$_{cG(qf)}$* that involve interference phenomena and a corresponding quantum potential. But it appeared that the $QM_{HD}$-representation of the state-ket of microstates $ms(unbound,1)_{cG(qf)}$ involves obstacles that *hinder the verification of the predictions on the outcomes of measurements of the* ***basic*** *momentum observable* ***P***. We have identified the source of these obstacles and this has led us to require a modified Hilbert-space representation of the state-ket of microstates $ms(unbound,1)_{cG(qf)}$ and a new sort of coding postulate for the momentum-observable ***P***, different from 'the method time-of-flight'; while inside $QM_{HD}$ it is presupposed – vaguely and implicitly – that a generalization of 'the method time-of-flight' is valid for measuring any observable on any sort of microstate. This led us to de Broglie's concept of guided momentum. Along the path opened up by the concept of guided momentum we achieved a convenient generalized equation (38') for the eigenfunctions and eigenvalues of a generalized concept $\boldsymbol{P_{vc\text{-}vd}}$ of momentum-observable. Therefrom emerged a new Hilbert-space representation (38')-(39)-(40) of the probabilistic predictions on the results of momentum-measurements on microstates $ms(unbound,1)_{cG(qf)}$.

Since any $QM_{HD}$-observable is a symmetrized function ***A(R,P)*** of the position-observable ***R*** and the momentum-observable ***P***, the mentioned result extends rigorously the assertion *Ass.2* and the equivalence (31) to also the microstates $ms(unbound,1)_{cG(qf)}$, for any observable defined inside the framework $[IQM\text{-}QM_{HD}\text{-}dB_{DS}(B)]$.

And finally it appeared that nothing hinders to apply the assertion *Ass.2* and (31) to also a *one* micro-state of two or several micro-systems, for any observable.

So inside (7.III).2 a fully intelligible *factually* rooted representation of the quantum measurements has revealed its whole contour; this is defined by the set of assertions and mathematical expressions

*[Ass.1, Ass.2,* (*2*3), (31), (38'), (39), (40) *]*

In (7.III).3 we have specified the peculiar conceptual position of the case of bound microstates, with respect to the representation of the quantum measurements constructed in (7.III).2.

In (7.III).4 we have shown that the factually rooted representation of the quantum measurements constructed in (7.III).2 permits to optimize the use of the Schrödinger equation, when it is available, and to replace it when it cannot be solved or even cannot be written.

# 8.III

## PRINCIPLES OF A FULLY INTELLIGIBLE SECOND QUANTUM MECHANICS: *QM2*

The representation of the quantum measurements constructed inside the two successive frameworks *[IQM-QM$_{HD}$]* and *[IQM-QM$_{HD}$-dB$_{DS}$(B)]* is the core of the fully intelligible Second Quantum Mechanics, *QM2*, researched in this work. This core is now alive. It beats already in the summarizing conclusion of the chapter III.7. By a somewhat uneasy act of conceptual surgery we shall now enact around it the essence of a whole abstract (and skeletal) organization that function coherently and meaningfully.

We *drop* now any scaffold and by just a few integrative gestures we try to constitute *QM2*.

### (8.III).1. THE THREE SOURCE-DOMAINS AND THEIR RESPECTIVE ROLES IN THE ORGANIZATION OF *QM2*

The prime matter for constructing *QM2* has been drawn out of *IQM*, *QM$_{HD}$* and the approach *dB$_{DS}$(B)*. The respective roles of these three source-domains will indicate the way in which they subsist inside the result to which they have contributed.

#### *(8.III).1.1. The Infra-(Quantum Mechanics) IQM.*

*IQM* has been constructed with the overt purpose to benefit of an infra-discipline from which to induce the whole conceptual-operational-methodological structure of *QM2*, by reference and immersion, in a sense similar to that in which the structure of a living being is induced by its genetic code. And indeed *IQM* has acted in this way. It has offered instructional constraints that have permitted to identify meanings, lacunae, potentialities, and to construct accordingly to a previously defined structure.

✚ *IQM* has revealed:

- The unacceptable character of the orthodox interdiction, in *QM$_{HD}$*, of a general model of a microstate.
- The total void inside *QM$_{HD}$* of an explicit and organized level of individual conceptualization; more specifically,
  - ÷ the absence inside *QM$_{HD}$* of an explicit definition of the unavoidable concept of a *factual* physical operation *G* 'of generation' of the microstate-to-be-studied;
  - ÷ the absence inside *QM$_{HD}$* of explicit definitions of the different sorts of 'microstates';
  - ÷ the absence inside *QM$_{HD}$* of explicit definitions of *coding*-measurement successions *[G.MesA]* founded on the posited general model of a 'microstate'.

✚ *IQM* has permitted to identify the general model of a microstate that – covertly – works inside the whole formalism of *QM$_{HD}$*.

✚ *IQM* has permitted to identify the coding-measurement-succession that is implicitly supposed to generally for any sort of act on measurement and any sort of microstate.

✚ *IQM* has permitted to identify the gross fallacies that vitiate the nowadays representation of quantum measurements, thereby founding a radical rejection of this representation.

✚ *IQM* has guided toward the refusal of the $QM_{HD}$ Hilbert-space representation of the microstates that involve a quantum-potential and toward a satisfactory alternative representation.

In short, throughout the chapters 6.II and 7.III of this work *IQM* has dictated *local* refusals or specifications, re-organizations, new constructs. While achieving these there emerged progressively zones of a reconstructed formalism ***where IQM got incorporated***. So *IQM* as a whole will be part of the substance of the integrated modified representation of the microstates that will be defined in (8.III).2.

### *(8.III).1.2. The nowadays Hilbert-Dirac formulation of Quantum Mechanics, $QM_{HD}$*

The $QM_{HD}$ representation of the microstates has been criticized throughout the chapters II.5 and II.6 of this work. In particular the $QM_{HD}$-theory of quantum measurements has been entirely rejected. And in the chapter III.7 the additive representation (15") of the state-ket of a microstate $ms(unbound,1)_{cG(qf)}$ has been found to lead to predictions that cannot be verified.

On the other hand, in consequence of Gleason's theorem (23) a Hilbert-space representation of the state-ket of *any* microstate is highly convenient for expressing formally the probabilistic predictions on the outcomes of quantum measurements, in connection with the relation (22). All the more so that inside $QM_{HD}$ – in association with the algebra of the observables that represent the mechanical qualifying concepts and with Dirac's bra-ket formalism – the Hilbert-space representation of the state-ket has become an organic part of a very complex and powerful network of mathematical tools. So the preservation of a non-restricted representation of any sort of microstate, by a state-ket from a Hilbert-space, constitutes a considerable practical purpose, notwithstanding the fundamental inadequacies from $QM_{HD}$ .

That is why in 7.III we have so decidedly researched an unquestionable new Hilbert-space representation for the unbound microstates that involve a quantum potential.

So the Hilbert-Dirac mathematical formalism is conserved inside *QM2*, though modified in a way that endows it with semantic consistency and with full verifiability of the asserted predictions.

### *(8.III).1.3. The de Broglie-Bohm approach, $dB_{DS}(B)$*

This approach started in Louis de Broglie's Thesis, with the Jacobi formulation of classical mechanics-and-optics where the conditions that restrict to the 'geometric approximation' have been *suppressed* so that a prolongation into arbitrarily small space-time dimensions emerged (de Broglie [1924], [1956], [1963], [1987]). Thereby – in continuity with the classical physics – de Broglie has proceeded toward a sort of mathematical representation of the Universal Substance in Spinoza's.

To this global mathematical model de Broglie's superposed an equation of evolution (de Broglie [1956)], [1957] [1987]). This equation is first written in very general and *relativistic* terms, with formal places explicitly secured for all the sorts of classical as well as quantum fields. Therefrom can then be drawn various particularizations or approximations[149].

As far as I know, up to now this is the unique available mathematical representation of the whole physical substance that covers the whole domain of possible space-time dimensions.

> The concept itself of a representation of this sort is very significant with respect to the purpose of an explicit unification of the sciences of matter; another concept that is at least as significant as this one is that of a general method of consensual conceptualization, predictive and verifiable.

Coming back to the $dB_{DS}(B)$ approach, the way in which one could make effective use of the general equation written by de Broglie – according to which criteria the *limiting conditions* are specified and how one could determine the *initial conditions* – is not considered. In fact it is implicitly supposed that this can always be performed *mentally* in an efficient way. Which is utterly false. It is the cognitive situation – that varies with the involved space-time dimensions – that entails what is to be performed mentally and what has to be performed factually, and what *conceptual and observational effects* are entailed by the sort of performance that is imposed by the cognitive situation.

That is one of the main reasons why each scientific discipline should founded in a corresponding infra-discipline that play with respect to it the role that IQM plays with respect to a theory of the microstates.

The $dB_{DS}(B)$ representation is entirely and passively referred to the current quantum mechanical formalism. It is conceived as the adjunction to the current quantum mechanics, of a mathematically expressed description of the physical substance that underlies it directly. The concept of a microstate is not individualized from inside the general model of a microstate, neither in a factually operational way nor only mentally. The conceptual-mathematical representation remains quintessentially *global and mental*. Local forms are just zoomed upon in imagination. Though individualizing words and ways of speaking do appear often in relation with a particularization or an approximation of the general equation, nevertheless – from the viewpoint of *IQM* – the conceptual distinction between individual or statistical features is very feeble, as well as the distinction between wave-features and mechanical features. Moreover features of these two different natures are often introduced inside one same descriptor. This is reminiscent of de Broglie's initial attitude in his Thesis where he initiated himself the use of the unhappy word 'particle', even though on the other hand he has quite explicitly presented his model as *exclusively* a wave endowed with a 'corpuscular-like' singularity; as he also defined 'a wave with a corpuscular phase' and he associated a *statistica*l 'wave-packet' with *one* 'particle'. For de Broglie such associations even seem to correspond to a personal esthetical preference. However, form a methodological point of view economies of this sort introduce a dangerously slippery slope, because the theoretical physicists tend to consider exclusively the mathematical expressions and to neglect the meanings that these carry.

In the deduction of the famous relation $p=h/\lambda$ the confrontation of the perceptions of two different observers plays an important role, but these are '*pure*' imaginary observers that do not accomplish physical operations. Etc.

Even de Broglie's theory of quantum measurements in de Broglie [1956)], [1957)] remains just an explanation of the *a priori and fully accepted* theory of measurements from $QM_{HD}$ *such as it stands*.

The theory as a whole is presented as a global, mental, that – directly – cannot be verified. But to this model is superposed an equation of evolution that is relativistic – which involves quintessentially conditions imposed to the processes of observation; moreover the equation cannot be solved – even in its most simplifying approximations – without limiting conditions and a posited initial state that presuppose an operation *G* of generation. *These are methodological inconsistencies*[150].

[149] The general view expressed in these works is what here we have denoted $dB_{DS}(B)$) because it integrates elements from Bohm's view, and later it has been associated with Viger-Bohm concept of sub-quantum fluctuations.

In short, Louis de Broglie has conceived his own representation like exclusively an interpretation of $QM_{HD}$ that is entirely accepted. He has never claimed another status for this representation [151].

But inside the framework *[IQM-*$QM_{HD}$*-*$dB_{DS}(B)$*]* the assignation to de Broglie's representation of an exclusively interpretive character appeared very likely to constitute a radical under-estimation. What acts in favour of this under-estimation is a general psychological submission to a formalism that has been developed under the intimidating pressure of a confrontation with a still very new and very cryptic cognitive situation. The physics of planets, stars and galaxies, like also that of the world of objects that we currently perceive everywhere around us, have started thousands of years ago. But a scientific exploration of a domain of physical facts where the space and time dimensions are so small that not only they exclude any sort of direct perception, but furthermore they cannot even be conceived, is a very recent feat.

The elaboration of *IQM* is a reaction to this situation. And inside the integration of *QM2* that follows just below the place inside Physics of the approach $dB_{DS}(B)$*]* is reconsidered drastically.

## (8.III).2. THE SKELETON OF *QM2*

Throughout the chapters 6.II and 7.III various features of what has been a priori named *QM2* have emerged in a scattered chaotic way. Now these features will just be summarized in a more organized way. The summary, moreover, will be synthetic in the extreme. Any elaboration will be banished. As stated from the start, inside this work the purpose is *not* to offer a fully achieved new theory of microstates; it is only to identify the conceptual loci in $QM_{HD}$ wherefrom spouts of un-intelligibility burst out, to suppress these, and to realize a well-defined new basic structure where is acting a *method* that guarantees full intelligibility and inner consistency for subsequent developments. We want to integrate only the *Principles*, literally, of a Second Quantum Mechanics.

Below we communicate this integration by an enumeration of the postulates followed by a brief verbal description of the semantic contents tied with these postulates.

### *(8.III).2.1. The postulates from QM2*

***1. The postulate of immersion in IQM*****:** *IQM* as a whole is conserved inside *QM2* in the role of an explicit pre-organized instructional epistemological-operational-methodological structure that constrains a priori the process of construction of *QM2*.

***2. The postulates from*** $MQ_{HD}$ ***are conserved, with the exceptions that appear at the points 4, 8, 9.***

***- The measurement-postulates are all suppressed.***

***3. The individual modelling postulate of a factually generated microstate:***

$$[MP(ms_{G,cw}) + (1')(G \leftrightarrow ms_{G,cw})]$$

---

[150] Nevertheless these contain suggestions for the development of *QM2*.

[151] The same is valid concerning Bohm's 1952 work.

is posited to be valid for ***any*** sort of factually generated microstate.

***4. The equation of evolution of the state-ket of a microstat:***

- The Schrödinger equation is maintained for the microstates ***without quantum-potential***.

- For microstates ***with quantum potential*** the general equation of evolution remains still an open problem to be solved on the basis of:
* de Broglie's approach ([1956], [1964], [1987]);
* *the results of the investigations that are developed by the Lisbon group* (cf. Croca&Araujo [2010]);
* the methodological constraints from *MCR* concerning chains of meta-descriptions (MMS [2002A], [2002B], [2006]) (in order to represent conveniently the superposed effects of classical macroscopic fields and of quantum fields interior to the studied microstate).

***5. Louis de Broglie's definition (33) of an evolving guided momentum-value p(r,t) – posited by him for ANY sort of microstate – is posited to be MEASURABLE.***

***6. The coding-measurement postulate (35)***

$$[G_t.Mes(\mathbf{r},\mathbf{p})] \quad \rightarrow_{trace} \quad (\mathbf{r}_k,\mathbf{p}_j)_t, \quad k=1,2,..K; \quad j=1,2,..J; \quad \forall ms_{Gt}$$

***is posited to be valid for* ANY *unbound microstate.***

***7. The coding-measurement postulate (20')***

$$[(G_{t1}\rightarrow\sigma_\Phi).MesA(\sigma_\Phi)] \quad \rightarrow_{H(A)} \quad (\text{marks registered in } (\Delta\mathbf{r}\Delta t)_j \simeq \text{'}a_j\text{'})$$

***is posited to be valid for the microstates ms(unbound,1)***$_{G(n-c)}$.

***8. The generalized representation* (38')**

$$\mathbf{P}_{vc\text{-}vd}.\Pi_m\big(a.exp((i/\hbar)\,(p_{jm}.\mathbf{k}_m(\mathbf{r}))_t.\mathbf{r}\big) = \mathbf{p}_j(\mathbf{r},t).\ a.exp\,\big((i/\hbar)\,(\textstyle\sum_m p_{jm}\,\mathbf{k}_m(\mathbf{r}))_t.\mathbf{r}\big),\ \forall j,\ m=1,2,..M$$

***of the concept $\mathbf{P}_{vc\text{-}vd}$ of a momentum-observable, with the corresponding basis of eigenket* (40)**

$$\big\{|\Pi_m\big(a.exp((i/\hbar)\,(p_{jm}.\mathbf{k}_m(\mathbf{r}))_t.r>\big\} = \big\{|a.exp\,\big((i/\hbar)\,(\textstyle\sum_m p_{jm}\,\mathbf{k}_m(\mathbf{r}))_t\,.\mathbf{r}\big)>\big\},\ \forall j,\ m=1,2,..M,$$

***replace the $QM_{HD}$-definition of the momentum observable.***

***9. The Hilbert-Dirac representation* (39)**

$$|\,\Psi(\mathbf{r},t)_{Gt(G1,G2,...Gm,...GM)}(\mathbf{r},t)>$$

***replaces the $QM_{HD}$ representation of the microstates ms(unbound,1)***$_{cG(q-f)}$.

---------

Via the postulates *3, 4, 5,* the $dB_{DS}(B)$ representation fuses with *QM2* whereby it gains access to the consensual, predictive and verifiable representation of the microstates from *QM2*.

### *(8.III).2.2. Remarks on the process of construction of QM2*

#### *A. On the constructive role of IQM*

* Throughout the chapters 6.II and 7.III, *IQM* has instilled an organizing distinction between the *individual* level of conceptualization, *the statistical-probabilistic* level, and the *meta-statistical level of correlations between whole probability spaces*.

* The *IQM* request of an explicit model of a microstate has led in (6.II).1 to the modelling postulate $MP(ms_{G,cw})$ associated with the re-definition (1') $G\leftrightarrow ms_{G,cw}$ of the concept of operation $G$ of factual generation of the microstate-to-be-studied and to the model (14) $ms_{G,cw}\equiv\{\sigma(ms_{G,cw})\}$ of such a microstate. In 7.III this model has then founded the coding-postulates (20') and (35) as well as the modified Hilbert-space representation (38')-(39)-(40) of the microstates $ms(unbound,1)_{cG(qf)}$..

* Via the basic assertions *Ass.1* and *Ass.2*, *IQM* has instilled into *QM2* a remarkable *fusion* expressed by the formula (31), between the *IQM*-definition (9") of the "primordial transferred description" of the studied microstate – the central descriptional concept from *IQM* – and on the other hand the Hilbert-space mathematical formalism from *QM2 such as it is perceived in the light of Gleason's theorem that* ***general*** *power of the Hilbert-space mathematics to represent* ***any*** *probability law in very convenient terms* that are different those that Kolmogorov had chosen.

* The construct (31) *doubles and controls factually* the purely mathematical outputs of the Schrödinger equation of the problem, when these can effectively be generated; while when these cannot be obtained, this construct offers inside *QM2* ***independence*** with respect to outputs of the equation of evolution. So inside *QM2* the construction and the solution of the Schrödinger equation of the problem cease to be a *necessity*. The unconditional acceptance of effects of approximations that cannot be foreseen, can be refused and compensated [152].

On the basis of Gleason's theorem *alone*, the whole essence of the $QM_{HD}$-algorithm (22)-(23) for prediction-and-verification is insured inside *QM2* by the equivalence (31).

#### *B. On the constructive role of $dB_{DS}(B)$*

The supply from the part of $dB_{DS}(B)$ is quite considerable:

** $dB_{DS}(B)$ connects *QM2* with the Hamilton-Jacobi formulation of the classical Mechanics *and Electromagnetic Optics* (i.e. with the whole essence of the classical Physics; while on the other hand, via the 'corpuscular-like-wave' model of a microstate, $dB_{DS}(B)$ roots *QM2* into the microphysical factuality.

** Via the channel of adduction into the *factually operational* representation from *QM2* (36') :

$$\Big[MP(ms_{G,cw}),\ (1')G_{cw}\leftrightarrow ms_{G,cw},\ (14)ms_{G,cw}\equiv\{\sigma(ms_{G,cw})\},\ (35)P(cod)_{\forall msG}),$$
$$(33')\boldsymbol{p_j}(\boldsymbol{r},t)=\boldsymbol{\nabla}.\beta(\boldsymbol{r})_t=\textstyle\sum_m(p_{jm}\boldsymbol{k_m}(\boldsymbol{r})_t\Big]$$

the metaphysical mathematical formalizations from $dB_{DS}(B)$ are transmuted into a rich reservoir of deeply conceived and carefully worked out representations, available for a future development inside the factual-operational and consensual representation from *QM2*. In order to illustrate this potentiality we add the following remarks.

---

[152] In the era of Moore's law (that still works and of which anyhow the already accumulated effects will subsist), of Big Data, and of the perspective of "the Singularity" in the human evolution, the realization of the contents of the equivalence (31) should not be a problem.

The approach $dB_{DS}(B)$ is quasi unanimously considered to be only an interpretation of $QM_{HD}$. But this sort of relation between $dB_{DS}(B)$ and $QM_{HD}$ might soon suffer a sort of inversion:

Both approaches $dB_{DS}(B)$ and $QM_{HD}$ define *any* classical dynamical quantity *A* as a function $A(\mathbf{r},\mathbf{p})$, like in the classical mechanics. On the other hand the coding-postulate (35) is posited to be valid for ***any*** sort of unbound microstate. So:

*Insofar that the guided momenta (33) can indeed be measured accordingly to (35) –* ***any*** *unbound microstate can be subjected to* ***simultaneous position-and-momentum measurements****.*

Thereby, inside *[IQM-*$QM_{HD}$*-*$dB_{DS}(B)$*]* the *statistical* notion of mutually incompatible mechanical qualifying-quantities can be circumvented. This entails that the probability-trees with mutually incompatible branches can be avoided and replaced by one-trunk probability-trees that – graphically – are similar to that from the Fig.3 from (3.I).1 but have a *different* content, namely:

*The physical-operational acts of measurement* $Mes(\mathbf{r},\mathbf{p})$ *are achieved accordingly to (35), by registration of the trace of de Broglie's guided momentum*; and the values of any dynamical quantity *A* (regarded like in the classical mechanics, as just a mechanical qualification *not as a quantum mechanical 'observable'*) are *calculated* ***after*** the performance of the physical-operational acts of measurement $Mes(\mathbf{r},\mathbf{p})$, from the obtained pairs of results $(\mathbf{r},\mathbf{p})$.

So the probability-tree from the Fig.2 reduces to only one physically-operational trunk that is common to *all* the classical mechanical quantities $A(\mathbf{r},\mathbf{p})$ and is topped by a crown of only ***conceptually*** worked out probability spaces mutually connected inside a meta-statistical level of correlations:

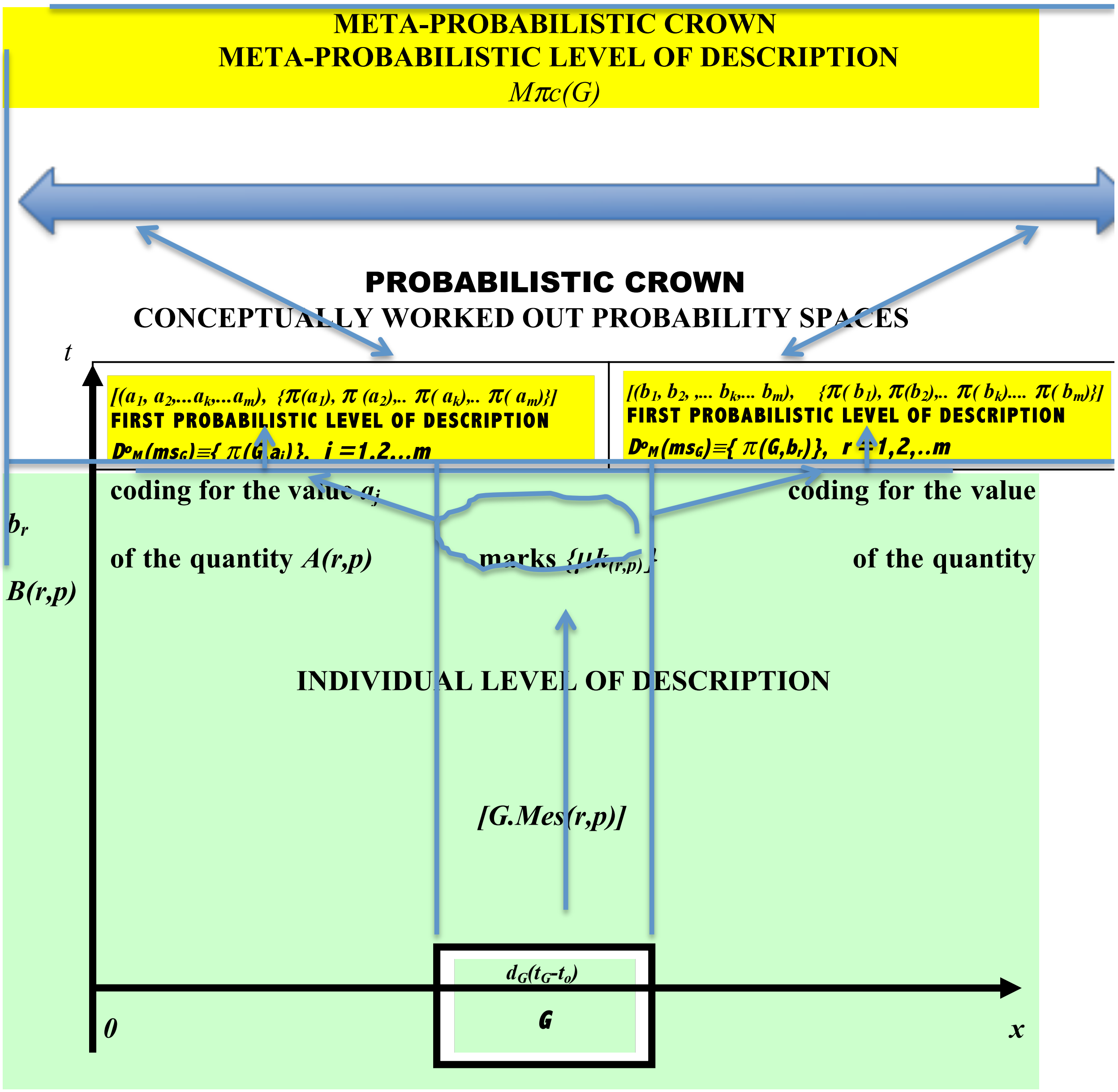

**Fig. 3'.** One-trunk probability-tree tied with the codingpPostulate (35), constructible for any sort of unbound microstate and any pair *((A(r,p), B(r,p))* of derived qualifying quantities.

In this way there appears a quite general category of physical operations and of corresponding representations that penetrate *deeper* into the a-conceptual physical reality than those tied with the quasi-classical microstates $ms(unbound,1)_{G(n\text{-}c)}$ for which the coding-postulate (20') is valid. The mentioned operations and representations concern the *radically* non-classical region of the a-conceptual physical factuality wherefrom are 'extracted' – in the sense of (1') – the microstates $ms(unbound,1)_{cG(qf)}$ of which the corpuscular-like singularities are immersed in 'inner' non-null quantum potentials that cannot be manipulated from the outside of the microstate, as it is required by the coding-postulate (20').

With respect to the generally possible representation from the Fig.3', the probability trees with mutually incompatible branches of the type represented in the Fig.2 appear now as a *particular* alternative possibility characteristic of *exclusively* the microstates $ms(unbound,1)_{G(n\text{-}c)}$, that are subjected to Heisenberg's principle of uncertainty *iff* the coding-postulate (20') *is* employed for them [153].

So inside the framework $[IQM\text{-}QM_{HD}\text{-}dB_{DS}(B)]$ the channel (36) entails the massive adduction into *'scientific' knowledge* – communicable, consensual, predictive and verifiable knowledge – of the new outputs of the *universal* possibility of one-trunk probability-trees represented in the Fig.3'.

Correlatively, inside the framework $[IQM\text{-}QM_{HD}\text{-}dB_{DS}(B)]$ the degree of 'communication' between the purely mental representations from $dB_{DS}(B)$ and the predictive representations from *QM2*, tends to *evolve*. For instance, consider the $[IQM\text{-}QM_{HD}\text{-}dB_{DS}(B)]$ probability $\pi(\mathbf{p_j})$ that the result of a momentum-measurement performed at a time $t$ on a microstate with state-ket $|\Psi_G(\mathbf{r},t)>$ be $\mathbf{p_j}$. In $dB_{DS}(B)$ terms this probability can be symbolized as

$$\pi(\mathbf{p_j}) = \int[\pi(\mathbf{r}).\pi[\nabla\beta(\mathbf{r},t)=\mathbf{p_j}]]d\mathbf{r} \;=\; \int[|\psi_G(\mathbf{r},t)|^2\pi[\nabla\varphi(\mathbf{r},t)]=\mathbf{p_j}]d\mathbf{r}$$

(the writing $\pi[\nabla\beta(\mathbf{r},t)=\mathbf{p_j}]$ is to be read: the probability that the guided momentum $\nabla\beta(\mathbf{r},t)$ possess the value $\mathbf{p_j}$). The purely conceptual (i.e. not physically operational) approach $dB_{DS}(B)$ does not define ways of verifying the factual truth of this symbolization. But conceptually, it permits to consider this symbolization as a $dB_{DS}(B)$ representation of Born's postulate:

$$|c(\mathbf{p_j},t)|^2 = \pi(\mathbf{p_j}) = \int[\pi(\mathbf{r}).\pi\,[\nabla\beta(\mathbf{r},t)=\mathbf{p_j}]]d\mathbf{r} \;=\; \int[|\psi_G(\mathbf{r},t)|^2\pi\,[\nabla\varphi(\mathbf{r},t)]=\mathbf{p_j}]d\mathbf{r}$$

Which acts as a $dB_{DS}(B)$ specification of the *significance* of Born's postulate. And furthermore inside the framework $[IQM\text{-}QM_{HD}\text{-}dB_{DS}(B)]$ the pertinence of this significance *can be verified factually* by use of the equivalence (31) constructed for a microstate $ms(unbound,1)_{G(n\text{-}c)}$ and, alternatively, accordingly to the coding-postulate (35) or accordingly to the coding-postulate (20'), taking into account the posit P2 from *(7.III).2.3.8*.

This illustrates how inside the framework $[IQM\text{-}QM_{HD}\text{-}dB_{DS}(B)]$ new possibilities arise of developing the consensual, predictive and verifiable representation of the microstates from *QM2*[154].

The investigations performed by the new Portuguese school (cf. Broca & Araujo [2010]) enforce notably this last assertion: Louis de Broglie's representation of a ***general*** – and *relativistic* – equation of propagation – that is *valid everywhere inside modern Physics, on all the various levels of space-time dimensions* – can be re-examined and stabilized, together with all its possible approximations or its particularizations to this or that cognitive situation; and it can be associated with the channel (36') of adduction into the *factually operational* representations from *QM2*.

This – together with the $dB_{DS}(B)$ mathematical representation of the Universal Substance, and with the generalization of *IQM* achieved in the Method of Relativized Conceptualization, *MCR* (cf. MMS [2002A], [2002B], [2006]), would constitute a genuinely general framework where, just by the specification of the mutual characteristics entailed for the pairs *(G,Mes.A)* by the acting cognitive situations, each given physical theory of a definite domain of physical facts would obtain a clear definition and a clear characterization with respect to any other theory of another definite domain of physical facts.

---

[153] It appears that that the non-boolean Birkhoff-voan Neumann 'logical' structure characterizes the way in which it is possible to code the results of measurerments, not the microstates themselves.

[154] Louis de Broglie's analyses of various cases of interference states where he makes use of the concept of variable 'quantum-mass (cf. dB [1987] *might lead to a genuinely scientific theory of teleportation.*

Which amounts to achieving *a* ***methodological*** *unification of Physics* that nothing seems to hinder. In 9.III we come back on this major question.

Globally considered, the preceding considerations point toward a genuine *inversion* of the conceptual *status* of the approach $dB_{DS}(B)$ with respect to that of $QM_{HD}$. The $dB_{DS}(B)$ representation tends to gain a basic conceptual status with respect to the un-intelligible $QM_{HD}$-algorithms for representing the quantum measurements.

***C. On the constructive role of*** $QM_{HD}$.

Notwithstanding the radical modification of the representation of the quantum measurements, the basic features of the $QM_{HD}$ mathematical representation via a Hilbert-Dirac state-ket, of the probabilistic predictions on the results of quantum measurements on a given microstate, is entirely conserved inside *QM2*. But it is conserved *cleaned* of all the conceptually unacceptable confusions brought forth by reference to *IQM*, and *modified* and/or extended in coherence with the criteria imposed by *IQM*.

### *(8.III).2.3. The main specificities of the second quantum mechanics QM2*

Consider now finally the global *inner* structure of the formalism from *QM2*.

This formalism satisfies strictly all the general requirements of *IQM*; these are recalled beneath:

° More or less deeply, the representations from *QM*2 are all rooted directly in the a-conceptual factual physical-operational reality. This establishes a zero-level of conceptualization. There*from* the process of conceptualization progresses constructively *bottom-up* along the vertical of human conceptualization, toward the classical Physics (cf. Fig.1).

° The individual level of conceptualization, and the statistical one, are explicitly and radically distinguished from one-another.

° The microstates are *classified* according to the definitions from (2.I)1.

° Each passage from the individual level of conceptualization, to the statistical one, is webbed factually by long series of repetitions of a given individual coding-measurement-succession *[G.MesA]*.

° What sort of acts of acts of measurement *MesA* is possible inside the coding-measurement-successions *[G.MesA]* – which inside the only general approach *IQM* remained undefined – is determined inside *QM2* for each class of microstates *separately*, on the common basis that consists of the general *model* posited for any microstate, namely de Broglie's model of *a wave with a corpuscular-like singularity in its amplitude*, and by taking into account also the specificities of the considered sort of microstate.

° The description of any microstate is ***a primordially transferred*** $[\varepsilon,\delta,N_0]$***-probabilistic description (9'')***.

° If the coding-postulate that is involved is of the type (20') – which according to *QM2* is possible only for the class of microstates $ms(free,1)_{G(n\text{-}c)}$ ***without quantum potential*** – then the construction of the corresponding primordially transferred probabilistic description can be inserted into a tree-like structure of the general type defined in (3.I).1 (Fig.2 and the Fig.9 below) of which the content varies in its details with the type, in the sense of the definitions from (2.I)1), of the considered microstate. (cf. Fig.2, Fig.3, Fig.4).

° If the involved coding-postulate is of the type (35) – which according to *QM2* is *possible* for ***ANY*** unbound microstate but is the ***unique*** possibility for the microstates from the class $ms(free,1)_{cG(qf)}$ – the corresponding primordially transferred probabilistic description can be represented by a one-trunk probability-tree (cf. Fig.3' and Fig.9 below).

Inside *QM2* these tree-like structures summarize in graphic language *a deep and complex unity of the representation of quantum measurements*.

Furthermore, in *MRC* it appears that the tree-like representation of a probability law is endowed with universality and constitutes a feature of a deep unity between *relativized and extended* versions of the three most basic sorts of human conceptualization, namely the logical conceptualization, the probabilistic one, and Shannon's informational conceptualization **[155]**.

We reproduce beneath a superposition of the Fig.2 and *a generalization of the Fig.3 entailed by the postulates 6 and 8*).

[155] The most complete account of this unity can be found in MMS [2006]. Less detailed and complete accounts can be found in MMS [2002A], [2002B]).

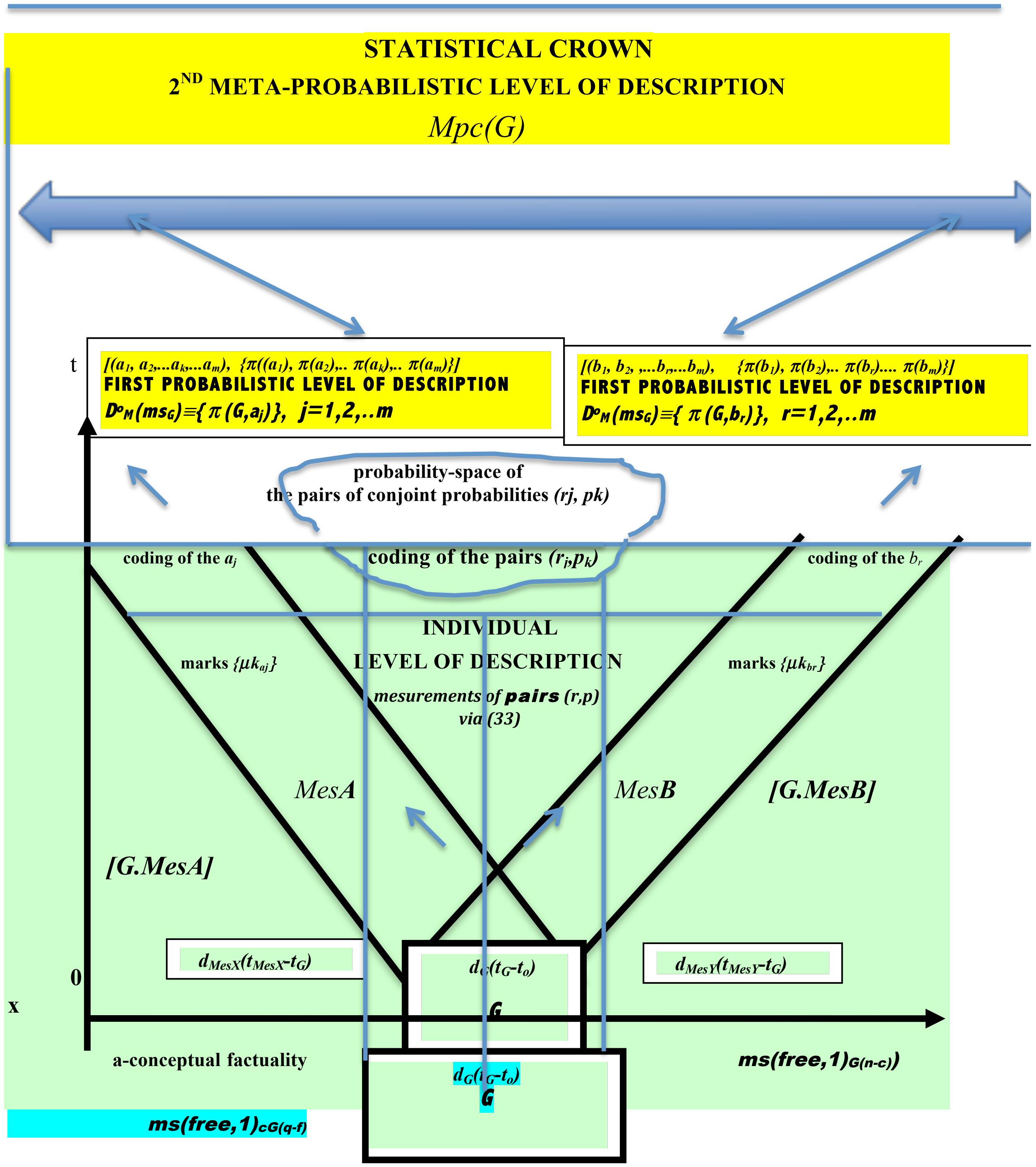

**Fig.9.** The one-trunk-tree consisting of simultaneous measurement of $\boldsymbol{r}$ and $\boldsymbol{p}$ surmounted by the two only conceptually worked out probabilistic crowns, corresponds to the coding postulate (35) that is *generally* valid. This one-trunk probability-tree of *any* microstate – whether a microstate $ms(free,1)_{G(n\text{-}c)})$ or a microstate $ms(free,1)_{cG(q\text{-}f)}$ – is rooted *deeper* into the a-conceptual factuality than any branches-probability-tree of a microstate $ms(free,1)_{G(n\text{-}c)})$.

### *(8.III).2.4. Global summary*

*QM2* is inserted in *IQM* and has been constructed under all the constraints imposed by *IQM*; whereby *IQM* has been infused in the structure of *QM2*.

*QM2* has been constructed with the purpose to offer a unified and intelligible Hilbert-Dirac mathematical representation of the quantum measurements, incorporating consensual prediction and verification. This has been realized via the assertions *Ass.1* and *Ass.2* associated with the predictive Hilbert-space algorithm (22)-(23) from $QM_{HD}$, re-considered in the light of Gleason's theorem.

*The representation that emerged in this way is in principle independent of the Schrödinger equation of evolution*.

*QM2* involves a radical re-structuration of the representation of the microstates $ms(unbound,1)_{cG(qf)}$ with non-null quantum potential. This restructuration has required recourse to the de Broglie-Bohm approach $dB_{DS}(B)$ that is derived from Hamilton-Jacoby formulation of the classical mechanics and optics. Thereby *QM2* emerges connected with the essence of the whole classical Physics. While accordingly to *IQM* the roots of *QM2* are directly implanted into the a-conceptual physical factuality. In consequence of this, and of the methodological structure instilled into *QM2* by the Infra-(quantum-mechanics) *IQM*:

> The representation of the microstates offered by *QM2* can naturally be connected with any domain of Physics. Which is a source of comparability, so also a source of an explicitly definable way toward a methodological unification of the whole Physics.

So *QM2* is, I think, fully intelligible from all the major points of view – genetic, factual, logical, mathematical, and contextual. It can now *work*. From now on *QM2* needs only growth, development (in particular in order to install in mathematical terms its requirement of effectiveness, so of finiteness).

# 9.III
# BRIEF FINAL CONSIDERATIONS

## (9.III).1. ON THE UNIVERSALITY OF QUANTUM MECHANICS

It is often felt that $QM_{HD}$ is endowed with a particular sort of universality and it is sometimes asserted that this is entailed by the fact that any material entity is a structure of microstates. But this belief is illusory, for at least two reasons, one epistemological and the other one formal.

***The epistemological reason***. Inside quantum mechanics a fragment of physical reality that is considered in order to create new scientific knowledge[156] on, specifically, this fragment, is placed *just* ***upon*** the extreme frontier between a still strict absence of any *such* knowledge, and on the other hand all that has already been previously conceptualized in scientific terms (Fig.1, Fig.2). This extreme character of the "position" inside the volume of the conceptualized, of the entities-to-be-studied, is not immediately perceivable *because inside $QM_{HD}$ the individual level of scientific conceptualization is occulted in consequence of a nearly exclusively top-down approach*. Only the statistical level is represented explicitly because it has been the *first* one encountered by the top-down progression. Nevertheless the individual level is irrepressibly involved and active. And the radically marginal, the limit-location of the origin the processes of creation of knowledge concerning factually generated microstates, imprints upon the emerging descriptions – the 'primordially transferred descriptions' that in *IQM* have been denoted $D/G,ms_G,V/$ – certain specific characters that are not apparent in the descriptions that are familiar to us.

Now, I hold that those who perceive universality in the formalism of quantum mechanics (cf. Aerts [1981]), in fact, more or less clearly, perceive:

- The specificities of 'primordially transferred descriptions' of factually generated microstates, with the unavoidable relativities that mark them.
- The statistical character that for such descriptions emerges irrepressibly.

They also feel more or less faintly that this sort of 'primordial' descriptions is *not confined* to the case of microstates; that the symbol '$ms_G$' can be replaced by the symbol of a quite general sort of entity, say '$œ_G$' (the 'object-entity-to-be-studied') that is generated *as such* by the operation of generation '$G$' [157]. They somehow perceive, though in a still unspeakable way, that the study of microstates introduces just an instance of certain ***general*** epistemological features that *call for a* ***method*** *for* ***startin****g a chain of knowledge at no matter what local but* ***total*** *relative zero of knowledge and for then developing the started chain into an unlimited process of creation of communicable knowledge that, by its structure, be consensual, predictive and verifiable*.

And indeed transferred descriptions emerge also, quite currently, inside the classical processes of conceptualization[158]. But in the case of factually generated microstates – because they are so totally unperceivable directly, and so difficultly accessible to deliberate interactions – *all* the involved descriptive features are filtered out *radicalized*, non-degenerate, mutually separable, pure; and that is why the concept of primordially transferred description – though universal – has revealed itself explicitly for

[156] That is, communicable, consensual and verifiable knowledge, not a subjective one, for instance imagined or metaphysical.

[157] cf. MMS [2002A], [2002B], [2006], [2014], etc.

[158] Henri Boulouet brought into evidence this very important conceptual fact, both in private communications and in his PhD thesis [2014], Univ. of Valenciennes.

the very first time only inside microphysics, and has only *therefrom* entailed new and striking questions of intelligibility as well as mathematical specificities.

***The formal reason***. Via Gleason's theorem the concept of Hilbert-vector-space revealed itself as a very expressive framework for just *lodging* inside it "*factual* probability laws" (MMS [2014]) that have been established *outside* this Hilbert-vector-space. Now, the concept of probability is omnipresent inside human thought.

But the mentioned possibility – little known and little understood – *has no necessary connection whatever with, specifically, microstates.* It is illusory to believe that there exists a direct logical relation between, for instance, social sciences in general, and on the other hand the concept of microstate; or even between the psychology of classical conceptualization, and microstates. In this sense expressions like "quantum social science" or "quantum cognitive science" are utterly misleading.

*The way of representing factual probabilities via vectors from a Hilbert-space,* ***has to be radically separated from the concept of microstate****.*

The *general* and basic construction of *relativized descriptions* and of the representational structures generated by these, constitute the object of a general discipline – a general Method of Relativized Conceptualization denoted *MRC* – that is ***independent*** of the study of microstates, notwithstanding that it has been suggested by a slow and long examination of the formalism of the nowadays quantum mechanics. I have developed *MRC* itself into an independent and rather complex whole where logic, probabilities and information theory become *unified*. And:

*This whole incorporates IQM as a particular application of MRC to the case of descriptions of factually generated microstates. But this does not in the least entail that MRC itself is a* **'quantum'** *method: in order to realize the mentioned application I had to construct from A to Z the* **bridge** *called IQM between, specifically, descriptions of factually generated microstates, and on the other hand, the general method of relativized conceptualization, MRC.*

For each application of *MRC* to a definite domain of facts, the construction of an analogous bridge is necessary, this cannot be economized without being captured in unpredictable confusions.

As for mathematical physics in general, and even for any mathematical scientific representation of a domain of "reality", we are still far from thoroughly understanding the conditions that restrict the acceptable, or fertile, or optimal association between, on the one hand this or that mathematical formal system, and on the other hand a given domain of what we call 'reality', physical or social or economical reality, etc.

This introduces the following point.

## (9.III).2. FACTS, MATHEMATICS, KNOWLEDGE, METHOD, TECHNIQUES, AND UNIFICATION

***Facts, cognitive features, mathematics***.

The approach developed in this work brings into evidence very general and fundamental questions concerning the relations, inside a mathematical theory of a domain of *physical* facts, between:

* The ***nature*** of the considered physical factuality;
* The descriptional ***purpose*** that acts.
* *The* ***cognitive*** *situation that is involved.*
* The mathematical tools that are made use of.

In this work, in order to generate in a guided manner an intelligible representation of the microstates, I have created *comparability* between two different and mutually independent sorts of descriptive systems – *IQM* and $QM_{HD}$ – and then, by the help of effective comparisons, I have explicated the conditions that permit to organize a mutual consistency between *IQM* and $QM_{HD}$ and I have constructed a realization of these conditions.

This process has permitted to perceive that, if one wants to insure in an intelligible way, consensual and verifiable prediction, it is an unavoidable necessity to pour into the mathematical descriptors well-defined factual contents, and the corresponding semantic, as it is *conceived by the human observers-conceptors*; indeed in the absence of such contents and such a semantic the mathematical tools that are claimed to "fit" the considered domain of physical entities, in fact remain largely disconnected from it; they simply do not.......‘*make sense*’ in a definite way. This is so because a mathematical system is a formal system, so a *closed* system; it is delimited, and the way of delimiting it – via axioms, definitions, well-formed expressions, laws of transformation of a well-formed expression – *involves an* ***own*** *semantic* (this is often occulted, if not even explicitly negated). So the closed semantic imported by the mathematical tools has to be quintessentially compatible with the factual semantic implied by the domain of facts to be represented, which is an open and unlimited semantic, fundamentally ***relative*** to the human ways of perceiving, of thinking, of "understanding" i.e. of experiencing a feeling of existence of "meaning". For this to happen it is necessary that the intersection (in the logical sense) between the own semantic of the mathematical tools and the semantic induced by the involved domain of facts be *non-null* and *identified*. For if the mentioned conditions of compatibility are not known and specifically fulfilled, the mathematical representations delivered by the mathematical system are not fully intelligible, and so they are exaggerated, or under-utilized, or distorted, and they push toward "interpretation problems" that attract more or less fantastic "solutions" (like that of "parallel universes" generated by each humble act of a quantum measurement). From a strictly logical point of view such a fantastic solution can be coherent, and *in this case, in the absence of a more definite [syntactic-semantic] point of view, it tends to last.* So the requirement of intelligibility can indefinitely stay violated.

Now, when inside a mathematical theory of a domain of physical entities there subsists some semantic *in*compatibility between the considered facts and the mathematical representation of these, the physicists always *feel* this. Usually they apprehend it as an unintelligibility of which the source cannot be located, nor specified. If this happens, the physicist's minds – nowadays at least – secrete a propensity to consider the mathematical formalism as if it were itself a physical reality of some superior essence, out of reach and immutable like a galaxy, which absolves from trying to change it. And so:

> A mathematical formalism that is applied to an important domain of physical entities without having previously worked out a thorough understanding of all the semantic features that are involved – those brought in by *the* ***cognitive features*** *generated by the involved* ***cognitive situation*** and those brought in by the mathematical tools – induces *unintelligibility, which induces a* ***reification of the formalism***.

The formalism is confounded with a *datum* and thereby it slips out of conceptual control. And often, in order to hold this back, inside many minds, the formalism undergoes a sort of divinisation: from a tool it is transmuted into an idol. (Jung and René Girard would have had much to say about this sort of effects of the modern collective unconscious).

Moreover this process also secretes a superposed penchant to *generalize to everything* the ill understood and divinised mathematical representation. It becomes *a general must*. This is what happens now with Einstein's General Relativity and the 'essential non-determinism' of Quantum Mechanics; with both in the same time. There exists a tendency to regard as general both these organizations of knowledge; which generates an artificial contradiction that leads to arbitrary, long and very difficult elaborations that are devoid of any sort of necessity.

Therefore I think that in the present stage of development of the human thought it has become urgent to focus much attention and effort upon the way in which, in *variable* cognitive situations, we generate a certain definite sort of *knowledge* that we agree to uniformly call "scientific" knowledge, i.e. communicable and consensually predictive and verifiable knowledge. This new urgency is entailed by the fact that we have only very recently reached a stage when we try to create scientific knowledge in the radical absence of the direct and ancestral basis of possibility of direct human perception, that has supported us since probably some 200.000 thousand years and has deeply shaped our "classical" laws of thought. In order to acquire the capacity to deal efficiently with this new challenge it has become acutely important to elaborate an explicit method for organizing the way in which is performed the passage from the scientific representations that are tied with *a given* cognitive situation, to any other scientific representation that is tied with *another* cognitive situation.

> There are differences between the systems of knowledge that are tied to different cognitive situations; but there are also *common features and these characterize the "scientific knowledge" as a whole*.

While:

> No sort of constructible ***knowledg*e** can be identified with a definition of *"the way in which the physical reality truly – absolutely – is in itself"*, because this is just an impossible notion, a basic and fragrant self-contradiction, the Fata Morgana of the nowadays concept of "scientific realism" that I call ***naïve realism***.

***Realism, knowledge, method, unification***

I have spent much time for the aim sketched out above. The results have been exposed elsewhere (MMS [2002A], [2002B], [2006]). I have called them a Method of Relativized Conceptualization, *MRC*. Here, in order to orient more precisely toward the meaning that I attach to the preceding very rapid considerations, I just add the following utterly synthetic illustration that concerns specifically the question of unification of the modern microphysics with the relativistic theory of gravitation.

- The physicists usually presuppose that a "*physical reality*" exists independently of our *knowledge* of it; if they did not they would be solipsists and solipsism is rare among physicists.

- But ***this posit of independent existence should be radically distinguished from the concept of scientific knowledge;*** it has to be explicitly and severely confined to the status of a strictly non-verifiable concept of merely mental existence. Kant, since already 1788, has confined this concept inside the metaphysics[159].

So let us stay resolutely inside the realm of scientific knowledge.

- Scientific knowledge is consensual, communicable, predictive and verifiable *descriptions*, by definition.

- Description is *qualification* of something, by definition.

- Any description involves an *object-entity-to-be-described* and *qualifications* of this entity, by definition.

The object-entity-to-be-described has to have been somehow separated from the whole of the physical reality, by some operation that has generated it *as* an object-entity-to-be-described, in this *role*; for if not it is not possible to think or to speak of "*it*" in a scientific sense, nor a fortiori to act on "*it*". Let us denote by $G$ this operation of generation (that can itself be purely mental, or factual-conceptual, or purely factual, or only selective, or also creative, or it can consist of some combination of such possibilities).

The object-entity-to-be-described generated by $G$ can be denoted by $æ_G$.

Consider now the qualifying-structure. We call it *a view*, we denote it by $V$, and posit that it possesses the formal structure of a union of *aspect-views*, $V=\cup_\alpha V_\alpha$ with $\alpha=1,2,....m$, where the index $\alpha$ designates a qualifying *aspect* (mental or conceptual-physical-operational (act of measurement) and each qualifying aspect $V_\alpha$ introduces a *semantic dimension of qualification* denoted $\alpha$ that carries a finite number $r_\alpha$ of freely chosen *values* $\alpha k$ of the aspect $\alpha$, with $k=1,2,....\ r_\alpha$ (this takes us very far away from the classical notion of "predicate")

> So any description involves an operation of generation $G$ that generates an object-entity-to-be-described $æ_G$, and a qualifying view $V$ that, acting on $æ_G$ via its aspect-views $V_\alpha$ , qualifies $æ_G$ in terms of qualifying values $\alpha k$. So any description involves a pair $(G,V)$: this is called the *epistemic referential* of the considered description. So the description itself can be symbolized by $D/G,æ_G,V/$. It is called *a relative description of* $æ_G$ *with respect to the qualifying view* $V$.

The general Method of Relativized Conceptualization mentioned above, *MRC*, constitutes a new sort of independent discipline that develops chains of hierarchically related relative descriptions. The chains do mutually connect in well-defined descriptional *knots* and then they separate again, thus generating progressively a rather complex network of radically relativized *scientific* ***knowledge***. While the postulate of mere *existence* of a "physical reality" – *without* specifying a mental model of it – that founds *MRC*, can pertinently be called a *postulate of* ***minimal realism***.

> The scientific knowledge constructed accordingly to *MRC* excludes by construction any false problems and paradoxes. And, in particular, it excludes by construction any *false* "impossibilities".

I say "false" problems, paradoxes, impossibilities, because inside *MRC* these dissolve; they are replaced by constructive *conditions* that avoid their emergence. The methodological rules of construction from *MRC* explicate the differences and the connections between the descriptions from the network of scientific knowledge, via the way in which the epistemic referential changes when one passes from a description from the network, to another one.

---

[159] The thing in itself (Ding an sich) is a concept Kantian concept that means "the such as it could be thought independently of any possible experimentation (in the $dB_{DS}(B)$ style).

As an example we consider the case of a description from the Quantum Mechanics (*QM*) and a description from the Relativistic Theory of Gravitation (*RTG*):

- Consider first a description *D/G,œ$_G$,V/* from the relativized version *QM2* of *QM*, that is subjected to *IQM*.

When *D/G,œ$_G$,V/* concerns an object-entity that consists of an ***unbound*** microstate *ms$_G$* (i.e. *œ$_G$≡ms$_G$*) then (in contradistinction to what happens for bound microstates that in general are pre-existing structures, atoms, molecules) the human observer has to first generate *factually* that unbound microstate in a way that make it factually available in the role of object-entity-to-be-studied.

And in this case, as expressed by (1), the involved operation *G* of factual generation unavoidably induces into the generated object-entity-t-be-studied *œ$_G$* a statistical-probabilistic character that cannot be eliminated on the basis of considerations of "imprecisions" in the acts of generation, because this character stems from the opposition between the finiteness of the possible definition of *G* and the unavoidably unlimited and unknown character of the physical reality on which *G* works.

The unique ***aim*** of the process of research is to establish consensually predictable and verifiable statistical-probabilistic distributions of the eigenvalues $a_j$ of the quantum observable ***A***.

And this aim is tied with the possibility to ***repeat*** as many time as one wants coding-measurement successions *[G.MesA]* that act on specimens of ***a microstate that nobody can observe directly*** and that is conceived to have the nature of a wave, while the unique feature from this wave that admits mechanical qualifications, consists of a very localized singularity in the amplitude of the wave; so it is ***interior*** to this wave, while the operation of generation *G* – without separating the whole wave from the rest of the physical reality – brings into the realm of the observable only marks produced by this singularity. (Whether this model is "really *tru*e", or not, is not the point here, the point in this context is that such a model is possible, and it is possible, by use of it, to produce consensual predictions and consensual verifications of these).

What is called "the observed eigenvalue" consists of registered observable marks that ***can wait an arbitrarily long time to be read by the human observer***.

Given the considered cognitive situation, the aim of the description, and the way in which this aim is realized, the *unique* ***condition of consensus*** that can be pertinently required is the following one: The statistical-probabilistic distributions of observable marks that code for the eigenvalues of the measured observable, that are obtained separately in various Laboratories via the same procedure applied to the same sort of microstate, must be all "identical" inside pre-specified limits of permitted difference. No light signals are involved, the state of movement of the observer and the Lorentz transformations play no role in the process.

The whole description – though only implicitly – is quintessentially relative to a definite sort of epistemic referential *(G,V)*; in particular, it is fundamentally dependent on *an operation of generation G that is* ***physically*** *operational.*

- Consider now *RTG*.

This is a theory of macroscopic and/or cosmic object-entities that in general *cannot be generated* ***factually*** *by the human observers-conceptors*. These object-entities are conceived to pre-exist. Inside the nowadays Physics the concept of operation of generation is not even conceived with respect to these object-entities. We are in presence of a totally classical conception that works in terms of "subjects" and "predicates" that just "are there" in the air of time, like in the current languages and the corresponding logic.

*The operations of generation G involved in RTG in general are* ***purely mental***, they consist of focusing the attention upon a certain conceptual representation that – with respect to some definite qualifying aspects – a *"View" V* in the sense of *MCR* – isolates *conceptually* the considered physical entity from inside the whole of the physical reality, *just like in the classical languages, grammars, logic*: ***the operations of generation G are realized via classical definitions***.

From the structure of descriptions that constitute *RTG*, it is tried to draw ***deductively*** certain observable consequences that in general are isolated events that *cannot be produced deliberately, so a fortiori they* ***cannot be repeated deliberately***. They can only be predicted to happen ***observably*** in certain definite conditions, in a way characterized in terms of values *gk* of the aspect-views $V_g$ from the view *V*.

The ***aim*** is to verify that these observable events do happen indeed, and such as they have been predicted.

The measurements that are involved *for this aim* must establish values *gk* of aspect-views $V_g$ that can concern very big mobiles that are conceived to mutually ***totally separated*** and that can be very distant form one another and from the human observers. The human observers also can be very distant from one another, carried by different ways of moving. So ***a same mobile or non-repeatable event*** *can be observed from different frames of reference that can have themselves various relative movements characterized by big velocities and sometimes also by accelerations*.

In such conditions the ***consensus*** on the results of the measurements made by these different observers concerning the same non-repeatable event or the same mobile, *do indeed involve in an essential role transformation laws and light signals*.

And when measurements of this sort are realized factually, the differences (in some definite sense) that can be detected in the results declared by different observers can in general be interpreted in terms of "imprecisions" that can be neglected in consequence of comparisons of orders of magnitude and of the admitted deterministic principle.

In Einstein's theories of relativity there is no trace of consciousness of the fact that, while knowledge is qualification, any qualification is *inextricably* ***relative as much to the way in which it emerges as to what is qualified***, which entails that knowledge of the reality as it "really is ", is not a scientific concept. *The basic conception RTG is* ***"naïvely realistic"***.

The cognitive situations considered in *QM2* and *RTG*, the descriptional aims, the procedures employed in order to realize these aims, are radically different. This – via descriptional relativities that cannot be suppressed – leads to radically different representations.

> *There is no contradiction between these theories.* They simply are tso radically different that they cannot even be pertinently compared *directly* (direct comparison presupposes some common features (MMS [2006])). But both are "scientific theories" in this sense that they both obey *MRC*. From a methodological point of view, the *specific* way in which each one of these theories is immersed in *MRC* via its own epistemic referential *(G,V)* can be analysed with as much detail as one wants. And ***these methodological features do permit comparability***.

In general terms now:

The unique sort of unity between the different physical theories that can be pertinently required, does not consist of a common mathematical representation; it consists of *a common* ***method*** *for representing* ***any*** *scientific theory of a physical domain of facts*, so that the methodological elements that are involved be clearly comparable. And *MRC* is such a method. For any scientific description it requires a corresponding pair *(G,V)* and it permits to explain the differences between the various scientific theories by the way in which these pairs have to be defined and to be made use of in order to satisfy the constraints imposed by the involved cognitive situation and by the descriptive aim. The unity stems from the fact that *any knowledge* is a structure of relative descriptions and that consequently all the *scientific* theories are subjected to *a same* *unique texture of* ***general legal relativizing structural features*** inside which there are ways to pass from any relativized representation of a given theory, to the relativized representation of any other relativized theory.

And furthermore, because inside *MRC any* consensual descriptional aim can find an adequate expression in terms of a network of chains of relativized descriptions, the methodological unification indicated above *can include the technical aims of construction of artefacts*. Which means inclusion of a relativized theory of systems and a relativized representation of engineering.

This has been the Ph.D. subject of the Thesis of Henri Boulouet [2014] (cf. www.mersyse.com/en/org_team.php).

# EPILOGUE

## UNIFICATION VERSUS TIM MAUDLIN'S VIEW

I feel uneasy to confess that I discovered only two weeks ago Tim Maudlin's book on Quantum Non-Locality & Relativity, and by chance. I have read this book in a continuously rising state of unexpected good surprise. Of course, such a state can stem only from a feeling of mutual agreement. To close this chapter 9 I just reproduce several brief extracts from TM, to which I add telegraphic comments.

From the page 22:

« 1. The quantum connection is unattenuated

The fall of a sparrow in Yougoslavia may have its effects in New Brunswick and on Saturn and beyond, but the effect becomes progressively smaller the farther away one goes. Since the gravitational force drops off as the square of the distance it eventually becomes negligible if one is concerned with observable effects.....

The quantum connection, in contrast, appears to be unaffected by the distance......... »

So TM detects *specificities* that – in our language – are tied with the "inside" and the "outside" of an entity-to-be-studied. And he draws attention on the fact that while according to the classical mechanics the "objects" that are studied are given as closed wholes that interact at distance only via "fields of forces" that lie entirely outside these "objects", in quantum mechanics two distinct entities that interact can be part of one same non-delimited whole, *inside* which they interact *otherwis*e than via "forces" in the classical sense (cf. the definitions from (I.2).1, and the Fig. 3 with its *IQM* meanings).

« 2. The quantum connection is discriminating

Gravitational forces affect similarly situated objects in in the same way .....

The quantum connection, however, is a private arrangement between our two photons. When one is measured its twin is affected , but no other particle in the universe need be......... »

So TM detects the specific effect of an operation *G* of generation!!!

« 2. The quantum connection is faster than light

..... For although no classical forces are unattenuated or discriminating, all were at least originally described as instantaneous......Any change in that global distribution would therefore have immediate effects on the forces felt everywhere .... »

So TM detects also consequences of the classical void of operations *G* of generation.

Pages 255-259:

**The Methodological Situation** (title)

The term "Methodological" is employed, which suggests that TM somehow requires also a methodological solution. Indeed on the page 256 he writes:

« ....We have somehow ended up in a world that seems , in its spatio-temporal aspect, to be relativistic, but also to be populated with matter governed by non-local laws.... »

A perfect characterisation of the mixed characteristics of the $dB_{DS}(B)$ approach (cf *(8.III).1.3.* in this work). And on the page 257 he writes:

« ....But the very radical falsity of the picture presents us with a methodological puzzle. Why, in the first place did we seek a theory that retains a relativistic account of space-time on the most fundamental level? Because that account of space-time structure has been so successful in making predictions that are checked at the macroscopic level. ....... This desire of a fully relativistic theory is, on the face of it, puzzling. Space and time do not come under our immediate experimental gaze........ »

And:

« ....The microscopic distribution of matter is not open to our direct inspection, but neither is the structure of space-time at any scale.... »

All these remarks stem from a hidden thrive toward a "minimal realism" and a methodological solution (in the sense specified above in (9.III).2. *Realism, knowledge, method, unification*); or at least, a potentiality to agree with such a composition.

# INSTEAD OF A GENERAL CONCLUSION

It is likely that the reader of this attempt has often been surprised and even repelled. In as much as this is so the reason, I think, is that it is very unusual in physics to make explicit use of epistemological and methodological arguments and constructions. Irrepressibly this is perceived as a noxious intrusion of "philosophy" into science.

But since now finally this work is entirely exposed, it has become possible to weigh whether the sort of philosophical intrusion that it involves has hindered or has enhanced the process of construction of a new representation of the microstates. So it has become possible to accept or to reject advisedly the procedure that has been employed, and its result.

In the present work *QM2* has reached the status of only a conceptual skeleton. But – as such – it might manifest force of life. In the realm of human representations this sort of miracle is current. The principles and the method are the spirit, and they can breathe life into the descriptional matter. If this does happen in the present case, then a whole new body of formal representation will grow for the skeleton constructed here, by its subsequent interactions with other minds.

**SHORT BIBLIOGRAPHY (provisional)**

Abbott, A., Calude, C., Conder, J., Svozil, K., [2012], “Strong Kochen-Specker theorem and incomputability of quantum randomness,” *arXiv:1207.2029* ; *Physical Review A 86, 062109*.

Aerts, D. (1981). *The one and the many: towards a unification of the quantum and the classical description of one and many physical entities* (Doctoral dissertation, PhD-thesis, Free University of Brussels).

Aharonov, Y., and Vaidman, L., (2007). "The Two-State Vector Formalism of Quantum Mechanics: an Updated Review". *Lecture Notes in Physics* **734**: 399–447. arXiv:quant-ph/0105101. doi:10.1007/978-3-540-73473-4_13.

Allori,V., Goldstein, S., Tumulka, R., Zanghi, N., [2008], "On the Common Structure of Bohmian Mechanics and the Ghirardi-Rimini-Weber Theory", *Br. J. Philos. Sci.* 59 353-389.

Atmanspacher , H. & Dalenoort, G.J., [1994], *Inside, Outside*, Springer Series in Synergetics, Springer

Bell [1964], J.S., “On the Einstein Podolsky Rosen Paradox”, *Physics* p.195.

Beltrametti, E. G., & Maczynski, M. J., [1991], On a characterization of classical and non-classical probabilities, *Journal of Mathematical Physics*, 32, 1280-1286.

Bitbol, M., [2010], *De l’intérieur du monde*, Flammarion, pp. 151-167.

Bohm, D., [1951], "Quantum Theory", Constable and Company Ltd..
Bohm, D., [1952], "A Suggested Interpretation of the Quantum Theory in Terms of "Hiden" Variables, I and II*", Phys. Rev. 85, 166-193.*

Bordley,F., [1989], "Reflection as an explanation of Bell's inequality paradox", Phys. Letts., Vol. 137, No 6, 1989.

Boulouet, H., [2014], *“Systémique Relativisée”*, Ph.D. Thesis, University of Valenciennes.

Broglie, L. de:
- [1963] *Recherches sur la théorie des quanta*, Masson &Cie, la ré-édition du texte de la Thèse [1924], Fac. Des Sciences de Paris. (La reproduction de l’Appendice 1 est pris de Annales de la Fondation Louis de Broglie, Vol. 17, No. 1, 1992).
- [1956], *Une tentative d’interprétation causale et non-linéaire de la mécanique ondulatoire (la théorie de la double solution)*, Gauthier-Villars.
- [1957], *La théorie de la mesure en mécanique ondulatoire (interprétation usuelle et interprétation causale))*, Gauthier-Villars.

- [1987], "Interpretation of quantum mechanics by the double solution theory", Annales de la Fondation Louis de Broglie, vol.12, no.4.

Cohen-Tannnoudji, C. ., Diu, B. et Laloë, [1973], *"Mécanique Quantique"*, Hermann. (dénoté *CTDL*)
Cohen-Tannoudji, C., Dupont-Roc, J & Grynberg,G., [1996], *Processus d'interaction entre photons et atomes*, EDP Sciences.
Croca, J.R., &J.E.F. Araujo, eds., [2010], *A New Vision on Physis, Eurhythmy, Emegence and Nonlinearity*, CFCUL.

Dirac, P.A.M., [1930], [1958], "The Principles of Quantum Mechanics", Oxford University Press (1958)

Dürr, D., Fusseder, W., Golgstein, S., Zanghi, N., [1993], Comment on "Surrealistic Bohm Trajectories", Z. Naturforsh, 48a, 1261-1262

Dürr & Teufel, [2009], *Bohmian Mechanics*, Springer pp. 131, 202.

Englert, B.G., Scully, M.O., Süssmann, G., Walther, Z, [1992], Surrealistic Bohm Trajectories., Z. Naturforschung, 47a.

Finkelstein, G., [1995 ] (12 July) "How to measure a beable", SJSU/TP 95

Gleason, A., [1957], "Measures on the closed subspaces of a Hilbert space". *Journal of Mathematics and Mechanics* **6**:885–893.

Goldstein, Norsen, Tausk and Zanghi, [2011], "Bell's theorem", *Scholarpedia* 6(10) 8378.

Goldstein,S., Zanghi,N., [2011], "Reality and the Role of the Wave-Function in Quantum Theory", arXiv:1101.4575.

Gudder [1976], “A generalized measure and probability theory”, in *Foundations of Probability Theory, Statistical Inference and Statistical Theories of Science*, Reidel, Dordrecht.

Hiley, B.J., Callagan, R.E., Marroney, O.J.E., [2000], Quntum Trajectories, Real, Surreal or an Approximation to Deeper Process?, arXiv:quant-ph/0010020v2

Hoffmann, Donald D. & Gefner Ananda, [2016], "Evolutionary Argument against Reality", *Quanta Magazine (Business Insider)*.

Hooft (G. van't), [2012], ["Quantum Mechanics from Classical Logic", *Journal of Physics: Conference Series* ***361*** 012024; [2007A], Emergent Quantum Mechanics and Emergent Symmetries, **arXiv:0707.4568v1; [2007B],** Entangled quantum states in a local deterministic theory, arXiv:0908.3408.

Kolmogorov, A., [1950], *Foundations of Probability*, Chelsea Publishing Company, New York.

Mackey [1963], *Mathematical Foundations of Quantum Mechanics*, Benjamin.

Maudlin, T., [2011], Quantum Non-Locality & Relativity, Metaphysical Intimations of Modern Physics,Willey-Blackwell, 3rd edition.

Mugur-Schächter, M., (MMS):
- [1963], "Sur la possibilité de trancher expérimentalement le problème du caractère « complet » de la Mécanique quantique", C.R. Acad. Sc., t. 256, p. 5514-5517, Groupe 4.
- [1980], Le concept nouveau de fonctionnelle d'opacité d'une statistique. Etude des relations entre la loi des grands nombres, l'entropie informationnelle et l'entropie statistique, Anns. de l'Inst. H. Poincaré, Section A, vol XXXII, no. 1, pp. 33-71.
- [1982], The Probabilistic-Informational Concept of an Opacity Functional, (en collab. Avec N. Hadjissavas), Kybernetes, pp.189-193, Vol. 11(3);
- [1987], "Locality, Reflection, and Wave-particle Duality", *Fonds. of Phys.,* Vol. 17, No 8.
- [1992], "Toward a factually induced Space-time Quantum Logic", *Founds. of Phys*., Vol. 22, No 7.
- [1993], "From Quantum Mechanics to Universal structures of Conceptualization and Feedback on Quantum Mechanics.
- [1994], "p=h/λ? W= h*v*? A Riddle Prior to any Attempt at Grand Unification", in Waves and Particles in Light and Matter, pp 541-569.
- [2002A], "Objectivity and Descriptional Relativities", *Fonds. of Science*.
- [2002B], "From Quantum Mechanics to a Method of Relativized Conceptualization", in *Quantum Mechanics, Mathematics, Cognition and Action*, Mugur-Schächter M. and Van Der Merwe A., Eds., Kluwer Academic.
- [2006], *Sur le tissage des connaissances*, Hermès-Lavoisier.
- [2010], "Kolmogorov's aporia and solution.....", arXiv:0901.2301 [quant-ph]
- [2013], "L'infra-mécanique quantique…..", arXiv:0903.4976**v3** [quant-ph] (***v3***), and [2017], "Infra-Mécanique Quantique, Indéterminisme, Non-localité", EUE.
- [2014], « On the Concept of Probability » in *Mathematical Structures in Computation and Information*, Cambridge Univ. Press, Volume 24, Special Issue 03 (Developments of the Concepts of Randomness, Statistic and Probability), Mioara Mugur-Schächter invited editor, Preface, contribution, organizer of a common debate.
- [2017A], "Principles of a Second Quantum Mechanics" arXiv:1310:1728v**4** [quantph]
- [2017B], "Infra-Mécanique Quantique, Indéterminisme, Non-localité.

Neumann, J. von, [1955], *Mathematical Foundations of Quantum Mechanics*, Princeton University Press.

Quadranti, P., [2007], *La raison constructrice,* Peter Lang.

Penrose, R., [1999], "The Emperor's New Mind", Martin Gardner.

Raichman, (in DEA-work (Diplome d'Etudes Approfondies)), [2003],: Destouches-Février [1946] et [1956], Ballentine [1973], Deutsch [1999]). Thereby these authors manifested non-perception of the unbridgeable abyss that separates logical-mathematical transformations, from the data that can be drawn only directly from facts, in an a-rational manner. On the contrary Anandan [2001]

**Rovelli, C.,** "Is Time's Arrow Perspectival?" **arXiv:1505.01125v2 [physics.hist-ph] 2015.**

Steinberg, A. M., [2011], "Observing the Average Trajectories of Single Photons in a Two-Slit Interferometer", Science 332, n° 6034.

Suppes, [1966], The probabilistic argument for a non-classical logic of quantum mec hanics, *Journal of Philosophy of Science*, **33**, 14-21.

Svozil, K.:

- [2012A] Unscrambling the Quantum Omelette;

- [2012B] "How much contextuality?" Natural Computing 11, 261–265, arXiv:1103.3980.

- with Josef Tkadlec, [1996], "Greechie diagrams, nonexistence of measures in quantum logics and Kochen–Specker type constructions," Journal of Mathematical Physics 37, 5380–5401 ;

Wigner, E.P., [1961], "Remarks on the mind-body question", in: I.J. Good, "The Scientist Speculates", London, Heinemann.